\documentclass[aps,prc,nofootinbib,onecolumn,floatfix,superscriptaddress]{revtex4-1}  
\usepackage{graphicx}  
\usepackage{dcolumn}   
\usepackage{bm}        
\usepackage{amssymb}   
\usepackage{amsmath}   
\usepackage[cal=euler,scr=zapfc,calscaled=1.0]{mathalfa}
\usepackage{verbatim}   
\usepackage{epstopdf}
\usepackage{appendix}
\usepackage{url}
\usepackage[f]{esvect}
\usepackage{float}
\usepackage{fixfoot}
\usepackage{multirow,makecell}
\usepackage{hhline} 
\usepackage{bookmark}
\usepackage{breakurl}
\usepackage{hyperref}
\hypersetup{breaklinks = true, colorlinks = true, citecolor = blue, linkcolor = blue, urlcolor = blue}
\usepackage[section]{placeins}

\allowdisplaybreaks  

\makeatletter
\newcommand{\colorcaption}[2][]{%
  \begingroup%
  \renewcommand{\@caption@fignum@sep}{ (color online). }%
  \caption[#1]{#2}%
  \endgroup%
}
\makeatother

\begin{document}

\title{A New Approach to Determine Radiative Capture \\Reaction Rates at Astrophysical Energies } 

\affiliation{Laboratory for Nuclear Science, Massachusetts Institute of Technology, Cambridge, Massachusetts 02139, USA}
\author{I.~Fri\v{s}\v{c}i\'{c}} \thanks{\href{mailto:friscic@mit.edu}
{\color{black}friscic@mit.edu}}   \affiliation{Laboratory for Nuclear Science, Massachusetts Institute of Technology,
Cambridge, Massachusetts 02139, USA}
\author{T.~W.~Donnelly} \affiliation{Laboratory for Nuclear Science, Massachusetts Institute of Technology,
Cambridge, Massachusetts 02139, USA}
\author{R.~G.~Milner} \affiliation{Laboratory for Nuclear Science, Massachusetts Institute of Technology,
Cambridge, Massachusetts 02139, USA}

\date{\today}

\begin{abstract}

Radiative capture reactions play a crucial role in stellar nucleosynthesis but
have proved challenging to determine experimentally.  In particular, the large
uncertainty ($\sim$100\%) in the measured rate of the $^{12}$C$(\alpha,\gamma)^{16}$O reaction is 
the largest source of uncertainty in any stellar evolution
model.    With development of new high current 
energy-recovery linear accelerators (ERLs) and high density gas targets, 
measurement of the $^{16}$O$(e,e^\prime \alpha)^{12}$C reaction close to threshold using detailed balance opens 
up a new approach to determine the $^{12}$C$(\alpha,\gamma)^{16}$O reaction
rate with significantly increased precision ($<$20\%). We present the formalism to relate photo- and electro-disintegration reactions and consider the design of an optimal experiment
to deliver increased precision. Once the new ERLs come online, an experiment to validate the
new approach we propose should be carried out. This new approach has broad applicability to radiative capture reactions in astrophysics.
\end{abstract}

\maketitle

\section{Introduction.} 

Radiative capture reactions, {\it i.e.}, nuclear reactions in which the incident projectile is absorbed by the target nucleus and $\gamma$-radiation is then emitted, play a crucial role in nucleosynthesis processes in stars~\cite{Rolfs1990}. For example, knowledge of their reaction rates at stellar energies is essential to understanding the abundance of the chemical elements in the universe.
However, determination of these reaction rates has proven to be challenging, principally due to the Coulomb repulsion between initial-state nuclei and the weakness of the electromagnetic force.  For example, the decay of unbound nuclear states by the emission of a particle of the same type as that captured, or by the emission of some other type of particle, is often $10^3 - 10^6$ times more probable than decay by $\gamma$-emission.

In stellar nucleosynthesis, at the completion of the hydrogen burning stage, the core of a massive 
star contracts and heats-up. When the temperature and the density 
of the core reaches sufficiently high values, the helium starts to burn
via the triple-$\alpha \rightarrow$ $^{12}$C process. Subsequently, the
$\alpha$ radiative capture reaction $^{12}$C$(\alpha,\gamma)^{16}$O
also becomes possible. The helium burning stage is fully 
dominated by these two reactions and their rates determine
the relative abundance of $^{12}$C and $^{16}$O, after the
helium is depleted. 
At helium burning temperatures, the rate of the 
triple-$\alpha$ process is known with an uncertainty of 
about $\pm$10\%, but the uncertainty of the $^{12}$C$(\alpha,\gamma)^{16}$O 
reaction rate is much larger. In fact, it is 
the largest source of uncertainty in any stellar evolution
model. Therefore, for many decades it has been the paramount experimental goal of nuclear astrophysics to determine the rate 
of $^{12}$C$(\alpha,\gamma)^{16}$O reaction at astrophysical energies with better 
precision~\cite{Woosley2003}. 

This task has been proven to be very difficult, not withstanding heroic 
experimental efforts for more than half a century. For the generic radiative capture reaction
\begin{equation}
A+B \rightarrow C \rightarrow D + \gamma \ ,
\end{equation}
the Coulomb repulsion is characterized by the Gamow factor (or Coulomb
barrier penetration factor) between $A$ and $B$ 
\begin{equation}
P_g = \exp{-\sqrt{E_g/E}} \ ,
\end{equation}
where $E_g \equiv 2 m_r c^2 (\pi \alpha Z_A Z_B)^2$ is the Gamow energy and $m_r =\frac{m_A m_B}{m_A+m_B}$ is the reduced mass.
The cross section $\sigma$ is then expressed~\cite{DFHMSa} as a product of $P_g$
and the astrophysical S-factor 
\begin{equation}
\sigma \equiv \frac{1}{E} \exp{[-2 \pi Z_A Z_B \alpha/v]} S(E) \ .
\end{equation}
$\sigma$ is further extrapolated to the Gamow energy, which is representative of stellar energies.   

At the helium burning
temperature $\sim 2 \times 10^8$ K and corresponding Gamow 
energy $E_g \sim 300$ keV, the cross section for the $^{12}C +\alpha \rightarrow \gamma + {}^{16}O$ 
reaction is $\approx 10^{-5}$ pb, which 
makes the direct measurement at stellar energies impossible. 
Unfortunately, the extrapolation is
not simple, since the structure of the cross section is complex.
It involves interference of the high-energy tail of the $J^{\pi} = 1^-$ 
subthreshold state in $^{16}$O (see \cite{Tilley1993}) at $7.12$ MeV 
and the broad $1^-$ resonance at $9.59$ MeV, and interference of 
the subthreshold state $2^+$ at $6.92$ MeV and the narrow $2^+$ 
resonance at $9.85$ MeV. 
Additionally, cascade transitions to the ground state of 
$^{16}$O need to be taken into account as well as the direct 
capture for the E2 amplitude.

\begin{figure}[h]
\begin{center}
\includegraphics[width=8.6cm]{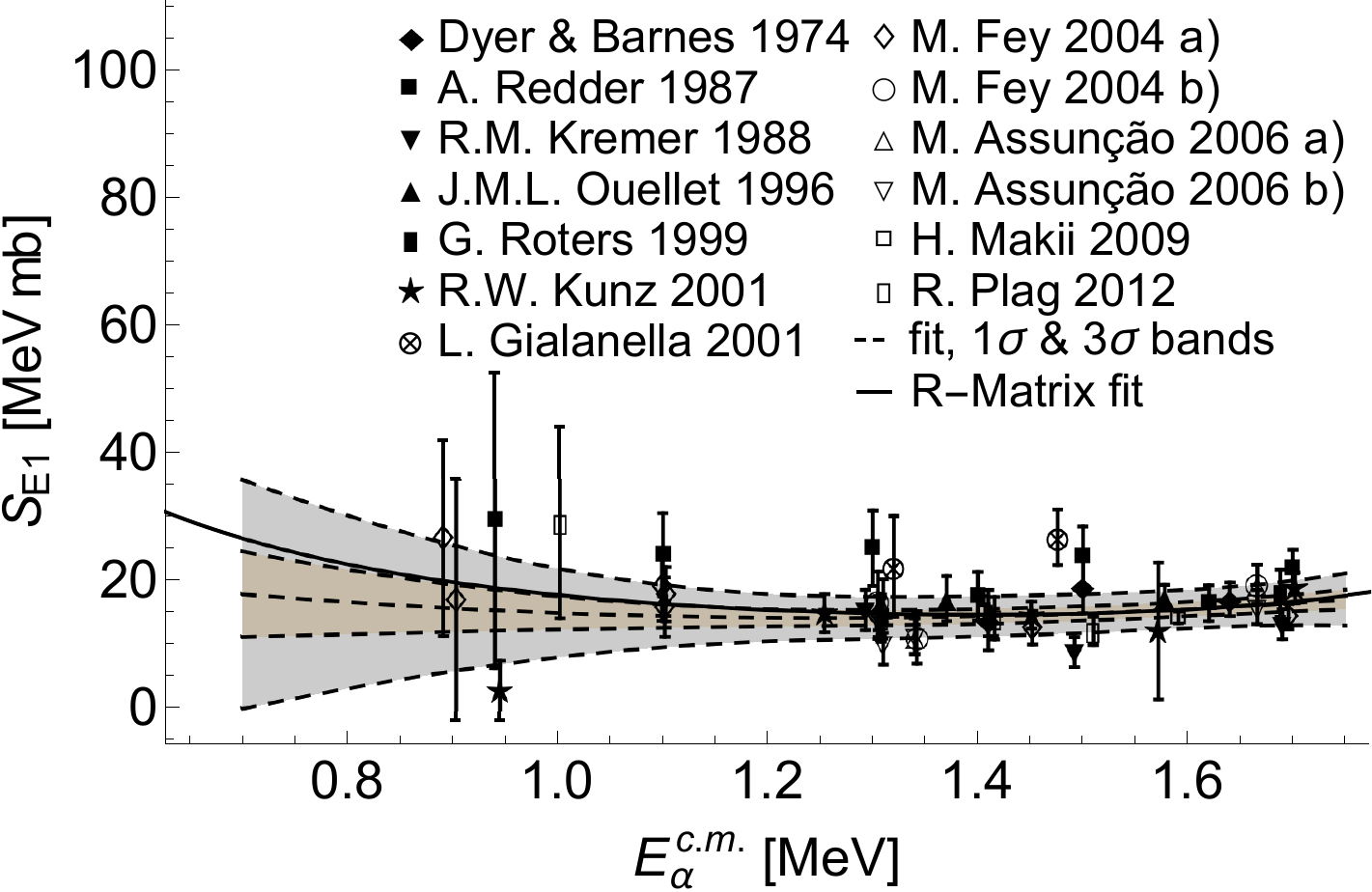}
\includegraphics[width=8.6cm]{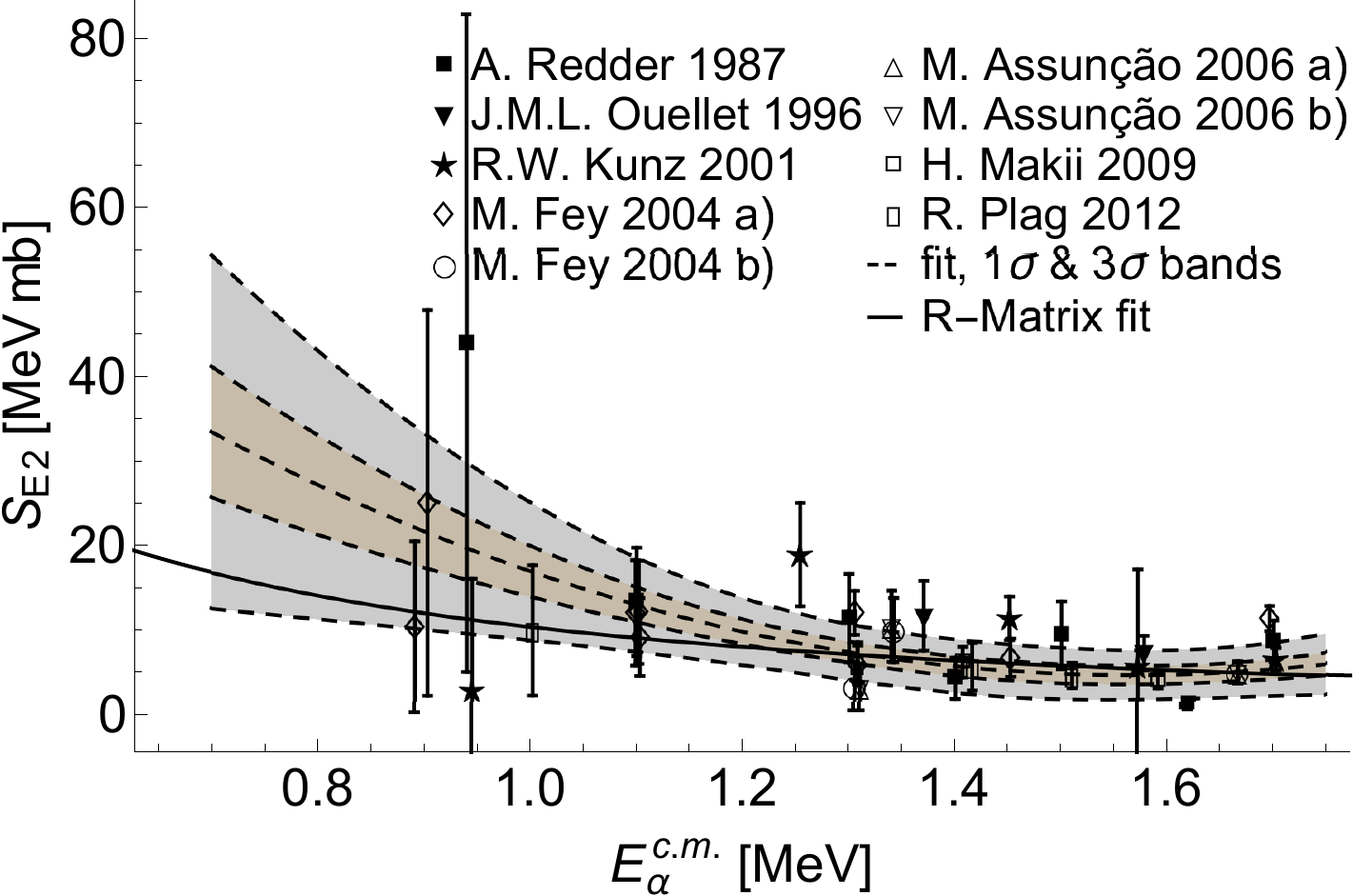} 
\caption{Measured astrophysical $S_{E1}$- and
$S_{E2}$-factors for $E_{\alpha} < 1.7$ MeV from Dyer 
and Barnes \cite{DyerBarnes1974}, A. Redder \cite{Redder1987}, R. M. Kremer \cite{Kremer1988}, 
J. M. L. Ouellet \cite{Ouellet1996}, G. Roters \cite{Roters1999},  R. W. Kunz \cite{Kunz2001, 
Kunz2002phd}, L. Gialanella \cite{Gialanella2001}, M. Fey a) turning table measurement and 
b) EUROGAM measurement \cite{Fey2004phd}, M. Assun\c{c}\~{a}o a) two parameters fit and b) 
three parameters fit \cite{Assuncao2006}, H. Makii \cite{Makii2009} and R. Plag \cite{Plag2012}.
An R-Matrix fit to the data, represented by the solid line, was performed using the AZURE2 code
\cite{Azuma2010}. \label{fig:SE12}}
\end{center}
\end{figure}

Through the years, different experimental approaches have been 
used to determine the rate of the $^{12}$C$(\alpha,\gamma)^{16}$O reaction.
These include measurements of the direct 
reaction \cite{DyerBarnes1974,Redder1987,Kremer1988,Ouellet1996,Roters1999,
Gialanella2001,Kunz2001,Kunz2002phd,Fey2004phd,Schurmann2005,Assuncao2006,
Makii2009,Schurmann2011,Plag2012}, $\beta-$delayed $\alpha$-decay of $^{16}$N \cite{Buchmann1993,Azuma1994,Tang2010} and elastic scattering 
$^{12}$C$(\alpha,\alpha)^{12}$C \cite{Plaga1987, Tischhauser2009}.
As described below, we have fit the world's data
in the region $0.7 \le E_{\alpha}^{c.m.} \le 1.7$ MeV, for both multipoles, where $E_{\alpha}^{c.m.}$ is the kinetic energy of the $\alpha$-particle in the 
center-of-mass ($c.m.$) of the ${}^{12}$C$- \alpha$ system.  The resulting
$S_{EJ}(E_{\alpha}^{c.m.})$
dependence was approximated by fitting the data to second-order polynomials, which are represented by the dashed curves in
Fig.~\ref{fig:SE12}.

However, due to the rapid decrease of the cross section in the region where $E_{\alpha}^{c.m.}$ falls below 2 MeV, 
the uncertainty in the S-factor experimental determination is increasingly dominated 
by the large statistical uncertainty.  Further, as $E_{\alpha}^{c.m.}$ decreases, 
the statistical uncertainties from the different experiments increase rapidly.  A comprehensive review of the experiments 
and methods developed so far, and the full list of astrophysical 
implications of the $^{12}$C$(\alpha,\gamma)^{16}$O rate can be found 
in~\cite{deBoer2017}.

In recent years, there have been new experimental approaches pursued.
One novel approach is based on a bubble chamber 
detector \cite{DiGiovine2015} where the number of photodisintegrations is counted and the total 
astrophysical $S$-factor could be measured even at very low
energies \cite{Holt2018}.  However, the isotopic impurities of $^{17}$O
and $^{18}$O have to be greatly suppressed~\cite{Ugalde2013}. Another $^{16}$O photodisintegration
experiment is based on the optical time projection chamber \cite{Gai2010}
where the angular distribution of $\alpha$-particles is measured and the $S_{E1}$- and
$S_{E2}$-factors can be determined. This approach works
well for higher $\alpha$-particle energies, but for
lower energies the density of the gas needs to be reduced.

In this paper, we present in some detail a new approach to the determination of radiative capture reactions
at stellar energies. We consider the inverse reaction initiated by an electron beam rather than a photon beam. 
The idea has been previously proposed~\cite{Tsentalovich2000}, but not measured, and more recently discussed by~\cite{Friscic2017,Lunkenheimer2017}.   
The theoretical formalism to relate electro- and photo-disintegration has been developed~\cite{Raskin1989}.
Most importantly, a new generation of high intensity ($\approx$ 10 mA) low-energy ($\approx$ 100 MeV) energy-recovery linear (ERL) electron accelerators
is under development~\cite{Hug2017,CBETAHoffstaetter2017} which, when used with state-of-the-art gas targets~\cite{GrieserKhoukaz2018}, can deliver luminosities of $\approx 10^{36}$ cm$^{-2}$ s$^{-1}$ for experiment~\cite{MIT2013}. In this way, the weakness of the electromagnetic force can be overcome.
Here, we have chosen to focus specifically on determination of the reaction rate of $^{12}$C$(\alpha,\gamma)^{16}$O
at stellar energies using this new approach.  However, our approach is generally applicable to all radiative capture reactions.

To provide a basis for the theory used to make estimations of event rates for the electrodisintegration reaction we have begun by revisiting what is typically done for photodisintegration. In the latter case, shell model or cluster model approaches have had some degree of success in yielding the general shape of the cross section, but fail to get its overall magnitude correct. On the one hand, since the electrodisintegration cross section demands even more of any modeling --- specifically, not only the energy dependence of the cross section, but also its momentum transfer behavior (see the following section) --- at present one cannot depend on typical modeling to provide reliable estimates of the cross section.  On the other hand, our focus is on very low energies (typically within an MeV or so of threshold) and relatively low momentum transfers (much smaller than a characteristic nuclear value of 200-300 MeV/c). This means that the form of the cross section as a function of the momentum transfer is tightly constrained. Indeed, as we show in the following sections, the momentum transfer dependence of the cross section can be characterized by a small number of constants, and, importantly, these few constants can be determined experimentally by making measurements at several values of the momentum transfer. In effect, at present it is possible to make reasonable estimates of the electrodisintegration cross section despite the lack of a satisfactory detailed model. Of course, our parametrization of the cross section has been designed to recover what is presently known about the photodisintegration cross section, namely, what must be recovered for the electrodisintegration cross section in the real-photon limit, as discussed in the next section.

We have considered the optimal experimental kinematics in terms of the incident electron energy, the oxygen gas target, the scattered electron spectrometer, and the final-state, low-energy $\alpha$-particle detection.  We have considered systematic uncertainties such as both isotopic and chemical contamination of the $^{16}$O; energy, angle and timing constraints of the final-state particles; energy loss in the gas jet and radiative corrections. 
Using realistic experimental assumptions, we propose an initial measurement of $^{16}$O$(e,e^\prime \alpha){}^{12}$C using an ERL with incident energy of order 100 MeV.  The experiment would take data at higher
$E_{\alpha}^{c.m.}$ where the reaction rates are relatively high and the running time is of order a month.  This initial measurement would aim to validate the extrapolation to photodisintegration and determine the contributions of different multipoles.  If successful, it would set the stage for a longer experiment (of order 6 months) with the highest electron intensity available to determine the $^{12}$C$(\alpha,\gamma)^{16}$O reaction rate with unprecedented precision in the astrophysical region. 

In Sect.~\ref{photo-electro}, the general relationship between electro- and photo-induced reactions is presented, while in 
Sect.~\ref{sec:two}, following the general formalism presented in \cite{Raskin1989}, these developments are applied to 
the exclusive $^{16}$O$(e,e^\prime \alpha)^{12}$C(g.s.) process in which all nuclear species have $J^\pi = 0^+$. In Sect.~\ref{sec:three} the multipole decomposition of 
the response functions involved is discussed, truncating the set of multipoles at the quadrupole response, and thus including C0, C1/E1 and C2/E2 multipoles\footnote{For completeness, the
multipole decompositions of the response functions up to C3/E3 are given in the Appendix.}. Following this general discussion, in Sect.~\ref{sec:four} the model adopted for the semi-inclusive electrodisintegration cross section is presented. Specifically, in Sect.~\ref{subsec:four-a} the present knowledge from studies of photodisintegration and radiative capture reactions is employed in a determination of the leading-order behavior of the C1/E1 and C2/E2 multipoles. Following this, in Sect.~\ref{subsec:four-b} our way of treating the next-to-leading order coefficients in expansions in $q$ is discussed, together with the approach taken for the C0 multipole. Section~\ref{subsec:four-c} concludes the discussion of the model with presentations of the electrodisintegration cross section for typical choices of kinematics in the desired low-$\omega$/low-$q$ region. Given the model, Sect.~\ref{sec:five} then continues with the central section of this paper in which it is shown that,  by making assumptions concerning the experimental capabilities that are projected to exist in the not-too-distant future, measurements of electrodisintegration of $^{16}$O appear to be feasible and that such measurements can be employed to significantly reduce the statistical uncertainties 
of the $S_{E1}$- and $S_{E2}$-factors in the $E_{\alpha}^{c.m.} < 2$ MeV
region. Additionally, in that section a discussion of how 
a smart choice of observable should allow one 
not only to identify the final-state $\alpha$, but also to
identify and remove background events such as $\alpha$-particles from
electrodisintegration of other oxygen isotopes ($^{17}$O and $^{18}$O)
or other ions emerging from electrodisintegration of impurities 
found in an oxygen gas target, {\it e.g.}, protons from $^{14}$N$(e,e'p)^{13}$C. 
We conclude with a summary and a perspective on the future in Section \ref{sec:six}.

\section{Relationships between Photo- and Electro-disintegration}\label{photo-electro}

We begin with a brief discussion of how studies of photodisintegration can be extended to those of electrodisintegration, focusing on the disintegration of $^{16}$O into the ground states of $^4$He (the $\alpha$ particle) and $^{12}$C. For the reader who is unfamiliar with the basic formalism that relates the two processes we can recommend the recent book involving two of the authors \cite{DFHMSa}, in particular Chapters 7 and 16,
including references therein.

\begin{figure}[h]
\begin{center}
\includegraphics[width=8.6cm]{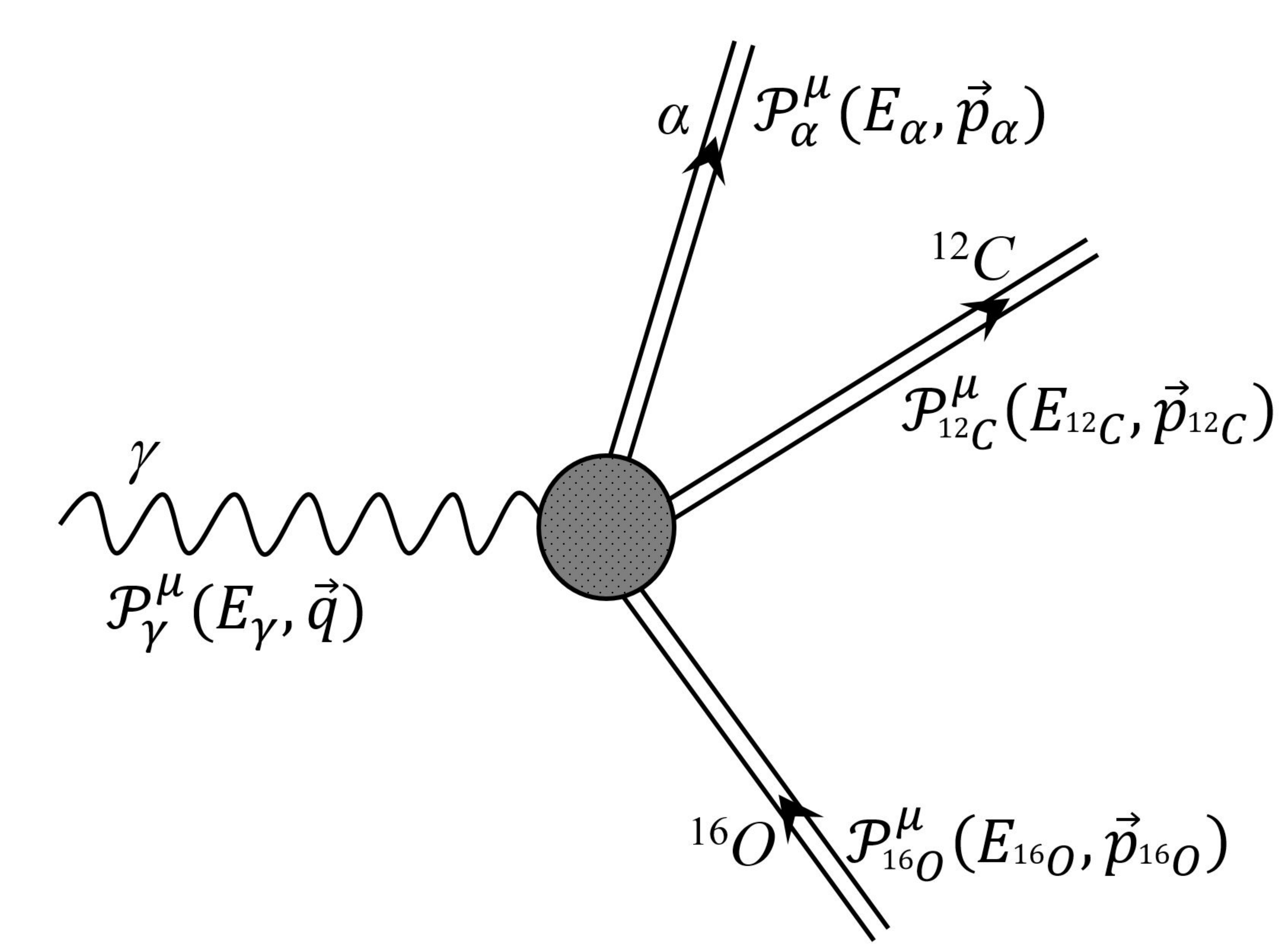}
\caption{Feynman diagram for the photodisintegration of $^{16}$O involving 
a real photon, $\gamma$, which requires that $q=\omega=E_\gamma$. The kinematic variables here will be discussed in more detail in Sect.~\ref{sec:two}. \label{fig:RelPh}}
\end{center}
\end{figure}

\begin{figure}[h]
\centering
\includegraphics[width=8.6cm]{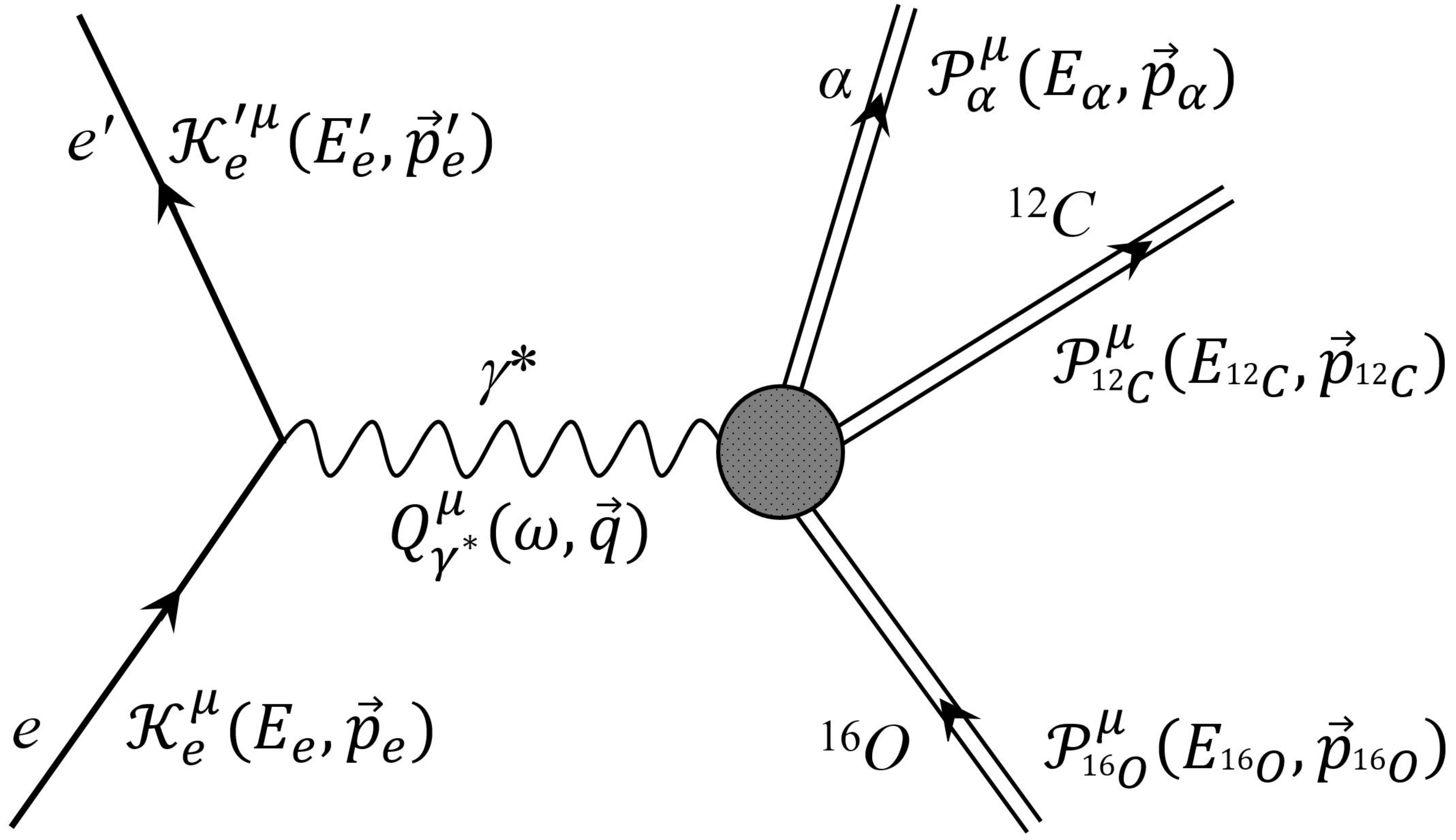} 
\caption{First-order Feynman diagram for the electrodisintegration of $^{16}$O
involving one virtual photon $\gamma\text{*}$ exchange, to be compared with Fig.~\ref{fig:RelPh}. Again, the kinematic variables here will be discussed in more detail in Sect.~\ref{sec:two}. \label{fig:VirtPh}}
\end{figure}

As discussed above, studies aimed at determinations of the alpha + carbon capture reaction $^{12}$C$(\alpha,\gamma)^{16}$O have made use of the inverse process, namely the {\em photodisintegration} of oxygen, $^{16}$O$(\gamma,\alpha)^{12}$C, together with detailed balance. In the present work we describe an extension of these ideas by focusing on the {\em electrodisintegration} reaction $^{16}$O$(e,e'\alpha)^{12}$C. Both photo- and electro-disintegration reactions are assumed to be exclusive, {\it i.e.,} to have the $\alpha$-particle in the final state detected. However, they differ in that the former involves real photons whose momenta $q$ must be equal to their energies $\omega=E_\gamma$, corresponding to  so-called real-photon kinematics, as illustrated in Fig.~\ref{fig:RelPh}\footnote{In most of this work we use natural units where 
$\hbar=c=1$, although later, when writing expressions for the cross sections, we include them to make the units explicit.}. In contrast, as illustrated in Fig.~\ref{fig:VirtPh}, in the one-photon-exchange approximation, which is generally good at the percent level for light nuclei, the latter involves virtual photon exchanges that may be shown to be spacelike, $q>\omega$. That is, by knowing the electron scattering kinematics it is possible to focus on a specific value of the excitation energy of the final-state $\alpha +^{12}$C system, for instance quite close to threshold, but to vary the three-momentum transfer $q$ for any value that keeps the exchanged virtual photon spacelike. Of course, the real-photon result is recovered by taking the limit where $q\to \omega$.

\begin{figure}[!h]
\centering
\includegraphics[width=8.6cm]{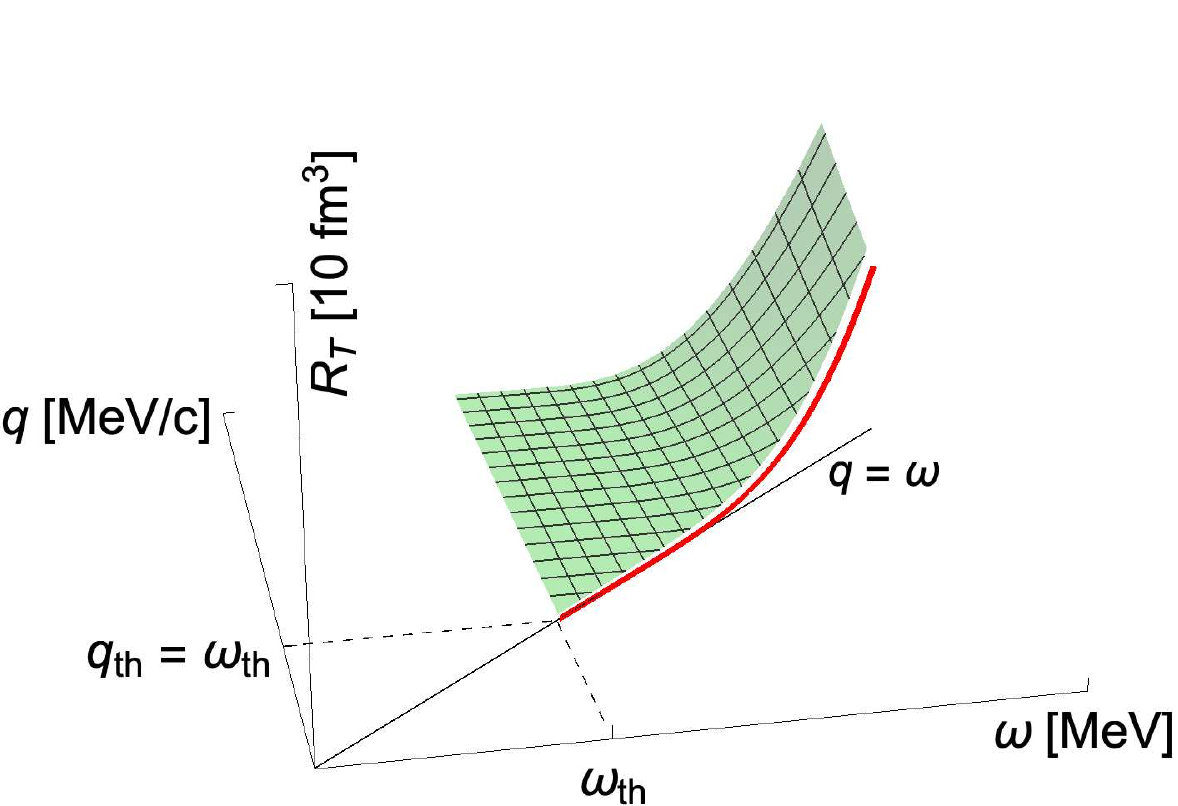} 
\colorcaption{Transverse response function $R_{T}$ as function of the photon energy $\omega$ and
the three-momentum transfer $q$, for the real photon case $q = \omega$ (solid line) and virtual photon case $q>\omega$ (surface plot), where $\omega_{th}$ denotes the value of the threshold photon energy for the reaction. \label{fig:sketch}}
\end{figure}

A sketch of the general landscape is given in Fig.~\ref{fig:sketch} which illustrates a typical response (see later sections of the present work for specifics) as a function of $q$ and $\omega$ together with the real-$\gamma$ line; here $\omega_{th}$ is the threshold value of $\omega$ for the reaction. The strategy in photodisintegration studies is to perform experiments at values of $\omega=E_\gamma$ where the cross section is large enough to be measured and then extrapolate along the real-$\gamma$ line to the very low energies of interest for astrophysics. The electrodisintegration reaction extends these ideas: now one can focus on small values of $\omega$ but have $q$ large enough to yield measurable cross sections. The extended strategy is then to extrapolate in both dimensions, namely, for the responses as functions of $q$ to approach the real-$\gamma$ line and as functions of $\omega$ to reach the interesting low-energy region. As will be discussed in the following sections, an advantage of having $q$ large enough is that one may work near threshold but have sufficient three-momentum imparted to the $\alpha$-particles in the final state that they can emerge from the target and be detected.

Both photo- and electro-disintegration reactions have in common that the angular distribution of the $\alpha$-particles in the final state can be measured. This yields information on the various multipoles that contribute to the process. We assume that $\omega$ is always quite small compared with a typical energy scale; in addition, for the electrodisintegration reaction we assume that $q$ is smaller than a typical scale for nuclear momenta, $q_0$, taken to be roughly of order 200--250 MeV/c. Given this, it is possible to limit the multipoles to a relatively small number. This is commonly done for the photodisintegration reaction near threshold where only $E1$ (electric dipole) and $E2$ (electric quadrupole) multipoles are assumed, although one can ask how important electric octupole $E3$ multipoles might be. Since the nuclear ground states involved are all $0^+$ states only electric multipoles can occur, and magnetic multipoles are absent. Here we have assumed that only the ground states of $^4$He and $^{12}$C are involved and that any excited states can be ignored by using the over-determined kinematics of the reaction. The electrodisintegration reaction is richer, as will be discussed in detail in the following sections of the paper. Since virtual photons are involved, now one has Coulomb $CJ$ as well as electric multipoles $EJ$; in the body of the paper we consider $C0$, $C1/E1$ and $C2/E2$ multipoles, although in an appendix we give some of the relevant formalism for a larger set that includes $C3/E3$ contributions.

At low values of the momentum transfer, $q\ll q_0$, each multipole is dominated by its low-$q$ behavior which enters as a specific power of $q$. For instance, later we show that the $CJ$ mulipole matrix elements go as $(q/q_0)^J$ at low $q$. Accordingly another advantage of electron scattering where $q$ may be varied while keeping $\omega$ fixed is that the balance of the multipole contributions can be varied. An example of this could, for instance, be the potential $C3/E3$ contributions: by increasing $q$ (still, of course, staying in the region where $q\ll q_0$) one may increase the relative importance of the octupole effects over the monopole, dipole and quadrupole effects to explore whether or not the former need to be taken into account.

Not only is there a richer set of mutipoles involved in the electron scattering case, but there are more response functions to be exploited. For real photons one has the transverse response $R_T$ at $q=\omega$ and potentially the transverse interference response $R_{TT}$ also at $q=\omega$ if linearly polarized real photons are involved (see Sect.~III for more discussion). For unpolarized electron scattering there are four types of responses, $R_T$ and $R_{TT}$ as for real photons but now with virtual photons and thus at $q>\omega$ and also $R_L$, the longitudinal/charge response and an interference between transverse and longitudinal contributions, $R_{TL}$, both at $q>\omega$. In the $R_T$ and $R_{TT}$ responses only $EJ$ multipoles enter, not simply squared but through interferences. The $R_{L}$ response contains only $CJ$ multipoles, again with interferences, while the $R_{TL}$ response has interferences between $CJ$ and $EJ$ mutipoles. All of this means that potentially one has more information with which to disentangle the various contributions. The angular distributions as functions of the alpha angles $\theta_\alpha$ and $\phi_\alpha$ (see the next section) will be discussed in detail. These may be written as expansions in terms of Legendre polynomials where the expansion coefficients that enter and may be determined experimentally contain valuable information on all bilinear products of the multipole matrix elements.

We now proceed to a summary of the kinematics and basic form of the semi-inclusive electron scattering cross section in the following section.

\section{Kinematics and the cross section}\label{sec:two}

We start this section with a brief discussion of 
exclusive-1 electron scattering A$(e,e'x)$, following 
\cite{Raskin1989}\footnote{An earlier version of the relevant formalism, based on the 
more general discussions presented in \cite{Raskin1989} 
was developed by Donnelly and Butler for a proposed measurement at the MIT-Bates Laboratory in 2000 \cite{Tsentalovich2000}; see also \cite{Donnelly2002}.}, although, in contrast to the more general study in \cite{Raskin1989}, here the discussion will be limited to the scattering of unpolarized electrons from an unpolarized
target nucleus, {\it i.e.,} polarization degrees of freedom will be neglected. 
We limit our consideration to the one-photon exchange contributions (lowest order, 
see Fig. \ref{fig:VirtPh}), and take the electron wave functions 
to be plane waves, namely, we invoke the plane-wave Born approximation (PWBA). 
The four-momenta of the incident and
scattered electrons are labeled $\mathcal{K}^{\mu}_e
(E_e,\vv{p}_{\mkern-4.mu e})$ and $\mathcal{K'}^{\mu}
_{\mkern-6.mu e}(E'_e,\vv{p}'_{\mkern-4.mu e})$,
respectively. $E_e$ and $E'_e$ are their energies, while 
$\vv{p}_{\mkern-4.mu e}$ and $\vv{p}'_{\mkern-4.mu e}$ 
are their three-momenta. The four-momentum transfer 
is defined by $Q^{\mu} \equiv (\omega,\vv{q}) = \mathcal{K}^{\mu}_e -
\mathcal{K'}^{\mu}_{\mkern-6.mu e} = \mathcal{P}^{\mu}_{^{16}O}
- \mathcal{P}^{\mu}_{^{12}C}-\mathcal{P}^{\mu}_{\alpha}$,
where $\mathcal{P}^{\mu}_{^{16}O}$, $\mathcal{P}^{\mu}_{^{12}C}$ 
and $\mathcal{P}^{\mu}_{\alpha}$ are the four-momenta of the
target nucleus $^{16}$O, residual nucleus $^{12}$C and 
exclusive nucleus $\alpha$. Also, $\omega =E_e - E'_e =
E_{\alpha}+E_{^{12}C} - E_{^{16}O}$ is the energy transfer and
$\vv{q}=\vv{p}_{\mkern-4.mu e}-\vv{p}'_{\mkern-4.mu e} =
\vv{p}_{\mkern-3.mu \alpha}+\vv{p}_{\mkern-3.mu ^{12}C}
-\vv{p}_{\mkern-3.mu ^{16}O}$  is the three-momentum transfer.

In order to identify the events belonging to the
electrodisintegration of $^{16}$O a scattered electron
needs to be detected in coincidence with a produced 
$\alpha$-particle and the four-momenta of both 
have to be measured. The remaining $^{12}$C nucleus 
does not need to be detected, since its final 
state can be reconstructed by using energy-momentum conservation. 
The variables typically used to characterize the semi-inclusive reaction are the following (see \cite{DFHMSa}): 
the missing momentum $\vv{p}_{miss}$ and
missing energy $E_{miss}$ are given by

\begin{align}
\vv{p}_{miss}&= \vv{q}-\vv{p}_{\alpha} \\
E_{miss}& =\omega + M_{^{16}O} - E_{\alpha} ,
\end{align}
and then the missing mass
\begin{equation}
m_{miss} = \sqrt{E_{miss}^2- \vv{p}_{miss}^2}
\end{equation}
may be calculated by subtracting the mass of the unobserved $^{12}$C nucleus,
$M_{^{12}C}$. One then obtains the excitation
energy of the $^{12}$C,
\begin{align}
E_{ex} = m_{miss}-M_{^{12}C} ,
\end{align}
where events which contribute to the astrophysical 
$S$-factor are those where one finds the $^{12}$C
nucleus in its ground state, that is, $E_{ex}=0$.

\begin{figure}[H]
\centering
\includegraphics[width=8.6cm]{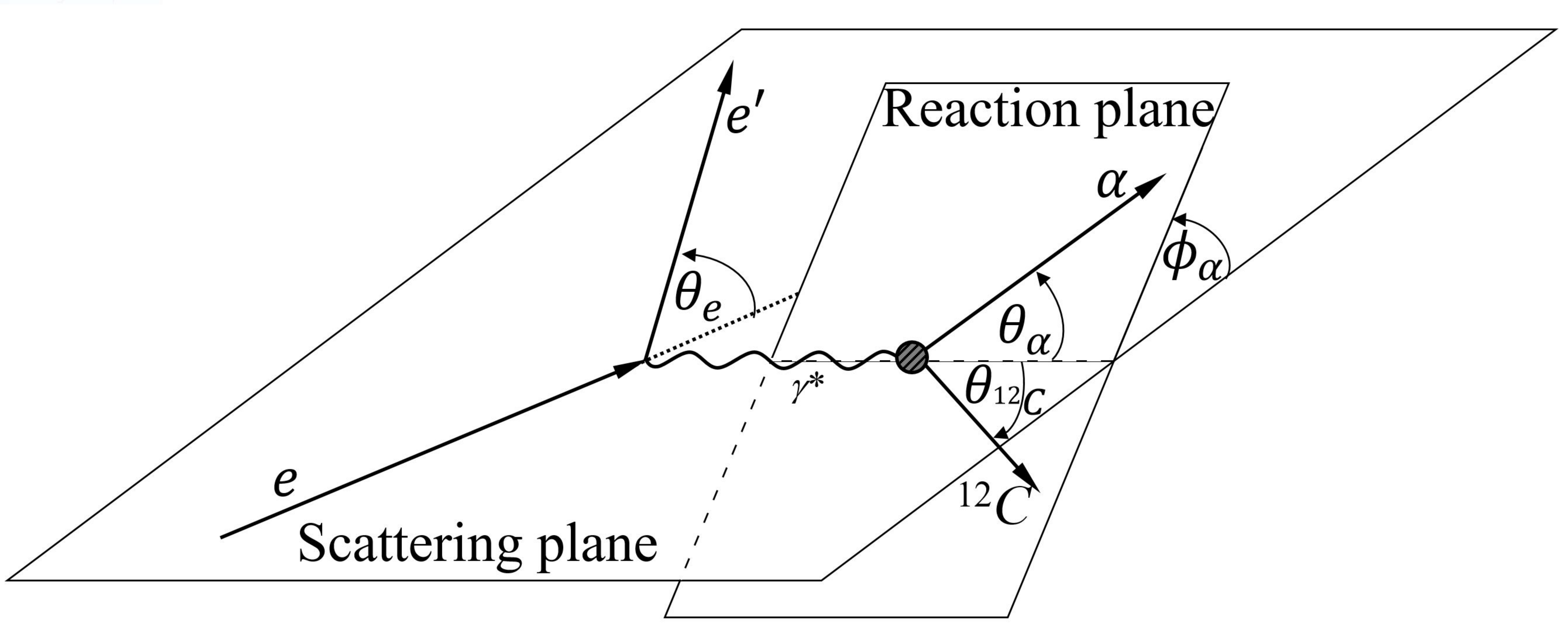} 
\caption{Kinematics of the exclusive $^{16}$O$(e,e^\prime \alpha)^{12}$C reaction.}
\label{fig2}
\end{figure}
The differential cross section in laboratory frame (where the target is at rest,  
$\mathcal{P}^{\mu}_{^{16}O}=(M_{^{16}O},0)$) is given by \cite{Bjorken1964}
\begin{widetext}
\begin{equation}
d\sigma=\frac{m_e}{E_e}\frac{1}{\beta_{e}}\overline{\sum\limits_{fi}}|\mathcal{M}_{fi}|^2\frac{m_e}{E'_{e}} \frac{d^3\vv{p}_{e}'}{(2\pi)^3}\frac{M_{\alpha}}{E_{\alpha}}\frac{d^3\vv{p}_{\mkern-3.mu \alpha}}{(2\pi)^3}\frac{M_{^{12}C}}{E_{^{12}C}}\frac{d^3\vv{p}_{\mkern-3.mu ^{12}C}}{(2\pi)^3}(4\pi)^4\delta^4(\mathcal{K}^{\mu}_{e}+\mathcal{P}^{\mu}_{\mkern-3.mu ^{16}O}-\mathcal{K'}^{\mu}_{\mkern-6.mu e}-\mathcal{P}^{\mu}_{\alpha}-\mathcal{P}^{\mu}_{^{12}C}),
\end{equation}
\end{widetext}
 
where $\beta_e = |\vv{p}_e|/E_e=|\vv{v}_e|$ and $\overline{\sum}_{fi}$
represents an average over initial states and sum over final states, under the assumption
that all particle states are normalized to unity. 
If we assume the momenta of the scattered electron and $\alpha$-particle to be measured 
but not the residual $^{12}$C nucleus, we need to perform an integration
over the recoil momentum $\vv{p}_{\mkern-3.mu ^{12}C}$:

\begin{widetext}
\begin{equation}
d\sigma=\frac{m_e^2 M_{\alpha} M_{^12{C}}}{E_e}\frac{1}{(2\pi)^5}\overline{\sum\limits_{fi}}|\mathcal{M}_{fi}|^2 \frac{p'^2_e dp'_e d\Omega_e p^2_{\alpha} dp_{\alpha} d\Omega_{\alpha}}{E'_e E_{\alpha} E_{^{12}C}}   \delta(E_e +E_{^{16}O} - E'_e - E_{\alpha}-E_{^{12}C}).
\end{equation}
\end{widetext}

We continue to integrate the energy-conserving delta-function of energy conservation over $p_{\alpha}$ 
and make use of the following formula
\begin{equation}
\delta(f(p))=\sum\limits_{i}\cfrac{\delta(p-p_i)}{\bigg|\cfrac{\partial f(p)}{\partial p}\bigg|_{p_i}},
\end{equation}
where $f(p_i)=0$ and
\begin{equation}
f(p_{\alpha})= \omega + M_{^{16}O}  - \sqrt{|\vv{p}_{\mkern-3.mu \alpha}|^2+M^2_{\alpha}}-\sqrt{(\vv{q}-\vv{p}_{\mkern-3.mu \alpha})^2+M^2_{^{12}C}}.
\end{equation}
After the integration we obtain
\begin{align}
\int & \frac{p^2_{\alpha} dp_{\alpha}}{E_{\alpha}E_{^{12}C}} \delta(E_e +E_{^{16}O} - E'_e - E_{\alpha}-E_{^{12}C})  \notag \\ 
    &= \frac{p_{\alpha}}{M_{^{16}O}}\cfrac{1}{\bigg|1+ \cfrac{\omega p_{\alpha} - E_{\alpha}|\vv{q}|\cos\theta_{\alpha}}{M_{^{16}O} \cdot p_{\alpha}} \bigg|}  \notag \\ 
 &=\frac{p_{\alpha}}{M_{^{16}O}} f_{rec}^{-1},
\end{align}
where $f_{rec}$ is the hadronic recoil factor and $\theta_{\alpha}$ is
the angle between $\vv{q}$ and $\vv{p}_{\alpha}$; see Fig \ref{fig2}. 
The cross section is now
\begin{equation}
d\sigma=\frac{m_e^2 M_{\alpha} M_{^{12}C}}{(2\pi)^5 M_{^{16}O}}\frac{E'_e \cdot p_{\alpha}}{E_e} f_{rec}^{-1} \overline{\sum\limits_{fi}}|\mathcal{M}_{fi}|^2 .
\end{equation}

The Lorentz-invariant matrix element $\mathcal{M}_{fi}$ is given by
\begin{equation}
\mathcal{M}_{fi}=\frac{ie}{Q^2}\Bigg(\frac{E_e E'_{e}}{m^2_e} \Bigg)^{\frac{1}{2}} j_e(\mathcal{K'}_{\mkern-6.mu e},\mathcal{K}_{e})_{\mu}J^{\mu}(\mathcal{P}_{^{12}C},\mathcal{P}_{\alpha};\mathcal{P}_{^{16}O})_{fi},
\end{equation}
where $j_e(\mathcal{K'}_{\mkern-6.mu e},\mathcal{K}_{e})_{\mu}$ is the electromagnetic electron current, $J^{\mu}(\mathcal{P}_{^{12}C},\mathcal{P}_{\alpha};\mathcal{P}_{^{16}O})_{fi}$ 
is the hadronic electromagnetic transition current and the square of
the four-momentum transfer in the extreme relativistic limit (ERL)
is given by $|Q^2|=4E_eE'_e\sin^2 (\theta_e/2)$, where $\theta_e$
is the electron scattering angle; see Fig. \ref{fig2}.
When we square $\mathcal{M}_{fi}$, sum over final states and average 
over initial states, we end up with
\begin{equation}
\overline{\sum\limits_{fi}}|\mathcal{M}_{fi}|^2=\frac{(4\pi\alpha)^2}{(Q^2)^2}\eta_{\mu\nu}W^{\mu\nu} ,
\end{equation}
where $\eta_{\mu\nu}$ is the leptonic tensor and $W^{\mu\nu}$ is 
the hadronic tensor. Note that the contraction of the leptonic and hadronic 
tensors is Lorentz invariant. Accordingly it can be evaluated in any frame and it is given by
\begin{equation}
\eta_{\mu\nu}W^{\mu\nu} =\frac{v_0}{4m_e^2}\sum\limits_{K}v_KR_K
\end{equation}
with $v_0=4E_e E'_{e}\cos^2(\theta_e/2)$. For the unpolarized exclusive
electron scattering we have four nuclear response functions $R_K$:
the longitudinal $R_L$ and transverse $R_T$ nuclear electromagnetic current
components (longitudinal and transverse with respect to the direction of the virtual photon $\vv{q}$), and
two interference responses, namely transverse-longitudinal $R_{TL}$ and 
transverse-transverse $R_{TT}$. In this notation $R_K$ will have
dimension of fm$^3$. The functions $v_K$ are electron kinematic factors and
in terms of ERL can be expressed as \cite{Raskin1989}
\begin{align}
v_L &=\rho^2  \notag \\
v_T &=\frac{1}{2}\rho+\tan^2\theta_e/2  \notag \\
v_{TL} &= - \frac{1}{\sqrt{2}}\rho\sqrt{\rho+\tan^2\theta_e/2}  \notag \\
v_{TT} &= - \frac{1}{2}\rho ,
\end{align}
where as usual $0 \le \rho \equiv |Q^2/q^2| = 1-(\omega / q)^2 \le 1 $. 
The most general discussion concerning the leptonic and hadronic tensor contraction,
which also includes polarization degrees of freedom, can be found in
\cite{Raskin1989, Donnelly1986}.

It is convenient to group variables to form the ERL Mott cross section
\begin{equation}
\sigma_{Mott}= \frac{\alpha^2 (\hbar c)^2 E'_e v_0}{(Q^2)^2 E_e} = \bigg(\frac{\alpha 
\hbar c \cos\theta_e/2}{2 E_e \sin^2\theta_e/2}\bigg)^2 .
\end{equation}
Note that here we include the factor $\hbar c = 197.327$ MeV\:\!fm so that $\sigma_{Mott}$ has dimensions of fm$^2$. Finally, the semi-inclusive electrodisintegration 
cross section for the reaction of interest in the laboratory frame takes the form 
\begin{widetext}
\begin{equation}
\Bigg[ \frac{d\sigma}{d\omega d\Omega_e d\Omega_{\alpha}} \Bigg]_{(e,e'\alpha)}=\frac{M_{\alpha}M_{^{12}C}}{8\pi^3 M_{^{16}O}}\frac{p_{\alpha}f^{-1}_{rec} \sigma_{Mott}}{(\hbar c)^3} \bigg(v_LR_L+v_TR_T+v_{TL}R_{TL}+v_{TT}R_{TT}\bigg) .
\end{equation}
\end{widetext}
Often it also very convenient to have an expression for the cross section in the center-of-mass ($c.m.$) frame, where the transformation between the frames involves a Lorentz boost along $\vv{q}$. We note  
that $W=\sqrt{(M_{^{16}O}+\omega)^2 -q^2}$ is the total invariant mass of the $\gamma + ^{16}$O and $\alpha + ^{12}$C systems, here evaluated in the incident channel laboratory frame with the $^{16}$O target nucleus at rest. Furthermore, $p^{c.m.}_{\alpha}=|\vv{p}\;^{c.m.}_{\alpha}|$ is the $\alpha$-particle 
three-momentum in the $c.m.$ frame, $R_K$ now represent quantities 
in the $c.m.$ frame and the lepton kinematic factors in the $c.m.$ 
frame are given by the following: 
\begin{align}
\widetilde{v}_L &=(W/M_{^{16}O})^2v_L,  \notag \\
\widetilde{v}_{TL} &= (W/M_{^{16}O})v_{T}, \notag \\
\widetilde{v}_T& =v_T,  \notag \\ 
\widetilde{v}_{TT} &=v_{TT} .
\end{align}
Finally, the cross section in the $c.m.$ frame can be written as
\begin{widetext}
\begin{equation}\label{eq:diffcross}
\Bigg[ \frac{d\sigma}{d\omega d\Omega_e d\Omega^{c.m.}_{\alpha}} \Bigg]_{(e,e'\alpha)}=\frac{M_{\alpha}M_{^{12}C}}{8\pi^3 W}\frac{p^{c.m.}_{\alpha}\sigma_{Mott}}{(\hbar c)^3} \bigg(\widetilde{v}_LR_L+\widetilde{v}_TR_T+\widetilde{v}_{TL}R_{TL}+\widetilde{v}_{TT}R_{TT}\bigg) .
\end{equation}
\end{widetext}
Note that $\phi_{\alpha} = \phi_{\alpha}^{c.m.}$, although $\theta_{\alpha} \neq 
\theta_{\alpha}^{c.m.}$. Again, we encourage the reader who is unfamiliar with these developments to look at \cite{DFHMSa}, especially Chapter 7 where the current matrix elements are discussed, multipole operators are introduced and the real-photon limit is briefly treated, as well as Chapter 16 where semi-inclusive electron scattering is the focus (there one also finds Exercises 16.4, 16.6 and 16.7 which are relevant for the present purposes, especially Exercise 16.7 where a problem involving the real-photon limit of semi-inclusive electron scattering is posed).

An analysis similar to the one in \cite{Raskin1989} can be performed
both for the photodisintegration process $^{16}$O$(\gamma,\alpha)^{12}$C
and for the radiative capture reaction $^{12}$C$(\alpha, \gamma)^{16}$O, where here for simplicity we take the real photons to be unpolarized (see the comment regarding linearly polarized photons in the next section). For the former reaction  
the differential cross section is given by
\begin{equation}\label{eq:photo}
\Bigg[\frac{d\sigma}{d\Omega^{c.m.}_{\alpha}}\Bigg]_{(\gamma,\alpha)}=\Bigg(\frac{M_{\alpha}M_{^{12}C}}{4\pi W}\Bigg)\frac{p^{c.m.}_{\alpha}}{\hbar c}\Bigg(\frac{\alpha}{E_{\gamma}}\Bigg)\mathcal{R}_{(\gamma,\alpha)},
\end{equation}
where $W=\sqrt{M_{^{16}O}(M_{^{16}O}+2E_{\gamma})}$ (that is, 
$q=\omega=E_{\gamma}$ above) and $\mathcal{R}_{(\gamma,\alpha)}$ 
is transverse response function having dimension of fm$^3$. Namely, one has the real-photon limit of the electrodisintegration result summarized above. The radiative capture cross section is then related by detailed balance and may be written in the form
\begin{equation}\label{eq:direct}
\Bigg[\frac{d\sigma}{d\Omega^{c.m.}_{\gamma}}\Bigg]_{(\alpha,\gamma)}=\Bigg(\frac{M_{\alpha}M_{^{12}C}}{2\pi W}\Bigg)\frac{E_{\gamma}}{\hbar c}\Bigg(\frac{\alpha}{p^{c.m.}_{\alpha}}\Bigg)\mathcal{R}_{(\alpha,\gamma)},
\end{equation}
where $W$ is the invariant mass above, which, in the incident channel laboratory frame where the $^{12}$C target is at rest, is equal to 
$W=\sqrt{M^{2}_{\alpha}+M^{2}_{^{12}C} + 2 M_{^{12}C} E^{Lab}_{\alpha}}$. 
As above, $\mathcal{R}_{(\alpha,\gamma)}$ is the transverse response function, here for real photons to be evaluated 
at $q=\omega=E_\gamma$; it has dimensions of fm$^3$. 

\section{Multipole decomposition of response functions involving spin $\bf 0$ nuclei} \label{sec:three}

Let us discuss the longitudinal-transverse decomposition a little further. 
For the specific initial and final nuclear states involved there 
are three independent current matrix elements, $\rho(\vv{q})$, 
$J^x(\vv{q})$ and $J^y(\vv{q})$, 
with $J^z(\vv{q})=(\omega/q)\rho(\vv{q})$ as required by current conservation. 
From them, we can obtain three independent quantities,
which transform as a rank-1 spherical tensor under rotations:
\begin{align}
J^{(0)}(\vv{q}) &\equiv J^{z}(\vv{q})
=(\omega/q) \rho(\vv{q}) \\
J^{(\pm1)}(\vv{q}) &\equiv \mp ( J^{x}(\vv{q}) 
\pm iJ^{y}(\vv{q}))/\sqrt{2} . 
\end{align}
The inverse relationships for Cartesian transverse projections are 
then given by $J^x_{fi} = -(J^{(+1)}_{fi}-J^{(-1)}_{fi})/\sqrt{2}$ and $J^y_{fi} = 
i(J^{(+1)}_{fi}-J^{(-1)}_{fi})/\sqrt{2}$.
Following \cite{Raskin1989} we define the generic quantity
\begin{equation}
X^{\lambda'\lambda}=J^{\lambda'\ast} J^{\lambda}
\end{equation}
and each structure function can be written in terms of the $X^{\lambda',\lambda}$.  
Furthermore,
\begin{align}
J^{(0)}(\vv{q}) &= \frac{\omega}{q}\sqrt{4\pi}\sum_{J\le0}[J]i^J \langle\alpha, ^{12}C|\widehat{T}_{J,0}|^{16}O \rangle\\
J^{(\pm1)}(\vv{q}) &= -\sqrt{2\pi}\sum_{J\le0}[J]i^J \langle\alpha, ^{12}C|\widehat{T}_{J,\pm1}|^{16}O \rangle
\end{align}
with notation $[J] \equiv \sqrt{2J+1}$ and the 
general form of structure functions can be now
written as
\begin{widetext}
\begin{equation}
X^{\lambda'\lambda}=4\pi\sum_{J',J}[J'][J](-i)^{J'}(i)^{J}\langle\alpha,^{12} C|\widehat{T}_{J',\lambda'}|^{16}O \rangle^{\ast}\langle\alpha,^{12} C|\widehat{T}_{J,\lambda}|^{16}O \rangle .
\end{equation}
\end{widetext}
Specifically, one has
\begin{eqnarray}
R_L &\equiv& X^{00} \\ \notag\\
R_T &\equiv& X^{11} + X^{-1-1} \\ \notag\\
R_{TT} &\equiv& X^{1-1} + X^{-11} \\ \notag\\
R_{TL} &\equiv& -2{\rm Re}\{X^{01} - X^{0-1} \} ,
\end{eqnarray}
where the transverse cases are labeled by the polarization that 
a photon would have in the real-$\gamma$ limit. In particular, 
it is clear that the $R_T$ response involves transverse projections 
of the current in a form corresponding to unpolarized photon exchange, 
while the $R_{TT}$ response enters when the photon is linearly polarized. Indeed, in the previous section where expressions for the real-$\gamma$ photodisintegration and radiative capture reactions were given we could have extended the analysis to include both $R_T$ and $R_{TT}$ contributions at $q=\omega=E_\gamma$ and thereby obtained expressions for linearly-polarized real-$\gamma$ processes. 

The responses are calculated from most general expressions Eqs.~(2.54 -- 2.58) in \cite{Raskin1989}. For the initial and final states $^{16}$O and $\alpha +^{12}$C, we have $J_{^{16}O}=J_{\alpha}=J_{^{12}C}=0$ 
which implies that $I_{^{16}O}=I_{\alpha}=I_{^{12}C}=0$. We have $S'=S=0$ which yields $\mathcal{J} = J=L$ and 
$\mathcal{J'} = J'=L'$. In the case of the completely unpolarized situation, Eqs.~(2.79--2.81) in
\cite{Raskin1989} yield
\begin{widetext}
\begin{eqnarray}
\widetilde{F} \; &\sim& \;\; 1 \\[6pt]
\widetilde{\mathcal{D}} \; &\sim& \;\;  \mathcal{D}^{(\ell)*}_{-\Lambda,0} (-\phi_x,\theta^c_x,\phi_x) =(-1)^{\Lambda}\sqrt{\frac{4\pi}{2\ell+1}}Y^{\Lambda}_{\ell}(\theta^{c}_x,-\phi_x) \\[6pt]
\widetilde{W}^{\lambda'\lambda}\; &\sim& \;\; (-1)^{J'+J+\ell+\lambda'}[J][J'][\ell]^2\begin{pmatrix}
  J & J' & \ell \\
  0 & 0 & 0
 \end{pmatrix}
 \begin{pmatrix}
  J & J' & \ell \\
  -\lambda& \lambda' & -\Lambda
 \end{pmatrix} ,
\end{eqnarray}
\end{widetext}
where here the 6-j symbols in \cite{Raskin1989} have been evaluated. The response functions 
will then involve sums of products of these elements 
\begin{align}
W^{\lambda'\lambda}\; &=\;\sum_{J'J}\widetilde{W}^{\lambda'\lambda}t^{\ast}_{J'(\lambda')}t_{J(\lambda)} \\
\overline{X^{\lambda'\lambda}_{fi}} \; &= \; \sum_{\ell}(-1)^{\Lambda}\sqrt{\frac{4\pi}{2\ell+1}}Y^{\Lambda}_{\ell}(\theta^{c}_x,-\phi_x)W^{\lambda'\lambda},
\end{align}
where $t_{J(\lambda)}$ are reduced matrix elements defined by
\begin{align}
t_{J,\lambda}&\equiv i^J  \langle J||\widehat{T}_{J,\lambda}  || J_i=0\rangle  \\
             &= \begin{cases}
             t_{CJ} & \text{ $\lambda = 0$}\\ 
             \frac{1}{\sqrt{2}}(t_{EJ,\lambda}+\lambda t_{MJ,\lambda} ) & \text{ $\lambda = \pm 1$} 
             \end{cases}
\end{align}
with the initial $^{16}$O state being $J_i =0$, and $J$ represents 
the total angular momentum of the partial wave of the final-state 
$\alpha$-particle plus $^{12}$C system. 
We note that this result is simplified enormously when the final 
state of $^{12}$C is the ground state, and not an excited state, and we shall assume that the kinematics of the reaction are well enough determined for this to be the case --- not an especially stringent requirement since the $2^+$ first excited state of $^{12}$C lies at 4.4389 MeV. 
This prevents excitation to unnatural parity states, in turn 
restricting the study to natural-parity $CJ$ and $EJ$ multipoles. 

Following the developments in \cite{Raskin1989} we can now describe the nature of the angular distributions 
themselves, accounting for both relative phases and magnitudes. 
In terms of the Coulomb and electric multipoles up to the quadrupole 
contribution $J=2$\footnote{For clarity, we restrict our attention in the body of the paper to $C0$, $C1/E1$ and $C2/E2$ multipoles; however, in the Appendix~\ref{app:ResOctu} we extend the analysis to include $C3/E3$ octupole multipoles. Additionally, for completeness there we also re-express the angular distributions in terms of sines and cosines of the angles involved, rather than in terms of Legendre polynomials as here.}, the responses may be written in terms of Legendre 
polynomials 
\begin{widetext}
\begin{flalign} \label{eq:RL}
R_L &= P_0 (\cos \theta_{\alpha}) \Bigg( |t_{C0}|^2 + |t_{C1}|^2 +|t_{C2}|^2\Bigg) \notag \\
    &+ P_1 (\cos \theta_{\alpha})  \Bigg(  2\sqrt{3} |t_{C0}||t_{C1}|\cos(\delta_{C1}-\delta_{C0}) + 4\sqrt{\frac{3}{5}} |t_{C1}||t_{C2}|\cos(\delta_{C2} - \delta_{C1}) \Bigg) \notag \\
    &+ P_2 (\cos \theta_{\alpha})  \Bigg(2|t_{C1}|^2  + \frac{10}{7} |t_{C2}|^2  +   2 \sqrt{5} |t_{C0}||t_{C2}|\cos(\delta_{C2}-\delta_{C0})  \Bigg) \notag \\
    &+ P_3 (\cos \theta_{\alpha})  \Bigg(6\sqrt{\frac{3}{5}} |t_{C1}||t_{C2}|\cos(\delta_{C2} - \delta_{C1}) \Bigg) \notag \\
    &+ P_4 (\cos \theta_{\alpha})  \Bigg(\frac{18}{7} |t_{C2}|^2  \Bigg) &&
\end{flalign}

\begin{flalign} \label{eq:RT}
R_T &= P_0 (\cos \theta_{\alpha}) \Bigg(|t_{E1}|^2 +|t_{E2}|^2  \Bigg) \notag \\
    &+ P_1 (\cos \theta_{\alpha})  \Bigg(\frac{6}{\sqrt{5}} |t_{E1}||t_{E2}|\cos(\delta_{E2}-\delta_{E1}) \Bigg) \notag \\
    &+ P_2 (\cos \theta_{\alpha})  \Bigg(-|t_{E1}|^2  + \frac{5}{7} |t_{E2}|^2   \Bigg) \notag \\
    &+ P_3 (\cos \theta_{\alpha})  \Bigg(-\frac{6}{\sqrt{5}} |t_{E1}||t_{E2}|\cos(\delta_{E2} - \delta_{E1})   \Bigg) \notag \\
    &+ P_4 (\cos \theta_{\alpha})  \Bigg(-\frac{12}{7} |t_{E2}|^2   \Bigg)   &&
\end{flalign}
\begin{flalign} \label{eq:RTL}
R_{TL} &= \cos\phi_{\alpha} \cdot \notag \\
    & \Bigg \{ P^1_1(\cos \theta_{\alpha}) \Bigg(  2\sqrt{3}|t_{C0}||t_{E1}|\cos(\delta_{E1}-\delta_{C0})  - 2\sqrt{\frac{3}{5}}|t_{C2}||t_{E1}| \cos(\delta_{C2}- \delta_{E1}) + \frac{6}{\sqrt{5}}|t_{C1}||t_{E2}| \cos(\delta_{C1} - \delta_{E2}) \Bigg)\notag \\
&+ P^1_2(\cos \theta_{\alpha}) \Bigg( 2|t_{C1}||t_{E1}|\cos(\delta_{C1}-\delta_{E1})+  2\sqrt{\frac{5}{3}}|t_{C0}||t_{E2}|\cos(\delta_{E2}-\delta_{C0})+\frac{10}{7\sqrt{3}}|t_{C2}||t_{E2}|\cos(\delta_{C2}-\delta_{E2})  \Bigg)  \notag\\
&+ P^1_3(\cos \theta_{\alpha}) \Bigg( 2\sqrt{\frac{3}{5}}|t_{C2}||t_{E1}|\cos(\delta_{C2}-\delta_{E1})+\frac{4}{\sqrt{5}}|t_{C1}||t_{E2}|\cos(\delta_{C1}-\delta_{E2})  \Bigg) \notag\\
&+ P^1_4(\cos \theta_{\alpha}) \Bigg( \frac{6\sqrt{3}}{7}|t_{C2}||t_{E2}|\cos(\delta_{C2}-\delta_{E2})  \Bigg)  \Bigg \} &&
\end{flalign}
\begin{flalign} \label{eq:RTT}
R_{TT} &= - R_T \cos(2\phi_{\alpha}) . &&
\end{flalign}
\end{widetext}
The $t_{(C,E)J}$ represent the Coulomb and electric reduced 
matrix elements, and are functions of $q$ and $\omega$. Similarly, 
the functions $\delta_{(C,E)J}$ represent the phases of the (in 
general complex) reduced matrix elements of each multipole current operator, and these 
too can be functions of $q$ and $\omega$. As expected only phase differences occur, and one overall phase may be chosen by establishing some specific phase convention.

It is now straightforward to obtain expressions for the angular distributions for specific choices of kinematics. For instance, assume that $\theta_{\alpha}=0^{\circ},180^{\circ}$. In this case, 
$\cos\theta_{\alpha}=\pm1$, so let $\beta=\pm1$ present these 
two possibilities. First $R_T=R_{TL}=R_{TT}=0$ in this case, and one has

\begin{eqnarray} \label{eq:RLspecial}
R_L &=&  |t_{C0}|^2 + 3 |t_{C1}|^2 + 5 |t_{C2}|^2 \notag \\
    &&+   2\sqrt{5} |t_{C0}||t_{C2}|\cos(\delta_{C2}-\delta_{C0}) \notag \\            
   &&+ \beta \Big(   2\sqrt{3} |t_{C0}||t_{C1}|\cos(\delta_{C1}-\delta_{C0}) \notag \\ 
    &&+ 2\sqrt{15}|t_{C1}||t_{C2}|\cos(\delta_{C2} - \delta_{C1})  \Big) .    
\end{eqnarray}
Or, consider the case where $\theta_{\alpha}=90^{\circ}$. Here
\begin{eqnarray}
R_L &=& |t_{C0}|^2 + \frac{5}{4}|t_{C2}|^2 \notag \\ &&-  \sqrt{5} |t_{C0}||t_{C2}|\cos(\delta_{C2}-\delta_{C0}) \\
R_T &=& \frac{3}{2}|t_{E1}|^2  \\
R_{TL} &=& \cos\phi_x \Big\{-  2\sqrt{3}|t_{C0}||t_{E1}|\cos(\delta_{E1}-\delta_{C0}) \notag \\
        &&+\sqrt{14}|t_{C2}||t_{E1}|\cos(\delta_{C2}-\delta_{E1})   \Big\}  \\
R_{TT} &=& - R_T \cos 2\phi_x . 
\end{eqnarray}

If we assume that the cross section is completely dominated by 
$R_L$, as is likely (see below), then there are as many unknowns as there 
are linearly independent Legendre polynomials in the expansion. 
One should also remember as noted above that, while we have stopped at $J = 2$ 
partial waves, there can be higher partial waves present. While these are
likely small for the kinematics of interest, any fit should test the convergence of these expansions by looking for higher-order Legendre polynomials.

We end this section with a discussion of our chosen parametrizations of the multipole matrix elements. These all depend on both $q$ and $\omega$ (which then determine the $c.m.$ energy 
of the final state); here we suppress the $\omega$-dependence, 
although one should remember that all functions written below should be taken 
to vary with $\omega$. Our focus is placed on kinematics where the excitation energies are near threshold and hence where $\omega$ is small, typically below a few MeV, and where $q$ is taken to be small compared with the typical nuclear scale for three-momentum denoted $q_0$. For $q_0$ we 
can use something like $2/b$, where $b$ is the oscillator
parameter (roughly 1.7 fm for our case, which yields 
$q_0 \cong 1.2 fm^{-1} \cong 230$ MeV/c). Accordingly, we can make use of the low-$q$ limits of the spherical Bessel functions involved in the definitions of the multipole operators, namely the fact that $j_J (qr) \to (qr)^J$ when $qr$ becomes small compared with unity. We may then with no loss of generality write the multipole matrix elements in a way that exposes the low-$q$ behavior which goes as $(q/q_0)^K$, where $K$ is some constant determined by 
the multipolarity of the transition (see below). For instance, 
the Coulomb multipole matrix elements may be parametrized 
in the form:
\begin{equation}\label{eq:tCJ}
t_{CJ}(q)\equiv \bigg(\frac{q}{q_0}\bigg)^J a'_{CJ}\Bigg[1+\bigg(\frac{q}{q_0}\bigg)^2 b'_{CJ}(q)\Bigg] e^{-(q/q_0)^2}
\end{equation} 
with $J\geq0$. Here $a'_{CJ}$ is independent of $q$ 
while $b'_{CJ}(q)$ depends on $q$; as noted above, 
they both depend on $\omega$. The powers of $q/q_0$ in 
the polynomial come from the nature of the spherical 
Bessel functions insofar as the leading power is fixed (the factor 
$(q/q_0)^J$) and the next term must begin two powers 
of $q/q_0$ higher, but otherwise, since $b'$ remains a general function of $q$, the expression is still completely general. The Gaussian factor is included to allow the results to have better behavior at high $q$ and may just as well be omitted if one wishes, since the entire focus here is on low-$q$ kinematics. Since we are assuming that $(q/q_0)\ll 1$, the multipoles are less and less important as the multipolarity J increases, in fact by $(q/q_0)^2$ for each additional increase in mutipolarity. This is a familiar result that leads one to characterize low-$q$ processes including real-$\gamma$ reactions by degrees of forbiddeness (see, for instance \cite{BlattWeisskopf1979}). The converse is also true: if $(q/q_0)\sim 1$ or larger, then one cannot order the multipoles by forbiddeness. A very old example --- from more than 50 years ago --- of this is provided by the first study of high-spin states in the giant resonance region where at values of $q$ of order $q_0$ M4 multipoles dominate over E1 multipoles \cite{Sick1969}.

It also proves useful to rewrite these expressions by letting
\begin{align}
a_{CJ}&\equiv a'_{CJ} \Big(1+(\omega/q_0)^2b'_{CJ}(q)\Big)e^{-(\omega/q_0)^2} \\
b_{CJ}(q)&\equiv b'_{CJ}(q)e^{-(\omega/q_0)^2} 
\end{align}
and then the parametrizations become
\begin{equation}
t_{CJ}(q)\equiv \bigg(\frac{q}{q_0}\bigg)^J a_{CJ} \Bigg[1+\bigg(\frac{|Q|^2}{q^2_0}\bigg) b_{CJ}(q)\Bigg] e^{-|Q|^2/q^2_0} .
\end{equation}
The electric multipole parametrizations may be written similarly:
\begin{align}\label{eq:tEJ}
t_{EJ}(q)&\equiv \bigg(\frac{\omega}{q}\bigg)\bigg(\frac{q}{q_0}\bigg)^J a'_{EJ}\Bigg[1+\bigg(\frac{q}{q_0}\bigg)^2 b'_{EJ}(q)\Bigg] e^{-(q/q_0)^2} \\
&\equiv \bigg(\frac{\omega}{q}\bigg)\bigg(\frac{q}{q_0}\bigg)^J a_{EJ} \Bigg[1+\bigg(\frac{|Q|^2}{q^2_0}\bigg) b_{EJ}(q)\Bigg] e^{-|Q|^2/q^2_0} ,
\end{align}
where now $J\geq1$ since there are no monopole electric multipoles, and where
\begin{align}
a_{EJ}&\equiv a'_{EJ}\Big(1+(\omega/q_0)^2b'_{EJ}(q)\Big)e^{-(\omega/q_0)^2} \\
b_{EJ}(q)&\equiv b'_{EJ}(q)e^{-(\omega/q_0)^2} .
\end{align}

From the continuity equation the long wavelength limit $(q\ll q_0)$ requires that
\begin{equation}
\lim_{q \ll q_0}\sqrt{\frac{J}{J+1}}t_{EJ}(q)= -\bigg(\frac{\omega}{q}\bigg) t_{CJ}(q),
\end{equation}
for $J \geq 1$, implying that 
\begin{equation}\label{eq:aEjaCj}
a'_{EJ}=-\sqrt{\frac{J+1}{J}}a'_{CJ} ,
\end{equation}
from which relationships involving the unprimed 
coefficients may be established.

For real photons all of the above parametrizations are to be evaluated at $q=\omega=E_\gamma$ and usually one invokes the above relationship between electric and Coulomb multipoles to employ the latter in real-$\gamma$ studies (see, for example, \cite{BlattWeisskopf1979}), although this is actually an approximation.


\section{Development of a Model for the Electrodisintegration Cross Section} \label{sec:four}

Having developed general expressions for the cross sections in Sect.~\ref{sec:two} and for the leading contributions to the angular distributions as functions of $\theta_\alpha$ in Sect.~\ref{sec:three}, here we proceed to make use of the still general parametrizations of the multipoles presented in Sect.~\ref{sec:three} and discuss our model for the electromagnetic response. We do this in two steps: first, we use the present knowledge of the real-$\gamma$ cross sections to constrain the leading-order behavior ({\it i.e.,} as functions of $q$) of the $E1$ and $E2$ multipoles. In the low-$q$ limit, current conservation then yields the leading-order behavior of the $C1$ and $C2$ multipoles. Second, we invoke ``naturalness'' --- to be explained below --- to model the next-to-leading order (NLO) dependences on $q$ in the $C1/E1$ and $C2/E2$ multipoles, which are not simply related by current conservation, as well as make an assumption concerning the behavior of the $C0$ multipole. Our goal is to develop a ``reasonable'' model and, using this model, to explore the feasibility of making electrodisintegration measurements in the interesting low-$\omega$/low-$q$ region. We emphasize that the model is used only to determine the feasibility of such experiments; in undertaking them the actual higher-order $q$-dependences will be measured and the region where the parametrizations are operative will be determined.


\subsection{Using Photodisintegration to Limit the Leading-order Behavior}\label{subsec:four-a}

The first step is to use the fact that the transverse response function
$R_T$ in electrodisintegration at $q=\omega$ is the same as the one in the real-$\gamma$ reactions and to establish the connection between our parameterization of electric
multipole matrix elements and the $E1$ and $E2$ astrophysical S-factors.
In the capture reaction $^{12}$C$(\alpha, \gamma)^{16}$O the radial distribution of the $\gamma$-rays is measured 
as a function of the $\alpha$-particle beam energy. The cross sections 
of the the $E1$ and $E2$ components, $\sigma_{E1}$ and $\sigma_{E2}$, are 
then extracted by fitting the data obtained to the differential 
cross section formula given in \cite{DyerBarnes1974}:
\begin{widetext}
\begin{eqnarray}
\Bigg[\frac{d\sigma}{d\Omega^{c.m.}_{\gamma}}\Bigg]_{(\alpha,\gamma)}&=&\frac{\sigma_{E1}}{4\pi} \Big(Q_0P_0-Q_2P_2(\cos\theta)\Big)+\frac{\sigma_{E2}}{4\pi} \Big(Q_0P_0+Q_2\frac{5}{7}P_2(\cos\theta)-Q_4\frac{12}{7}P_4(\cos\theta)\Big) \\
&+& \frac{\sqrt{\sigma_{E1}\sigma_{E2}}}{4\pi}\cos(\phi_{12})\frac{6}{\sqrt{5}}\Big(Q_1P_1(\cos\theta)-Q_3P_3(\cos\theta) \Big) ,
\notag
\end{eqnarray}
\end{widetext}
where $Q_{l}$ are attenuation factors \cite{Rose1953} 
determined by the geometry of $\gamma$ detectors. This is just a rewriting of Eq. (\ref{eq:RT}). Furthermore,  
$\phi_{12} = \delta_{E2} -\delta_{E1}$ is the phase between the $E1$ and $E2$ components (sometimes also used as a third fitting parameter). 
From multilevel R-Matrix theory \cite{Barker1991} 
the phase $\phi_{12}$ can be expressed as
\begin{align}\label{eq:phase12}
\phi_{12} &= \delta_{d} -\delta_{p} + \arctan{\eta/2}; \notag \\
 \eta &= \frac{e^2 Z_{\alpha}Z_{^{12}C}}{\hbar c}\sqrt{\frac{M_{\alpha} M_{^{12}C}} {M_{\alpha}+M_{^{12}C}} \frac{1}{2 E_{\alpha}^{c.m.}}} ,
\end{align}

where $\eta$ is the Sommerfeld parameter, while $\delta_p$ 
and $\delta_d$ are p- and d-wave phase shifts from elastic
$\alpha$ scattering on carbon. Barker derived first equation \ref{eq:phase12} for single-level R-Matrix \cite{Barker1974}, and later Barker and Kajino for multi-level R-Matrix \cite{Barker1991}. 
For the general case Knutson \cite{Knutson1999}, used 
Watson’s theorem \cite{Watson1954} to show that the 
phase shifts of the radiative capture data at low energy 
can be related to elastic scattering phase shifts. This also holds for elastic 
$^{12}$C$(\alpha,\alpha)^{12}$C and radiative capture $^{12}$C$(\alpha,
\gamma)^{16}$O phase shifts. The final step is to convert 
the extracted $\sigma_{E1}$ and $\sigma_{E2}$ into S-factors
 $S_{EJ}(E_{\alpha}^{c.m.})= E_{\alpha}^{c.m.}\cdot \sigma_{EJ}\cdot 
e^{2\pi\eta}$ as shown of Fig. \ref{fig:SE12}. Note that here and below, following common practice in studies of photodisintegration, we assume that the nuclear phase difference is small and therefore that the complete phase difference arises largely from the term containing the Sommerfeld parameter. However, a word of caution should be inserted here: the result above may be either as written or could be $\pi$ minus that result. Said another way, the $E1/E2$ interference term may have the sign as written or might have the opposite sign. Upon fitting the angular distributions in photodisintegration it was found that typically in the kinematic region of interest the sign is as written above \cite{Brune2001}. We shall discuss this in more depth below for the case of electrodisintegration.

We will go in opposite direction: by using earlier 
obtained differential cross sections for the real-$\gamma$ reaction, 
Eq.~(\ref{eq:direct}), and parameterization of the electric 
multipole matrix elements in the real-photon limit, $t_{EJ}(\omega) = (\omega/q_0)^J a'_{EJ}$, we can express the leading coefficients 
$a'_{E1}$ and $a'_{E2}$ in terms of S-factor data:
\begin{align}\label{eq:aE1aE2}
a'_{EJ} &= \bigg(\frac{q_0}{\omega}\bigg)^J \! \sqrt{\frac{\hbar c \;  p^{c.m.}_{\alpha}    W  }{2\alpha   \; \omega  \; M_{\alpha}M_{^{12}C}}\frac{S_{EJ}(E_{\alpha}^{c.m.})  e^{-2\pi \eta(E_{\alpha}^{c.m.})}}{E_{\alpha}^{c.m.}}}; \notag \\ \quad J &=1,2 .
\end{align}
For the sake of simplicity we did not perform an R-Matrix fit
on the S-factor data. Instead, for both multipoles, the $S_{EJ}(E_{\alpha}^{c.m.})$
dependence was approximated by fitting the data to second-order polynomials, which are represented by the dashed curves in
Fig.~\ref{fig:SE12}.

The feasibility of performing measurements will be discussed in detail in the next section. For the present purposes we assume typical values for the kinematics of interest and postpone their justification for later. In this section we shall assume an electron beam energy of $E_e = 114$ MeV and work in the region $0.7 \le
E_{\alpha}^{c.m.} \le 1.7$ MeV. Accordingly the other kinematic variables must lie in relatively narrow ranges. Specifically, the scattered electron energy is found to lie roughly in the range $105 < E'_e < 107$ MeV for the assumed value of $E_e$ and the electron scattering energy loss $\omega = E_e - E'_e$ then falls in the range $7 < \omega < 9$ MeV. The electron three-momenta are, as usual, given by $p_e=\sqrt{E_e^2 + m_e^2}$ and ${p'}_e=\sqrt{{E'}_e^2 + m_e^2}$ (see Fig.~\ref{fig:VirtPh}), from which one can obtain the square of the three-momentum transfer
\begin{equation}
  q^2 = p_e^2 + {p'}_e^2 - 2 p_e {p'}_e \cos{\theta_e} .  
\end{equation}

\begin{figure*}[h]
\centering
\includegraphics[width=7.6cm]{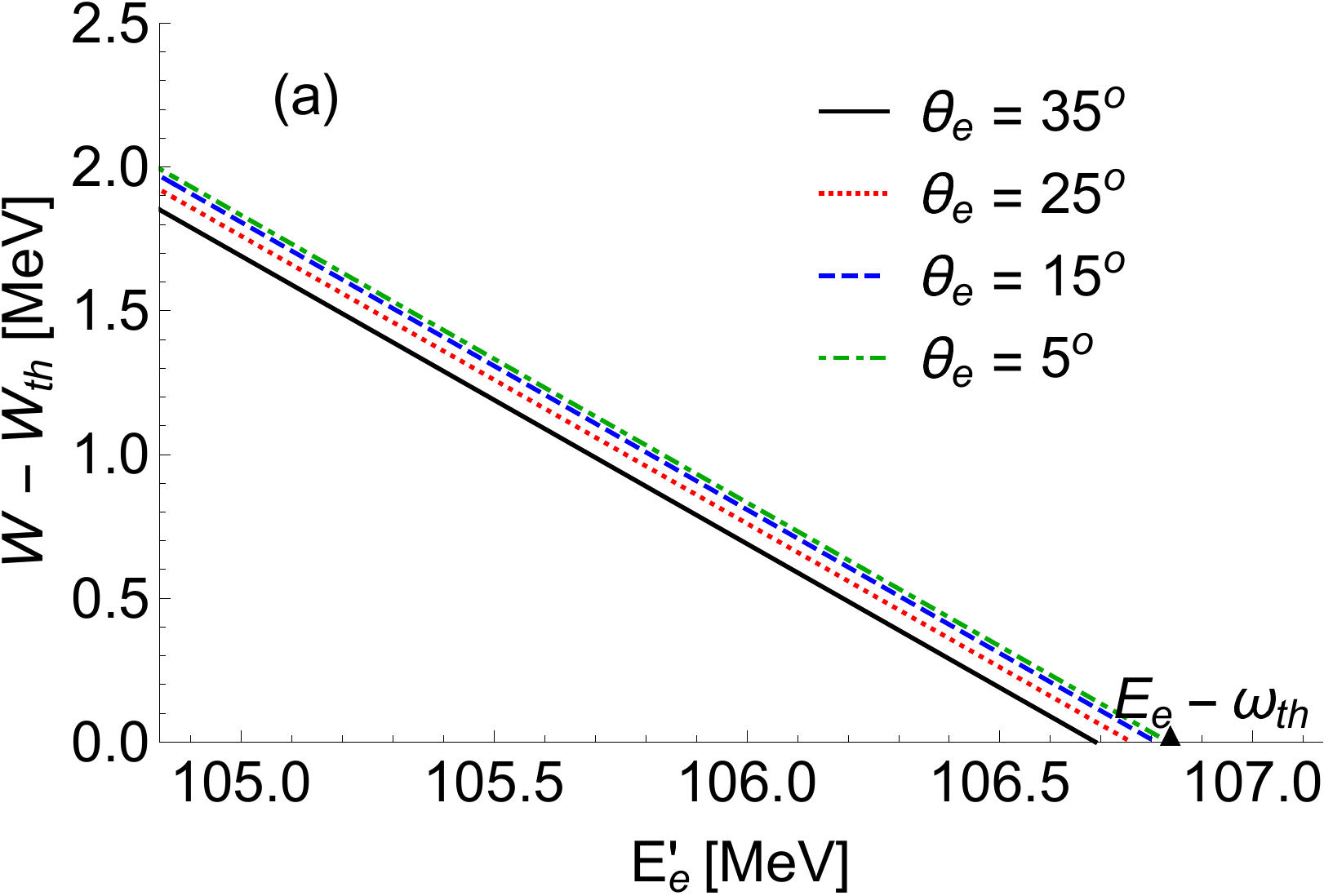}
\includegraphics[width=7.6cm]{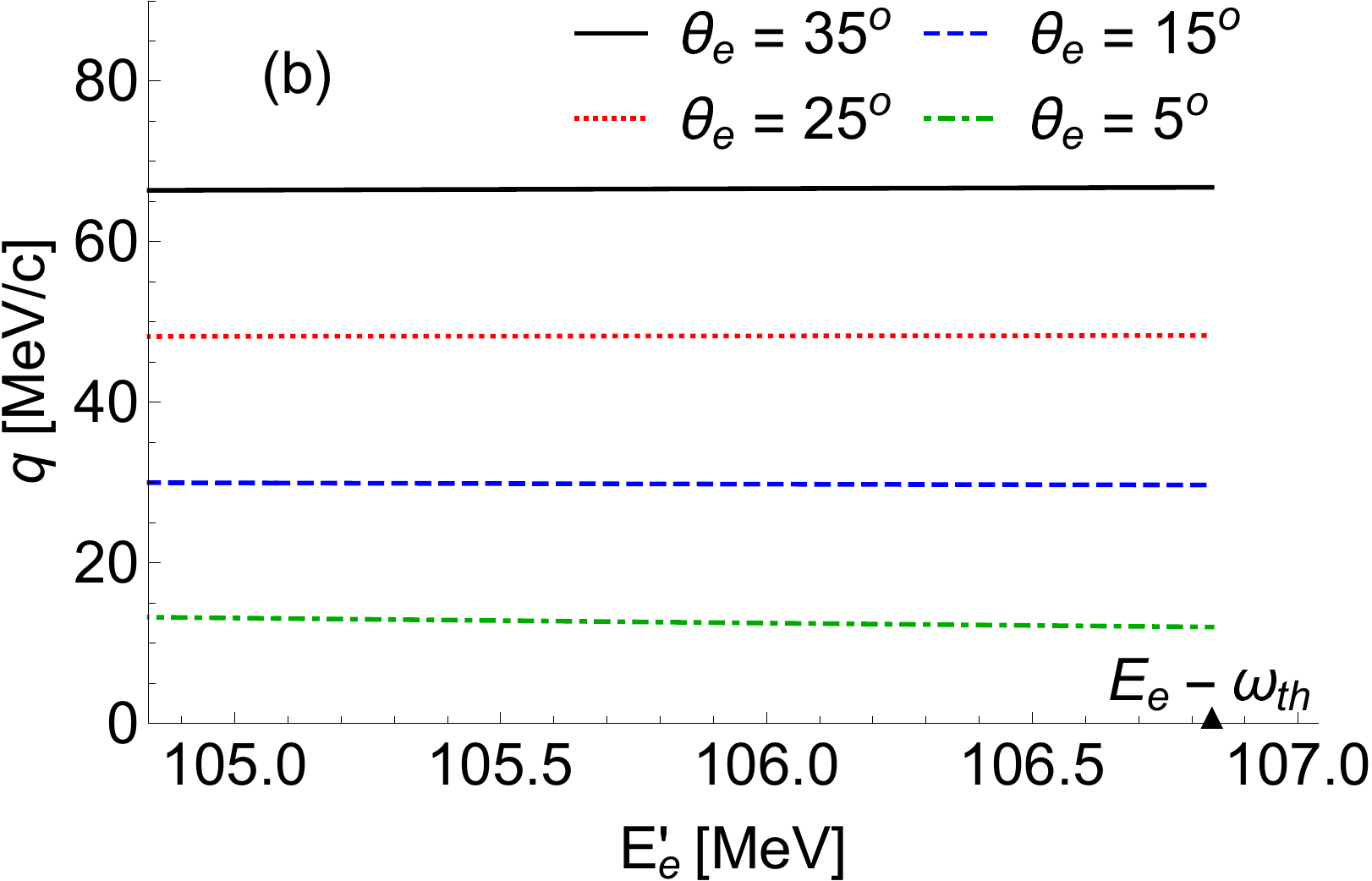} 
\includegraphics[width=7.6cm]{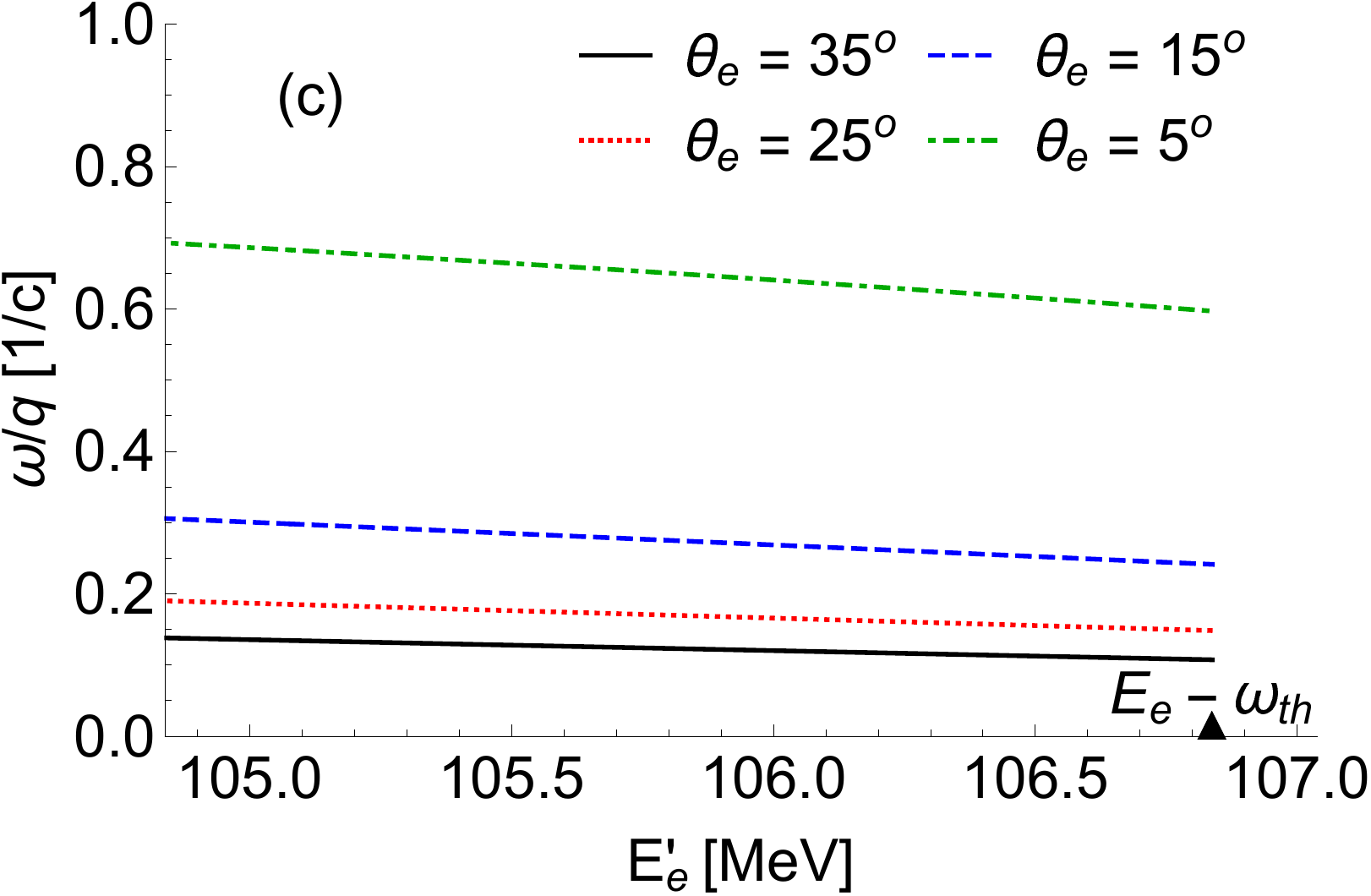}
\includegraphics[width=7.6cm]{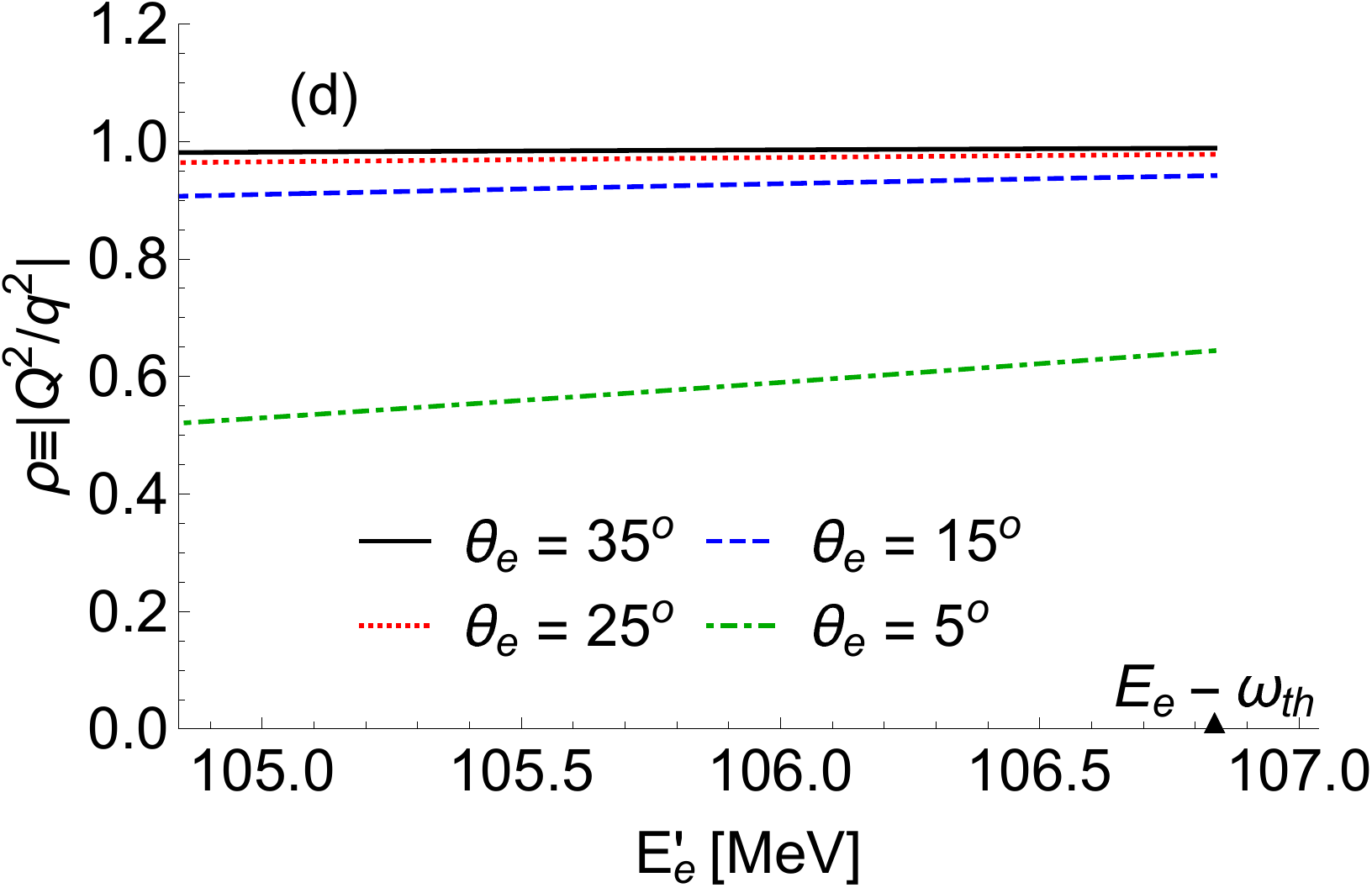}
\colorcaption{The dependences of the invariant mass above threshold $W - W_{th}$ (a), of the transferred three-momentum $q$ (b), of the ratio of the transferred energy to the transferred three-momentum $\omega/q$ (c), and of the virtuality of the exchanged photon $\rho \equiv |Q^2/q^2|$ (d), on the scattered electron energy $E'_e$ for kinematics corresponding to within 2 MeV above threshold. \label{fig:EprPlots}}
\end{figure*}

In Fig.~\ref{fig:EprPlots} (a) we show $W-W_{th}$ versus $E'_e$ for the range discussed here for three typical values of the electron scattering angle, from which we see that $E'_e$ goes from about 105 MeV to 107 MeV when $W-W_{th}$ goes from 0 to 2 MeV, as stated above. Within this range one finds that $q$ behaves as shown in the (b) panel. Clearly $q$ is nearly, but not exactly, constant as a function of $E'_e$ for the chosen kinematics. The two lower panels illustrate the virtuality of the electron scattering reaction, showing the ratio $\omega/q$ in the (c) panel and $\rho \equiv |Q^2/q^2|$ in the (d) panel. Each varies both as a function of $E'_e$ and $\theta_e$, as shown. As is clearly seen in the $\omega/q$ ratio plot, one can go from rather virtual photons ($q$ significantly larger than $\omega$; larger angles) towards real-$\gamma$ kinematics ($q$ comparable to $\omega$; smaller angles). And the invariant mass above threshold (effectively the excitation energy of the $\alpha + ^{12}$C system) has a nearly linear relationship with $E'_e$. 

\begin{figure}[h]
\centering
\includegraphics[width=7.6cm]{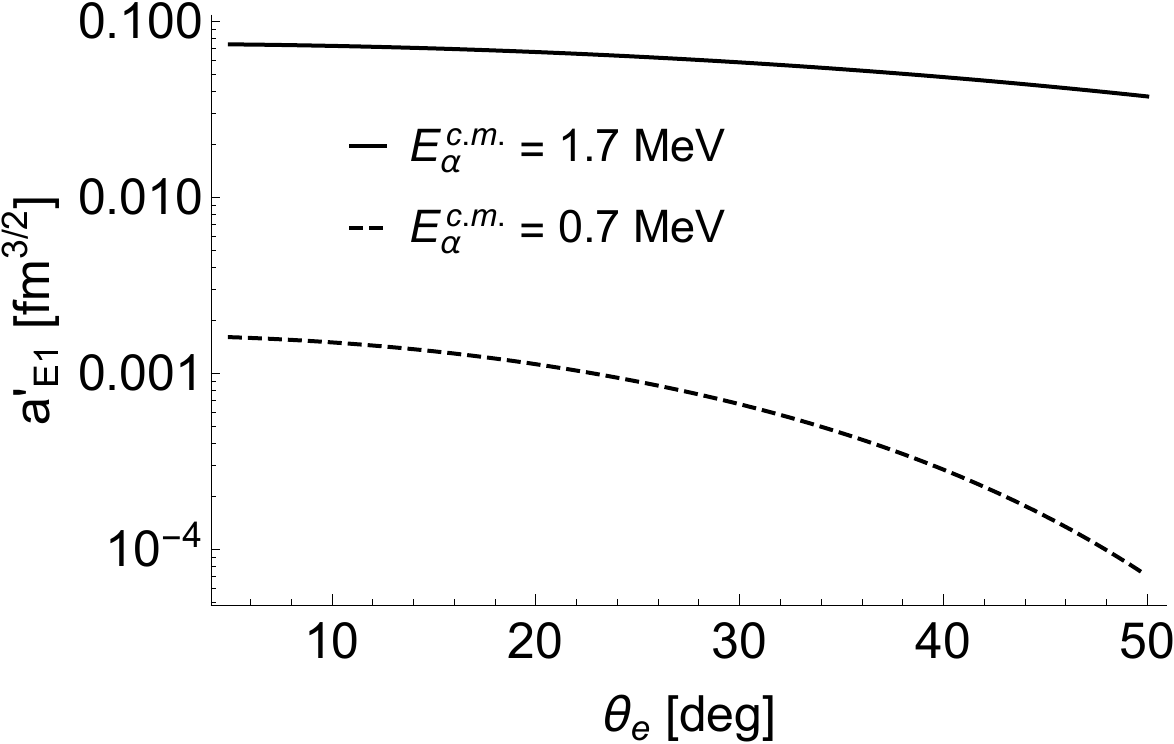}
\includegraphics[width=7.6cm]{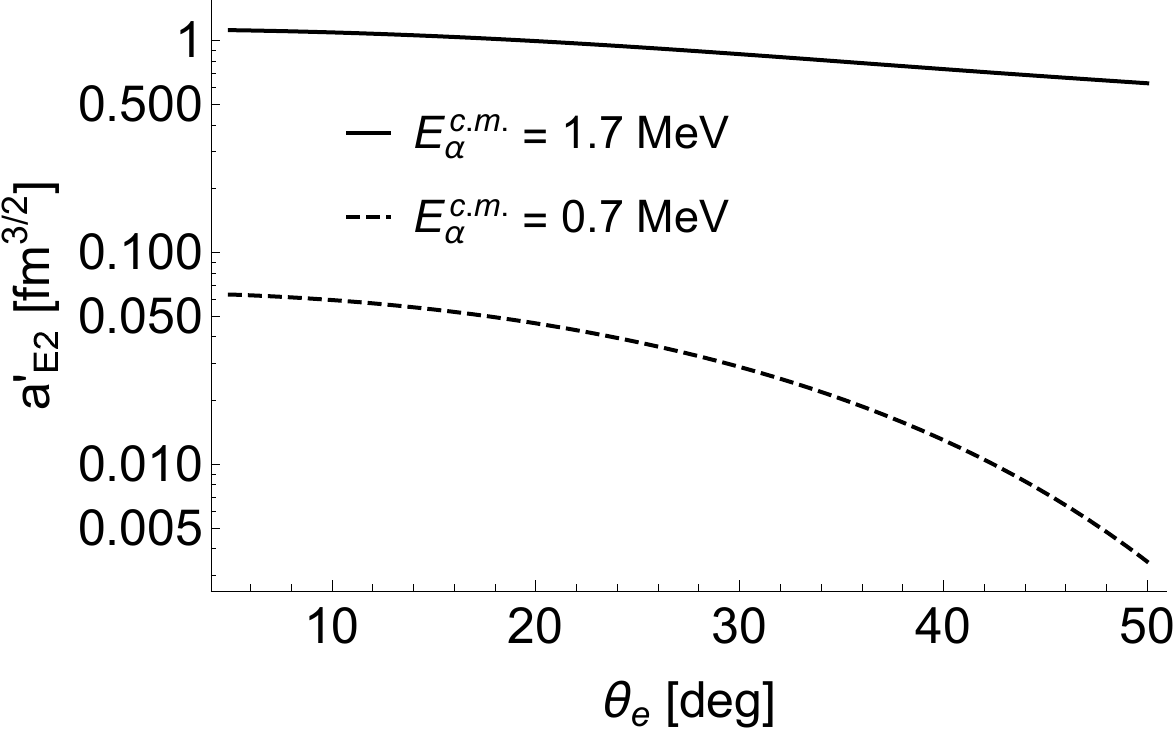} 
\caption{Leading-order coefficients $a'_{E1}$ and $a'_{E2}$ as 
functions of electron scattering angle $\theta_e$ at a beam energy $E_e$ of 114 MeV.  \label{fig:aE1aE2} }
\end{figure}

Given these choices of kinematics in Fig.~\ref{fig:aE1aE2} we then present the leading-order E1 and E2 coefficients, $a'_{E1}$ and $a'_{E2}$, as functions of $\theta_e$ for two values of the $\alpha$-particle $c.m.$ kinetic energy
$E_{\alpha}^{c.m.}$, 0.7 and 1.7 MeV (one should remember that these coefficients are constants as functions of $q$ but still depend on $\omega$). One sees that 
the values of both leading-order coefficients decrease over 
almost two orders of magnitude when $E_{\alpha}^{c.m.}$ changes
from 1.7 to 0.7 MeV, reflecting the steep falloff of the
cross section when approaching threshold.

Note that, in the case of the radiative capture 
reaction, $E_{\alpha}^{c.m.}$ denotes the kinetic energy in the 
center-of-mass frame of the relative motion of the $\alpha$ 
and $^{12}$C pair in the incident channel and can be expressed as
$E_{\alpha}^{Lab}M_{^{12}C}/(M_{\alpha}+M_{^{12}C})$, where 
$E_{\alpha}^{Lab}$ is the $\alpha$-particle kinetic energy in the laboratory frame. 
For the electrodisintegration 
of $^{16}$O, $E_{\alpha}^{c.m.}$ is the difference between 
the invariant mass $W$ and its value 
$W_{th}$ at threshold. 

Finally, from the continuity equation and in the long wavelength 
limit $(q \ll q_0)$ we know how to relate electric  $a'_{EJ}$
and Coulomb $a'_{CJ}$  coefficients (Eq. (\ref{eq:aEjaCj})):
\begin{equation}\label{eq:aCjaEj}
a'_{CJ}= -\sqrt{\frac{J}{J+1}}a'_{EJ} .
\end{equation}


\subsection{Next-to-leading Order $q$-dependences}\label{subsec:four-b}

Presently we have no information 
concerning the next-to-leading order contributions in our general parametrization of the multipoles, $b'_{CJ,EJ}(q)$ with $J=1,2$. These are independent functions of $q$, {\it i.e.,} cannot be related via current conservation as can the leading-order contributions. It should be remembered that, at this higher order in $q$, even the way real-$\gamma$ processes are traditionally treated is an approximation, since the electric multipole matrix elements are typically computed as Coulomb matrix elements using the current conservation assumption. Since $\omega/q_0=E_\gamma/q_0$ is not zero, but is small, one is actually making an assumption when following this procedure. In the virtual-$\gamma$ case that occurs with electron scattering the expansion is via higher-order contributions in $q/q_0$, and, since $q$ can take on any value where the virtual photon is spacelike, $q>\omega$, as stated earlier, one now has a different situation where when $q\ll q_0$ these NLO terms are likely safely negligible; however, if $q$ is allowed to become too large compared with the scale $q_0$, then the form taken by these NLO functions may not be simple.

Accordingly, we now make the basic assumption involved in our parametrization of the $C1/E1$ and $C2/E2$ multipoles, namely, we shall assume that the general functions of $q$, $b'_{CJ,EJ}(q)$, are in fact constants. When measurements are made these constants will be determined experimentally using the $q$-dependences inherent in the semi-inclusive cross sections. And, with fine enough measurements, one may look for evidence of $q$-dependences that involve even higher powers of $(q/q_0)^2$ to validate the truncations of the expansions.

This strategy is what can be followed when making measurements of the semi-inclusive electrodisintegration cross section as a function of both $q$ and $\omega$. For the present, lacking such measurements, our approach is to make ``reasonable'' assumptions for these NLO coefficients. Since the multipole matrix elements were parameterized to 
reflect the nature of spherical Bessel functions, it is reasonable to expect that they are of order unity and accordingly the simplest approximation at present 
is to assume that $|b'_{CJ,EJ}| \approx 1$ for $J=1,2$, and thus the $C1/E1$ and $C2/E2$ multipole matrix elements will be parametrized as
\begin{align}
t_{CJ}(q) &\approx -\sqrt{\frac{J}{J+1}} \bigg(\frac{q}{q_0}\bigg)^J a'_{EJ}\Bigg[1 \pm \bigg(\frac{q}{q_0}\bigg)^2 \Bigg] e^{-(q/q_0)^2};\notag \\
 t_{EJ}(q) &\approx  \bigg(\frac{\omega}{q}\bigg)\bigg(\frac{q}{q_0}\bigg)^J a'_{EJ}\Bigg[1 \pm\bigg(\frac{q}{q_0}\bigg)^2 \Bigg] e^{-(q/q_0)^2} .
\end{align}

A special case involves the monopole Coulomb matrix element $t_{C0}$: 
there the leading dependence, which from above would appear 
to be $(q/q_0)^J$ with $J=0$, cannot occur due to the orthogonality of 
the initial and final nuclear wave functions 
and in fact the leading behavior of 
$t_{C0}(q)$ at low-$q$ is proportional to $(q/q_0)^2$.
Again, there are no experimental data which would fix the value
of the product $c'_{C0} \equiv a'_{C0}\cdot b'_{C0}$. Therefore, in our feasibility study we investigated the 
contribution of the  $t_{C0}$ to the rate of the $^{16}$O$(e,e^\prime \alpha)^{12}$C 
reaction by setting the $|b'_{C0}| = 1$ and replacing  $a'_{C0}$
first with $a'_{E2}$, denoted Case A, and then with $0.5 \cdot a'_{E2}$, denoted
Case B. In general, when dealing with experimental data from 
electrodisintegration of $^{16}$O, $c'_{C0}$ needs 
to be handled as a fit parameter, as with the coefficients $b'_{CJ,EJ}$ for $J=1,2$ discussed above. As noted above, we do not know the sign of the $C0$ multipole ({\it i.e.,} with respect to the other multipoles) and so could have either choice of sign for all interferences between $C0$ and the other multipoles (see below). When presenting results in the following, for the sake of simplicity we have usually chosen the sign to be positive, although both sign choices have been investigated. The detailed angular distributions that result from changing the sign of the $C0$ multipole are found to be comparable but clearly different and accordingly the sign can be determined from the data, as was the case for the $E1/E2$ interference contribution in photodisintegration (see above).

\begin{figure}[t]
\centering
\includegraphics[width=7.6cm]{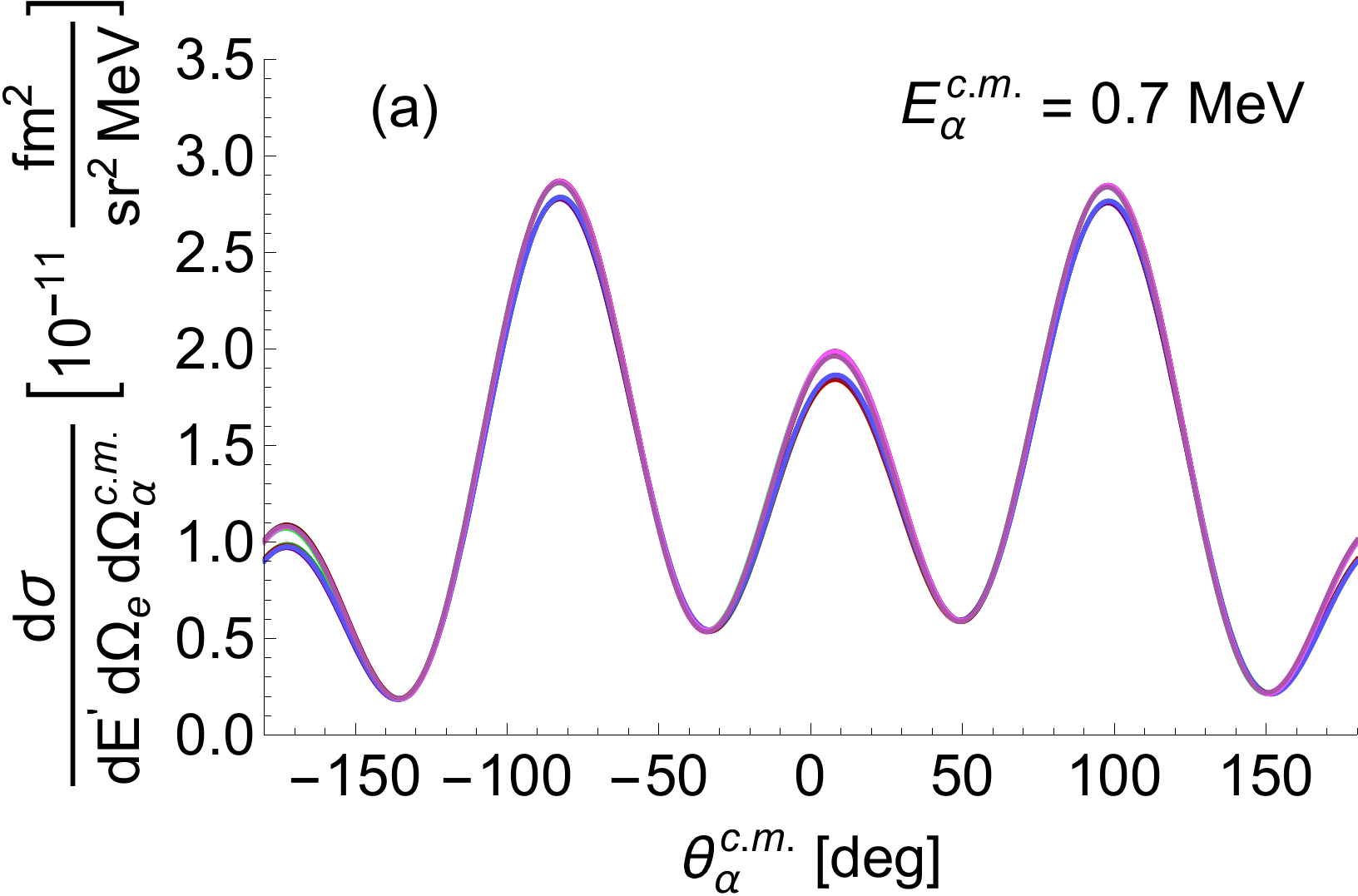}
\includegraphics[width=7.6cm]{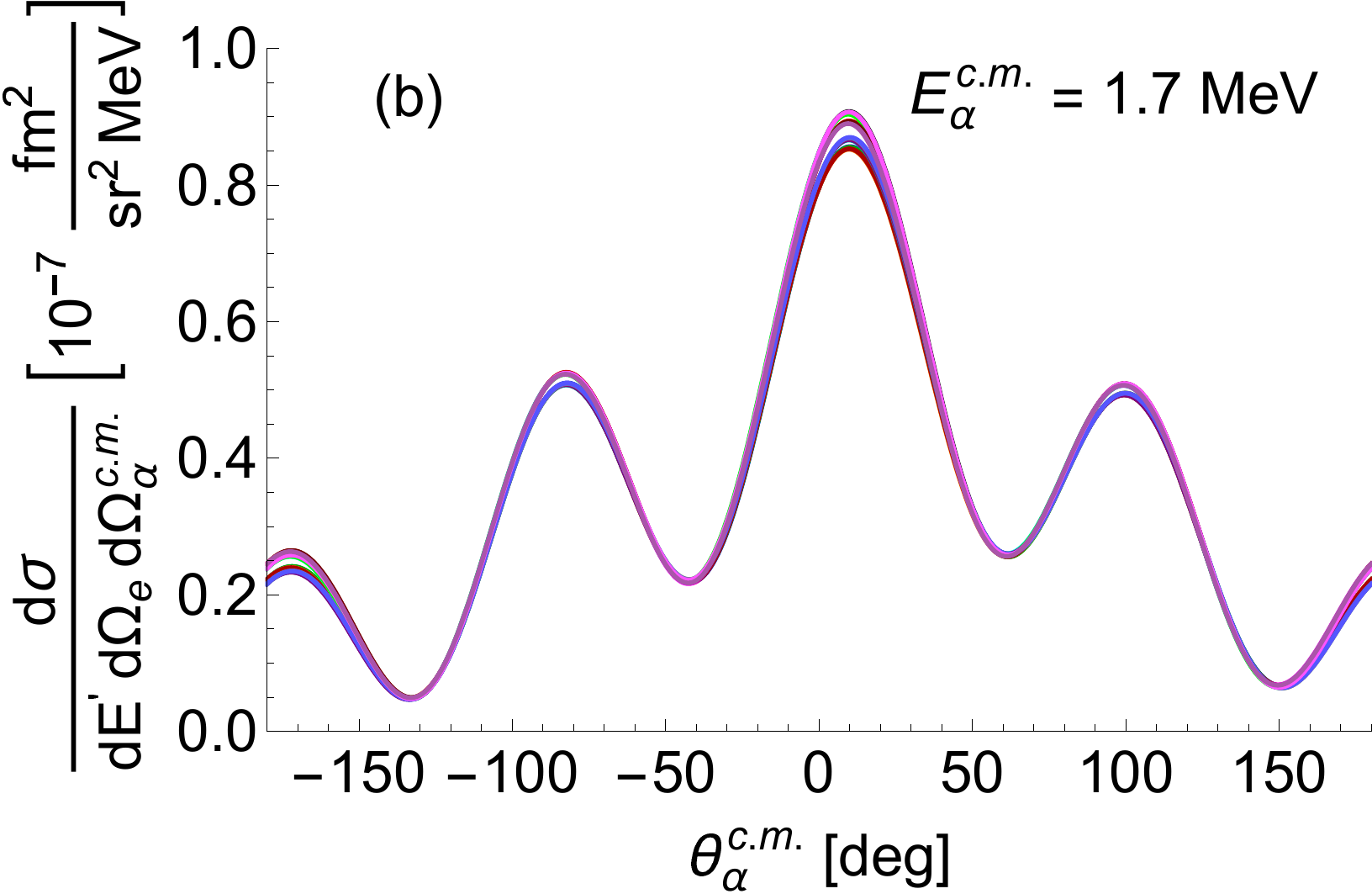} 
\colorcaption{The semi-inclusive electrodisintegration differential cross section as a function of $\theta_{\alpha}^{c.m.}$ at a beam energy $E_e$ of 114 MeV
and an electron scattering angle $\theta_e$ of 15$^{\circ}$, for $E_{\alpha}^{c.m.} =$ 0.7 MeV (a) and $E_{\alpha}^{c.m.} =$ 1.7 MeV (b). There are 16 curves
on each plot corresponding to ``$+$" and ``$-$" sign choices for each of the four next-to-leading order coefficients $b'_{C1}$, $b'_{C2}$, $b'_{E1}$ and $b'_{E2}$.
The difference introduced by the change of the sign is so small that the most of the lines are overlapping.
At each of the local maxima, it is possible  to distinguish two groups of lines which correspond to $+b'_{C2}$ and $-b'_{C2}$ contributions. Here all interferences involving the $C0$ multipole have been taken to have a plus sign together with the choices of phase differences discussed in the text. Alternatively, all such inteferences could enter with a minus sign. \label{fig:bCombinations}}
\end{figure}

\begin{figure*}[b]
\centering
\includegraphics[width=7.6cm]{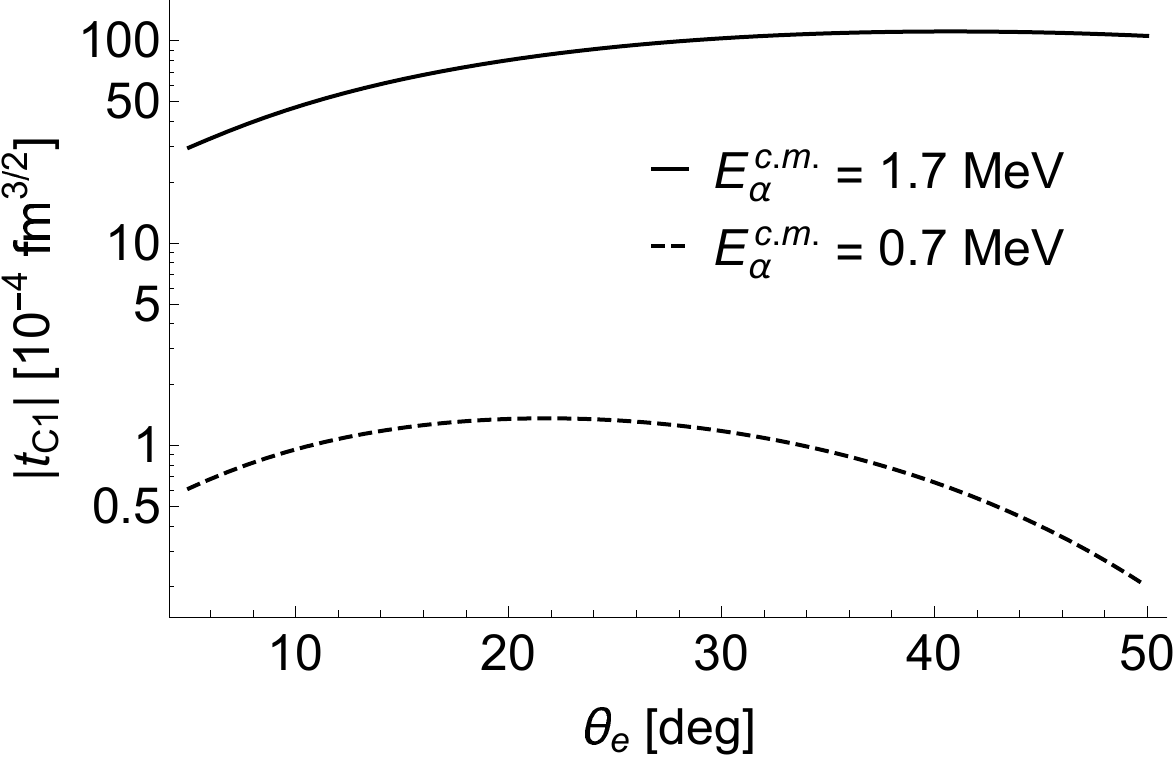}
\includegraphics[width=7.6cm]{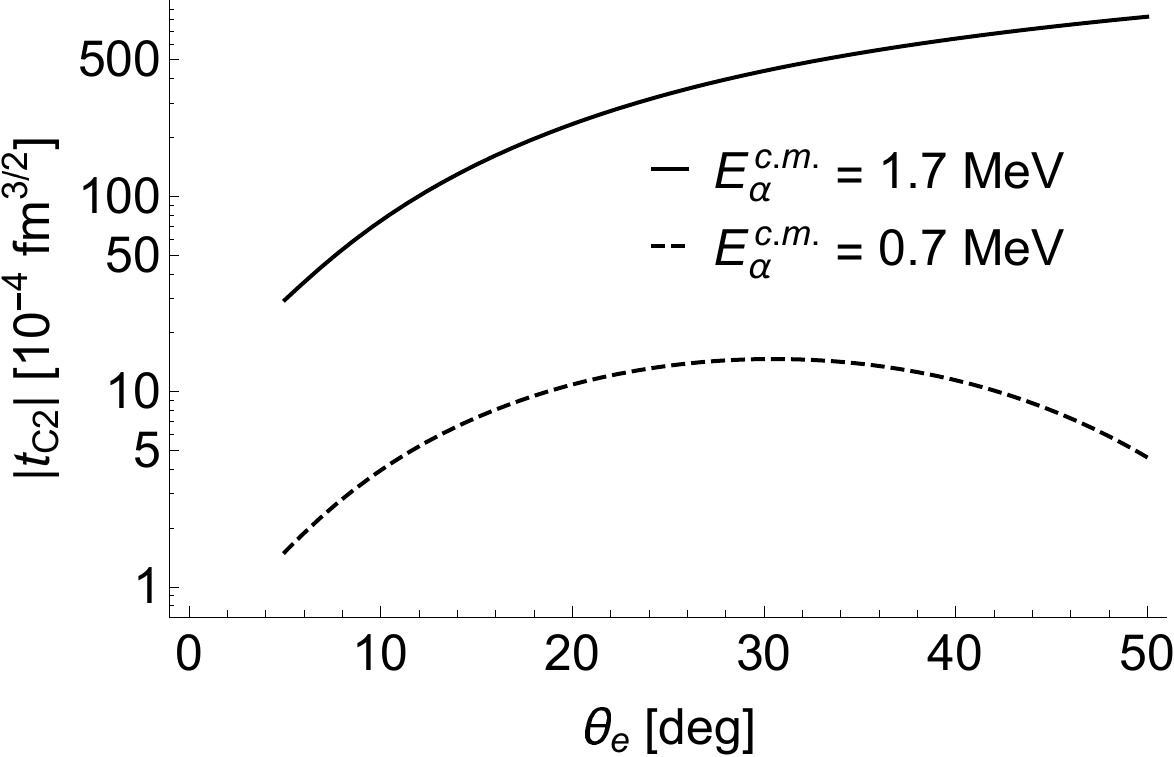}
\includegraphics[width=7.6cm]{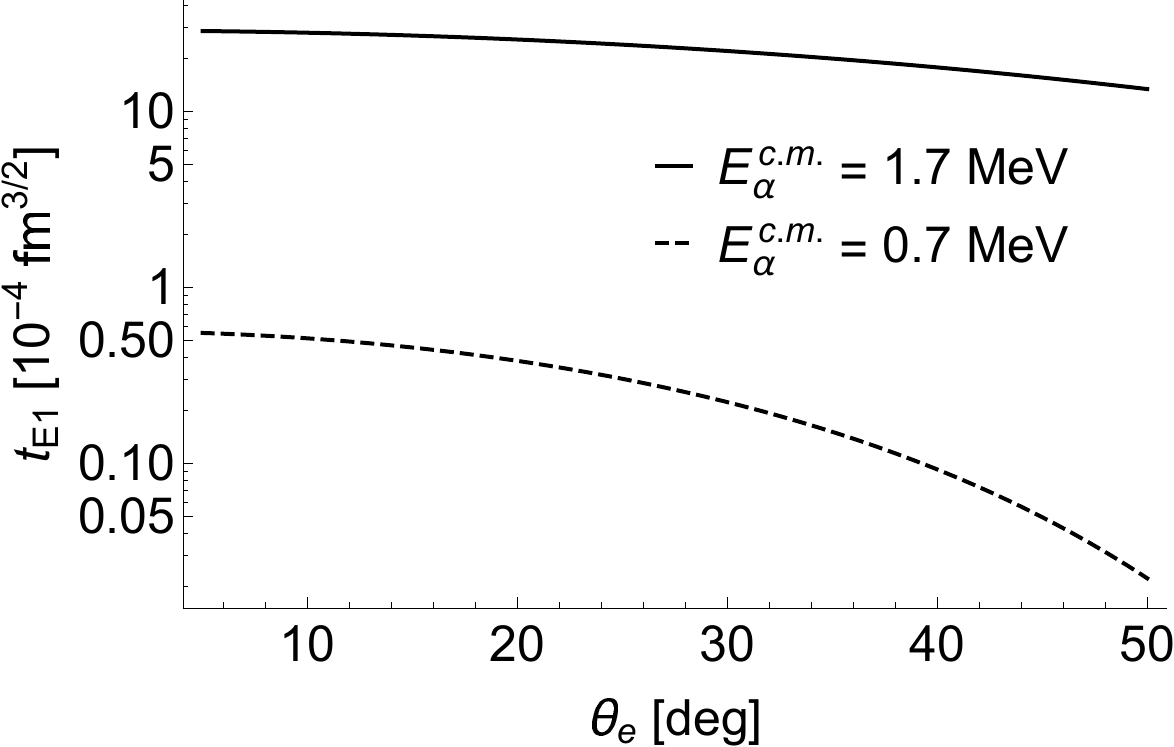}
\includegraphics[width=7.6cm]{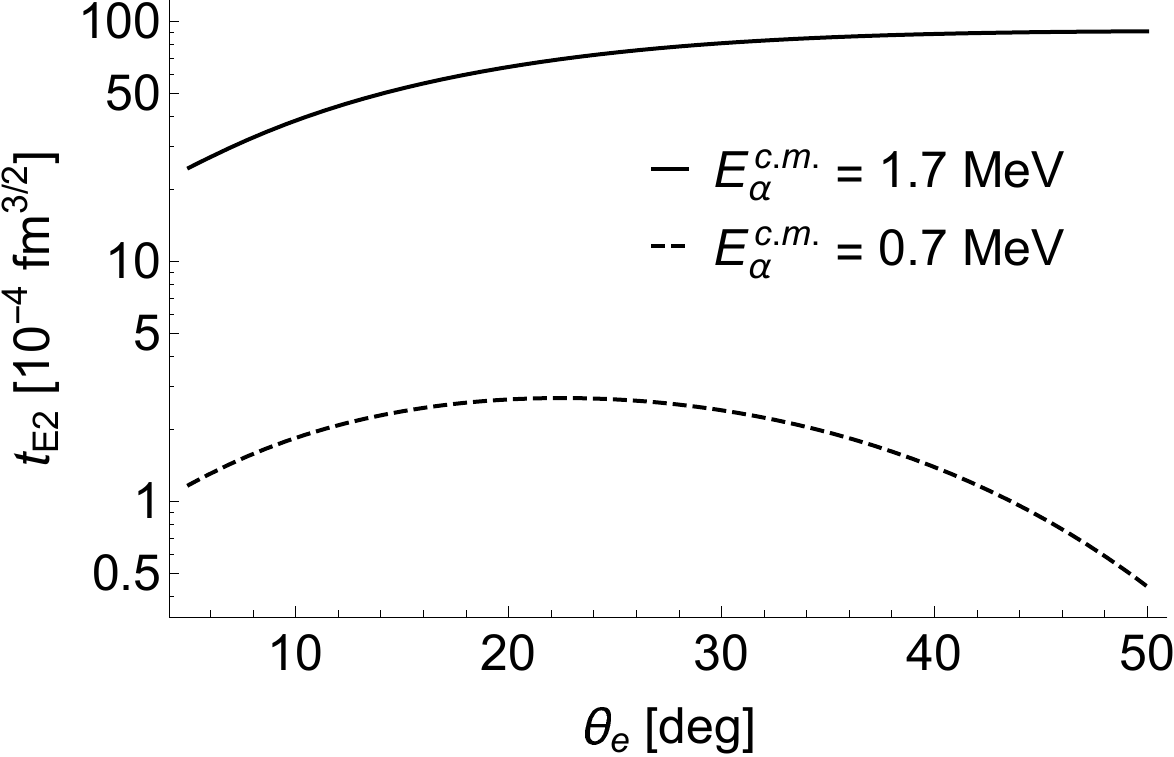} 
\caption{Electric ($t_{E1}$, $t_{E2}$) and Coulomb ($|t_{C1}|$, $|t_{C2}|)$ 
multipole matrix elements as functions of the electron scattering angle $\theta_e$ 
at a beam energy $E_e$ of 114 MeV.}
\label{fig:tCE1aCE2}
\end{figure*}

In Figs.~\ref{fig:bCombinations} ((a) and (b)) we show the semi-inclusive electrodisintegration cross section as a function of $\theta_{\alpha}^{c.m.}$ for two values of $E_{\alpha}^{c.m.}$, at a beam energy $E_e$ of 114 MeV and an electron scattering angle $\theta_e$ of 15$^{\circ}$. In each case there are 16 curves corresponding to the two sign choices for each of the next-to-leading order coefficients. Clearly, for the selected kinematics these higher-order effects are quite small, typically less that 6.4\%. We again stress that this is not the limiting factor in making such measurements, since, in any actual experiment, the slight extra dependence on $q$ will be determined by varying the kinematics. Having found that the next-to-leading order effects are small, for simplicity henceforth we make the choice $b'_{CJ,EJ} = +1$ for $J=1,2$, and, given this choice, Fig.~\ref{fig:tCE1aCE2} then shows the $\theta_e$ dependence
of the electric $t_{EJ}$ and Coulomb $|t_{CJ}|$ multipole 
matrix elements for the selected kinematics.

Finally, we note that in the region of interest  $0.7 \le E_{\alpha}^{c.m.} \le 1.7$ MeV,
the elastic phase shifts of the $s$-, $p$- and $d$-waves are almost equal 
to zero \cite{Plaga1987, Tischhauser2009} and therefore we neglected them in our calculation of the rate. The only contribution to the phase shift then comes from the Coulomb field, which is equal to the difference 
of the Coulomb phase shift $\sigma_{l} = \arg \Gamma (1 + l + i \eta)$ 
of partial wave $l$ and the phase shift of the Coulomb monopole 
$\sigma_{0} = \arg \Gamma (1 + i \eta)$ \cite{Lane1958}:
\begin{equation}\label{eq:CoulPhase}
\omega_{l} \equiv \sigma_{l} - \sigma_{0} = \sum_{n=1}^{l}\arctan \frac{\eta}{l} .
\end{equation} 
We see that the last term in $\phi_{12}$, Eq.~(\ref{eq:phase12}), follows from the general expression, Eq.~(\ref{eq:CoulPhase}). At the end, we will assume that the phase-shift differences that occur in the electrodisintegration
response functions written above, $\delta_{Cl}-\delta_{C0}$ and $\delta_{El}-\delta_{E0}$, are both
equal to $\omega_{l}$ for the corresponding partial wave $l$.

Having chosen to use these for the phase-shift differences, as noted above, we must allow for either plus or minus signs to enter for the interferences between the various mutipoles. For the $E1$ and $E2$ cases we follow the lead from photodisintegration and choose the relative sign to be positive. The low-$q$ relationships between $CJ$ and $EJ$ multipoles then fix the signs of the $C1$ and $C2$ multipoles relative to the $E1$ and hence $E2$ multipoles. However, we do not have any information concerning the relative sign of the $C0$ multipole compared with the $C1$, $C2$, $E1$ and $E2$ multipoles. Hence, all terms involving interference with the $C0$ multipole could occur with either sign. During the rest of what is presented in this study usually we arbitrarily choose the sign to be positive, although we have examined what happens when the opposite sign choice is made: the detailed angular distributions change, although are roughly of similar sizes. When measurements are made the appropriate sign choice should be clear following what was done in studies of photodisintegration.

\subsection{Electrodisintegration Cross Section Predictions}\label{subsec:four-c}

Having specified the model, we employ this to make projections of the electrodisintegration cross sections in 
the low-$\omega$ and low-$q$ region and to explore these projections for a range of kinematics, and, in the following section, to provide estimates for the uncertainties that might be expected in practical experiments in extrapolating towards the real-$\gamma$ line and towards threshold. These estimates will then be used to make projections for the desired astrophysical S-factors.


\begin{figure*}[h!]
\centering
\includegraphics[width=8.6cm]{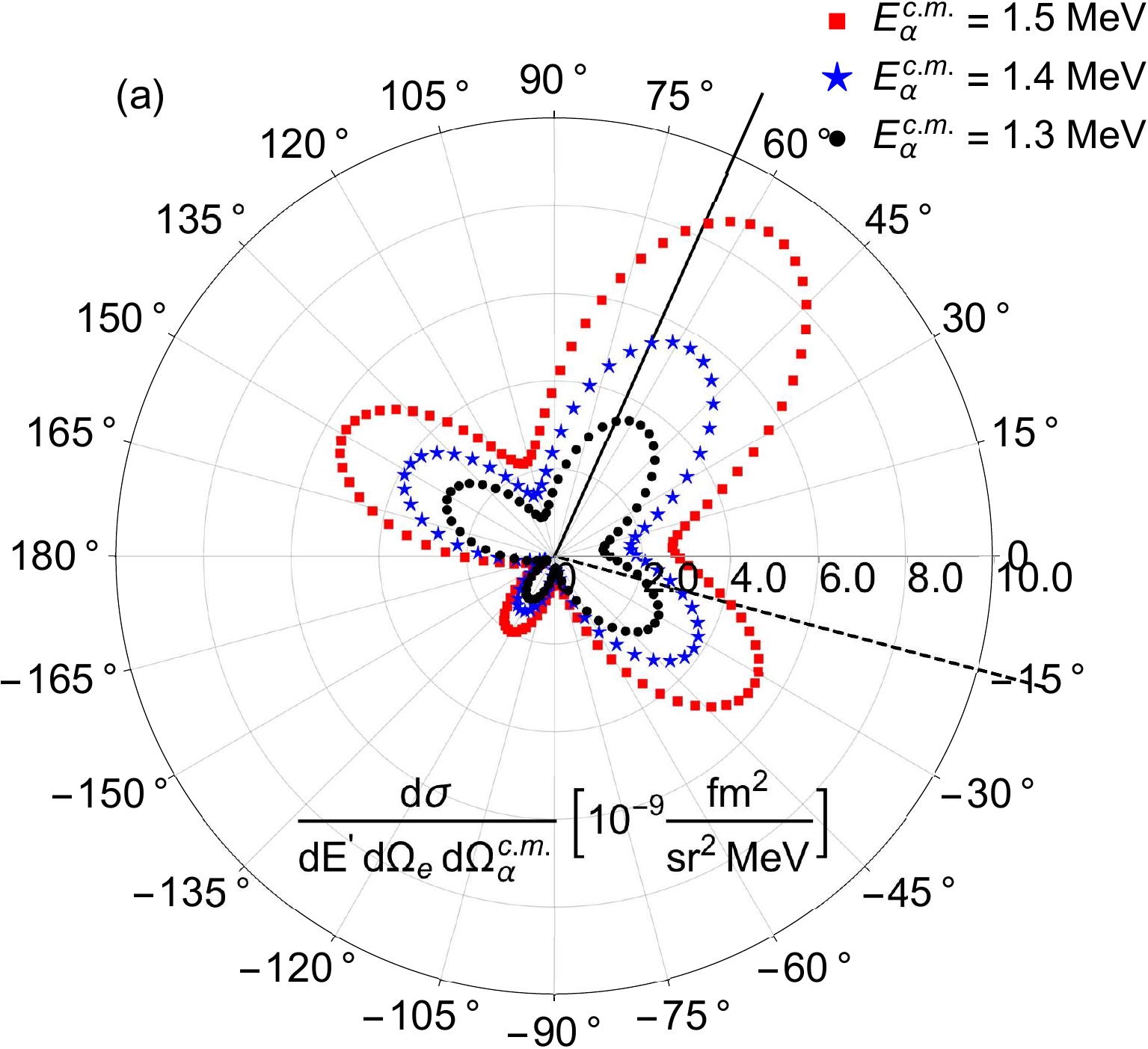}
\includegraphics[width=8.6cm]{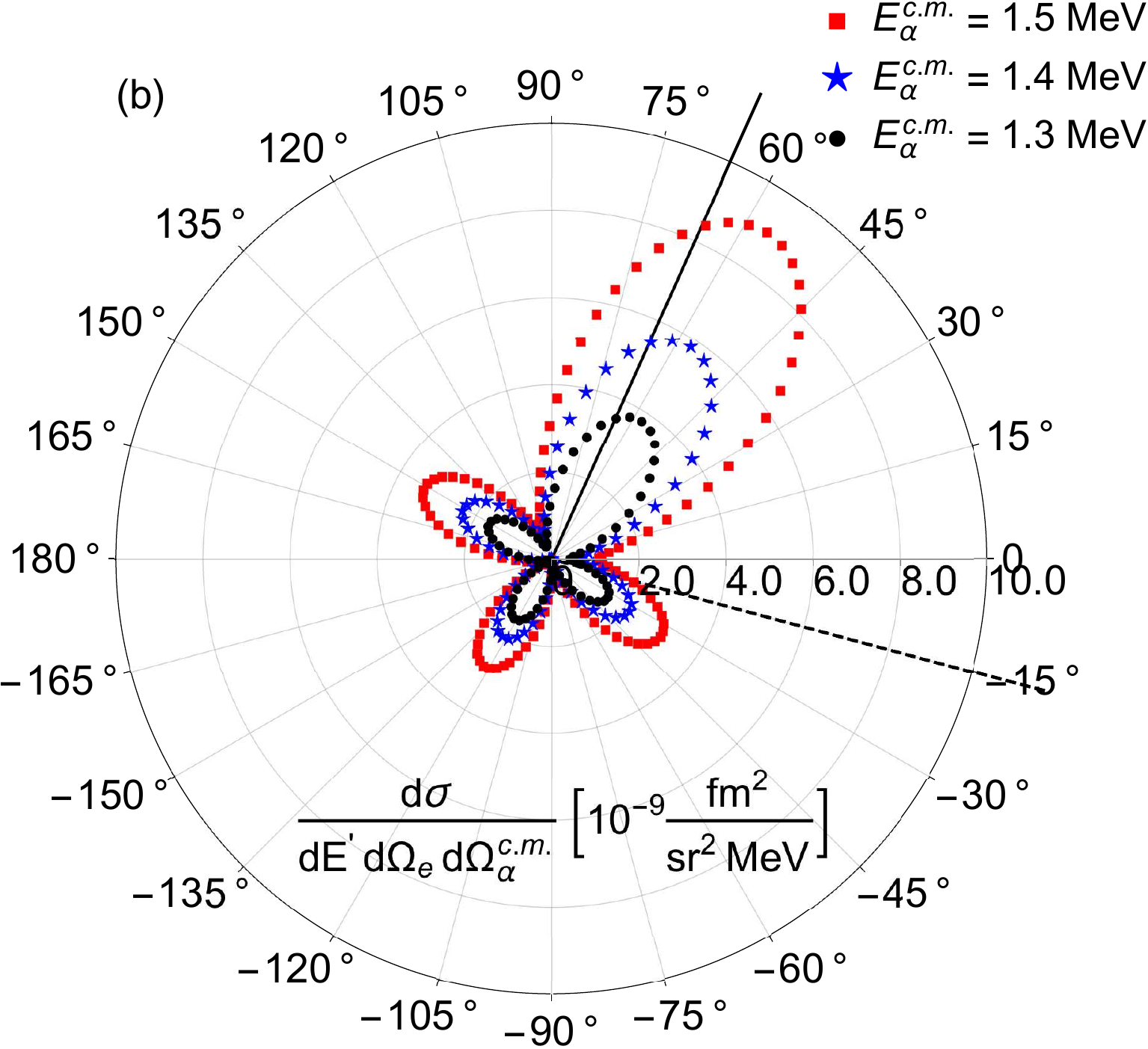} 
\colorcaption{Angular distribution of the  $^{16}$O$(e,e^\prime \alpha)^{12}$C  
differential cross section for beam energy of $E_{e} = 114$ MeV, electron 
scattering angle $\theta_e =15^{\circ}$ and $\alpha$-particle $c.m.$ kinetic 
energies $E_{\alpha}^{c.m.}=$ 1.3, 1.4 and 1.5 MeV. The electron beam lies on
a ray from 180$^{\circ}$ to 0$^{\circ}$, the direction of the scattered
electron is represented by a dashed line, and the direction of the 
virtual photon by a solid line. The results were calculated 
as a function of $\alpha$-particle $c.m.$ production angle $\theta^{c.m.}_{\alpha}$,
but plotted with respect to the direction that the virtual photon
has in the laboratory system. Figure (a)
shows $t_{C0}$ Case A and (b) Case B. \label{fig:diffradar}}
\end{figure*}

Figure \ref{fig:diffradar} shows polar plots of the differential 
cross section for $^{16}$O electrodisintegration as a function of the 
$\alpha$-particle $c.m.$ production angle $\theta^{c.m.}_{\alpha}$ 
with respect to the direction of the virtual photon for A (a) 
and B (b) cases for the choice of $t_{C0}$ discussed above. A very rapid fall-off of the differential cross
section can be observed as the $c.m.$ kinetic energy of the 
$\alpha$-particle decreases. By comparing the (a) and (b) panels in the 
figure, we see that the choice of the $C0$ coefficient 
influences to some extent the shape of  the differential cross section 
around the virtual photon direction and its contribution is more important around $\pm90^{\circ}$ with respect to the virtual photon direction (see later discussion of what impact the monopole contributions have on the extraction of the astrophysical S-factors).

\begin{figure*}[h!]
\centering
\includegraphics[width=8.6cm]{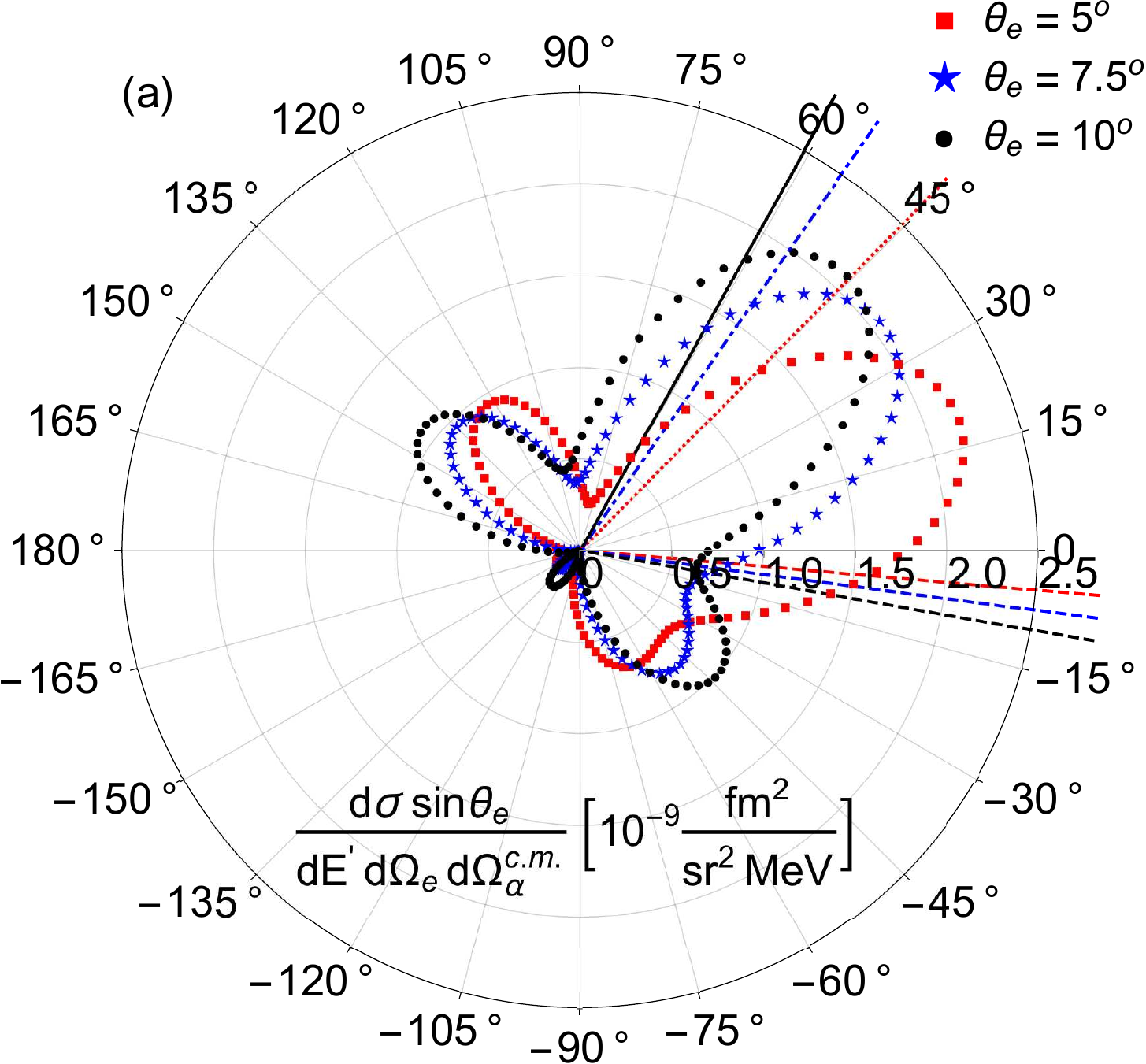}
\includegraphics[width=8.6cm]{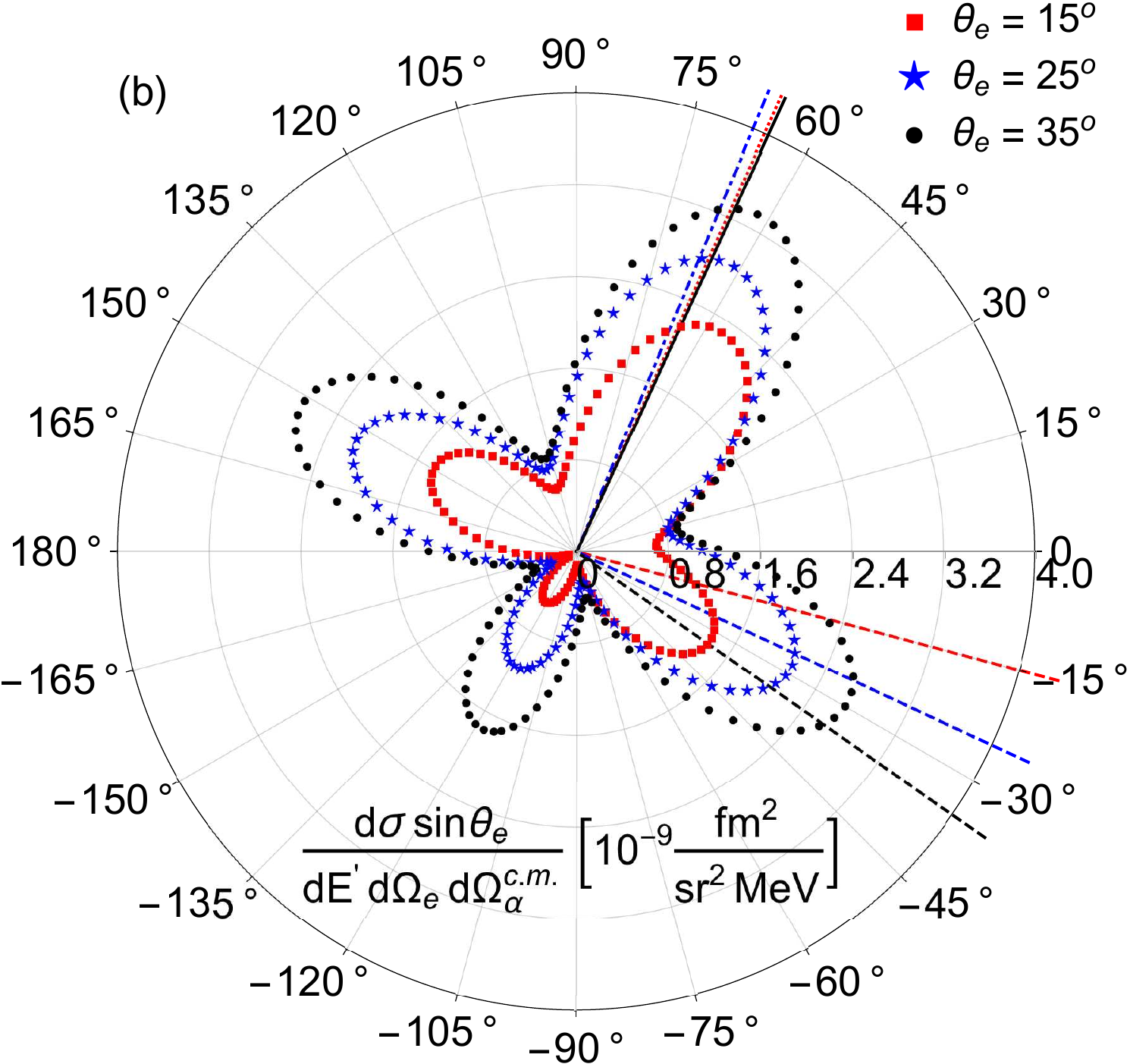} 
\colorcaption{Angular distribution of the  $^{16}$O$(e,e^\prime \alpha)^{12}$C  
differential cross section multiplied by $\sin\theta_e$ for beam energy of $E_{e} = 114$ MeV, 
$E_{\alpha}^{c.m.}=$ 1.5 MeV, for different electron 
scattering angles $\theta_e = 5^{\circ}, 7.5^{\circ}, 10^{\circ} ({\rm a}); 
15^{\circ}, 25^{\circ}, 35^{\circ} ({\rm b})$. The electron beam lies on
a ray from 180$^{\circ}$ to 0$^{\circ}$, the directions of the scattered
electrons are represented by dashed lines, and the other type of lines on the 
positive angle side represents the direction of the virtual photon in
the laboratory system. \label{fig:diffradar1}}
\end{figure*}

\begin{figure*}[b!]
\centering
\includegraphics[width=8.6cm]{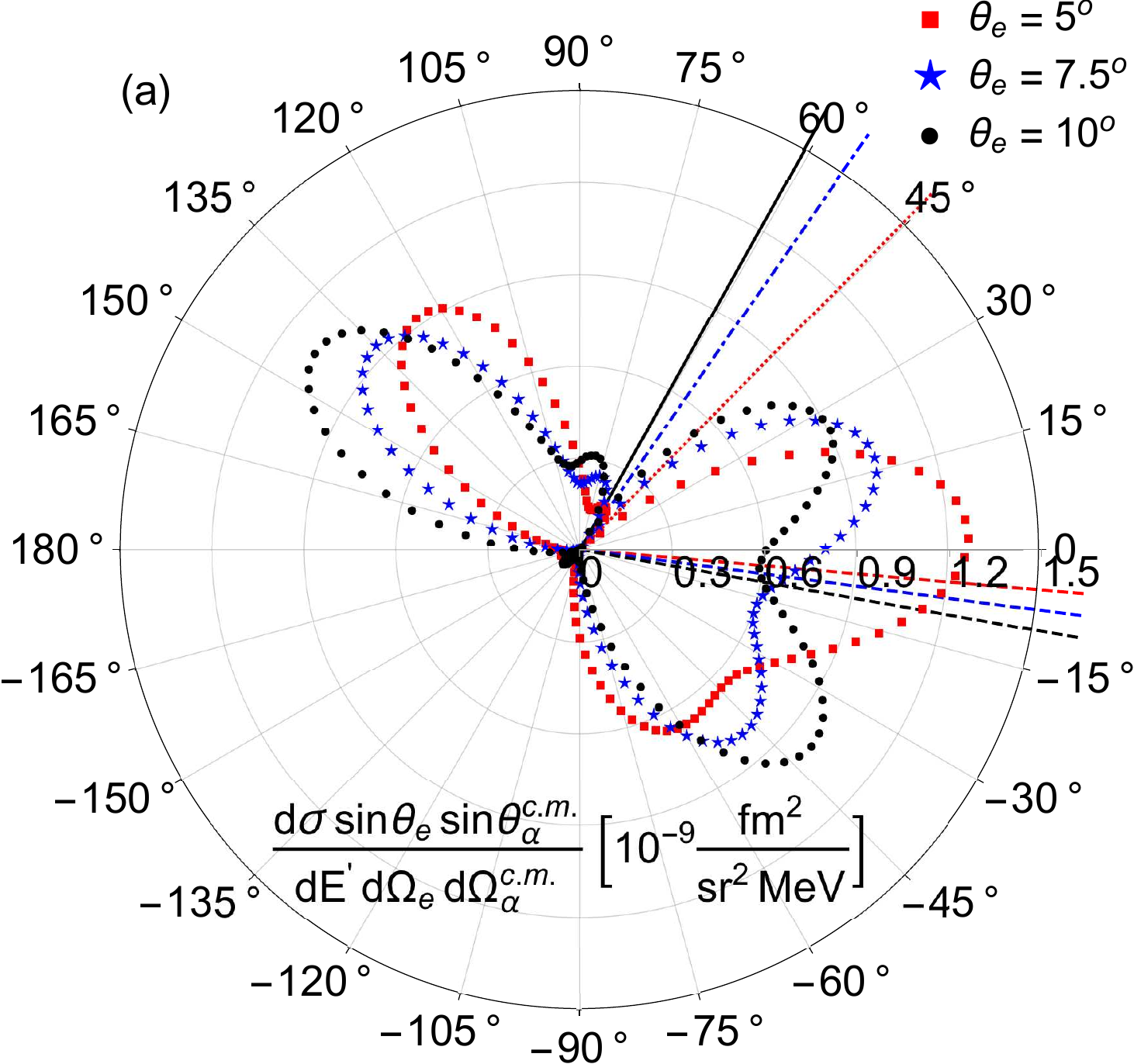}
\includegraphics[width=8.6cm]{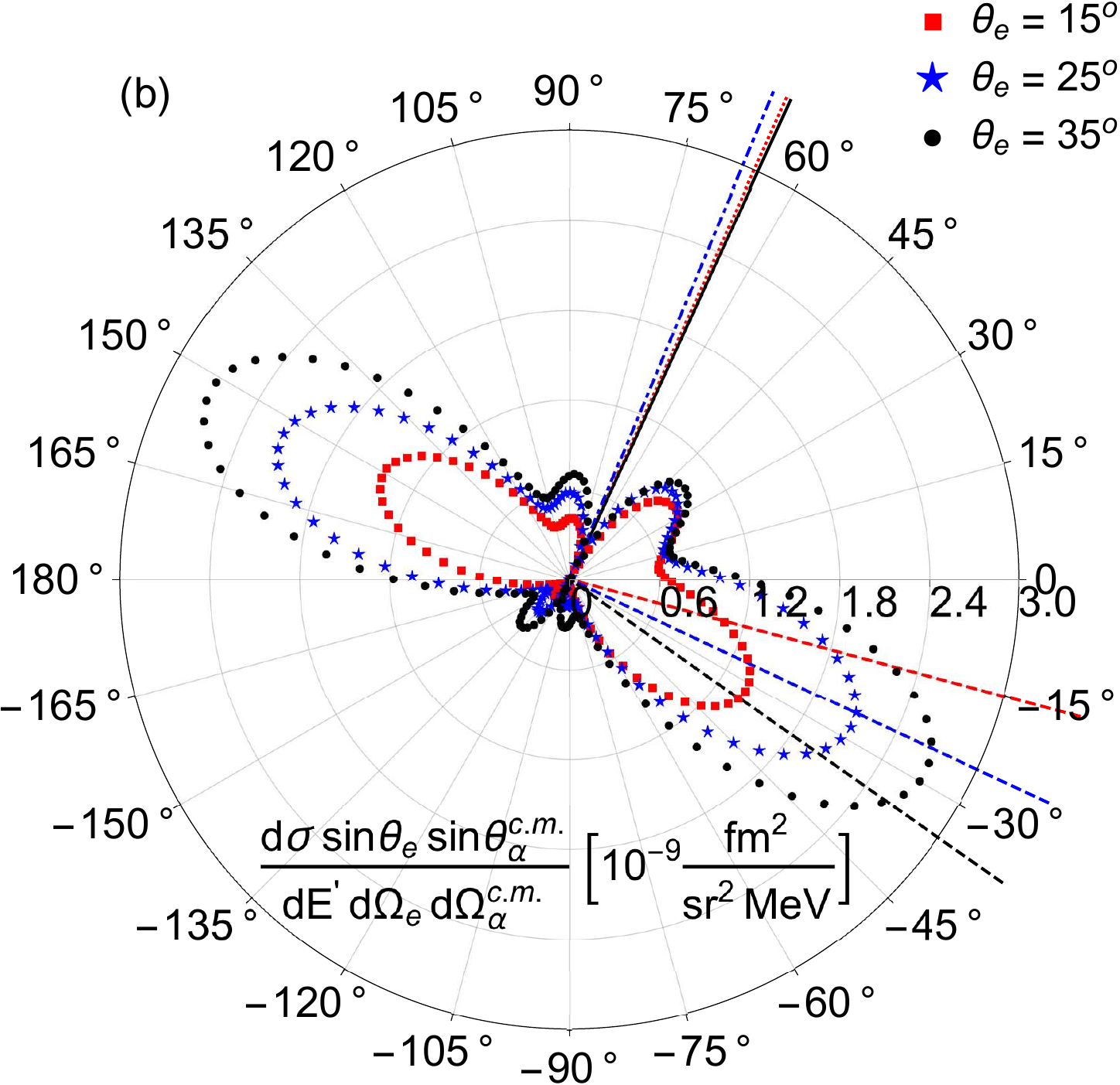} 
\colorcaption{Angular distribution of the  $^{16}$O$(e,e^\prime \alpha)^{12}$C  
differential cross section multiplied by $\sin\theta_e \cdot \sin\theta_{\alpha}^{c.m.}$ 
for beam energy of $E_{e} = 114$ MeV, $E_{\alpha}^{c.m.}=$ 1.5 MeV, for different electron 
scattering angles $\theta_e = 5^{\circ}, 7.5^{\circ}, 10^{\circ} ({\rm a}); 
15^{\circ}, 25^{\circ}, 35^{\circ} ({\rm b})$. The electron beam lies on
a ray from 180$^{\circ}$ to 0$^{\circ}$, the directions of the scattered
electrons are represented by dashed lines, and the other type of lines on the 
positive angle side represents the direction of the virtual photon in
the laboratory system. \label{fig:diffradar2}}
\end{figure*}

Figure \ref{fig:diffradar1} shows the product of the differential cross section and
the electron's solid angle factor $\sin\theta_e$ as a function of $\alpha$ production
angle $\theta_{\alpha}^{c.m.}$ for several values of electron scattering angle $\theta_e$. 
The plots suggest that there is no advantage to reaching very low values of $\theta_e$, 
since the product saturates and only increases in magnitude when increasing the electron 
scattering angle $\theta_e$.  The increase in magnitude comes 
from the response functions -- at fixed beam energy $E_e$ larger $\theta_e$ means larger
$q$, that is larger values of the response functions. In addition, one needs to
keep in mind that a finite sized collimator in a typical electron spectrometer accepts 
larger angular phase space ($\sin\theta_e d\theta_e d\phi_e$) at smaller electron
scattering angle $\theta_e$. 
Later, we will make clear that these two competing effects, for specific experimental
conditions, influence the final coincidence rate and, consequently, the statistical uncertainty.

The polar plot of the product of the differential cross section and the solid angle
factor $\sin\theta_e \sin\theta_{\alpha}^{c.m.}$, shown in Fig. \ref{fig:diffradar2},
indicates the values of  $\theta_{\alpha}^{c.m.}$ for which we can expect the maximum 
rate of $\alpha$-particle production. For $\theta_e \geq$ 15$^{\circ}$ the maximum
rate is around $\pm$90$^{\circ}$ with respect to the direction of the virtual photon. 
At energies ($E_{\alpha}^{Lab}\ge 2$ MeV) this can be a good guide to where to place
an $\alpha$-particle detector, but  at lower energies the placement 
of the $\alpha$-particle detector will be governed by the minimization of the
energy loss and the angular spread of the $\alpha$-particles when traveling
through the target material.

\begin{figure*}[h]
\centering
\includegraphics[width=8.6cm]{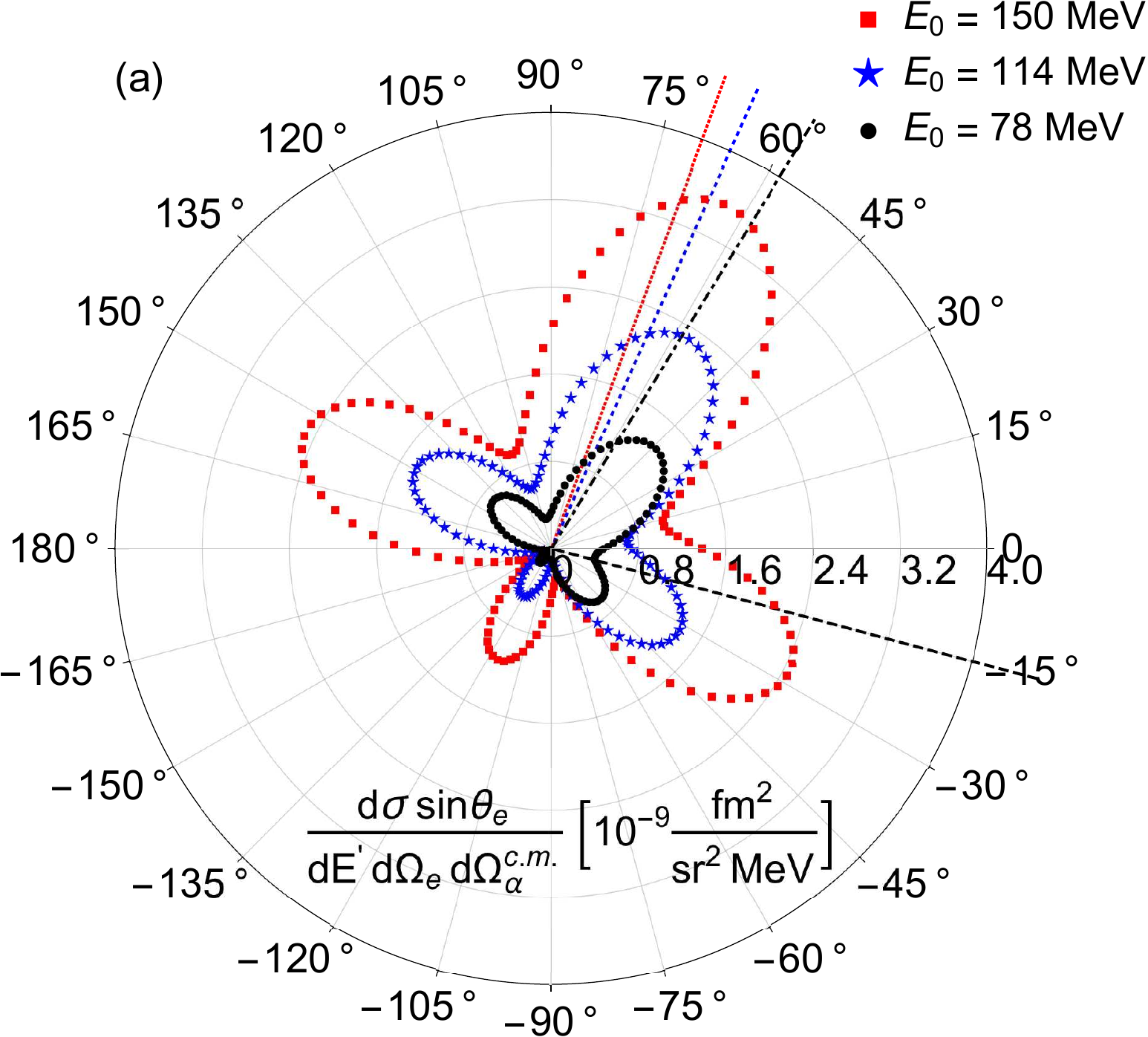}
\includegraphics[width=8.6cm]{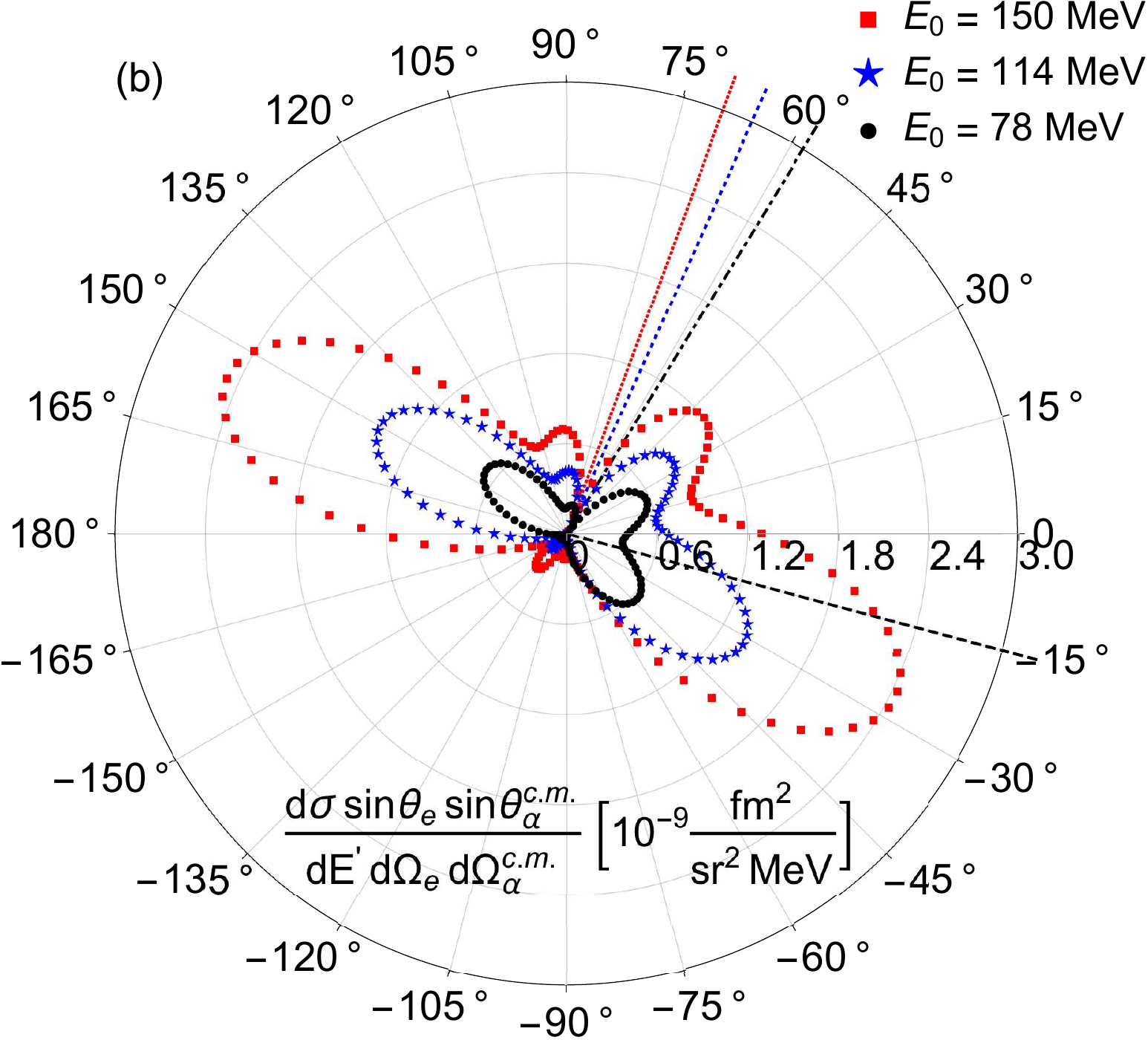} 
\colorcaption{Angular distribution of the  $^{16}$O$(e,e^\prime \alpha)^{12}$C  
differential cross section multiplied by $\sin\theta_e$ (a)
and $\sin\theta_e \cdot \sin\theta_{\alpha}^{c.m.}$ (b)
for beam energies of $E_{e} =$ 78, 114 and 150 MeV, $E_{\alpha}^{c.m.}=$ 1.5 MeV and electron 
scattering angle of $\theta_e = 15^{\circ}$. The electron beam lies on
a ray from 180$^{\circ}$ to 0$^{\circ}$, the directions of the scattered
electrons are represented by dashed lines, and the other type of lines on the 
positive angle side represents the direction of the virtual photon in
the laboratory system.\label{fig:diffradar3}}
\end{figure*}
Figure \ref{fig:diffradar3} shows the product of the differential cross section and 
the solid angle factors $\sin\theta_e$ and $\sin\theta_e \sin\theta_{\alpha}^{c.m.}$,
and illustrates that at fixed electron scattering angle $\theta_e$ one can increase 
the magnitude of the product by increasing the electron beam energy $E_e$.

\section{Consideration of an Experiment to Measure the $^{\bf 16}$O$({\bf e},{\bf e}^{\bf \prime} {\boldsymbol \alpha})^{\bf 12}$C Reaction in the Astrophysically Interesting Region} \label{sec:five}

Alpha-cluster knockout in the $^{16}$O(e,e$^\prime \alpha)^{12}$C reaction has been previously studied~\cite{DeM2001} at 615 and 639 MeV incident electron energy.  The shape of the measured missing-momentum distribution is reasonably well described by shell-model and cluster-model calculations, but the theoretical curves over-predict the data by a factor of three to four.  
However, at present, there exists no dedicated set-up for measuring the electrodisintegration of $^{16}$O at lower energies with the astrophysical goals above in mind. Assuming the availability of high intensity energy-recovery linacs (ERLs) in the near future~\cite{Hug2017,CBETAHoffstaetter2017}, we here develop a conceptual experiment based on these new, advanced accelerator technologies.  In doing so, there are nevertheless practical constraints on what is likely to be possible, and these are discussed below.

\subsection{Experimental Considerations}

\subsubsection{Electron Detection}

The detector system suitable for measuring the 
four-momentum of the scattered electron is a high 
precision, focusing magnetic spectrometer, equipped with
focal plane detectors, capable of achieving a momentum 
resolution $\Delta p_{e} / p_{e} $ better than 
$\le 10^{-4}$ and an in-plane scattering angle resolution  
$\Delta \theta_{e}$ better than $\le 0.5^{\circ}$.
Spectrometers of this type are standard in  
electron scattering nuclear research, but they differ
in angular and momentum acceptance ranges, and 
in the type of focal plane detector systems used.  

\subsubsection{Isotopic and Chemical Contamination \label{IandCC}}

\begin{figure}[h!]
\centering
\includegraphics[width=7.6cm]{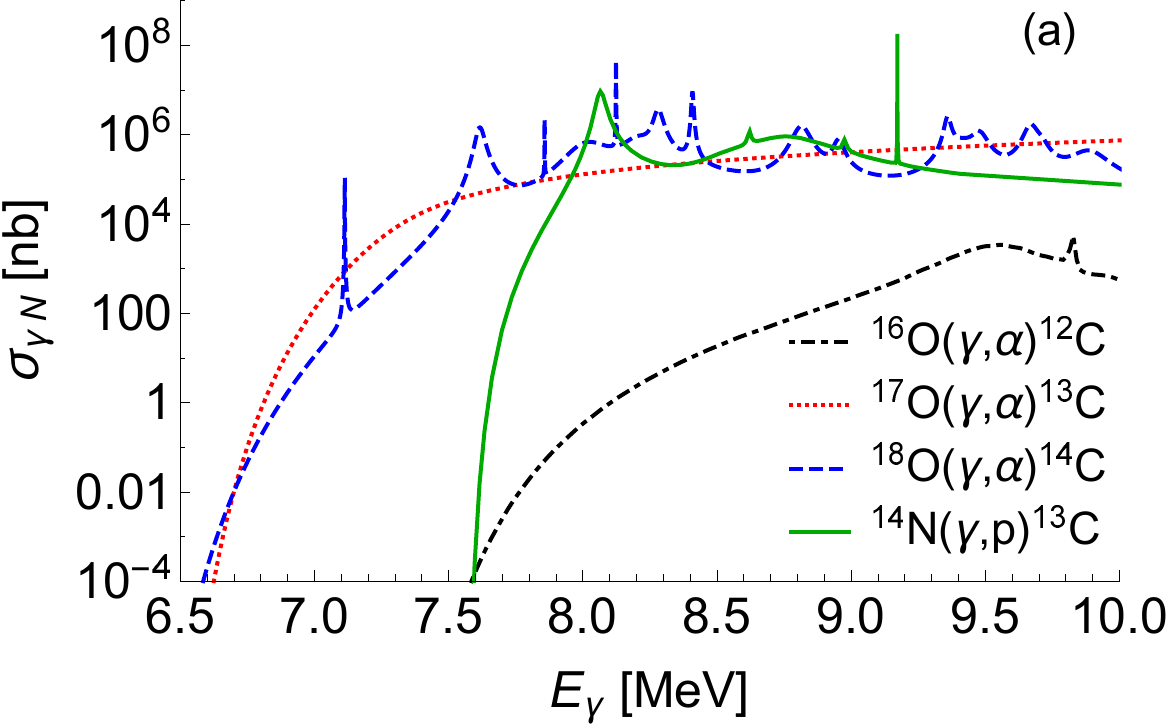}
\includegraphics[width=7.6cm]{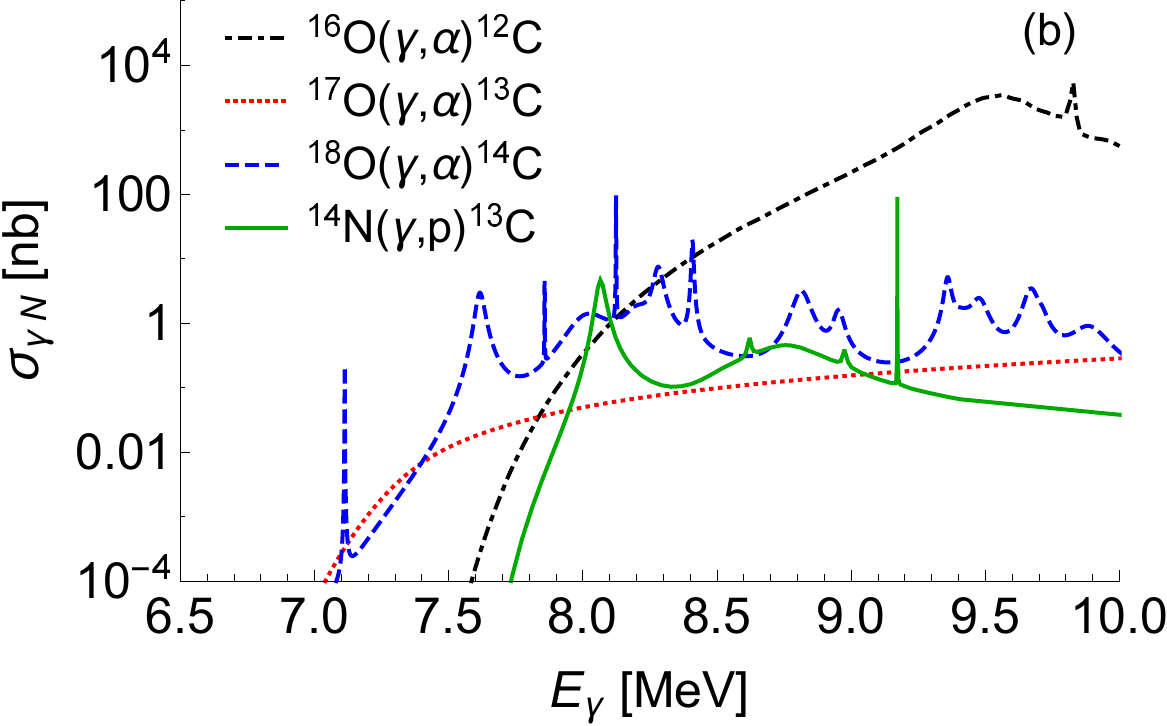} 
\colorcaption{(a) the theoretical photo-nuclear cross section 
$\sigma_{\gamma N}$ as a function of the gamma energy $E_{\gamma}$ 
for the signal reaction $^{16}$O$(\gamma,\alpha)^{12}$C, and 
background reactions  $^{17}$O$(\gamma,\alpha)^{13}$C, $^{18}$O$(\gamma,\alpha)^{14}$C and $^{14}$N$(\gamma,p)^{13}$C, from \cite{PhotoNucCross}. The same curves are
shown on (b), but now the cross sections of the oxygen isotopes were normalized under the assumption that the natural abundances of  $^{17}$O and $^{18}$O were depleted by a factor of 1000, and that oxygen gas is contaminated with 5 ppmv of $^{14}$N. 
\label{fig:PNC}}
\end{figure}

When dealing with the photodisintegration of $^{16}$O into 
an $\alpha$-particle and $^{12}$C one needs to take into 
account a large background coming from $\alpha$-particles produced
on $^{17}$O and $^{18}$O.  The average isotopic abundances of 
the oxygen isotopes are 99.7570\% for $^{16}$O, 0.03835\% for 
$^{17}$O and 0.2045\% for $^{18}$O \cite{Meija2016}. 
The cross sections for photodisintegration of $^{17}$O and $^{18}$O
into an $\alpha$-particle and corresponding carbon isotope 
are several orders of magnitude larger than for the
case of photodisintegration of $^{16}$O; see Fig. \ref{fig:PNC}(a).
Further, there is always some finite amount 
of nitrogen present in the oxygen gas (depending on the
vendor usually 5 ppmv or less).  This will give rise to 
protons from the photodisintegration reaction $^{14}$N$(\gamma,p)^{13}$C
and also contribute to the background.  Even if one
depletes the $^{17}$O and $^{18}$O by a factor of 1000,
and normalizes the cross sections accordingly as shown in the (b) panel of Fig.~\ref{fig:PNC}, 
in the region of interest ($E_{\gamma} = E^{c.m.}_{\alpha} +$ 7.162 MeV)
$E_{\gamma} \le 8.5$ MeV, photodisintegration of $^{17}$O 
significantly contributes to the background and 
the contributions of $^{18}$O and $^{14}$N are comparable 
or at some energies even larger. The same problem
can also be expected in the case of electrodisintegration.  
\begin{figure*}[h]
\centering
\includegraphics[width=7.6cm]{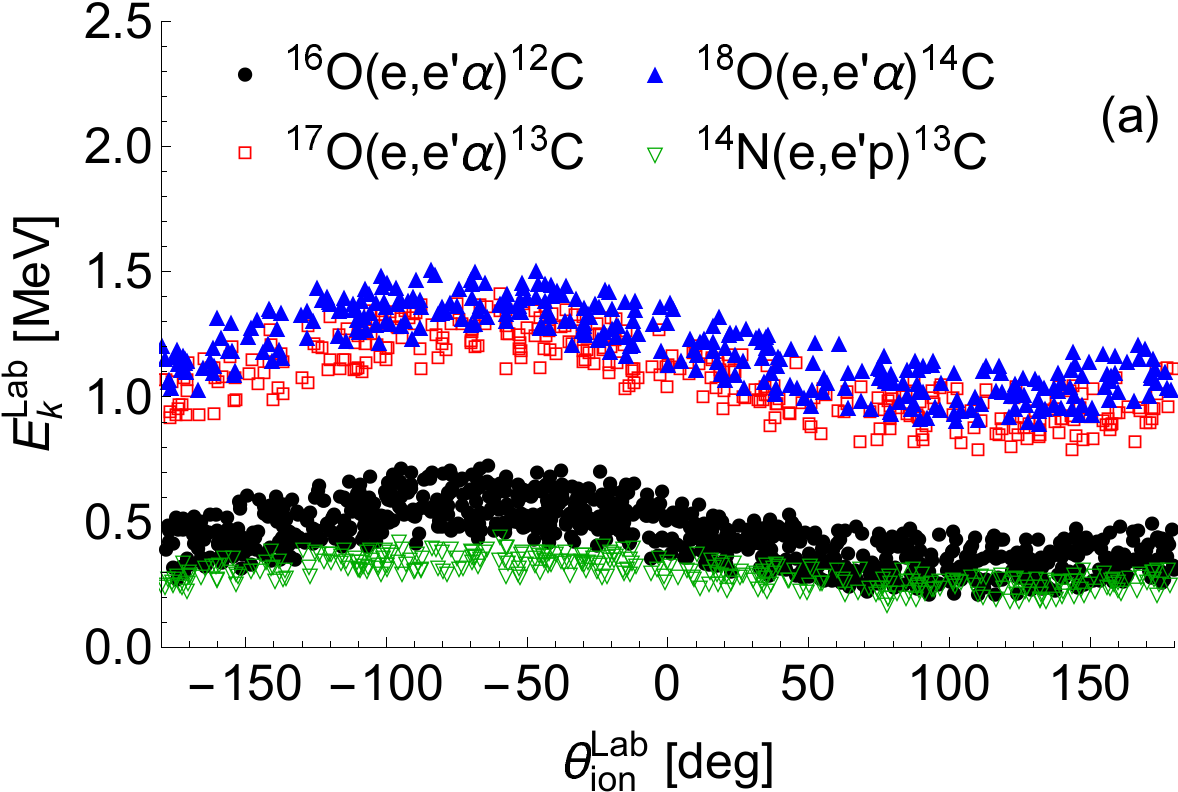}
\includegraphics[width=7.6cm]{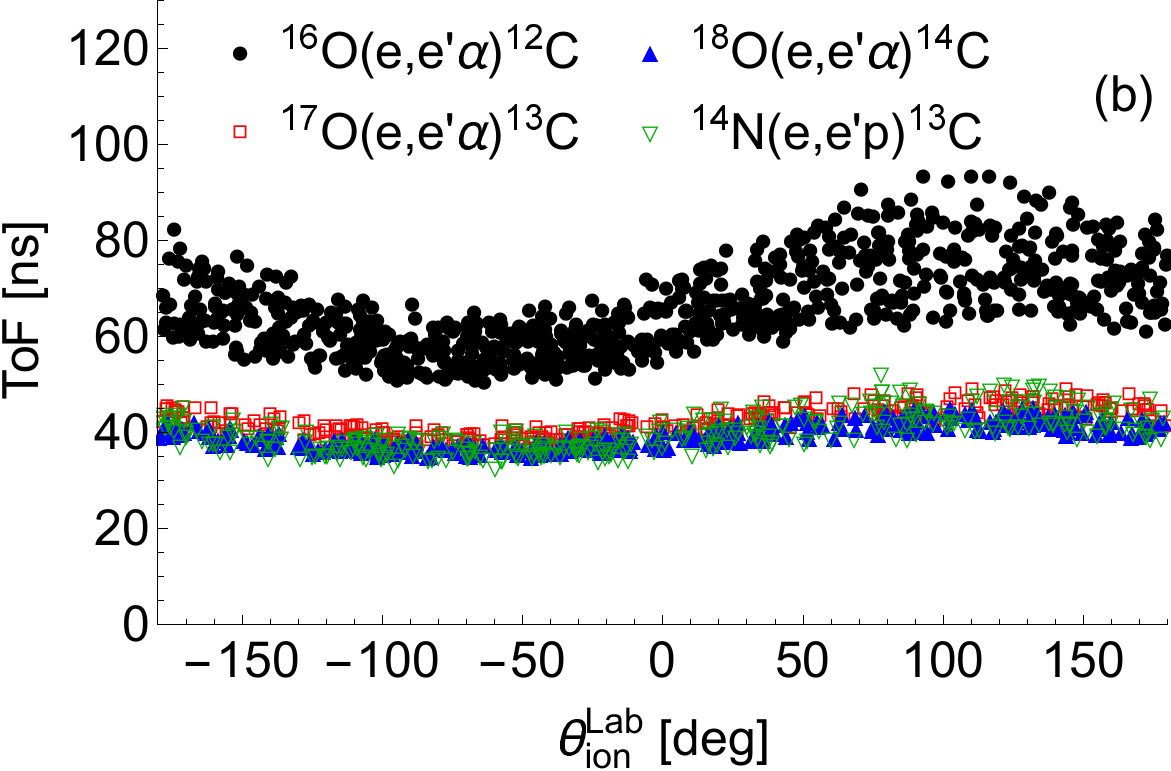} 
\includegraphics[width=7.6cm]{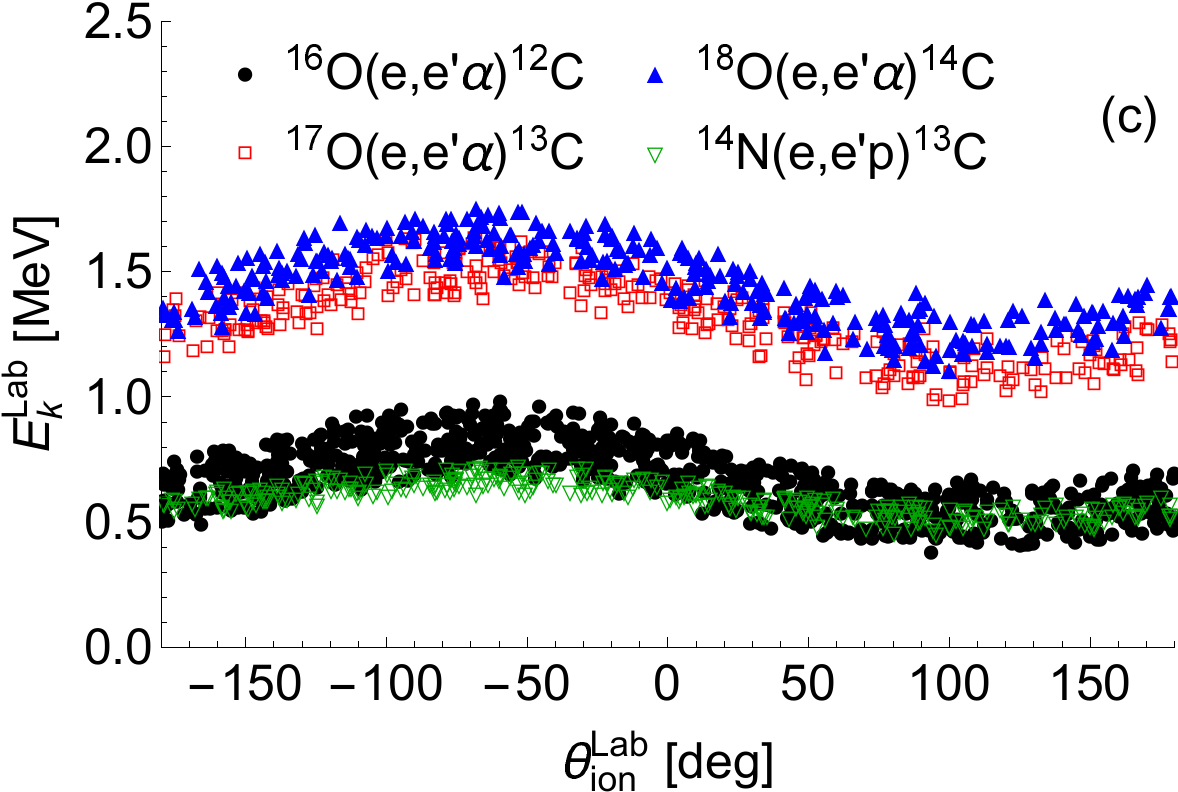}
\includegraphics[width=7.6cm]{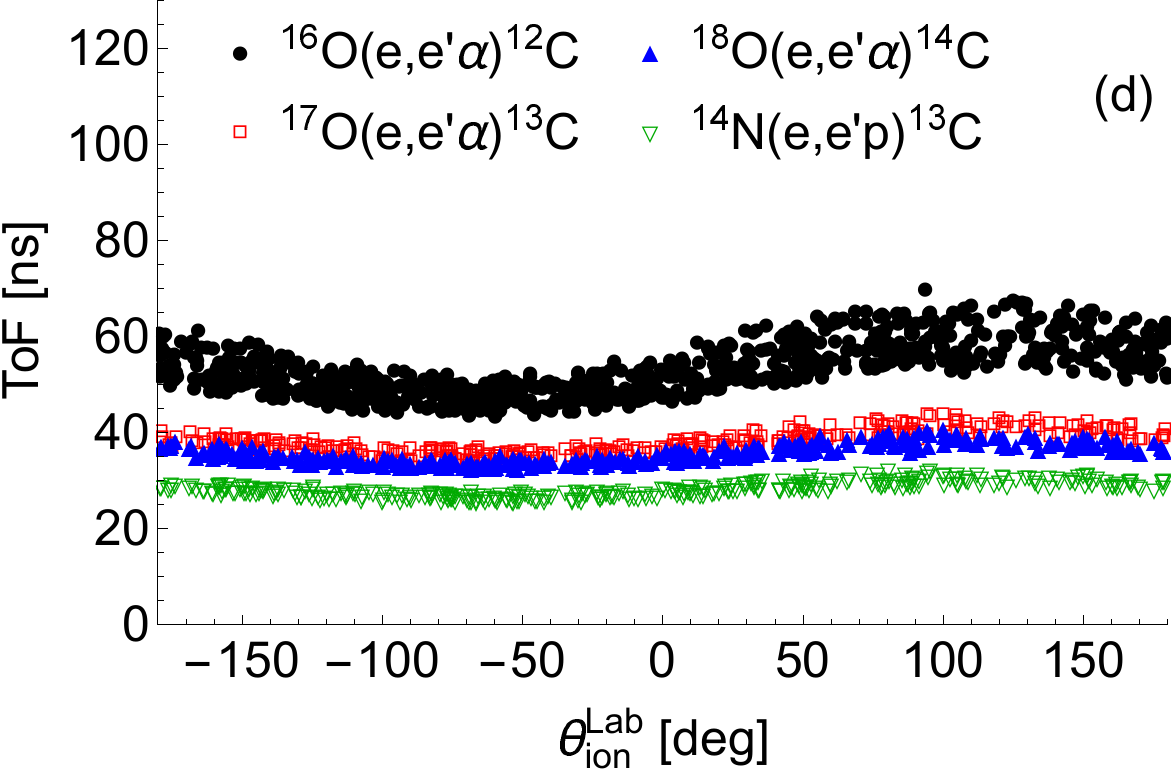} 
\colorcaption{Energy-loss corrected kinetic energy $E_{k}^{Lab}$ and time-of-flight ToF as functions of laboratory ion production angle
$\theta^{Lab}_{ion}$ assuming that the ions were produced by electrons involved in the 
electrodisintegration of $^{16}$O, at $E_{e} =$ 114 MeV and 
$\theta_e =$ 15$^{\circ}$: (a) and (b) cut on $0.7 \le E_{\alpha}^{c.m.} \le 0.8$
MeV, (c) and (d) cut on $1.0 \le E_{\alpha}^{c.m.} \le 1.1$ MeV. \label{fig:back}}
\end{figure*}

The modern photodisintegration experiments, \cite{DiGiovine2015,Ugalde2013} and 
\cite{Gai2010}, address these isotopic and chemical contamination issues.   
Here, we investigate how the background
problems can be mitigated in an electrodisintegration experiment
with a gas jet target.

In the presented study, SRIM-2013 simulation software \cite{Ziegler2010, SRIM} 
was used for calculation of the average energy loss of the $\alpha$-particle or proton at a given kinetic energy in a 2 mm wide $^{16}$O gas jet having a density of 6.65$\times$10$^{-4}$ g/cm$^3$. The full electrodisintegration kinematics calculation was performed
for oxygen isotopes and $^{14}$N target nuclei, and the data were
sorted by selecting the electrons having momenta capable of
producing $\alpha$-particles on $^{16}$O in a given 
$E_{\alpha}^{c.m.}$-range. The kinetic energy of selected $\alpha$-particles 
and protons was corrected for the energy loss
assuming that these particles are created at different positions
inside the gas jet. The maximum correction was applied 
when the particle is created at the edge of the jet and
needs to travel through the full extension of the gas jet. 
In this way, the corrected kinetic energies where converted 
to time-of-flight (ToF), assuming a flight path of 30 cm
between the gas jet and the ion detectors. Figure \ref{fig:back}
shows the energy-loss corrected kinetic energies and ToF of the
$\alpha$-particles and protons for two $E_{\alpha}^{c.m.}$-ranges.
In both cases, we see that the kinetic energy can be used to distinguish 
the signal from the background $\alpha$-particles. 
However, to distinguish between protons and $\alpha$-particles from $^{16}$O,
the ToF observable is the most effective. It allows a
clear background identification and removal from the 
collected events for all $E_{\alpha}^{c.m.}$ of interest. 
Furthermore, it is easier to determine the final state of
low-energy ions by measuring the ToF and not their kinetic
energy $-$ this method is very well known in experimental 
nuclear physics. Most importantly, such detectors can be
designed to be electron blind.

\begin{figure}[h]
\centering
\includegraphics[width=7.6cm]{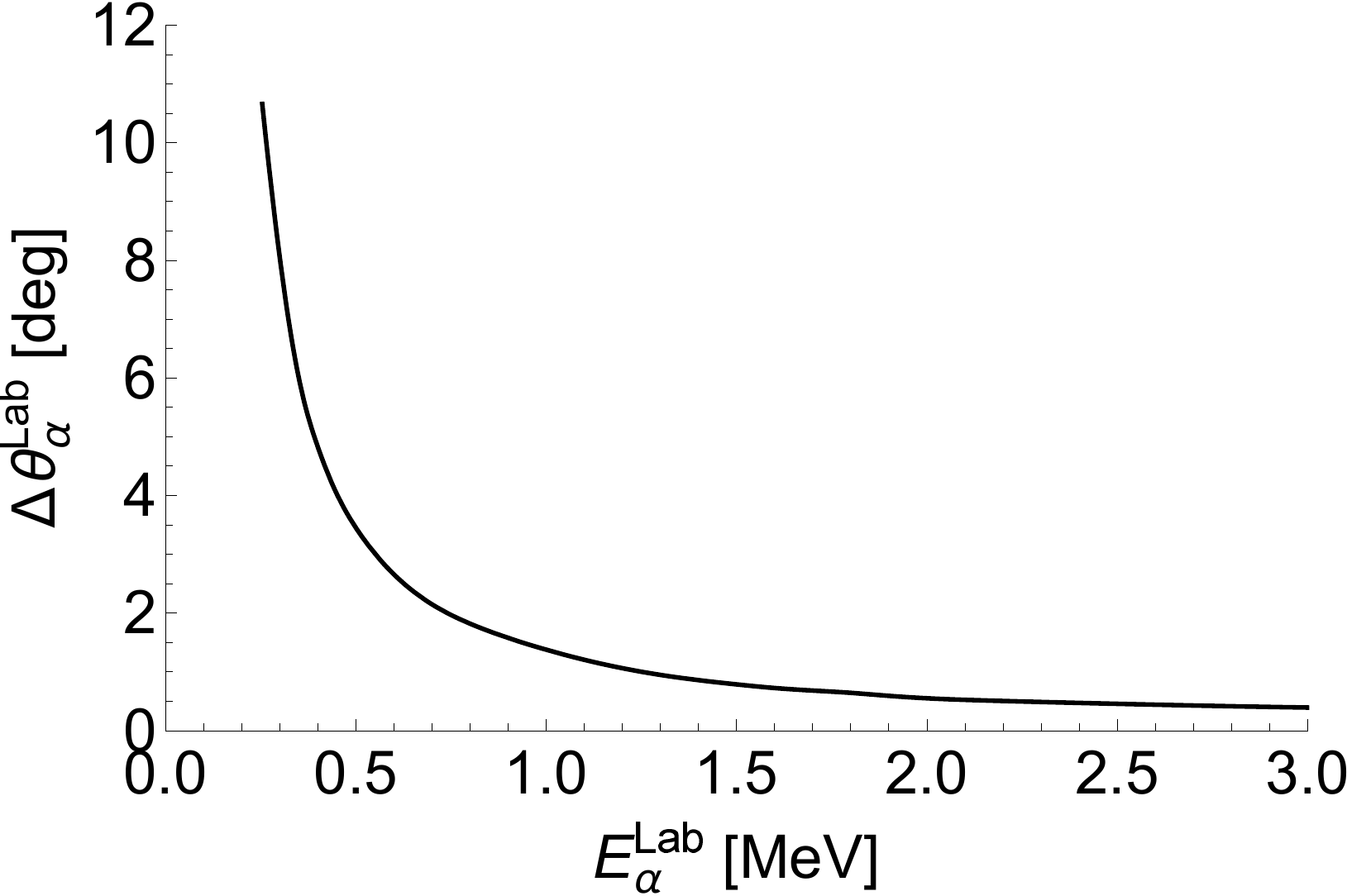} 
\caption{Standard deviation of the $\alpha$-production angle
$\Delta\theta^{Lab}_{\alpha}$ as a function of $\alpha$-particle
kinetic energy $E^{Lab}_{\alpha}$ for an $\alpha$-particle passing 
through a 2 mm wide $^{16}$O gas jet with a density of 6.65$\times$10$^{-4}$ g/cm$^3$. \label{fig:sigmaAng}}
\end{figure}

Very close to the reaction threshold one has to deal with
$\alpha$-particles having very small kinetic energies, {\it i.e.}, so small that the target material itself can smear 
their angular resolution significantly. To quantify the
angular smearing by the target material, we used data
obtained from a  SRIM-2013 simulation to calculate the 
standard deviation of the $\alpha$-production angle
$\Delta\theta^{Lab}_{\alpha}$ as a function of the $\alpha$-particle
kinetic energy $E^{Lab}_{\alpha}$; see Fig. \ref{fig:sigmaAng}.
At kinetic energy of $E^{Lab}_{\alpha} =$ 0.7 MeV, the 
standard deviation of the $\alpha$-production angle is 
already equal to $\Delta\theta^{Lab}_{\alpha} =$ 2.1$^{\circ}$ and,
with deceasing $E^{Lab}_{\alpha}$, the $\Delta\theta^{Lab}_{\alpha}$
starts to increase even faster.

\begin{figure}[h]
\centering
\includegraphics[width=7.6cm]{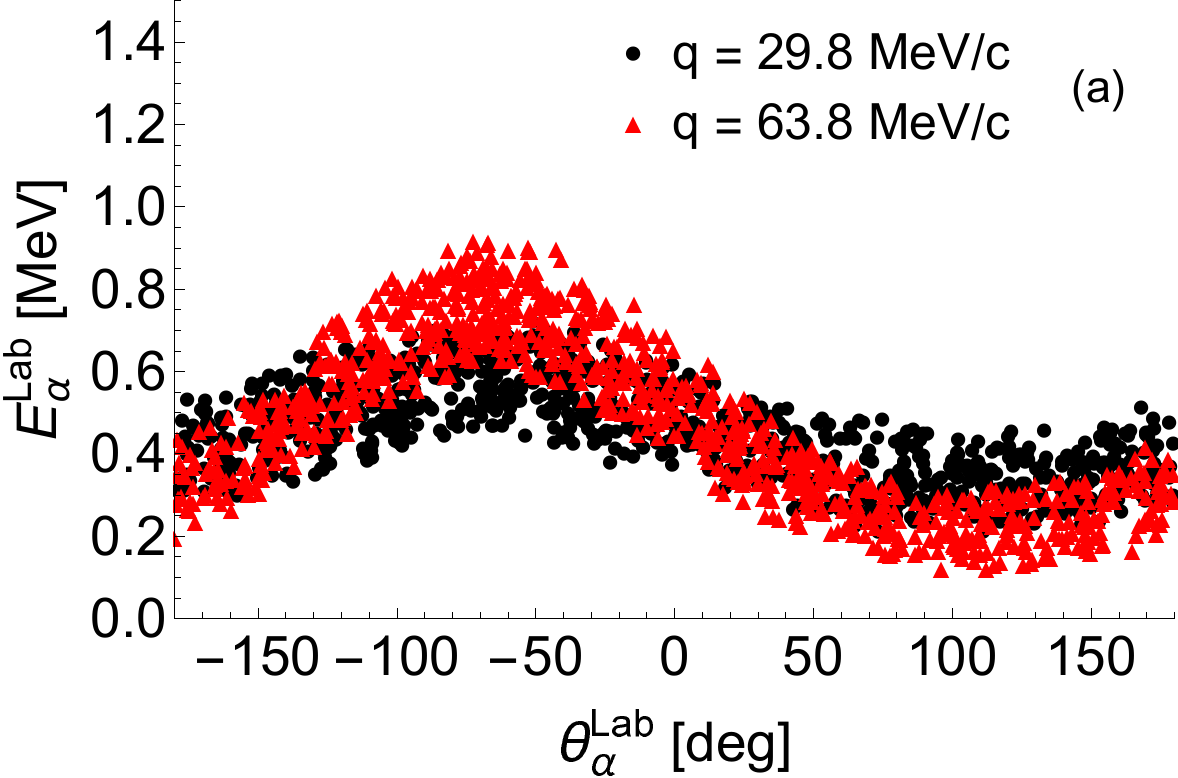}
\includegraphics[width=7.6cm]{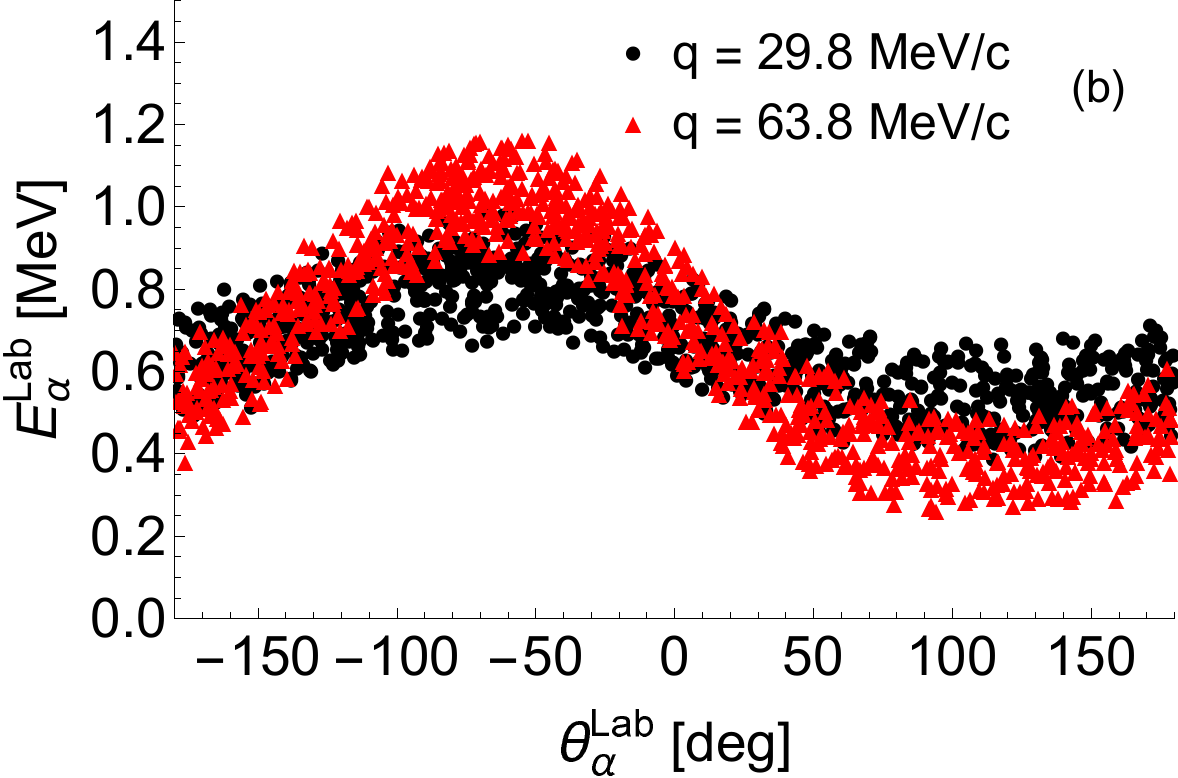}  
\colorcaption{Energy-loss corrected kinetic energy of $\alpha$-particles $E_{\alpha}^{Lab}$  
as functions of laboratory $\alpha$ production angle $\theta^{Lab}_{\alpha}$ for two
values of transferred 3-momentum of the virtual photon $q$. The (a) panel shows
$\alpha$-particles in the range of $0.7 \le E_{\alpha}^{c.m.} \le 0.8$ MeV, 
and in the range of $1.0 \le E_{\alpha}^{c.m.} \le 1.1$ MeV for the (b) panel. \label{fig:virtualq}}
\end{figure}

Since the circular profile of the jet can be easily 
changed to a different one, like demonstrated in  \cite{Kohler2012}. For the fixed luminosity, the problem of the multiple scattering inside the jet can be minimized by extending the jet in the direction of the beam. But in this case, the electron spectrometer will
need to have good spatial resolution to be able to reconstruct the position of the vertex along the extended gas jet. 
Another option, to partially solve this problem, is to make use of the virtual photon properties: at
fixed $\omega$ one can independently dial the value of 
the transferred 3-momentum $q$. Figure \ref{fig:virtualq}
shows examples of angular distributions of the $\alpha$-particle 
kinetic energy for fixed $\omega$ but for two different 
values of $q$. For larger-$q$, around the direction of the
virtual photon ($\sim-$67$^{\circ}$), the kinetic energy of the
$\alpha$ particles is larger compared with the lower-$q$ case. 
In the opposite direction the larger-$q$ $E^{Lab}_{\alpha}$
is decreased.  Ultimately, the measurement close to threshold will need to be 
performed at an optimized value of $q$, with a gas jet having an
optimized density and shape. 

\subsubsection{Alpha-particle Detection}

The $\alpha$-particle detector system has to be able 
to cover the maximum possible solid angle
around the beam-target interaction.  Further, the detectors have to be blind to electrons, positrons 
and $\gamma$-rays, due to high rates from elastic, inelastic,
and M{\o}ller electrons, gammas from radiative processes, and
positrons and electrons from radiative pair production. 
In the region of interest ($0.7 \le E_{\alpha}^{c.m.} 
\le 1.7 $ MeV) it is straightforward to measure the time-of-light
of the $\alpha$-particle to obtain its energy. Thus, these
detectors should have a good timing resolution. 

Measuring the time-of-light has a crucial advantage
since it can be used for ion identification purposes,
as well as for distinguishing the $\alpha$-particles
coming from different oxygen isotopes. 

We have given some consideration to the choice of $\alpha$-particle detector, which is required to detect  ions with kinetic energies of about 1 MeV.   At 1 MeV kinetic energy, 
the range in silicon is about 1 mg/cm$^2$. Silicon has a density of 2.33 
g cm$^{-2}$ and so this corresponds to a thickness of 4.3 microns.  The count rate for the 
$(e,e'\alpha)$ process is low, $\approx 1-10$ 
Hz. It is required:
\begin{itemize}
\item to measure the total energy of the $\alpha$ to 
about $\approx \pm 10\%$
\item to distinguish between protons, $\alpha$-particles and $^{12}$C
\item to measure the position to $\sim$mm and the timing to a few nsec
\item that the ion detection system be blind to scattered electrons and photons.
\end{itemize}

There are a several different detector possibilities:
\begin{itemize}
\item \textbf{Silicon detector}  \cite{Kordyasz2015} \\
Silicon detectors have a high position resolution in tracking
charged particles but are expensive and require cooling to reduce 
leakage currents. They also suffer degradation over time from 
radiation; however, by cooling them to low temperatures, this 
effect can be significantly reversed.
\item  \textbf{Micro-channel-plate electron (MCP) detector} \cite{Friedman1988} \\
A micro-channel plate is a slab made from highly resistive material 
of typically 2 mm thickness with a regular array of tiny tubes or 
slots (microchannels) leading from one face to the opposite, 
densely distributed over the whole surface. The microchannels
are typically approximately 10 microns in diameter (6 micron 
in high resolution MCPs) and spaced apart by approximately 15 
microns; they are parallel to each other and often enter the plate 
at a small angle to the surface ($\approx 8^{\circ}$ from normal).

The gain of an MCP is very noisy, meaning that two identical particles 
detected in succession will often produce wildly different signal magnitudes. 
The temporal jitter resulting from the peak height variation can be removed 
using a constant fraction discriminator. Employed in this way, MCPs are 
capable of measuring particle arrival times with very high resolution, 
making them an ideal detector for mass spectrometers.

\item \textbf{Parallel-plate avalanche counter (PPAC)} \\
The PPAC detector consists of two parallel thin electrode films separated 
by 3--4 mm and is filled with 3--50 Torr of gases such as isobutane (C$_4$H$_10$) 
or perfluoropropane (C$_3$F$_8$). When a voltage gradient corresponding to a few hundreds 
of volts per millimeter is applied between the anodes and cathodes, ionized 
electrons from incident heavy ions immediately cause an electron avalanche. 
Because there is no time delay before the avalanche occurs and the electrons 
move at high mobile velocity (mobility), the resulting signals have good 
timing properties, with rise and fall times of a few nanoseconds, as compared 
with other types.

A PPAC detector has been developed at RIKEN RIBF in Japan \cite{Kumagai2013} 
that has a sensitive area of 240 mm $\times$ 150 mm, and the position 
information is obtained by a delay-line readout method. Being called a {\it double 
PPAC}, it is composed of two full PPACs, each measuring the particle locus 
in two dimensions. High detection efficiency has been made possible by the 
twofold measurement using the double PPAC detector. The sensitivity 
uniformity is also found to be excellent. The root-mean-square position 
resolution is measured to be 0.25 mm using an $\alpha$ source, while the
position linearity is as good as $\pm$0.1 mm for the detector size of 
240 mm.

\item \textbf{Time Projection Chamber} \\
A time projection chamber (TPC) is a type of particle detector that 
uses a combination of electric and magnetic fields together with 
a sensitive volume of gas or liquid to perform a three-dimensional 
reconstruction of a particle trajectory or interaction. 

A Micromegas TPC is under development \cite{Iguaz2014} for the detection 
of low-energy heavy ions. The first prototype consists of a 10 $\times$ 10 $\times$ 
10 cm$^3$ gaseous vessel equipped with a field shaping cage and a Micromegas 
detector. With 1 atm of gas, the energy resolution for 6 MeV $\alpha$-particles is 
about 10\%. The window is 10 $\mu$m of Mylar (polyethylene terephthalate) which has a thickness 
of 1.4 mg cm$^{-2}$.

The DMTPC detector technology has been developed at MIT \cite{Deaconu2017} to search for dark matter.
It consists of a TPC 
filled with low pressure CF$_4$ gas. Charged particles incident on the gas 
are slowed and eventually stopped, leaving a trail of free electrons and 
ionized molecules. The electrons are drifted by an electric field 
toward an amplification region. Instead of using MWPC endplates for 
amplification and event readout, as in the traditional TPC design, 
the DMTPC amplification region consists of a metal wire mesh separated 
from a copper anode with a high electric field between them. This 
creates a more uniform electric field in order to preserve the shape
of the original track during amplification. The avalanche of electrons 
also creates a great deal of scintillation light, which passes through 
the wire mesh. Some of this light is collected by a charge-coupled device (CCD) camera located 
outside the main detector volume. This results in a two dimensional 
image of the ionization signal of the track as it appeared on the 
amplification plane. Information about the charged particle, including 
its direction of motion within the detector, can be reconstructed
from the CCD readout. Additional track information is obtained from 
readout of the charge signal on the anode plane. The largest existing 
prototype detectors each have a total of 20 liters of CF$_4$ gas 
within the drift region, where measurable events can occur. Recoil 
$^{19}$F and $^{12}$C nuclei with energies from 20 keV to 200 keV and 
$\alpha$-particles from an $^{241}$Am source have been detected in DMTPC~\cite{Deaconu2017}.

\item \textbf{Low-Pressure Multistep Detector for Very Low 
Energy Heavy Ions} \cite{Astabatyan2012} \\
A large-area timing and position-sensitive multistep gaseous detector 
designed for the detection of very low energy heavy ions has been 
developed \cite{Breskin1984}. It consists of a preamplification stage 
operating as a parallel plate avalanche chamber directly coupled to a 
multiwire proportional chamber. The multistep avalanche counter (MSC) was 
tested with $\alpha$-particles, fission fragments and heavy ions. The detector 
operates at a pressure range of 1-4 Torr isobutane, with very thin ($\sim$ 
50 $\mu$g cm$^{-2}$) polypropylene window foils. It has a high gain and good time resolution (better than 180 
ps fwhm) and a position resolution better than 0.2 mm (fwhm). Its 
efficiency for low-energy, high-mass ions was tested with $^{160}$Gd 
ions and found to be 93\% down to kinetic energies of 1.3 MeV. In 
its original design, the MSC does not provide $\Delta$E information. 
Information concerning the energy loss, in addition to timing and 
localization, can be obtained by adding an independent wide-gap 
collection and low-gain element.
\end{itemize}

We note that:

\begin{itemize}
\item A large area, thin, silicon detector with adequate position resolution and with threshold set so that
minimum ionizing particles do not trigger, is an attractive option.

\item The first stage could be a thin gas detector, {\it e.g.}, 10 
cm length of gas at 5 Torr. For isobutane (C$_4$H$_{10}$), the 
thickness is 0.15 mg cm$^{-2}$. The energy lost by a 1 MeV $\alpha$-particle 
in such a detector is of order 0.2 MeV. 

\item This gas detector must be contained in the vacuum 
system of the gas target. The detector gas volume can be 
isolated from the gas jet volume by a thin window, {\it e.g.},
50 $\mu$g cm$^{-2}$ of polypropylene. The energy loss of 
the $\alpha$-particle will be small in this window.

\item The energy lost by a minimizing particle (stopping 
power $\sim$2 MeV/(g cm$^{-2}$)) will be of order 0.5 keV 
so the gas detector will be blind to scattered electrons.

\item The detailed technical aspects of the gas detector ({\it e.g.}, charge 
collection mechanism, amplification, transverse size, gas type, 
{\it etc.}) need to be considered in detail. The gas pressure could 
be high enough to stop the $\alpha$ or it could 
be thin enough to have another detector ({\it e.g.}, thin silicon) behind 
it. Note that a higher detector gas pressure will require a thicker 
entrance window. Until the details of the gas detector are specified, 
it is hard to characterize the energy, position and time resolutions.

\item Finally, the possibility to integrate the oxygen gas target and 
the $\alpha$-detector by using the oxygen gas as the ionizing gas for
the detector is worthy of consideration.
\end{itemize}

Below we continue with the calculation of the $^{16}$O$(e,e'\alpha)^{12}$C
reaction rate and perform an estimate of the statistical uncertainties by using established parameters 
of existing cluster gas-jet targets \cite{GrieserKhoukaz2018} and expected performance of electron accelerators
(MESA \cite{Hug2017} and CBETA \cite{Trbojevic2017}) under construction. In the rate calculations, we identify and
consider the most significant sources of systematic uncertainty.   Furthermore, systematic effects due to scattering in the gas jet target can be
reduced by extending the profile of the jet and/or by increasing the transferred $q$-value; 
optimization here needs to be carried out experimentally. 
Nevertheless, in calculation of the rate, we will use what we have learned in this section and 
restrict the accepted range of the $\alpha$-production angle $\theta_{\alpha}^{c.m.}$ 
and the accepted $\alpha$-particle kinetic energy $E_{\alpha}^{Lab}$ to reasonable values.

\subsection{Proposed Experiment Concept}

\begin{figure}[b!]
\centering
\includegraphics[width=7.6cm]{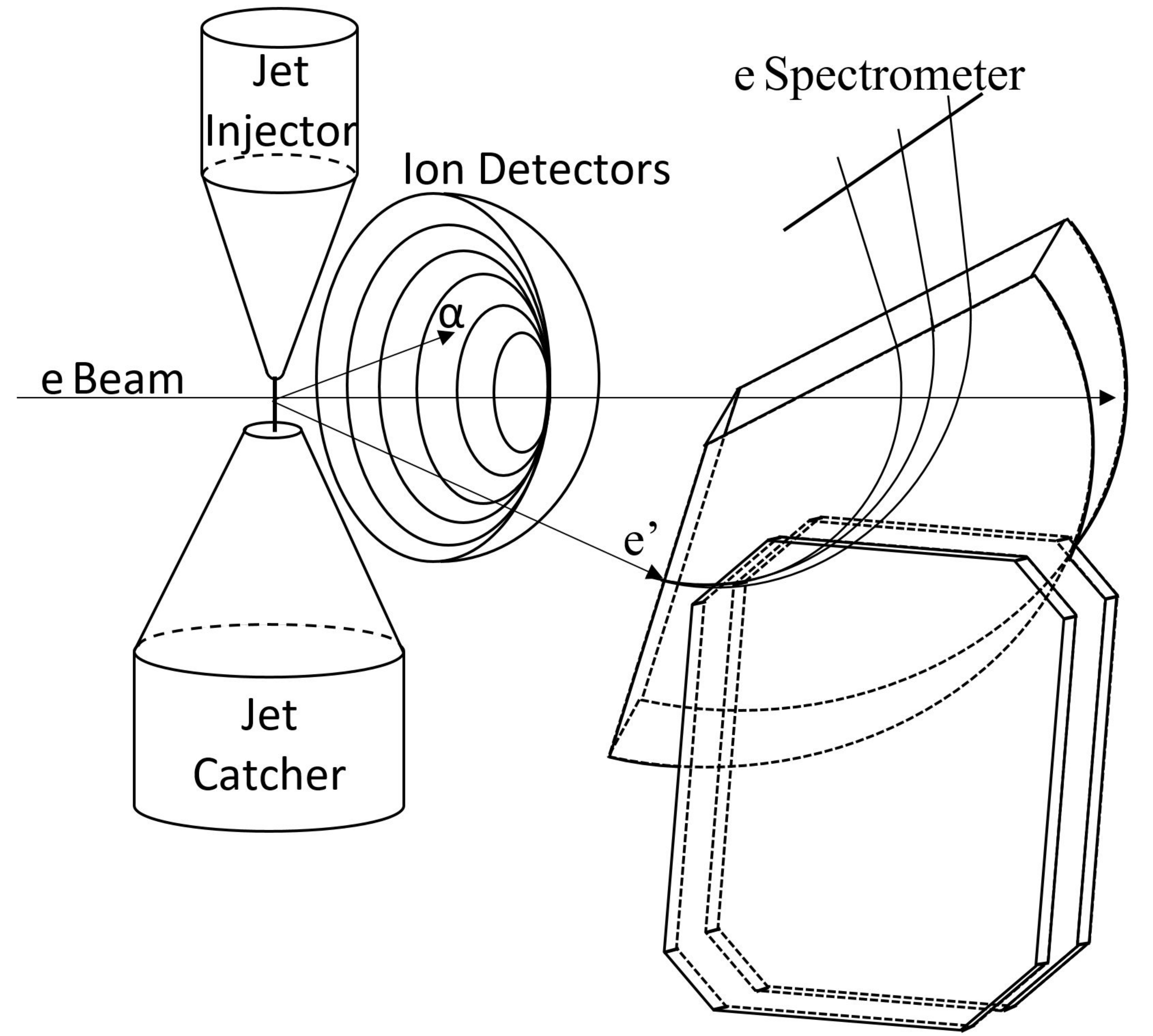} 
\caption{Schematic layout of our proposed $^{16}$O$(e,e^\prime \alpha)^{12}$C experiment:
 $^{16}$O, inside a gas cluster-jet target, is disintegrated 
by the electron beam into $\alpha$-particles and $^{12}$C nuclei. 
The scattered electron is detected in an electron spectrometer 
and the produced $\alpha$-particle in the ion detectors. \label{fig:setup}}
\end{figure}

First, having made exploratory projections using our model, we have 
come to the conclusion that the luminosity should be larger than $10^{35}$ cm$^{-2}$s$^{-1}$, 
but that the density of the target oxygen has to be low enough  
to allow the $\alpha$-particles that exit the target to be detected. 
A suitable target design here is a windowless oxygen 
cluster-jet target, like the one described in \cite{GrieserKhoukaz2018}.
The areal thickness of 2.4$\times$10$^{18}$ atoms/cm$^2$
was measured for ($\approx$ 2 mm wide) hydrogen jet at a gas temperature of 40 K and
gas flow of 40 $l/$min. For our purposes, we will assume one has
an oxygen cluster-jet target capable of achieving an areal thickness 
of  5$\times$10$^{18}$ atoms/cm$^2$, which for a 2 mm wide jet
corresponds to a density of 6.65$\times$10$^{-4}$ g/cm$^3$. 
We also require an electron accelerator which can deliver a beam energy of about 100 MeV and a beam current of at least 10 mA. Two suitable electron accelerators are currently being constructed, namely,
MESA, which should deliver a beam current of 10 mA \cite{Hug2017} 
and CBETA which should be able to go up to 40 mA 
\cite{Trbojevic2017} for beam energies of 42, 78, 114 and 150 MeV (any energy in between should also be possible).
In what follows, we assume a beam current of 40 mA and a jet target
as described above, which is equivalent to a luminosity of 1.25$\times$10$^{36}$ cm$^{-2}$s$^{-1}$.

\begin{figure}[h!]
\centering
\includegraphics[width=7.6cm]{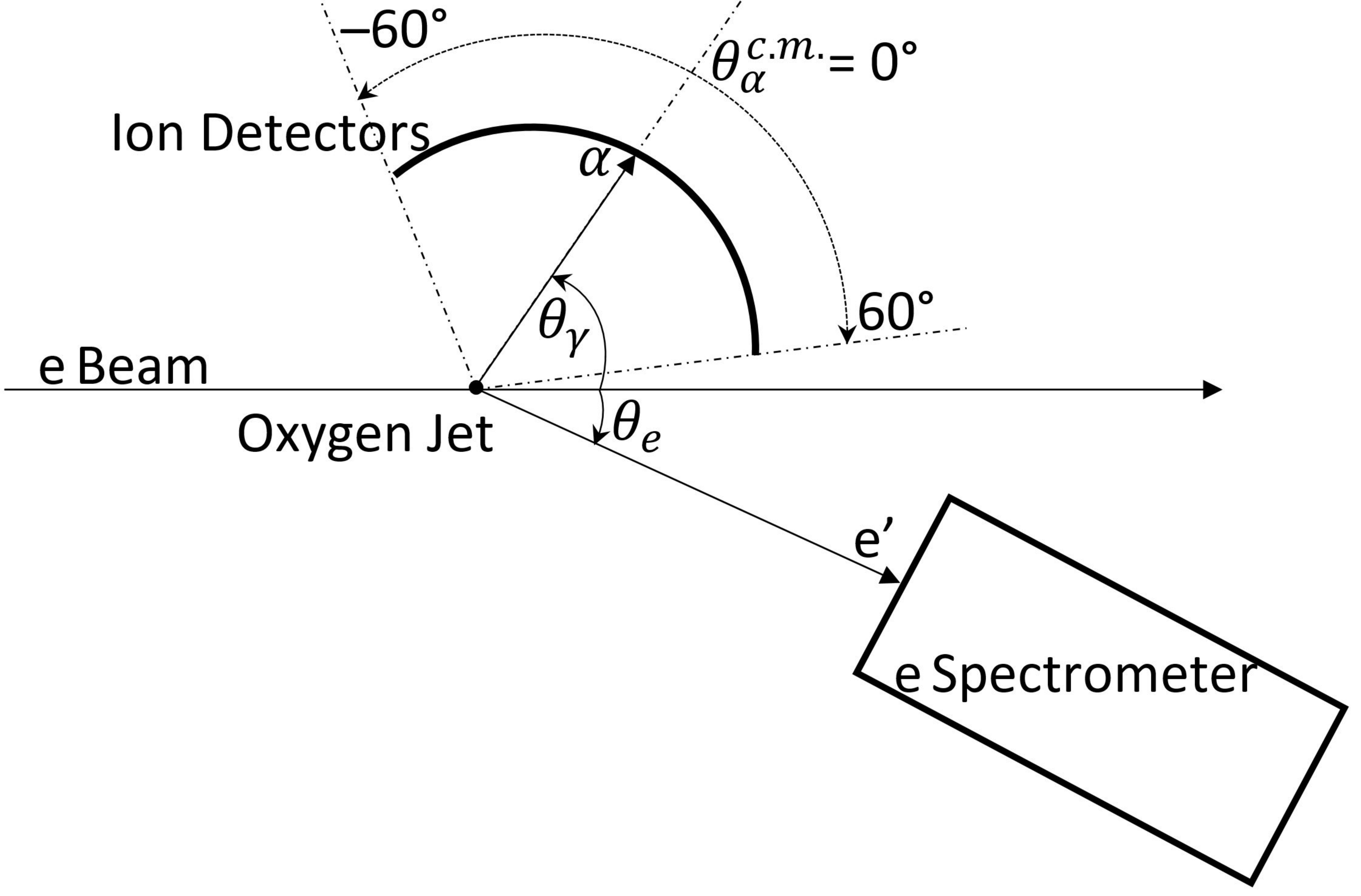} 
\caption{Top-view layout of proposed $^{16}$O$(e,e^\prime \alpha)^{12}$C experiment, showing
the in-plane angular acceptance of the ion detectors. \label{fig:topview}}
\end{figure}

To identify events belonging to the $^{16}$O$(e,e^\prime \alpha)^{12}$C 
reaction we need to detect the scattered electron in coincidence
with the produced $\alpha$-particle. Fig. \ref{fig:setup} shows a schematic layout 
of a possible experiment.

A high precision magnetic spectrometer is suitable for detection of 
the scattered electron. For the purpose of defining electrons accepted
by the electron spectrometer, we will assume that the spectrometer has 
an in-plane acceptance of $ \pm$2.08$^{\circ}$ and out-of-plane acceptance 
of $\pm$4.16$^{\circ}$; this amounts to a solid angle of 10.5 msr. 
\renewcommand{\arraystretch}{1.3}
\begin{table}[h!]
\caption{Summary of experimental parameters for the rate calculation.} 
\centering
\begin {tabular} {l|c|r}
\hline
\multicolumn{3}{c}{Parameters } \\\hline
\multirow{2}{*}{Oxygen Target} & Thickness  &  5$\times$10$^{18}$ atoms/cm$^2$ \\
       & Density   & 6.65$\times$10$^{-4}$ g/cm$^3$   \\ \hline
\multirow{2}{*}{Electron Beam}   & Current  &  40 mA  \\
       & Energies &  78, 114, 150 MeV \\ \hline
\multirow{3}{*}{\shortstack[l]{Electron arm \\ acceptance}} & In-plane  &  $ \pm$2.08$^{\circ}$  \\ 
                      & Out-of-plane  & $\pm$4.16$^{\circ}$ \\ 
                      & Solid angle  &  10.5 msr \\ \hline  
\multirow{3}{*}{\shortstack[l]{$\alpha$-particle arm \\acceptance}} & In-plane  &  60$^{\circ}$  \\ 
                      & Out-of-plane  & 360$^{\circ}$ \\ 
                      & Solid angle  &  3.14 sr \\ \hline
\multicolumn{2}{l|}{Luminosity }   &   1.25$\times$10$^{36}$ cm$^{-2}$s$^{-1}$ \\ \hline
\multicolumn{2}{l|}{Integrated Luminosity (100 days)} & 1.08$\times$10$^7$ pb$^{-1}$ \\ \hline
\multicolumn{2}{l|}{Central electron scattering angles} &  15$^{\circ}$, 25$^{\circ}$, 35$^{\circ}$\\ \hline
\multicolumn{2}{l|}{$E^{c.m.}_{\alpha}$-range of interest}   &   $0.7 \le E_{\alpha}^{c.m.} \le 1.7$ MeV \\ \hline 
\end{tabular} 
\label{Tab1}
\end{table}

Since we want to obtain S-factors close to the Gamow energy (300 keV), we will need to
deal with $\alpha$-particles having very low kinetic energy $E_{\alpha}^{lab}$, (see Fig. \ref{fig:virtualq}),
where the energy loss in the target and the multiple scattering in the target material
play important roles as shown in Fig. \ref{fig:sigmaAng}. In order to select $\alpha$-particles
with reasonable energy and angular spread one should either reduce the density
of the gas jet or set a cut on the minimum accepted kinetic energy $E_{\alpha}^{Lab}$,
{\it i.e.}, to accept $\alpha$-particles within a certain range
around the direction of the virtual photon. We decided to go with the second 
option and set a cut to accept $\alpha$-particles having a kinetic energy 
$E_{\alpha}^{Lab} \geq 0.55$ MeV. This cut also imposes a limit on the maximal accepted
in-plane scattering angle $\theta^{c.m.}_{\alpha}$, and to cover all settings listed in
Table~\ref{Tab1} within an equal angular range, only $\alpha$-particles having an in-plane scattering 
angle $\theta^{c.m.}_{\alpha}$ in the range from 0$^{\circ}$ to 60$^{\circ}$ were accepted. 
For the out-of-plane angle $\phi_{\alpha}$, the full acceptance
from 0$^{\circ}$ to 360$^{\circ}$ was assumed. Note that by selecting the full range of  $\phi_{\alpha}$, the integral of interference response functions $R_{TT}$ and 
$R_{TL}$  over $\phi_{\alpha}$ will be equal to zero and
only longitudinal $R_{L}$ and transverse $R_{T}$ response functions will contribute to the total cross section.  
Figure~\ref{fig:topview} shows a top-view layout of the experiment.

Table \ref{Tab2} summarizes the assumptions for the parameters used in the
differential cross section (Eq.~(\ref{eq:diffcross})) used for the calculation of the
rate and subsequent statistical uncertainties.

\begin{table}[h!]
\caption{Summary of theoretical assumptions for the rate calculation.} 
\centering
\begin {tabular} {l|l|l}
\hline
\multicolumn{3}{c}{Assumptions} \\\hline 
\multirow{2}{*}{$b'_{CJ,EJ}$} & Value & $ \approx 1$ for $J=1,2$ \\ 
                             & Sign  & "$+$"  for $J=1,2$  \\  \hline
\multirow{3}{*}{$c'_{C0} \equiv a'_{C0}\times b'_{C0}$} & Value of $b'_{C0}$ &  $ \approx 1$  \\ \hhline{*{1}~|--}
                             & \multirow{2}{*}{Value of $a'_{C0}$}  & $= a'_{E2}$, Case A \\ \hhline{*{2}~|-}
                             &                                     & $=0.5 a'_{E2}$, Case B \\ \hline
$t_{C0}$ & Sign & "$+$" \\  \hline
\multicolumn{3}{l}{In $E^{c.m.}_{\alpha}$-region of interest only the Coulomb phase } \\
\multicolumn{3}{l}{ contributes.} \\\hline 
\end {tabular} 
\label{Tab2}
\end{table}

\subsection{Estimation of Event Rates}

\begin{figure}[h!]
\centering
\includegraphics[width=7.6cm]{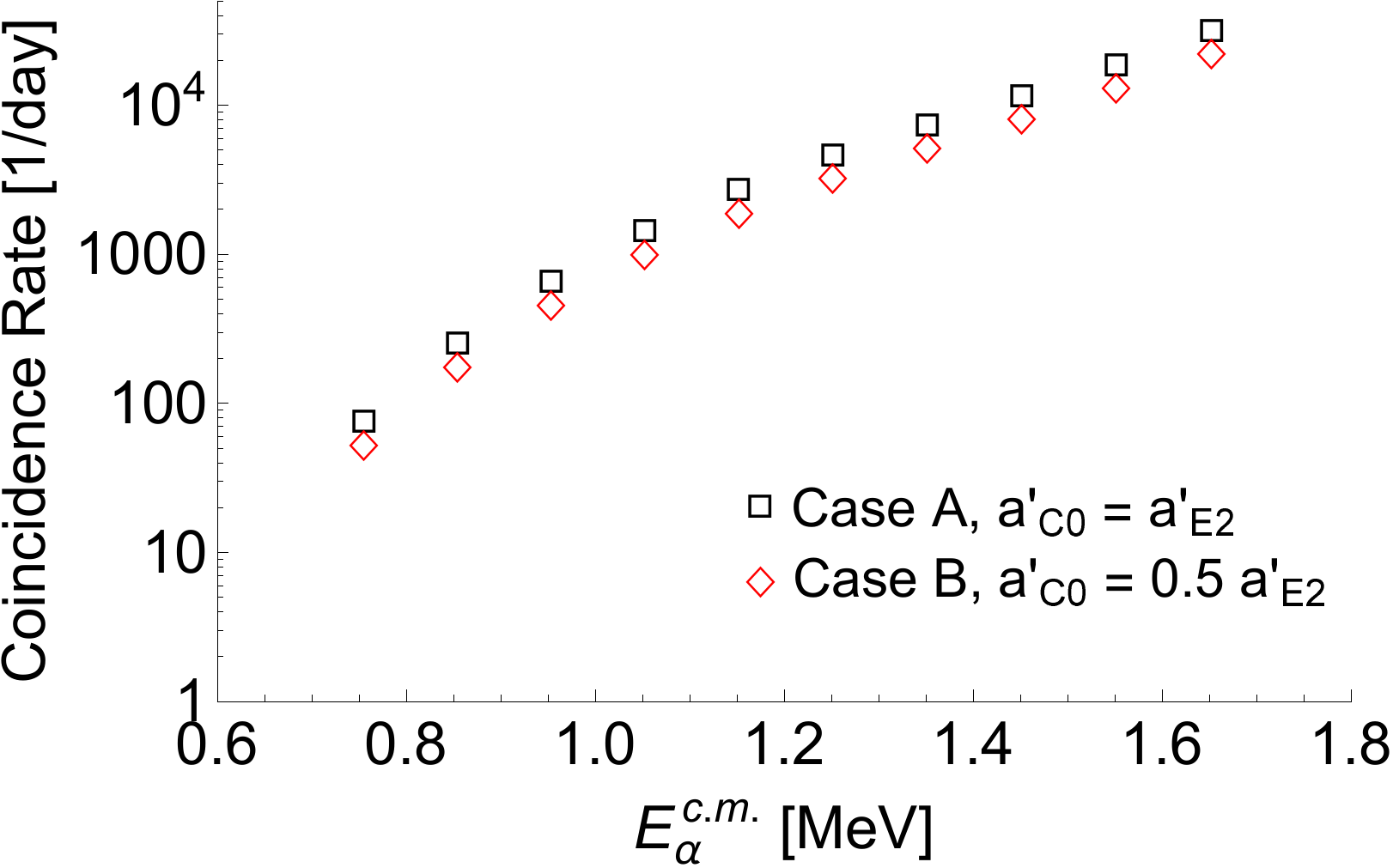} 
\colorcaption{Coincidence rate per day for $t_{C0}$ Cases A and B, 
for electron beam energy of $E_{e} = $ 114 MeV, central electron scattering
angle of $\theta_e = $ 15$^{\circ}$, electron spectrometer acceptance 
of 10.5 msr, $\alpha$-particle detector acceptance of  3.14 sr   
and luminosity of 1.25$\times$10$^{36}$ cm$^{-2}$s$^{-1}$.\label{fig:crpd}}
\end{figure}

Since it is difficult to calculate the rate analytically,
we have carried out a numerical simulation of the conceptual experiment illustrated in Fig.~\ref{fig:setup}. By using Monte Carlo integration and explicit experimental parameters (see Table~\ref{Tab1})
and theoretical assumptions (see Table~\ref{Tab2}), 
we have estimated the rate of the coincidences
per day in the energy range  $0.7 \le E_{\alpha}^{c.m.} \le 1.7$ MeV
divided into 100 keV wide bins; see Fig.~\ref{fig:crpd}. 
For $t_{C0}$ Case A the coincidence rate ranged from 73 day$^{-1}$ up to 
30602 day$^{-1}$, and for Case B from 55 day$^{-1}$ up to 23123  day$^{-1}$.
In total, the coincidence rate of $t_{C0}$ Case A is
$\approx$32\% larger than for Case B.

\begin{figure*}[t!]
\centering
\includegraphics[width=6.90cm]{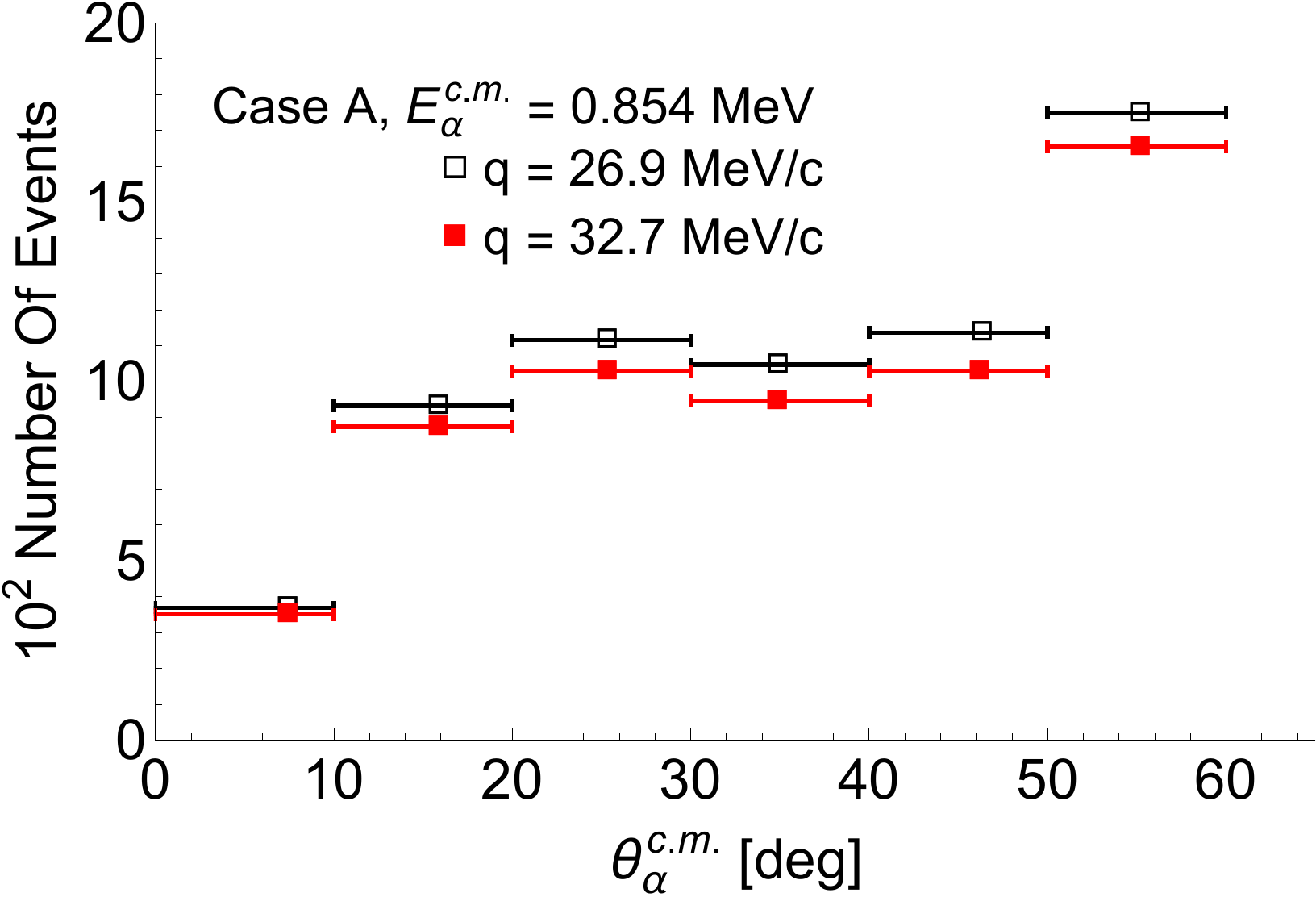}
\includegraphics[width=6.90cm]{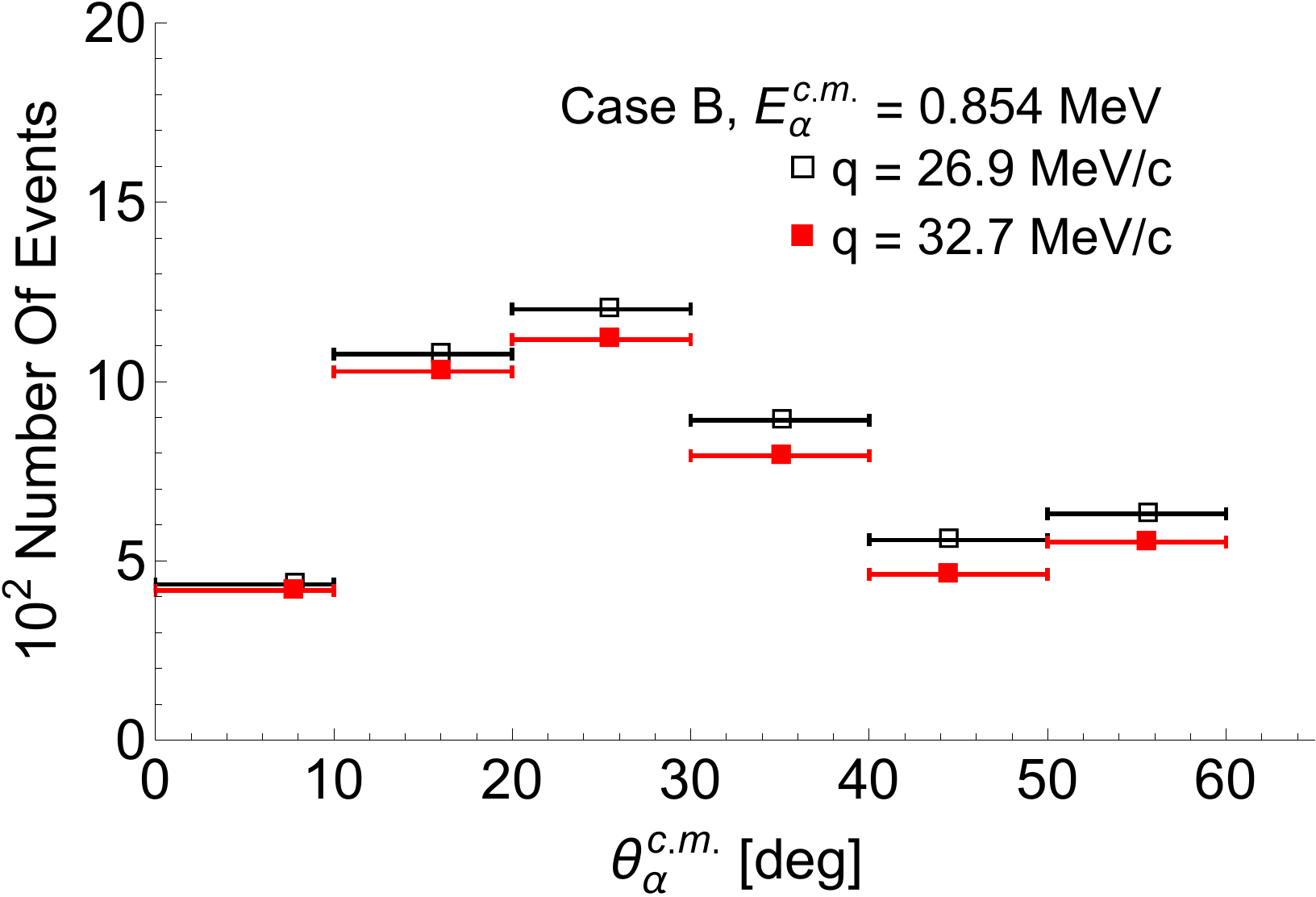}
\includegraphics[width=6.90cm]{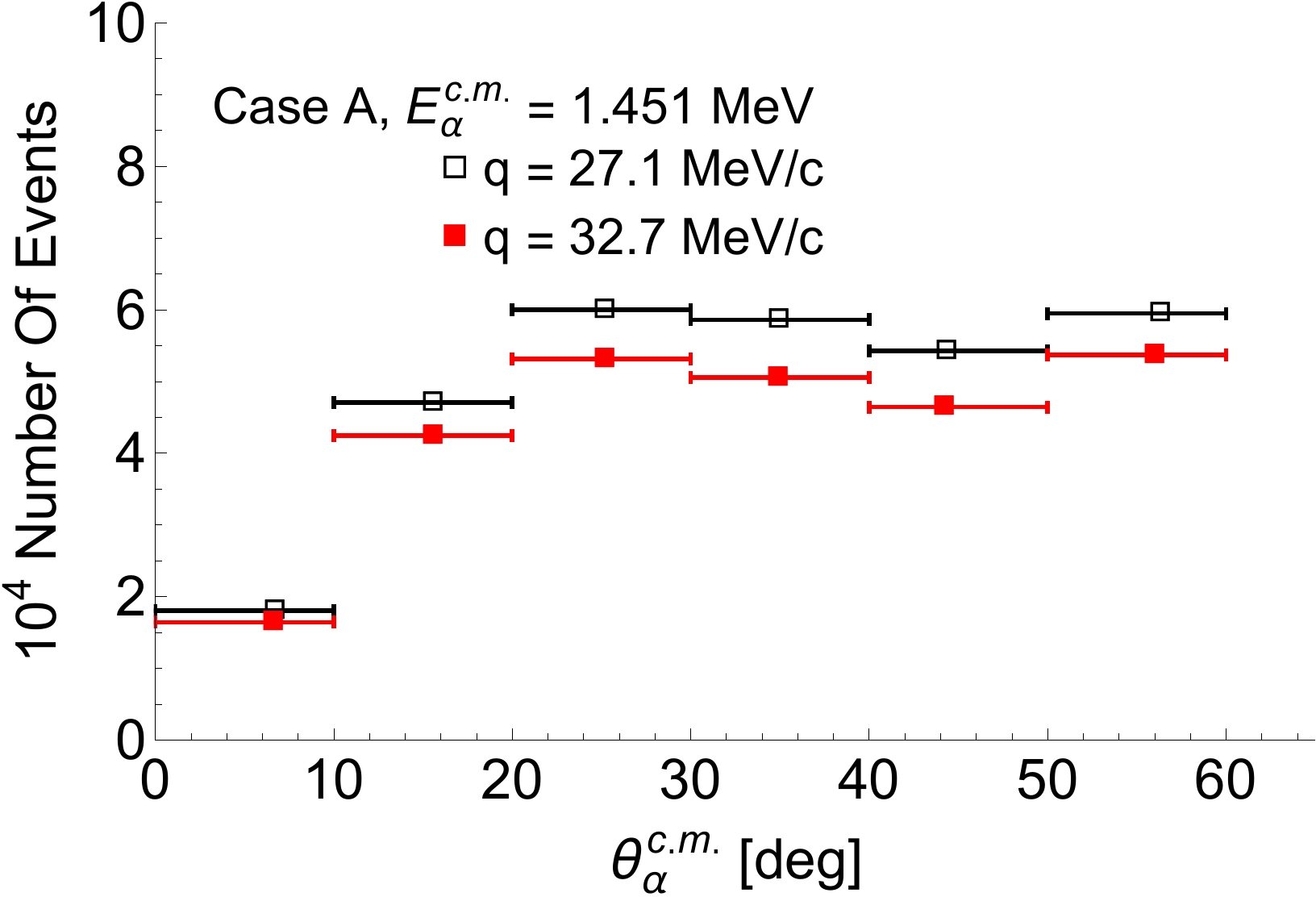}
\includegraphics[width=6.90cm]{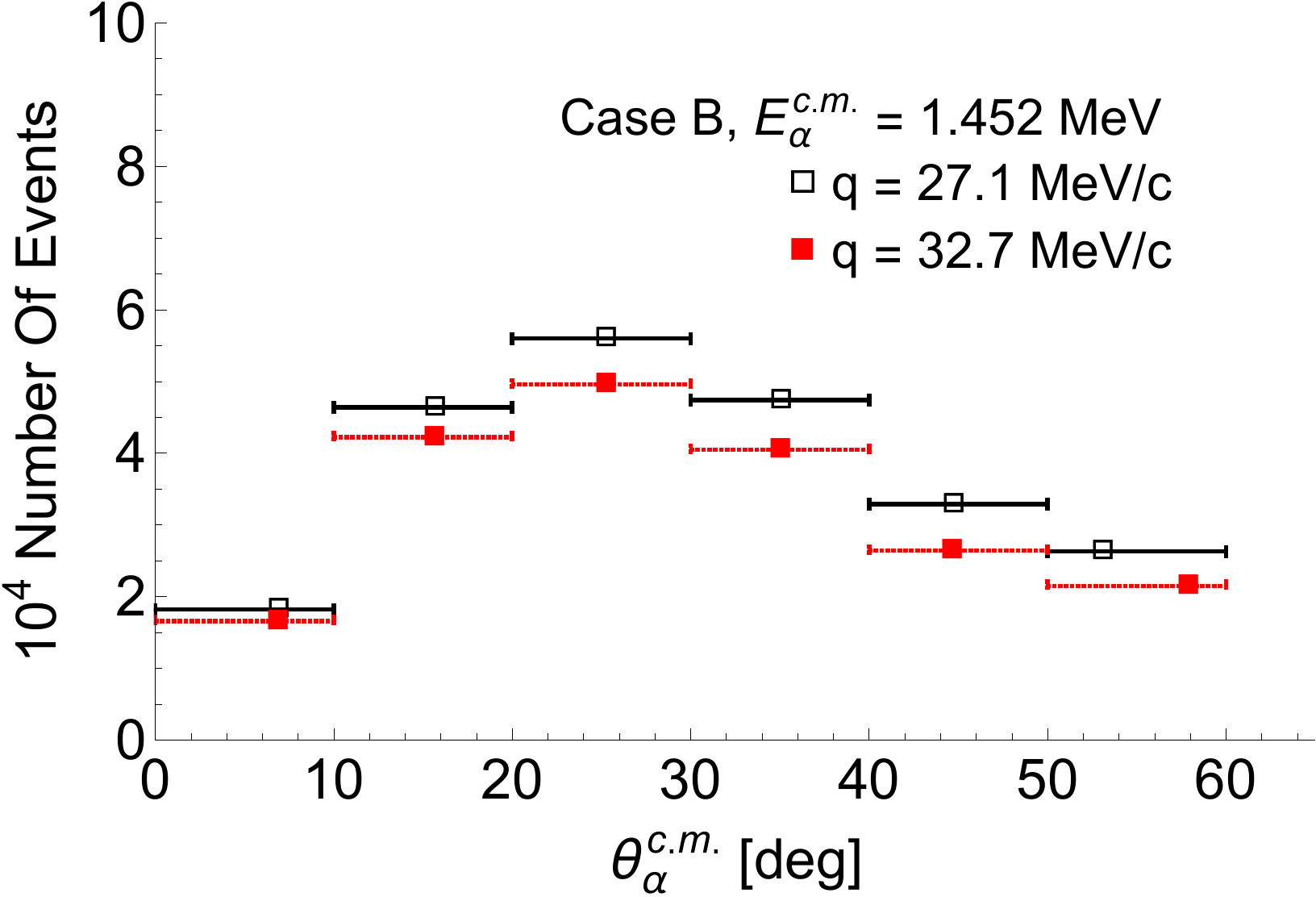} 
\vspace{-0.3cm}
\colorcaption{Number of events as a function of  $\theta^{c.m.}_{\alpha}$ assuming
100 days of data taking at the luminosity of $2.5\times 10^{36}$ cm$^{-2}$s$^{-1}$.
The left column represents $t_{C0}$ Case A and right one Case B. 
Horizontal bars denote the width of the $\theta^{c.m.}_{\alpha}$-bin
which here is equal to 10$^\circ$. Horizontal placement of the data
point within a bin was done according to the procedure recommended 
in \cite{Lafferty1995}. The $q$-bins are 1.91 MeV/c wide and
$E_{\alpha}^{c.m.}$-bins 100 keV.  For all events inside a particular $q$- and $E_{\alpha}^{c.m.}$-bin, the 
specified $q$-values represent the average value, and the 
stated $E_{\alpha}^{c.m.}$-values represent the average of the expected and averaged $E_{\alpha}^{c.m.}$-value.\label{fig:noe}}
\end{figure*}
\begin{figure*}[t!]
\centering
\includegraphics[width=6.90cm]{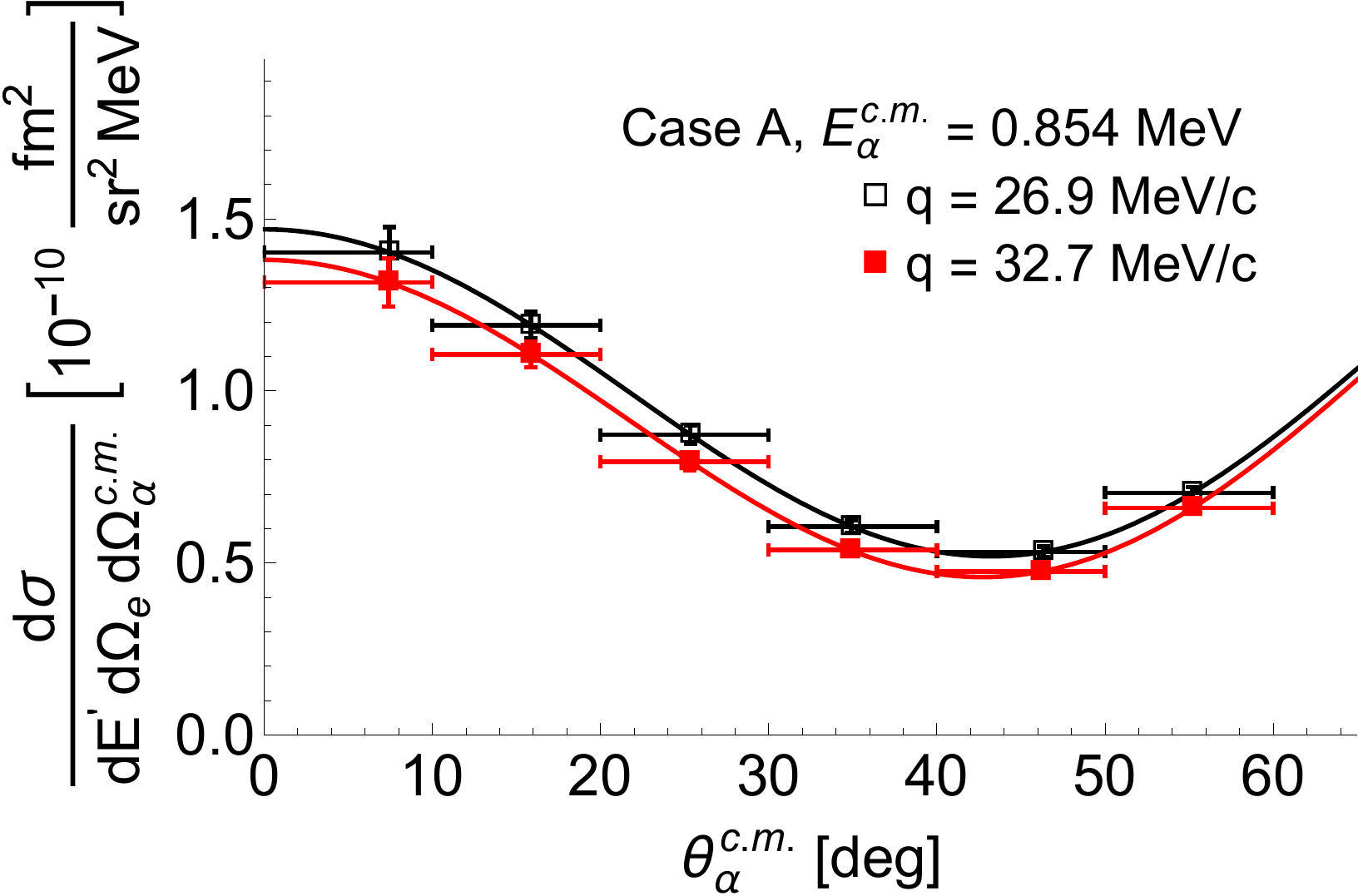}
\includegraphics[width=6.90cm]{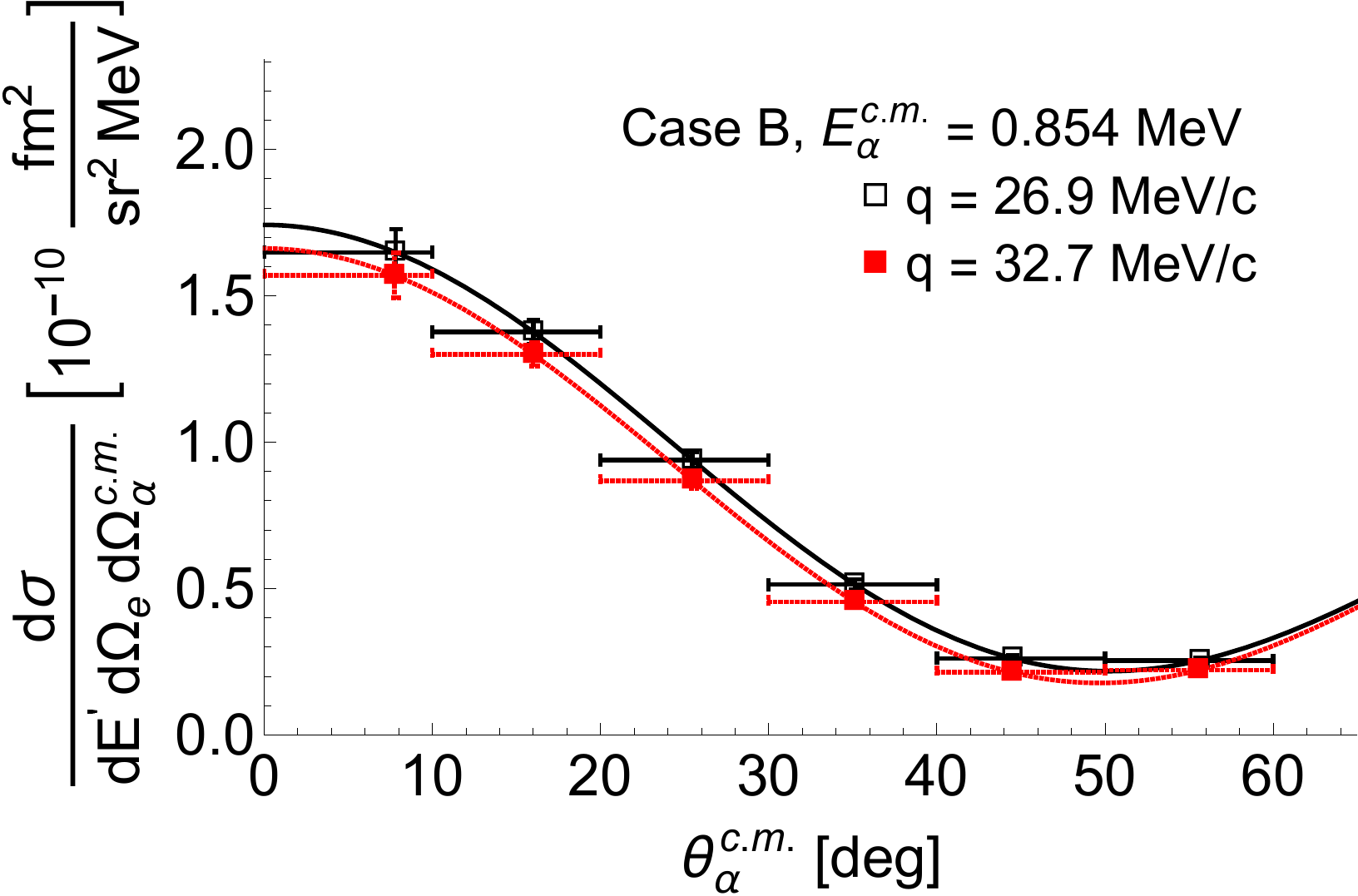}
\includegraphics[width=6.90cm]{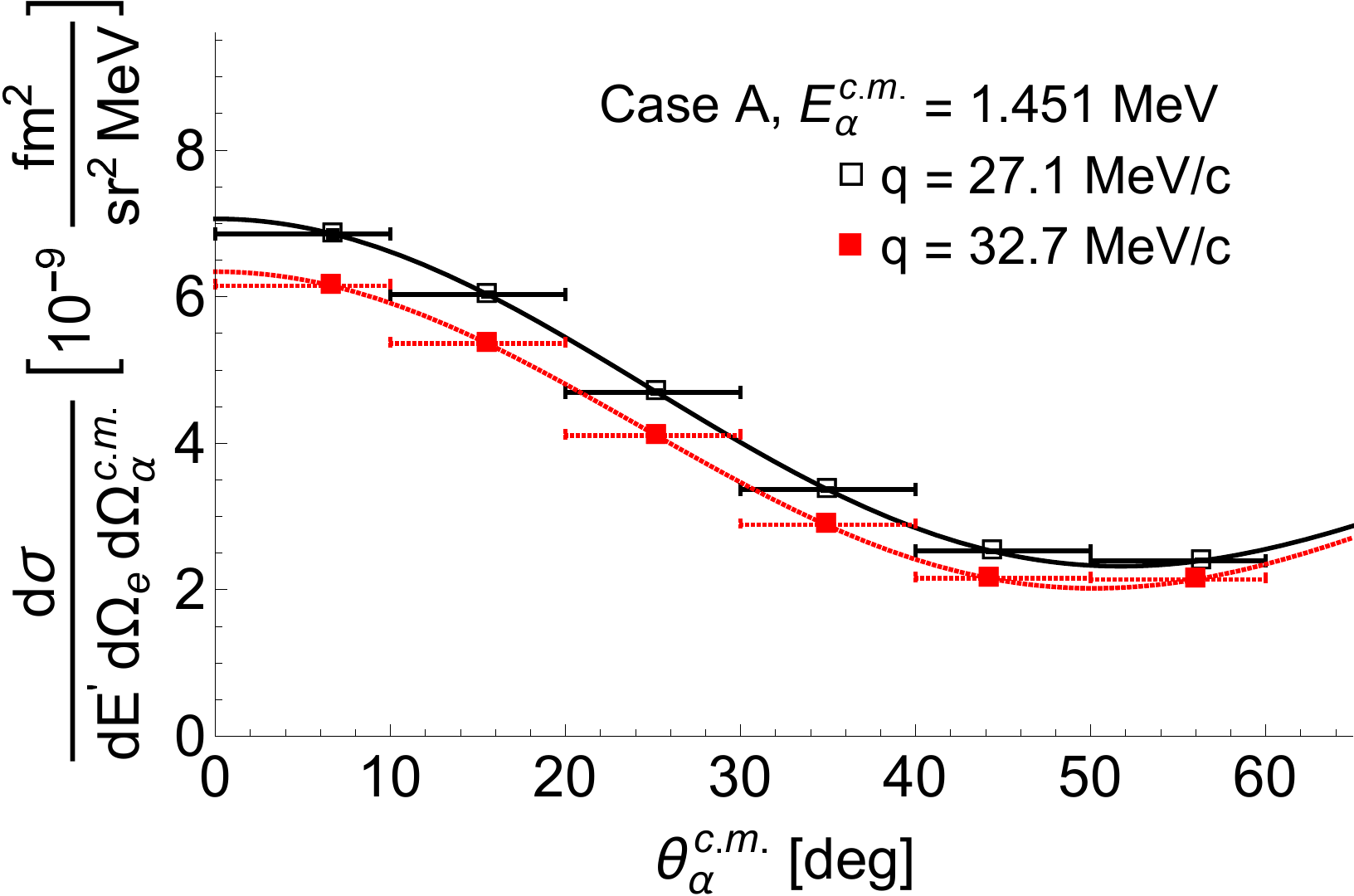}
\includegraphics[width=6.90cm]{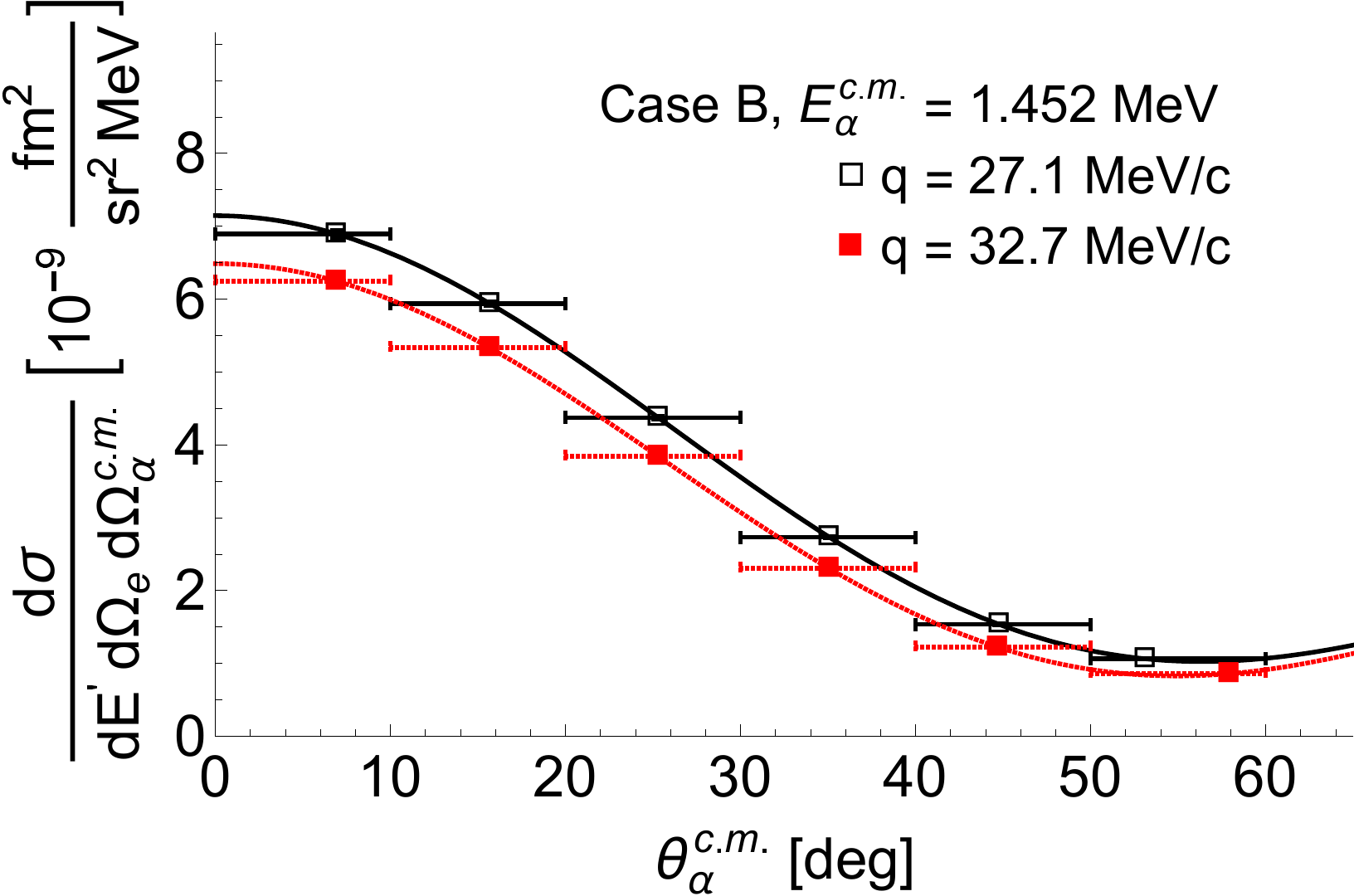} 
\vspace{-0.3cm}
\colorcaption{Differential cross section as function of  $\theta^{c.m.}_{\alpha}$.
The left column represents $t_{C0}$ Cases A and right Case B. 
Vertical bars correspond to statistical uncertainties 
assuming 100 days of data taking at the luminosity of $2.5\times 10^{36}$ cm$^{-2}$s$^{-1}$.
The same binning procedure was performed as described in caption of Fig. \ref{fig:noe}.
\label{fig:diffstat}}
\end{figure*}

In order to mimic the data treatment in a real experiment,
the accepted events in the energy range $0.7 \le
E_{\alpha}^{c.m.} \le 1.7$ MeV were placed in 100 keV wide bins,
for which, as shown in figure \ref{fig:back}, it is possible
to identify the $\alpha$-particles from electrodisintegration
of $^{16}$O and fully separate them from the background. 
Additionally, the full range of accepted electron scattering 
angle $\theta_e$ was divided into four bins corresponding to four 
different $q$-values, and events in each $\theta_e$-bin 
were finally sorted into six $\theta_{\alpha}^{c.m.}$-bins ranging from 
0$^{\circ}$ to 60$^{\circ}$. An example of the
sorting can be seen in Fig. \ref{fig:noe}. The rate 
was converted into the number of events collected over
100 days by multiplying it with the integrated luminosity of 
1.08$\times$10$^7$pb$^{-1}$. 

The number of events per bin was used to calculate the corresponding
statistical uncertainty and this is the quantity for which
we performed the above described procedure, since it determines
how large an advantage one might have measuring the electrodisintegration 
of $^{16}$O compared with previous experiments.

\subsection{Estimated Uncertainties in Determination of Astrophysical S-Factors}

Now that we have determined the angular distribution of the number of events,
we can proceed to predict the astrophysical S-factors, with associated uncertainties.
First, the event distribution is converted back
into the differential cross section distribution by dividing it with 
the Monte Carlo integrated phase space covered by each bin
and the integrated luminosity, but now including the statistical uncertainties;
see Fig. \ref{fig:diffstat}. 
\begin{figure*}[t]
\centering
\includegraphics[width=7.0cm]{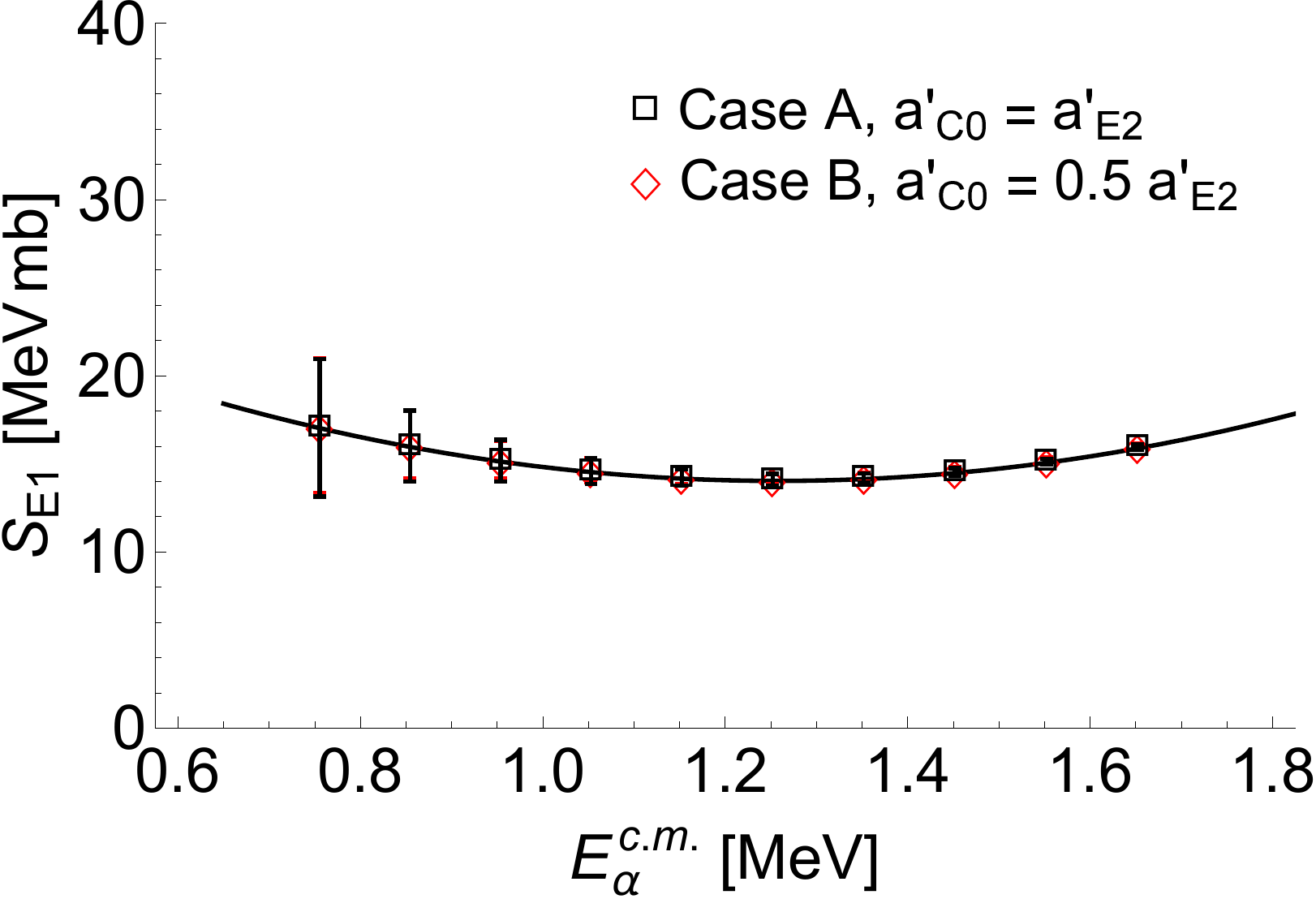}
\includegraphics[width=7.0cm]{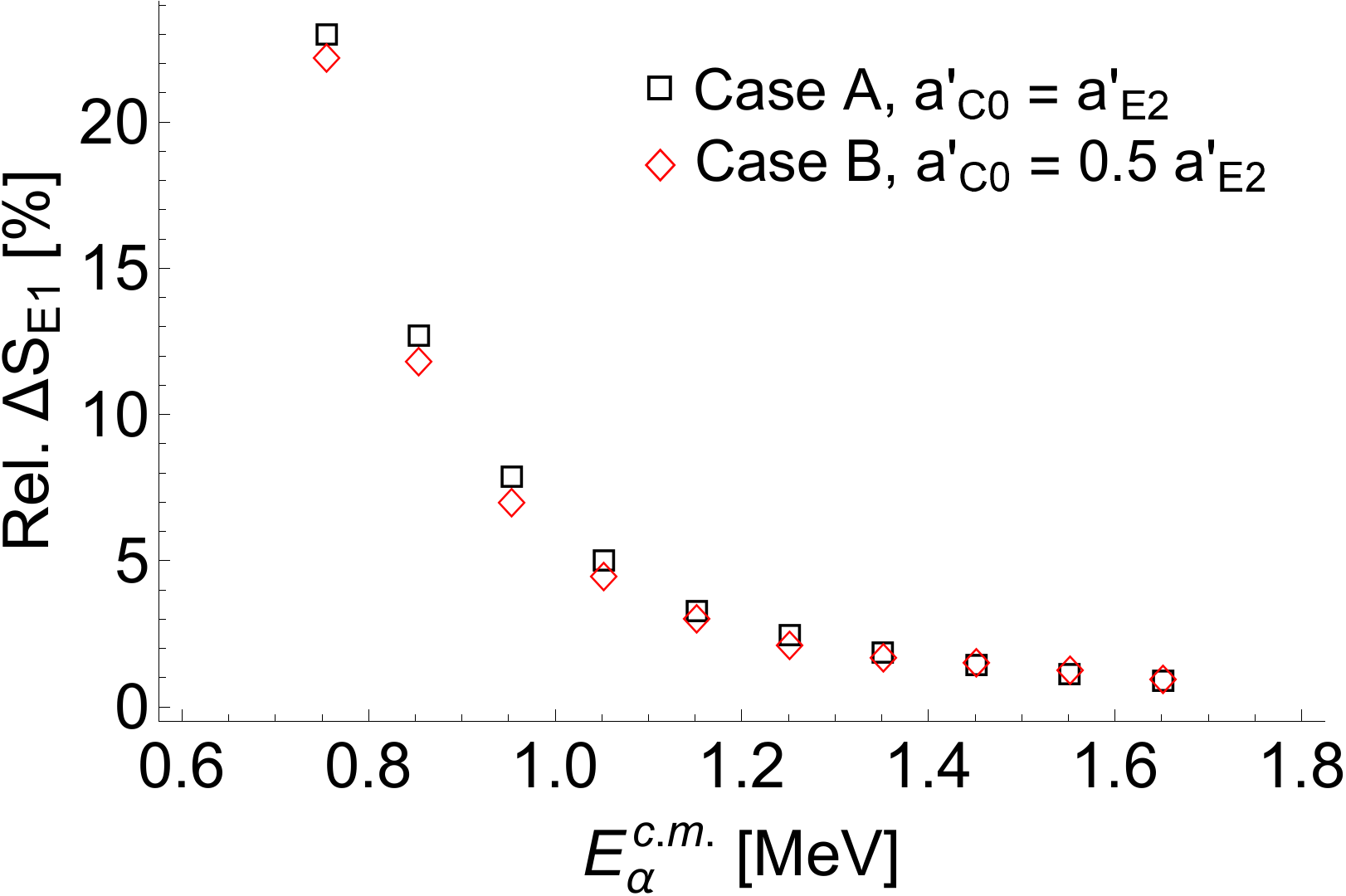}
\includegraphics[width=7.0cm]{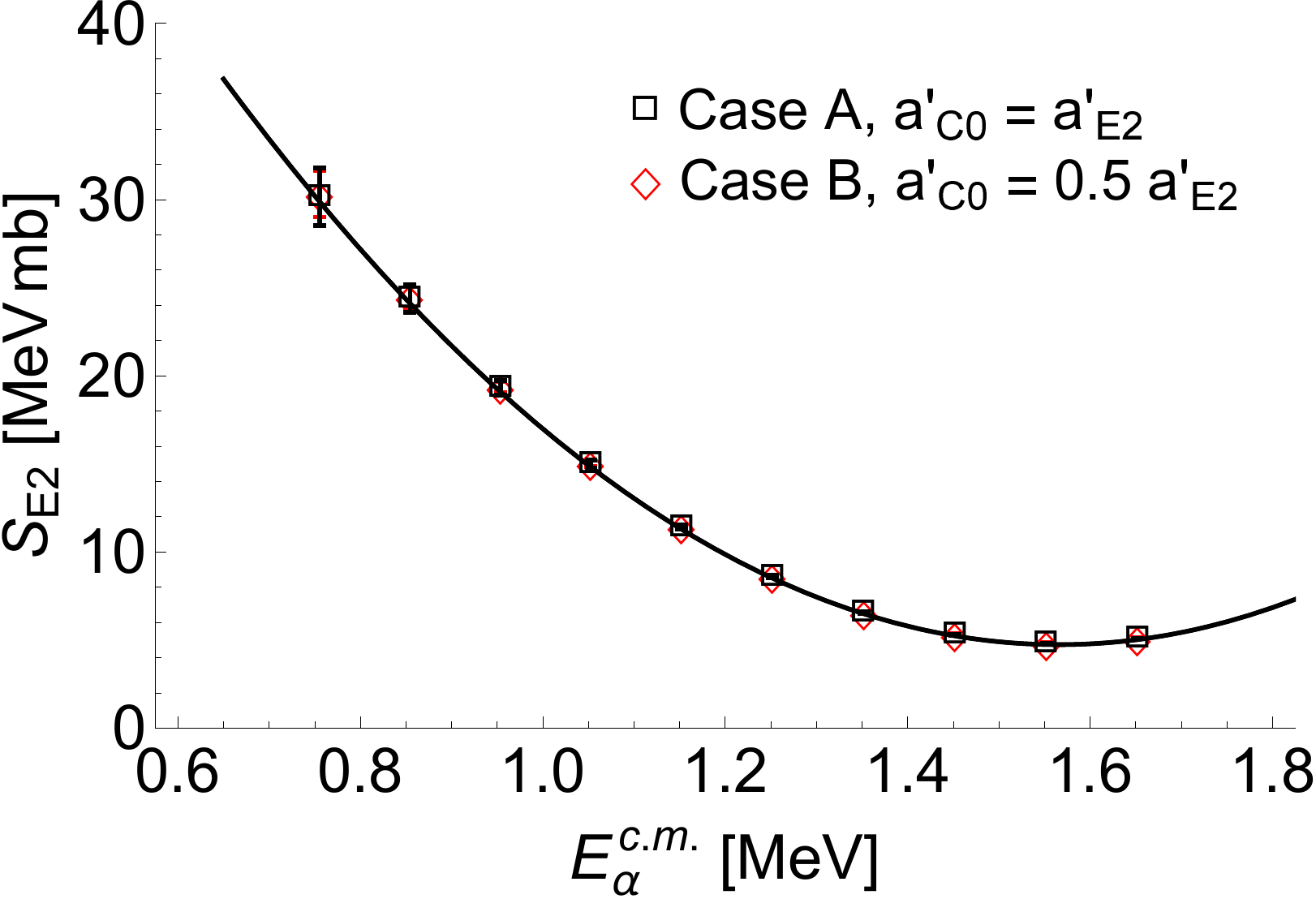}
\includegraphics[width=7.0cm]{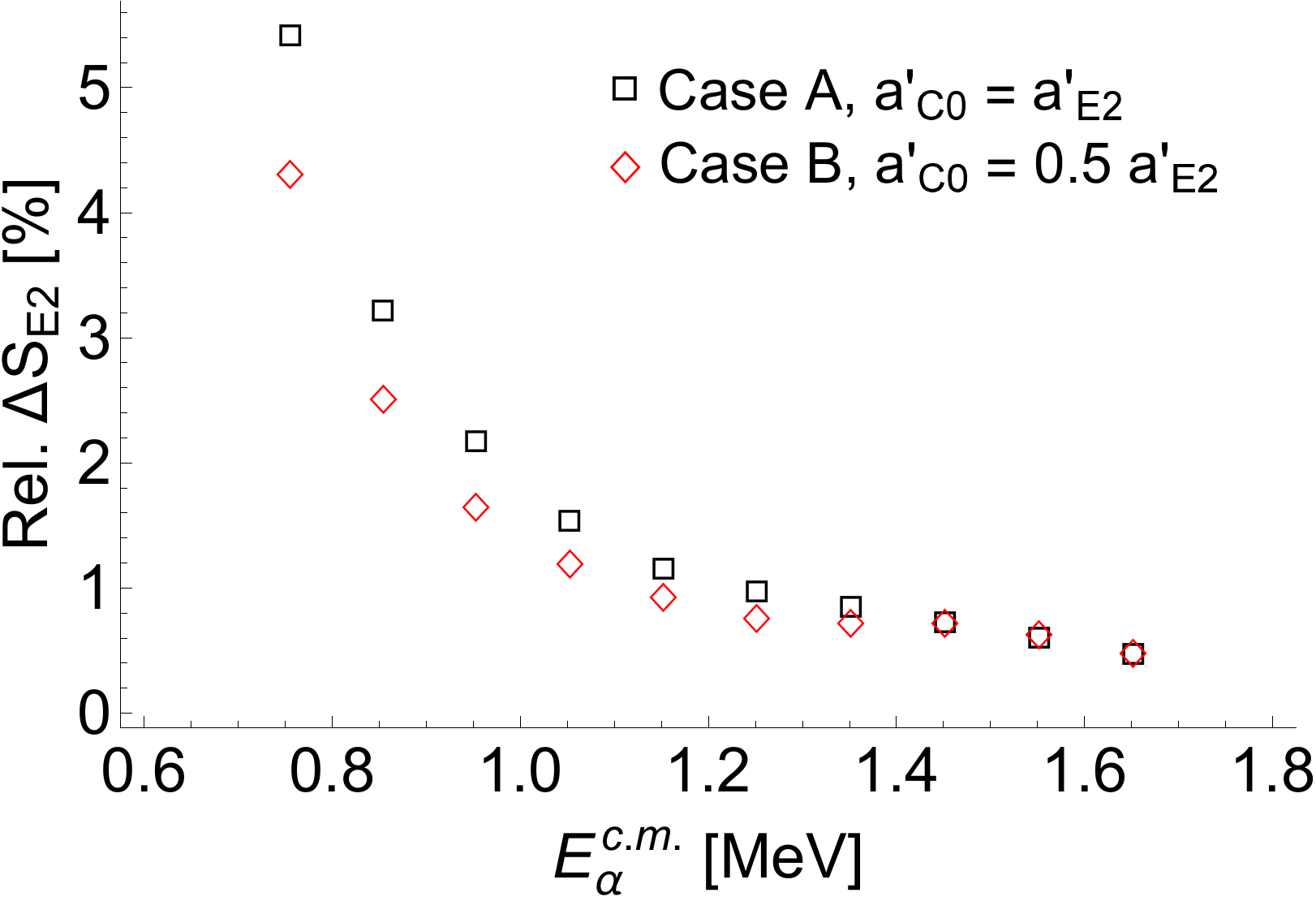}
\includegraphics[width=7.0cm]{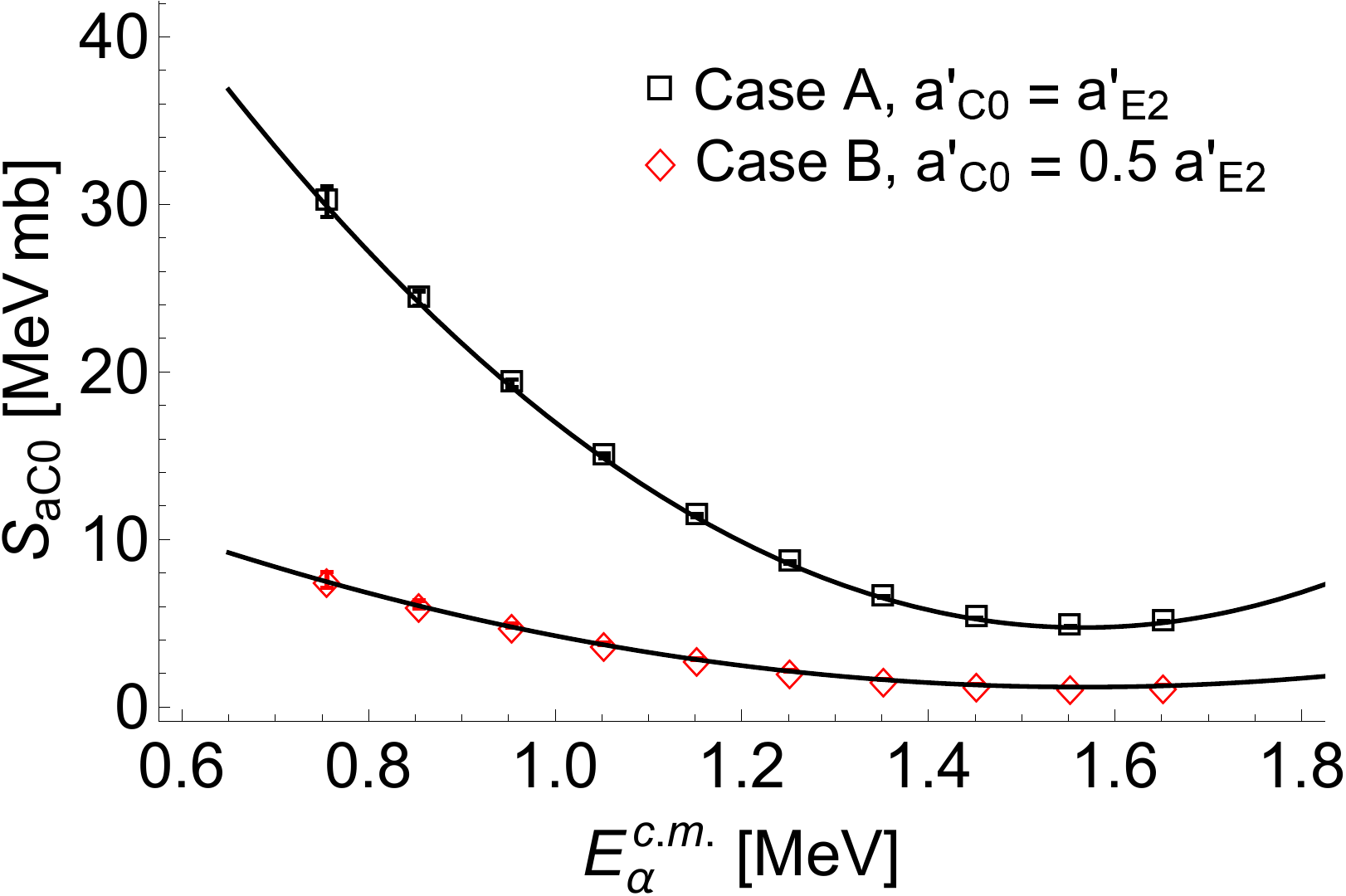}
\includegraphics[width=7.0cm]{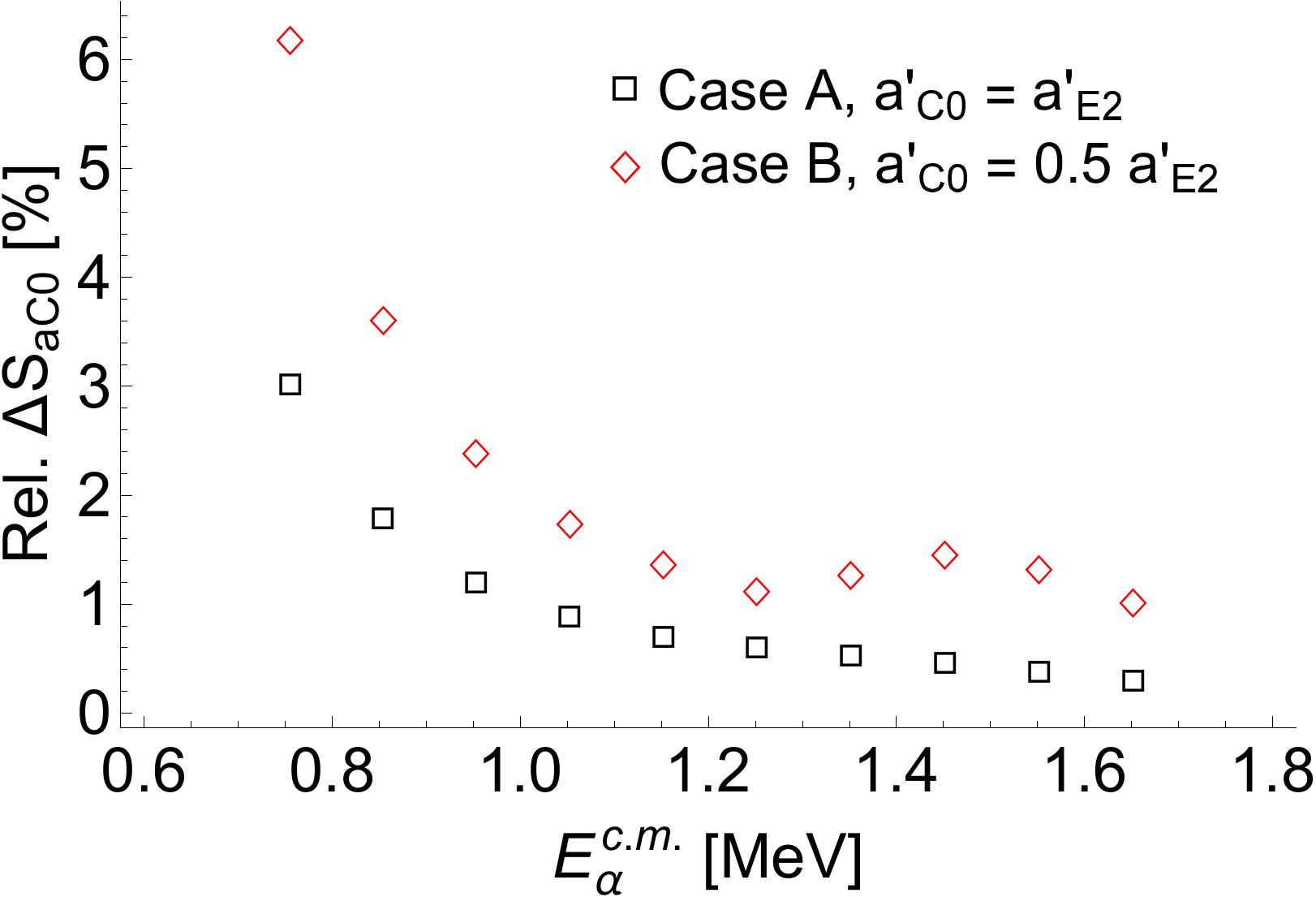}
\colorcaption{Reconstructed astrophysical $S_{E1}$- and $S_{E2}$-factors, showing their absolute (left column) and relative uncertainties (right column)
for $t_{C0}$ Case A and B, $E_{e} = $ 114 MeV and $\theta_e = $ 15$^{\circ}$. $S_{aC0}$ does not have an astrophysical
counterpart and it just a conversion of the third fitting parameter
into an S-factor and corresponding uncertainty in order to put it in
perspective with $S_{E1}$ and $S_{E2}$. \label{fig:Srecon}}
\end{figure*}

The Levenberg-Marquardt method was used to extract three fitting parameters $a^{'}_{E1}$,  $a^{'}_{E2}$
and  $a^{'}_{C0}$ (from the Coulomb monopole) from the data, as well as their uncertainties. 
At given $E_{\alpha}^{c.m.}$-bin we obtained four values for each fitting parameter
originating from four $q$-bins, which were combined together by taking the average 
value of each parameter and by calculating their total uncertainty.  
The last step is to invert Eq.~(\ref{eq:aE1aE2}) and for each 
$E_{\alpha}^{c.m.}$-bin calculate the $S_{E1}$- and $S_{E2}$-factors and their
uncertainties; see Fig. \ref{fig:Srecon} as an example for $t_{C0}$ Case A and Case B
at  $E_{e} =$ 114 MeV and $\theta_e =$ 15$^{\circ}$. When we compare $t_{C0}$ Cases A and B, the 
value of $a'_{C0}$ has a minor effect ($\sim$3\%) on the uncertainties in
$S_{E1}$. For the same comparison, the relative uncertainties in $S_{E2}$ are
approximately 25\% larger in Case A, but the uncertainties in 
$S_{aC0}$ in Case B are twice as large as in Case A. The "bump" in the
relative uncertainties of $S_{aC0}$ Case B is caused by fluctuation 
in $\theta_{\alpha}^{c.m.}$-position of the data point inside the last bin
$50^{\circ}<\theta_{\alpha}^{c.m.}<60^{\circ}$.

\begin{figure}[t]
\centering
\includegraphics[width=8.0cm]{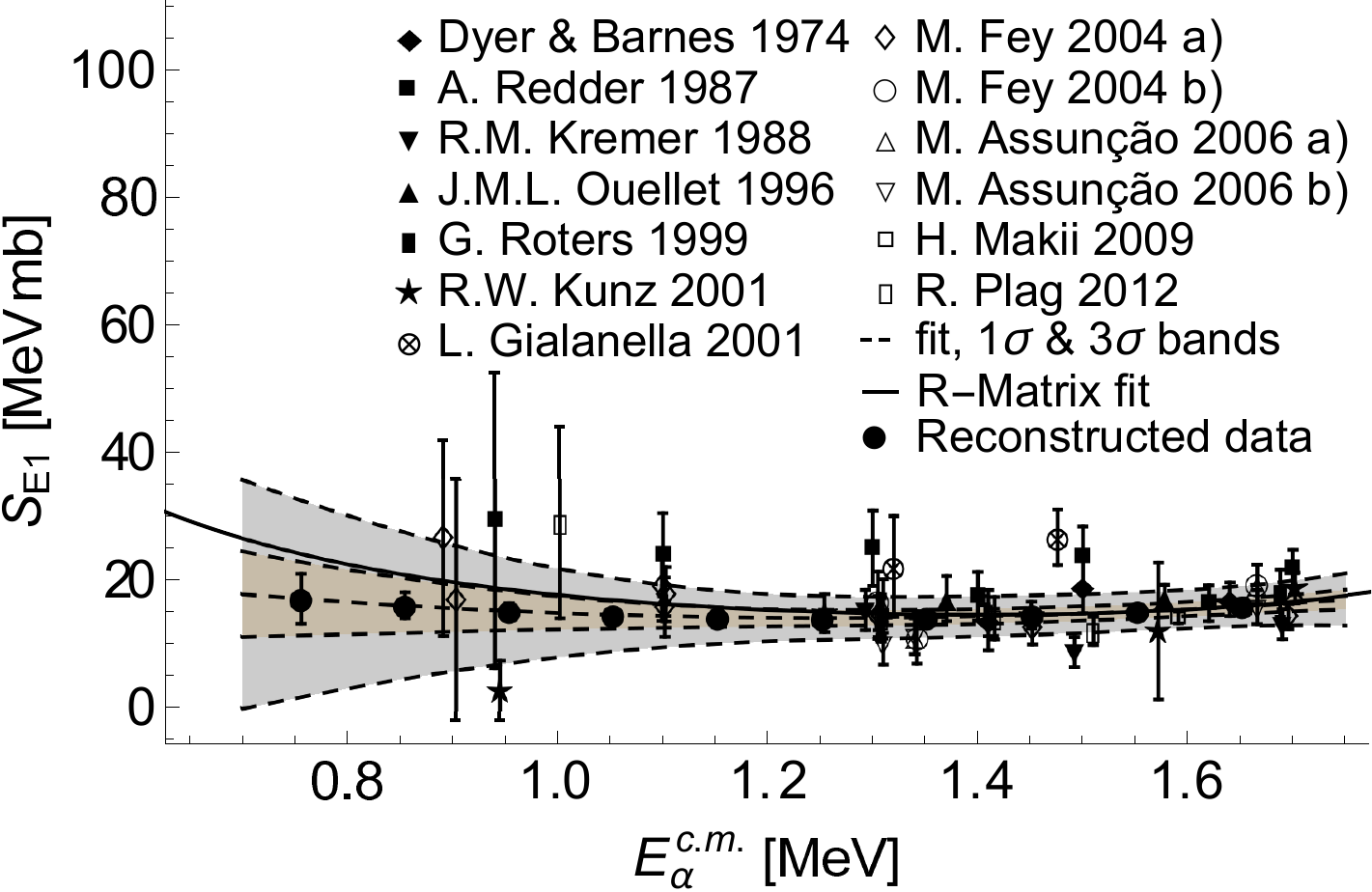}
\includegraphics[width=8.0cm]{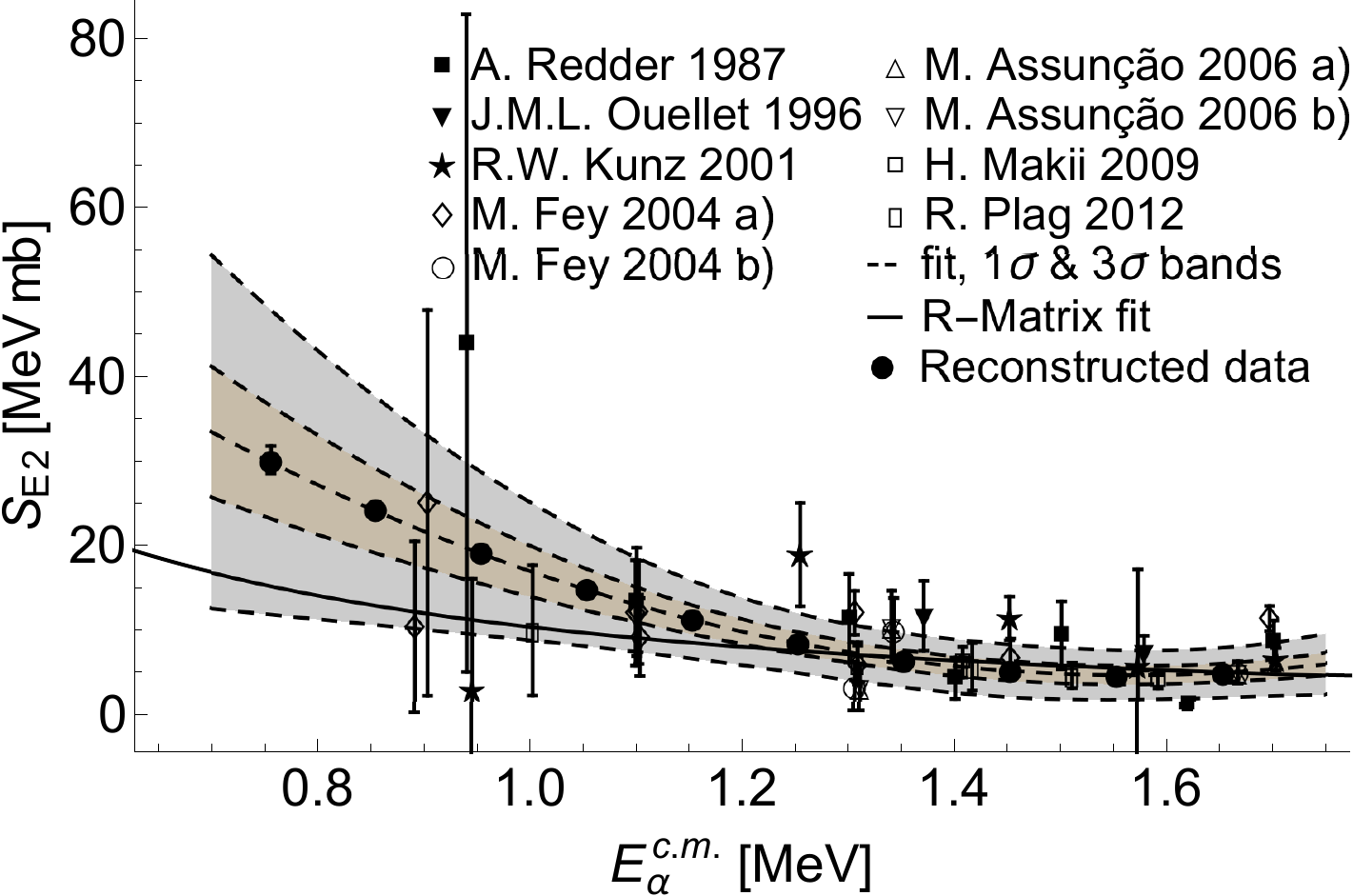} 
\caption{Reconstructed astrophysical $S_{E1}$- and
$S_{E2}$-factors with statistical 
error bars (represented by solid circles) from our calculation
for $E_{e} = $ 114 MeV, $\theta_e = $ 15$^{\circ}$, $t_{C0}$ Case A, together with experimental data from \cite{DyerBarnes1974, Redder1987, Kremer1988,
Ouellet1996,Roters1999,Kunz2001,Kunz2002phd,Gialanella2001,
Fey2004phd,Assuncao2006,Makii2009,Plag2012} and Azure2 R-Matrix fit \cite{Azuma2010}. \label{fig:SECalc}}
\end{figure}

\begin{figure}[t!]
\centering
\includegraphics[width=8.0cm]{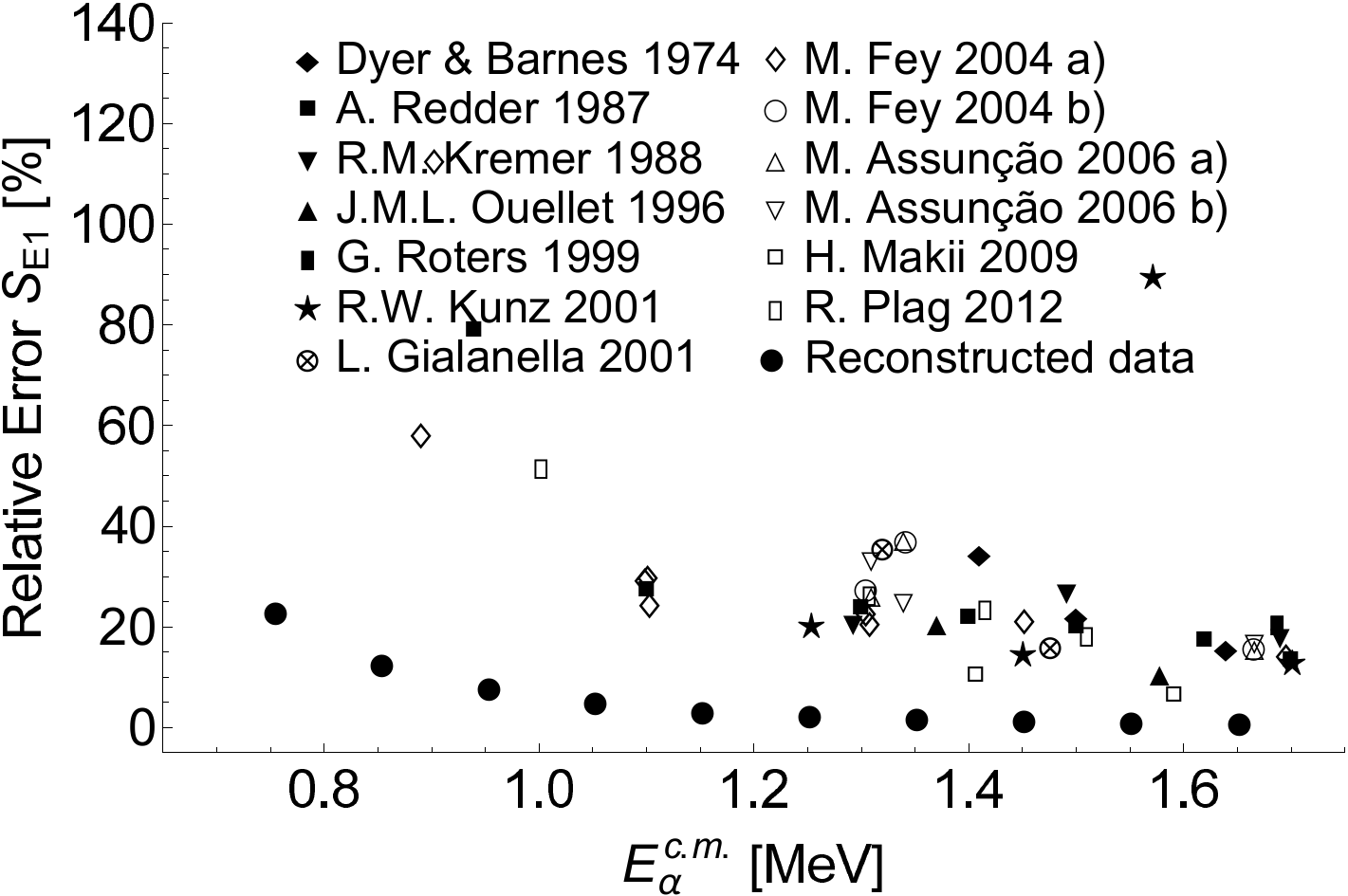}
\includegraphics[width=8.0cm]{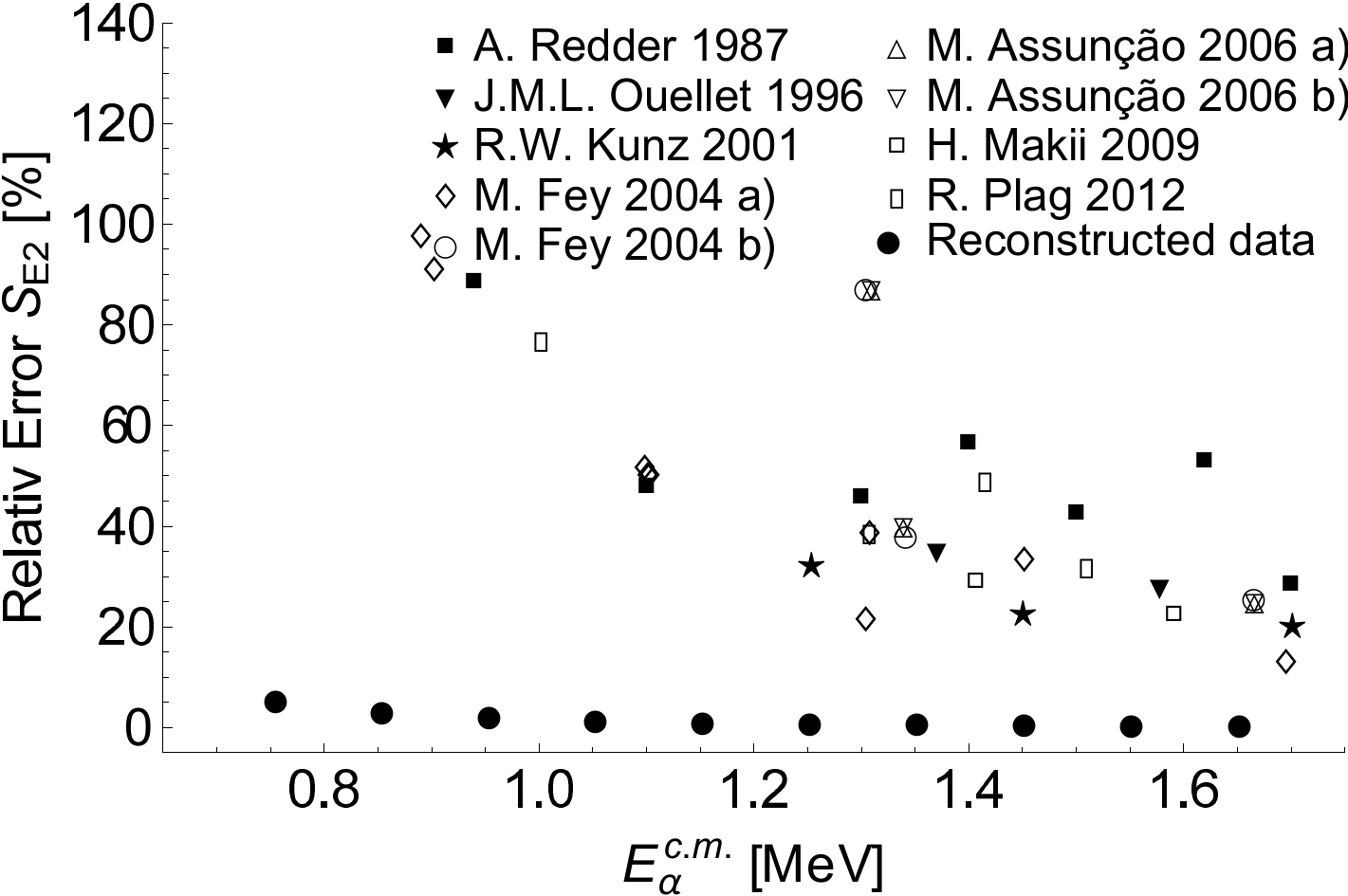} 
\caption{Relative uncertainties of reconstructed the astrophysical $S_{E1}$- and
$S_{E2}$-factors (represented by solid circles) from our calculation
for $E_{e} = $ 114 MeV, $\theta_e = $ 15$^{\circ}$, $t_{C0}$ Case A
and relative errors from \cite{DyerBarnes1974, Redder1987, Kremer1988,
Ouellet1996,Roters1999,Kunz2001,Kunz2002phd,Gialanella2001,
Fey2004phd,Assuncao2006,Makii2009,Plag2012} experiments. Data points
with uncertainties larger then 140\% are not shown. \label{fig:SErelative}}
\end{figure}

Figure \ref{fig:SECalc} shows one example of the calculated $S_{E1}$- 
and $S_{E2}$-factors with projected statistical uncertainties for parameters
 $E_{e} =$ 114 MeV, $\theta_e =$ 15$^{\circ}$ and $t_{C0}$ Case A, as well as
data from past experiments. These results are also plotted in terms
of relative uncertainties in Fig. \ref{fig:SErelative} to point
out a clear advantage of measuring the $^{16}$O$(e,e^\prime \alpha)^{12}$C
reaction for several $E_{\alpha}^{c.m.}$ energies. Compared with the most 
accurate measurements from \cite{Fey2004phd} and \cite{Makii2009}, 
the relative uncertainties in $S_{E1}$ and $S_{E2}$ at a given energy 
are improved at least by factors of $\times$5.6 and $\times$23.9, respectively. 

In the following Figs.~\ref{fig:DeltaS}$-$\ref{fig:DeltaSHalf}, we 
summarize the calculation of the projected relative uncertainties 
in the $S_{E1}$-, $S_{E2}$- and $S_{aC0}$-factors as functions of the 
beam energies $E_{e} =$  78, 114 and 150 MeV and the electron 
scattering angles $\theta_{e} =$ 15$^{\circ}$, 25$^{\circ}$
and 35$^{\circ}$. Even for values of $E_{e}$, $\theta_e$ and $t_{C0}$, which give 
the worst projected statistical uncertainties, improvements
in the relative uncertainties of $S_{E1}$ and $S_{E2}$ at a given energy,
compared with previous experimental data from \cite{Fey2004phd} and 
\cite{Makii2009}, are at least $\times$2.6 and $\times$15.5, respectively.

In general, with increasing electron beam energy $E_e$
all uncertainties are reduced. This can be easily 
understood, because at fixed central electron scattering 
angle $\theta_e$ the accepted angular phase space of the
electron is also fixed, but at larger beam energy $E_e$ 
we also get a larger $q$ value, and thus the coincidence 
rate is larger. 

If we vary the central electron scattering angle $\theta_e$
at fixed  beam energy $E_e$, the uncertainty in $S_{E1}$
is smaller at smaller values of angle $\theta_e$, which 
favors the kinematic setting having a larger accepted
electron angular phase space, thus having the larger rate 
for fixed $E_e$. The uncertainty in $S_{E2}$ behaves 
the opposite way, favoring the kinematic setting with larger
$q$ value at fixed $E_e$.

\begin{figure}[h]
\centering
\includegraphics[scale=0.410]{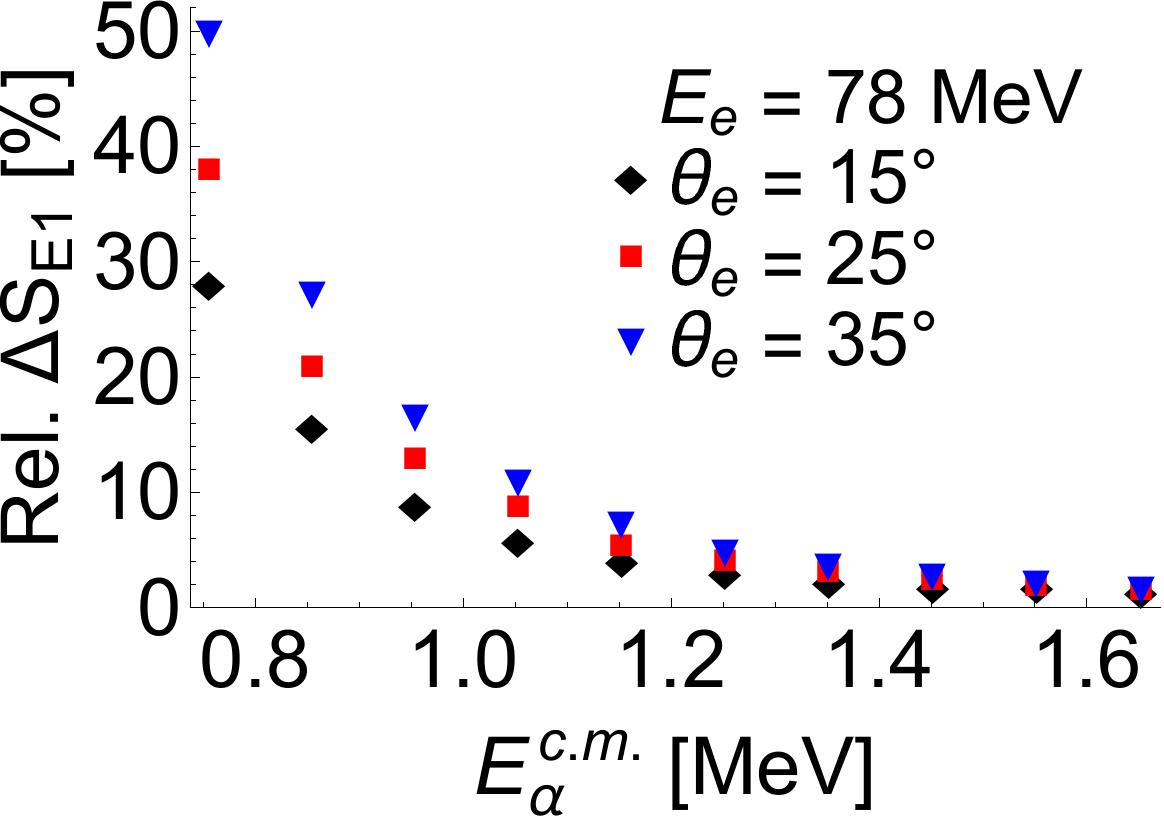}
\includegraphics[scale=0.410]{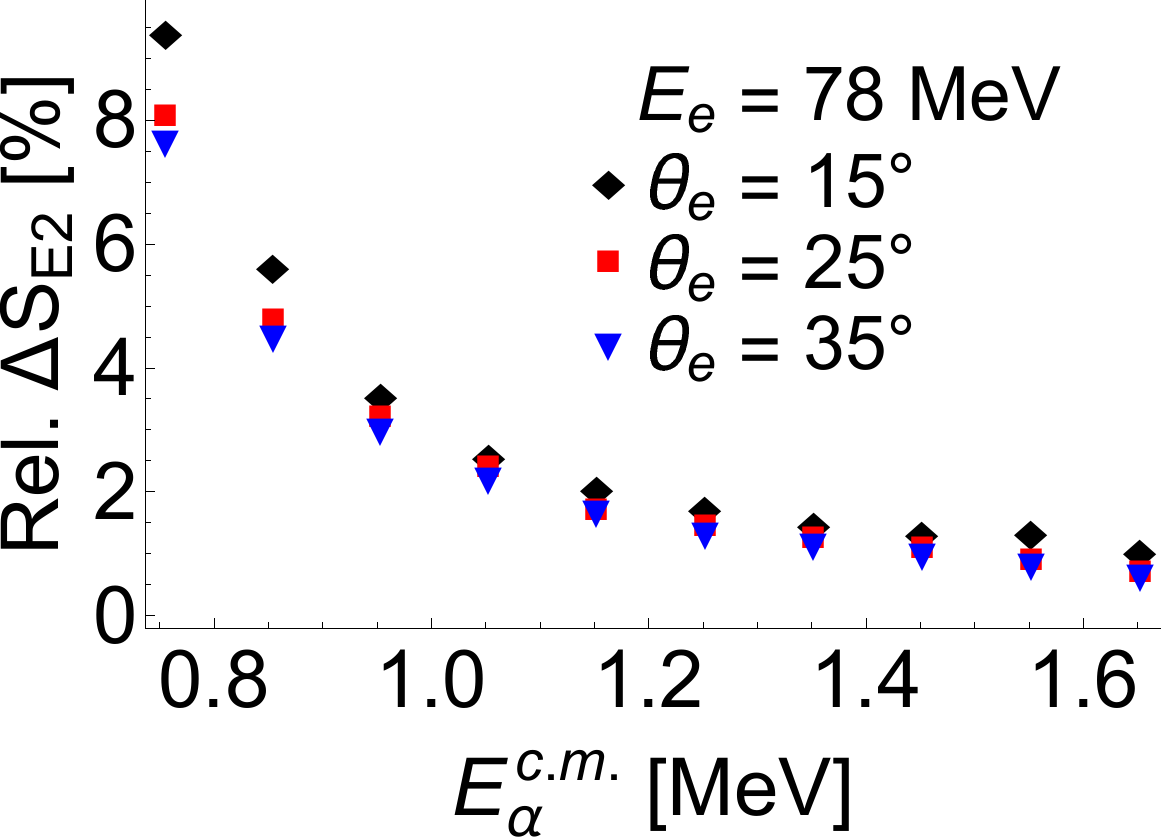}
\includegraphics[scale=0.410]{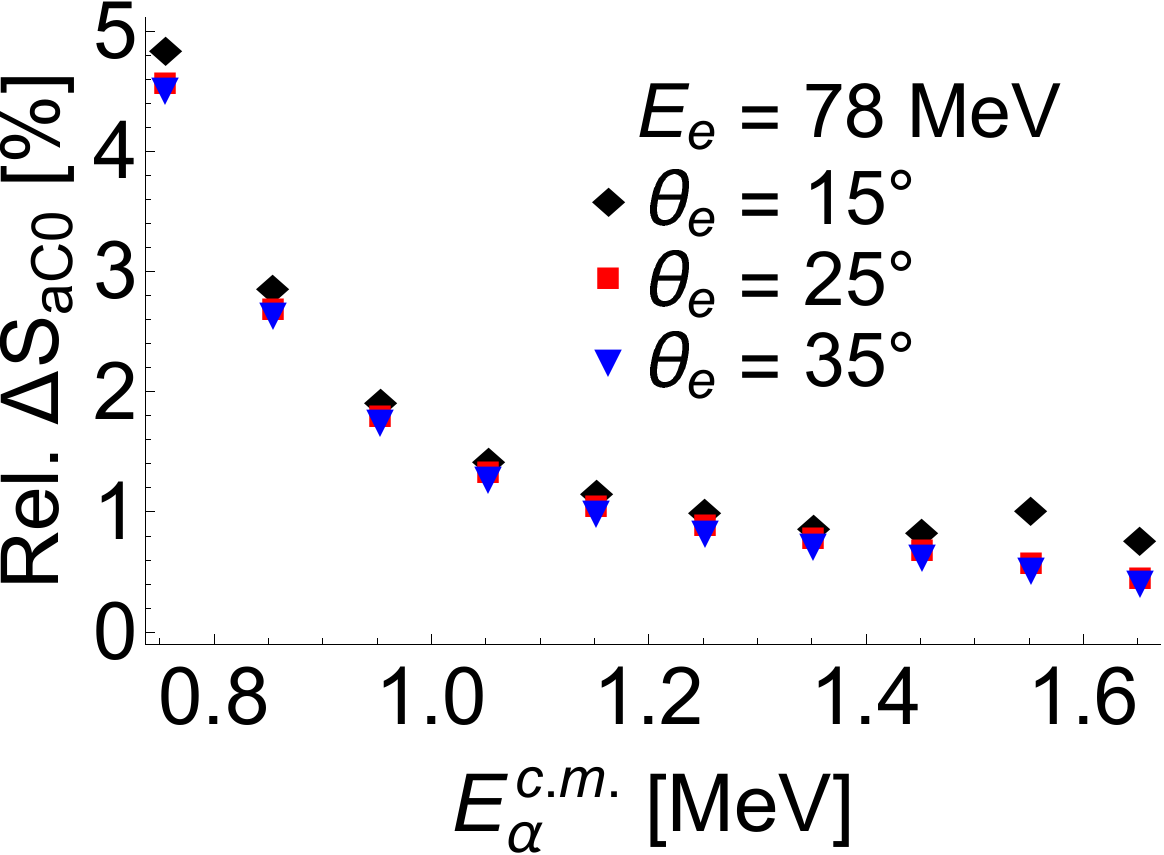}
\includegraphics[scale=0.410]{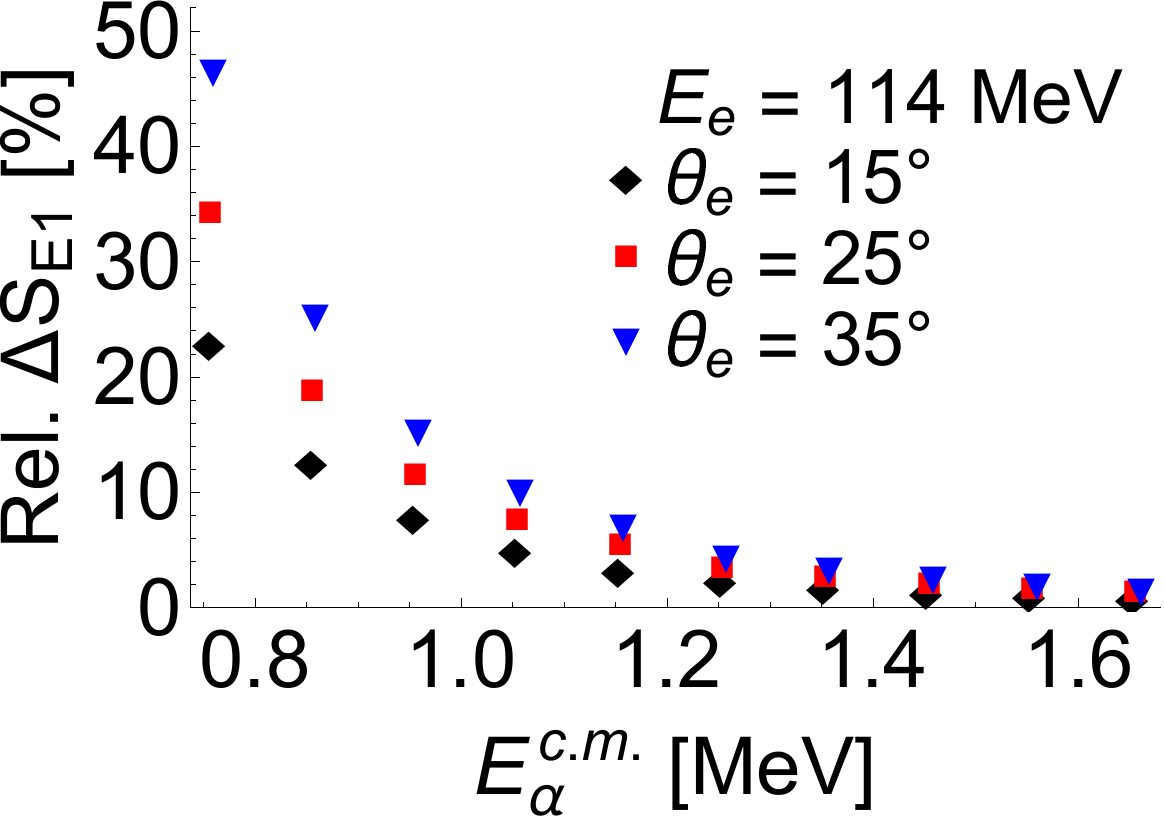}
\includegraphics[scale=0.410]{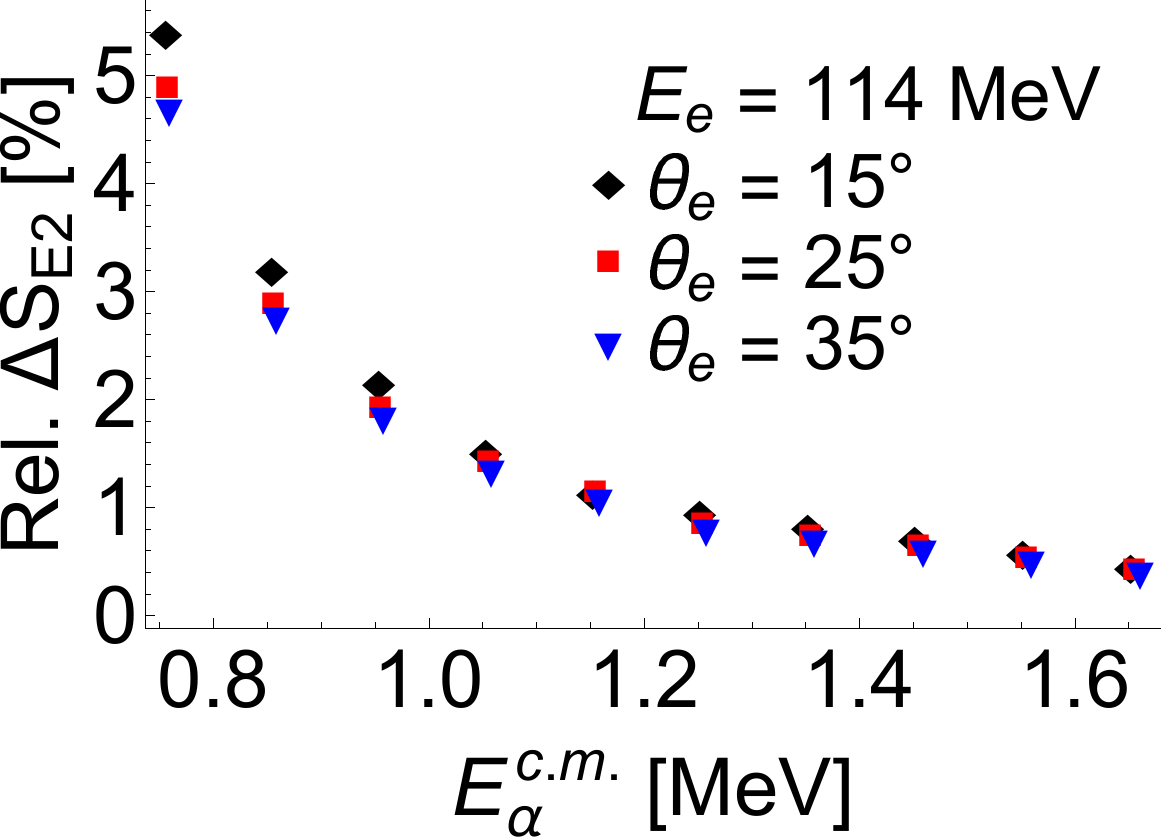}
\includegraphics[scale=0.410]{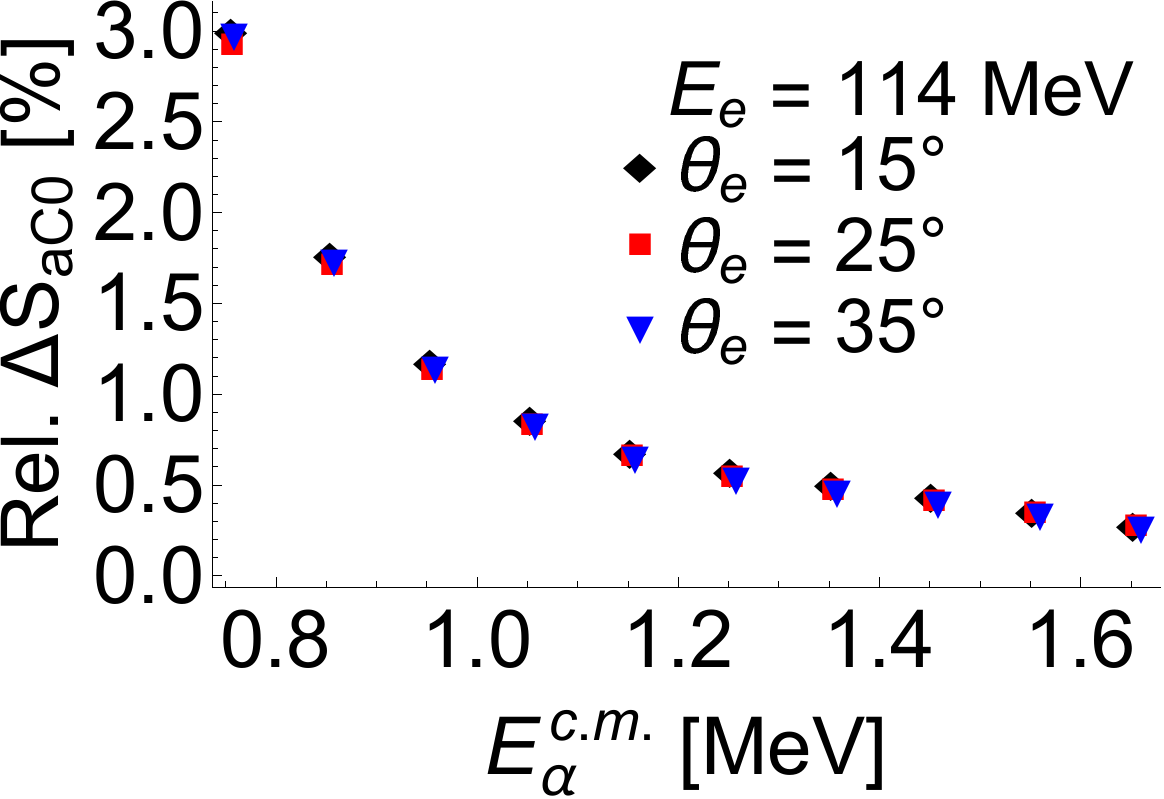}
\includegraphics[scale=0.410]{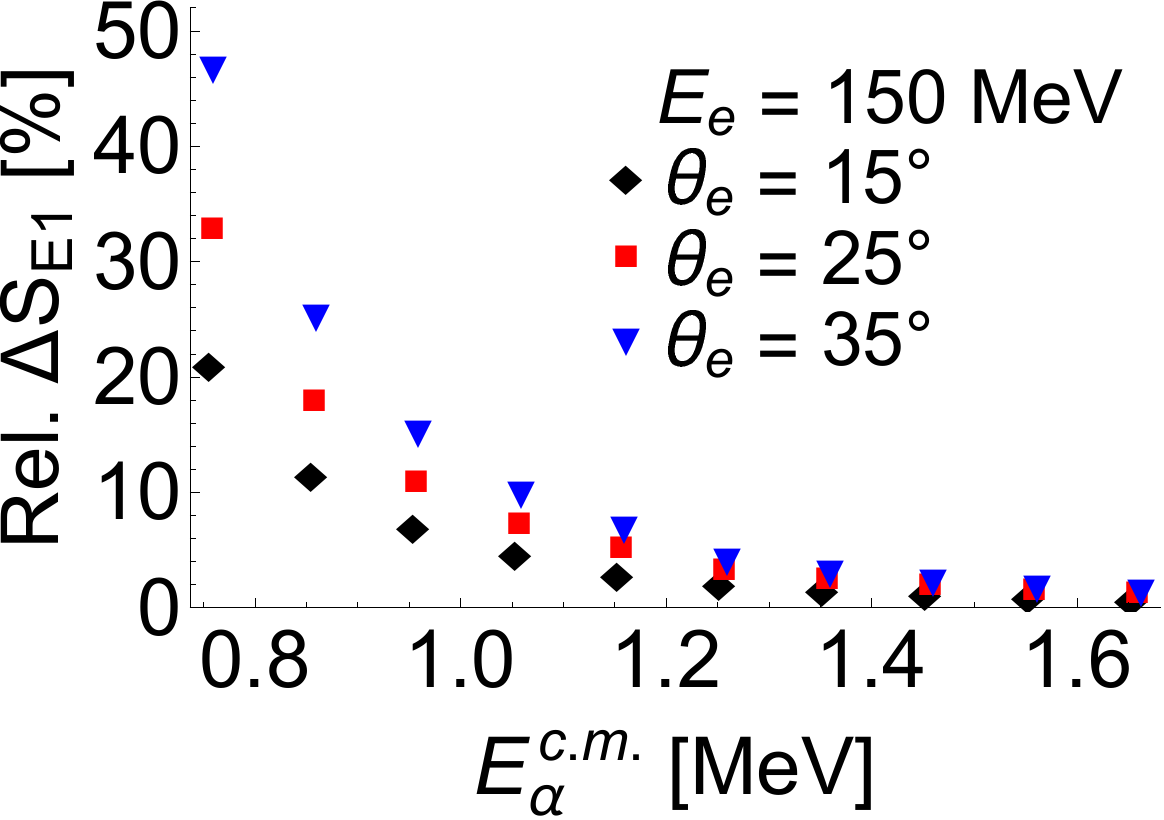}
\includegraphics[scale=0.410]{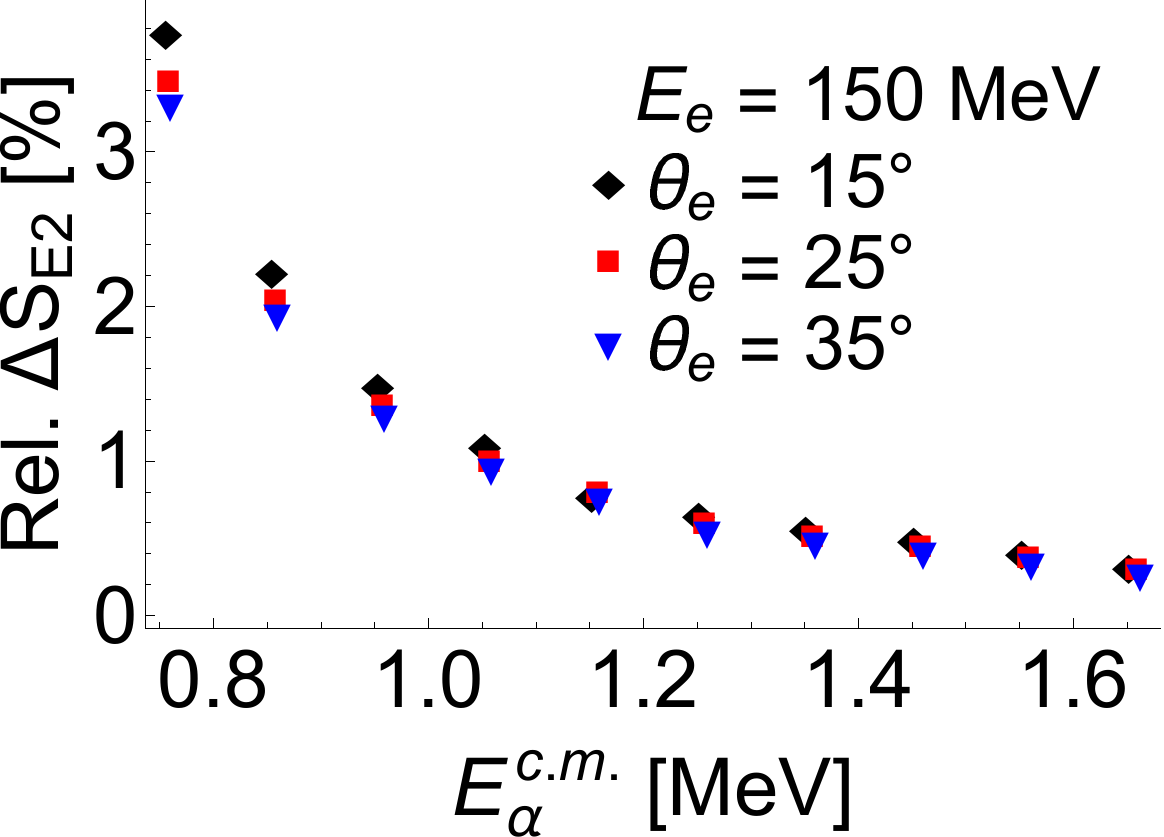}
\includegraphics[scale=0.410]{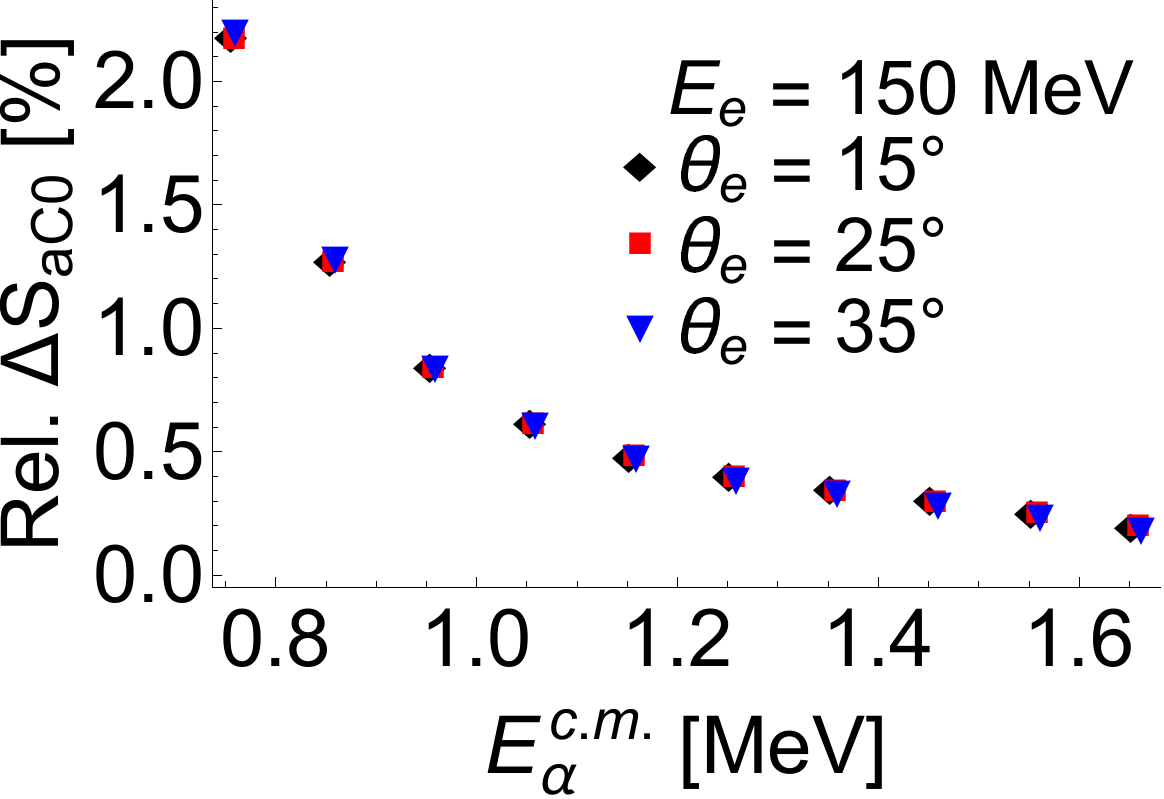}
\vspace*{-0.2cm}
\colorcaption{Relative uncertainties of the $S_{E1}$-, $S_{E2}$- and
$S_{aC0}$-factors for several values of beam energies $E_{e}$,
electron scattering angles $\theta_e$ and for  $t_{C0}$ Case A. 
The $S_{aC0}$ does not have an astrophysical counterpart and 
it just a conversion of the third fitting parameter
into an S-factor and corresponding uncertainty in order to put it in
a perspective with $S_{E1}$ and $S_{E2}$. \label{fig:DeltaS}}
\vspace*{0.2cm}
\includegraphics[scale=0.410]{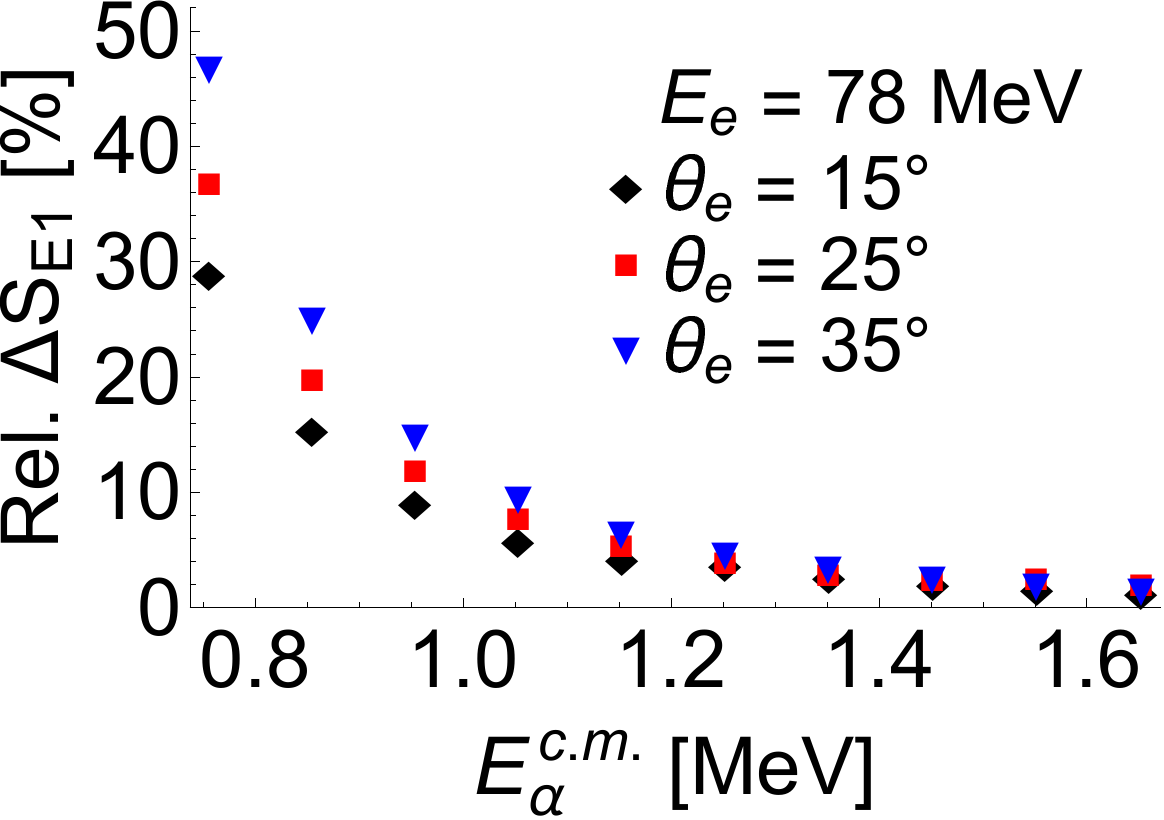}
\includegraphics[scale=0.410]{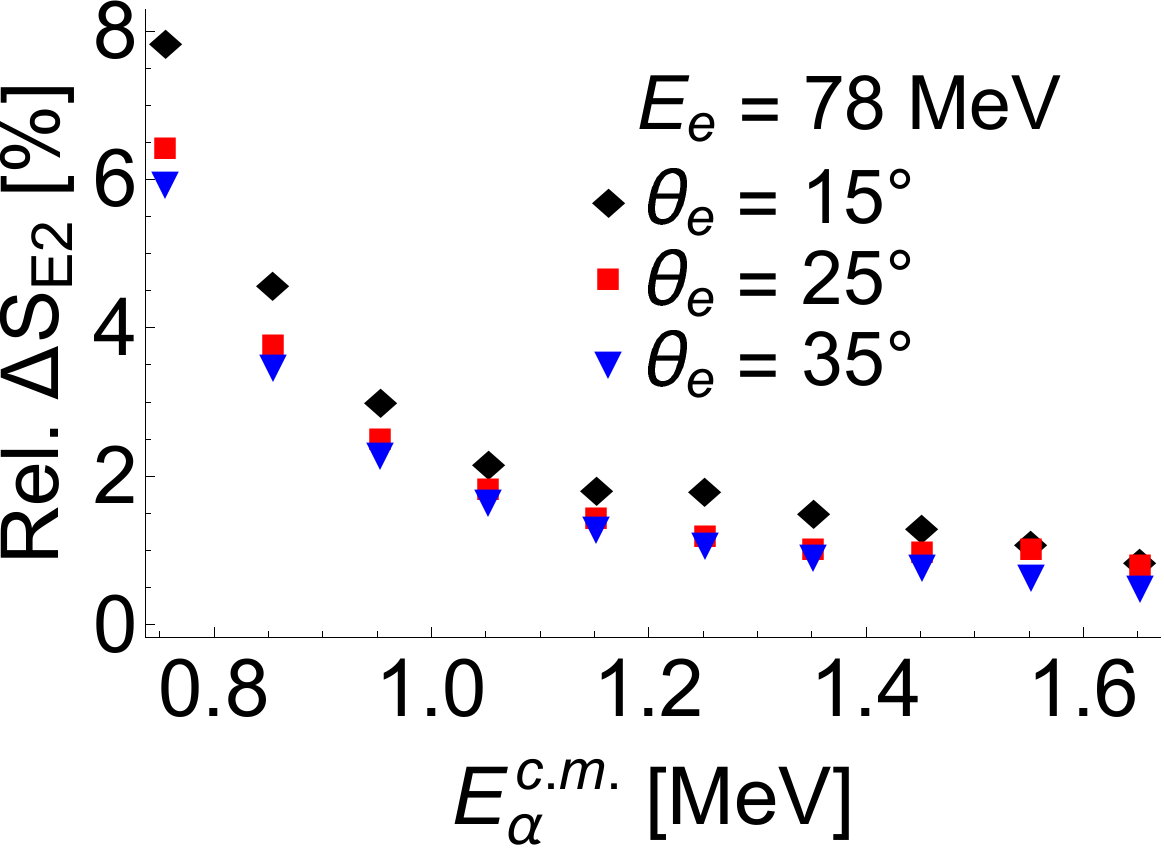}
\includegraphics[scale=0.410]{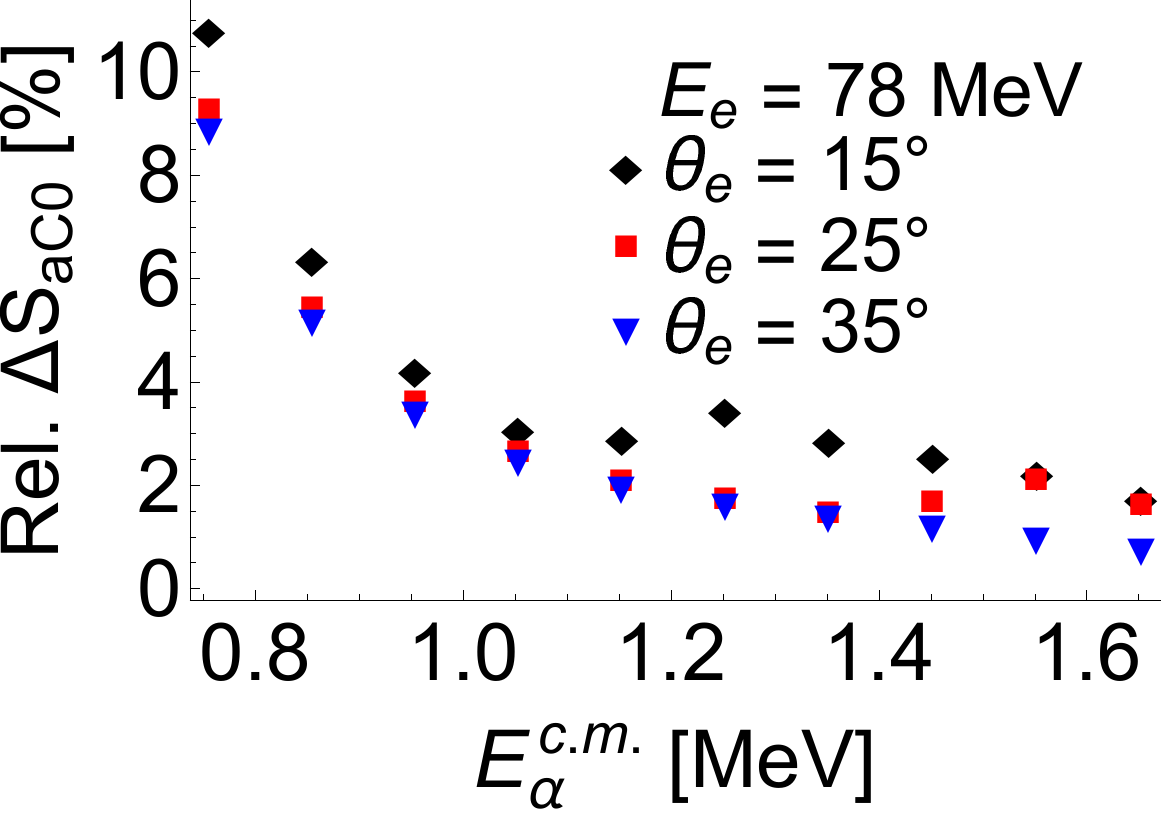}
\includegraphics[scale=0.410]{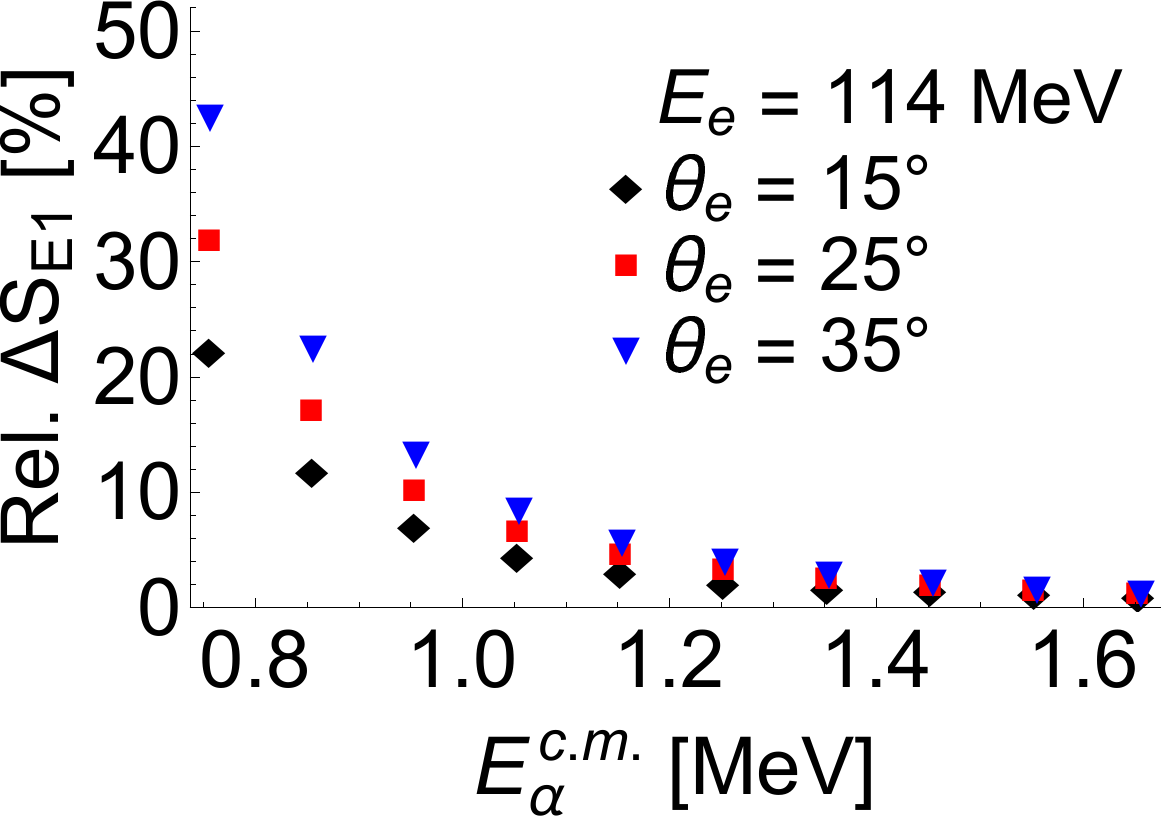}
\includegraphics[scale=0.410]{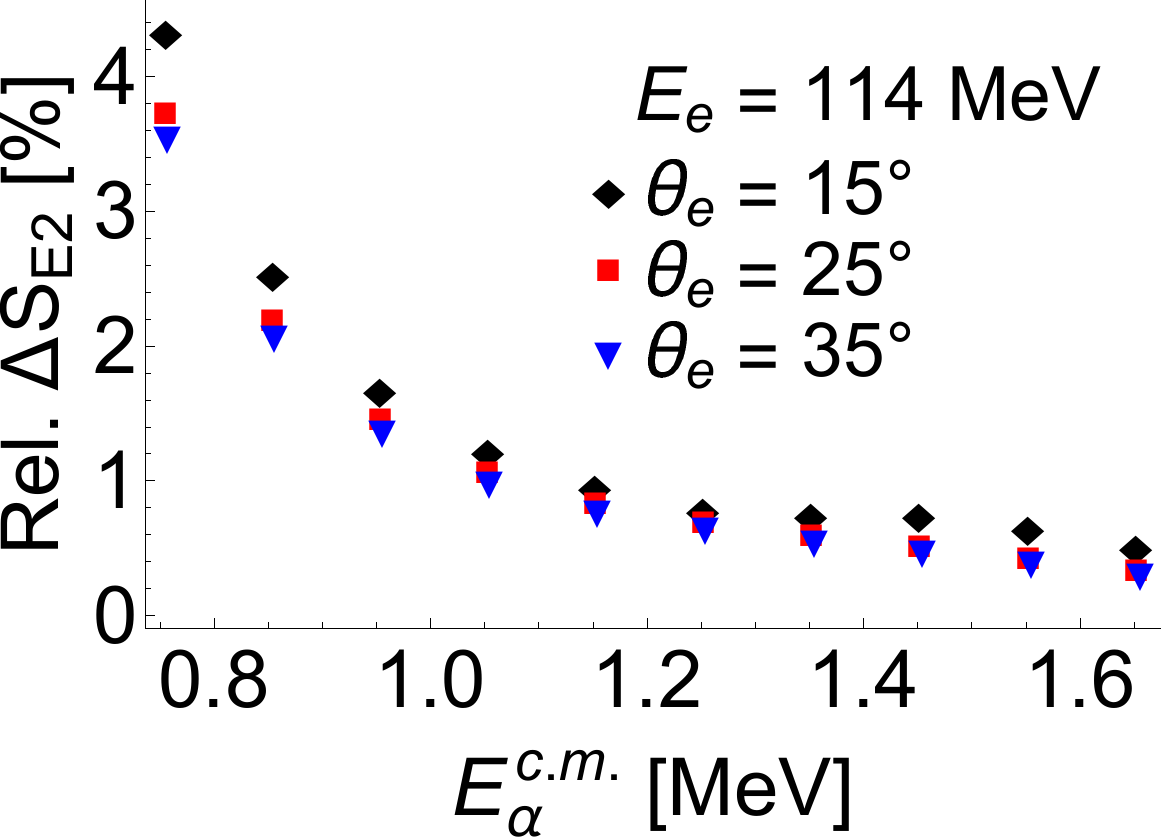}
\includegraphics[scale=0.410]{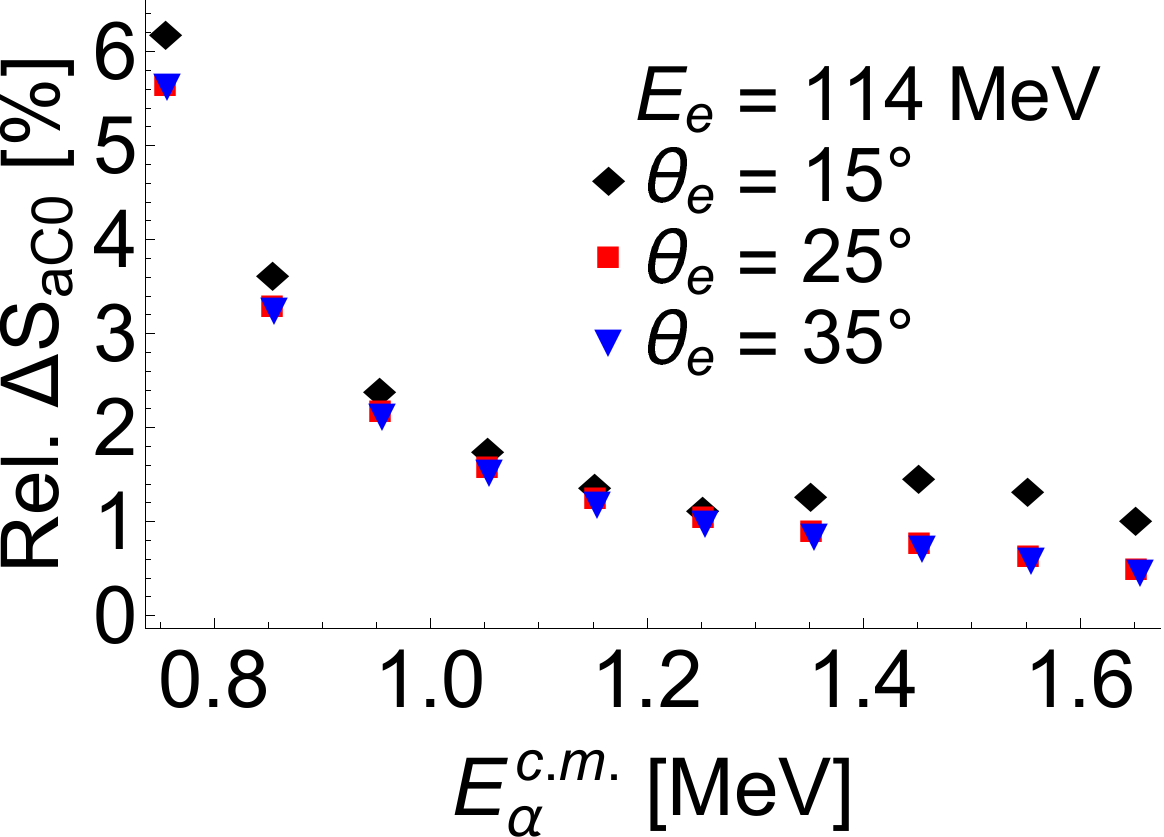}
\includegraphics[scale=0.410]{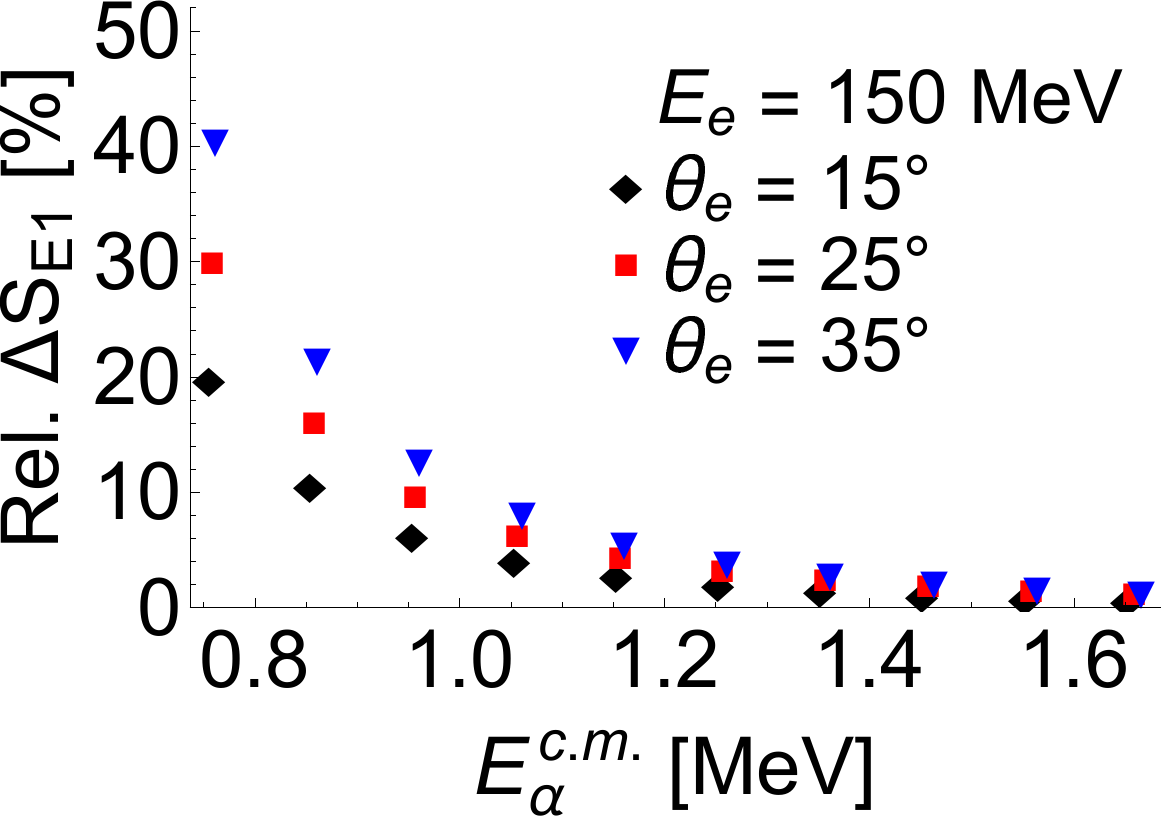}
\includegraphics[scale=0.410]{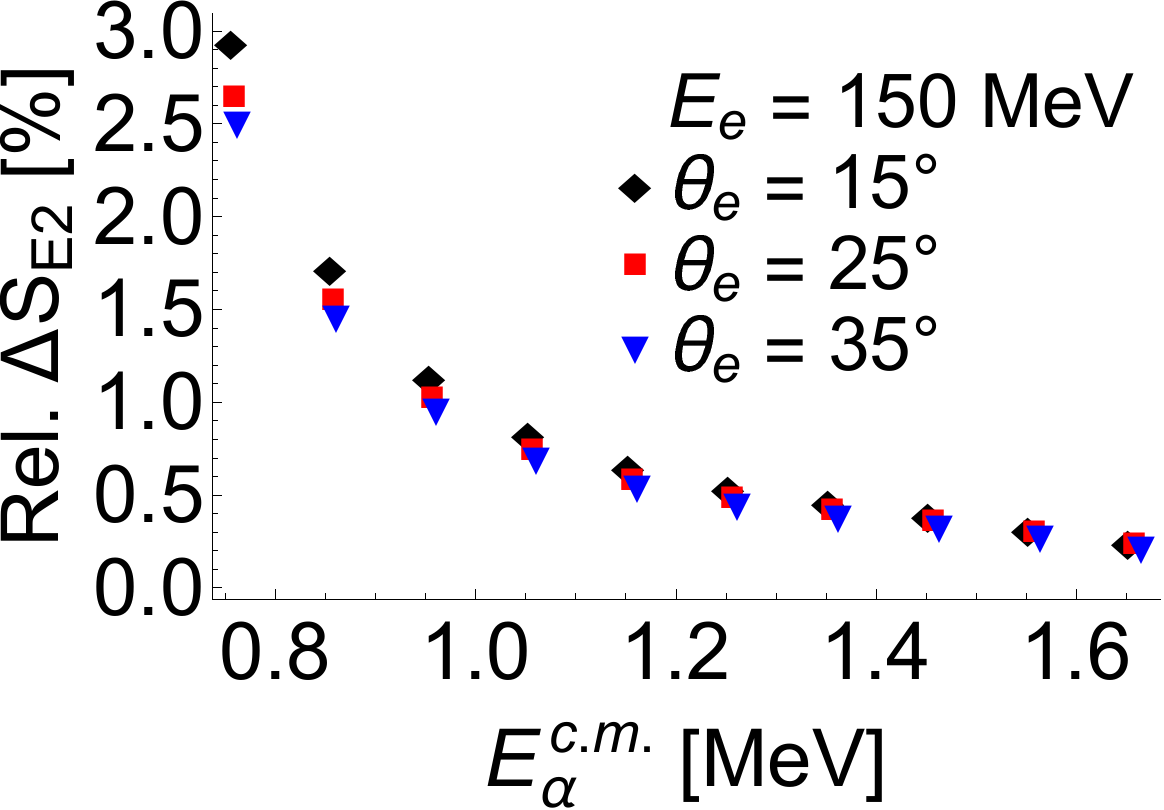}
\includegraphics[scale=0.410]{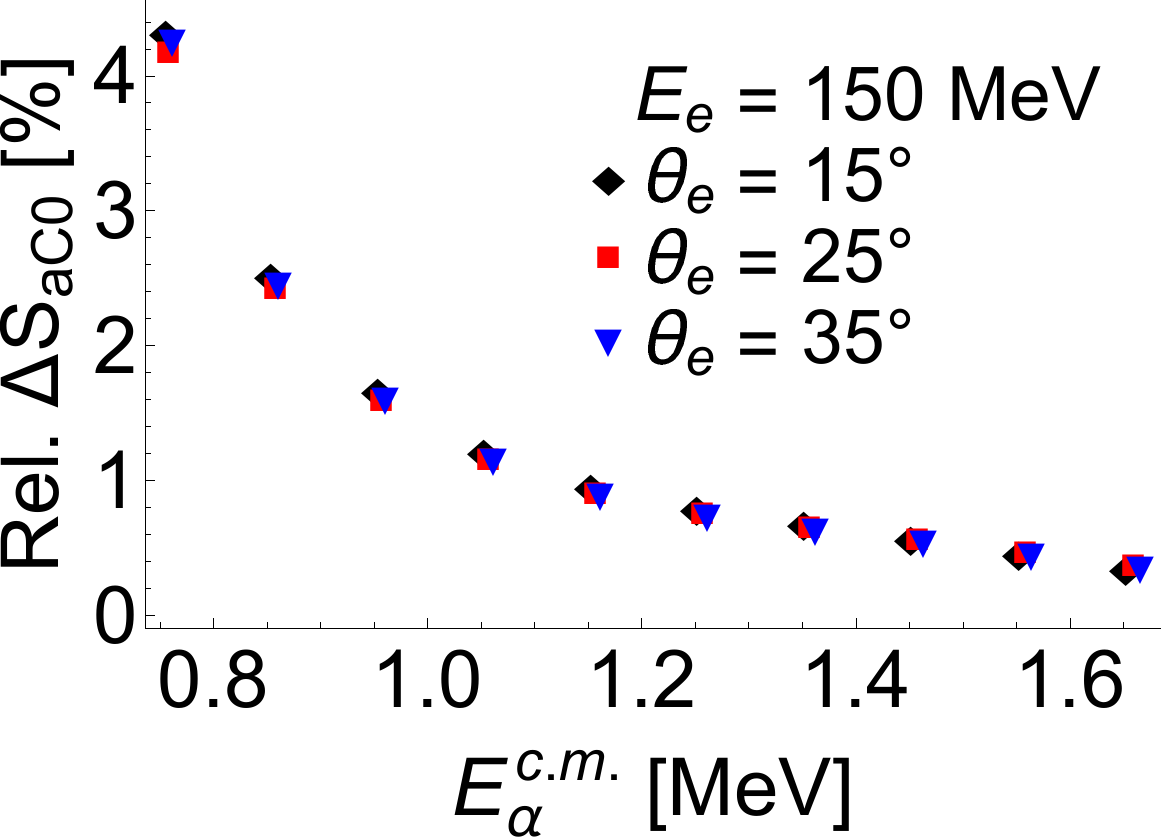}
\vspace*{-0.2cm}
\colorcaption{Same as in the caption of Fig. \ref{fig:DeltaS}, but for $t_{C0}$ Case B. \label{fig:DeltaSHalf}}
\end{figure}

\subsection{Discussion of Results}

The results summarized in Figs.~\ref{fig:SECalc}$-$\ref{fig:SErelative} offer significant potential that the $S$-factors associated with the $E1$ and $E2$ multipoles 
of the $^{16}$O$(e,e^\prime \alpha)^{12}$C reaction at astrophysical energies can be determined with significantly
reduced uncertainties.  We would like to emphasize important details about the assumptions we have made:
\begin{itemize}
\item Although we obtained excellent results in reducing the statistical
      uncertainties, note that our calculation does not include detailed consideration of systematic
      uncertainties, which are always present in experimental data. However, we are not aware of any
      systematic effect that can reduce the large improvement in the determination of the radiative capture 
      reaction with our new approach.
      At fixed beam energy $E_e$ and electron scattering angle $\theta_e$, the electron 
      spectrometer detects scattered electrons in a narrow range of electron momenta 
      $E_{e}^{'}$ and the $\theta_e$. Therefore, one can expect that the systematic 
      uncertainty connected with the detection of the electron  will not vary significantly 
      over these ranges. As discussed in section \ref{IandCC}, the systematic uncertainties
      related to the detection of the $\alpha$-particles are very energy dependent. They
      increase rapidly as the $\alpha$-particle kinetic energy  $E_{\alpha}^{Lab}$ 
      decreases; see for example Fig. \ref{fig:sigmaAng}. However, we note that the kinetic energy of
      the $\alpha$-particle can be controlled by the transferred momentum $q$ and the
      thickness of the jet-target traversed by the $\alpha$-particle can be reduced 
      by extending the shape of the jet's profile. This optimization needs further consideration.
\item One significant source of systematic uncertainty which needs to be considered is
      the uncertainty of the electron beam energy $E_e$, which is especially important
      at low $E_{\alpha}^{c.m.}$-values. In a coincidence measurement of the electrodisintegration 
      of $^{16}$O, the kinematics are over-determined.  Thus, an attractive method to determine the 
      electron beam energy would be to reconstruct the energy of the electron beam $E_e$ for each coincidence 
      $e' \alpha$ pair separately.
\item The Coulomb and electric multipole matrix elements have only 
      been expanded up to the NLO, see Eq. (\ref{eq:tCJ})
      and (\ref{eq:tEJ}), and for the corresponding NLO  
      coefficients we assumed $b'_{CJ,EJ} \approx 1$. In general these 
      coefficients are functions of $q$ and, when dealing with experimental
      $^{16}$O$(e,e^\prime \alpha)^{12}$C data, their magnitude and $q$-behavior
      will have to be verified by including them as four additional fitting parameters.
      If values of $b'_{C0,C1,C2,E1,E2}$ are smaller than unity for large
      range of $q$, truncating the expansion of multipole matrix 
      elements at the NLO term is justified. But, if the values are larger 
      than 1, we may need to include the third-order in the expansion 
      with corresponding coefficients $c'_{CJ,EJ}$.  Which order in this
      expansion needs to be included can easily be verified by measuring 
      the rate of electrodisintegration of $^{16}$O at several larger
      $q$-points. 
\item The calculations here were focused on the $E_{\alpha}^{c.m.}$-range
      from 0.7 to 1.7 MeV, but a typical electron spectrometer has at 
      least a momentum acceptance of 10\% and for $E_{e} =$ 78 MeV 
      the full available $E_{\alpha}^{c.m.}$-range would be from 
      0.0 to 6.7 MeV, or for $E_{e} =$ 150 MeV from 0.0 to 13.6 MeV.
      By choosing the appropriate beam energy, a single $^{16}$O
      electrodisintegration measurement could cover the 
      $E_{\alpha}^{c.m.}$-range of almost all previous experiments,
      and crosscheck their results. Furthermore, at higher $\alpha$-energies,
      multipoles $E3$ and $C3$ could start to significantly contribute 
      to the cross section (although this was not yet observed \cite{deBoer2017}).
      Because of this we have provided the multipole decomposition of the response
      functions up to octupole terms in the Appendix \ref{app:ResOctu}.

      \item For all choices of the parameters $E_e$ and $\theta_{e}$ we obtained
      a smaller uncertainty for $S_{E2}$ compared to $S_{E1}$. There are two reasons for this result. 
       Firstly, the $C2/E2$ matrix elements which enter in the response functions $R_{L,T,TL,TT}$
      differ from $C1/E1$ matrix elements by a factor $q/\omega$. This is the dominant contribution, 
      which does not exist in the case of real photon experiments, since $q/\omega =$ 1. Secondly,
      a minor contribution comes from the $\theta^{c.m.}_{\alpha}$-distribution
      of the relevant multipoles. In the $\theta^{c.m.}_{\alpha}$-range 
      from 0$^{\circ}$ to 60$^{\circ}$, the magnitudes of the $C2/E2$ 
      $\theta^{c.m.}_{\alpha}$-distributions are larger compared with
      those of the $C1/E1$. The same behavior can also be observed for
      the $E1$ and $E2$ multipoles in the case of real-photon experiments; for example this 
      is shown in Fig. 5 in \cite{deBoer2017}. 
      
      \item In section \ref{IandCC} we have
      considered the most probable sources of background and demonstrated how to identify the $\alpha$-particles
       from the electrodisintegration of $^{16}$O. 
       If one takes a closer look at figures 
       \ref{fig:PNC}  and \ref{fig:back}, in the same experiment, we can also identify the proton
       and measure the rate of the $^{14}$N$(e,e'p)^{13}$C reaction. Furthermore, the photodisintegration cross
       section of $^{18}$O is much larger compared with that of $^{17}$O and, with further work,
       it would also be possible to extract the rate of the $^{18}$O$(e,e'\alpha)^{14}$C reaction. 
       Note that with minor modifications, the same formalism presented in this paper
       for electrodisintegration of $^{16}$O can also be be applied to electrodisintegration of $^{18}$O. 
 \end{itemize}

\section{Conclusion and outlook} \label{sec:six}

In summary, we have considered in some detail a new approach to determine radiative capture reactions at astrophysical energies.
Using detailed balance, we consider the inverse electron-induced disintegration process.  Specifically, in this paper we have focused on the $^{16}$O(e,e$^\prime \alpha$)$^{12}$C reaction as a means to determine the astrophysically crucial radiative capture process $^{12}$C($\alpha, \gamma$)$^{16}$O. We have applied a multipole decomposition constrained to fit existing data together with some reasonable theoretical assumptions to extrapolate from the electrodisintegration process to the photodisintegration reaction. We have developed a Monte Carlo simulation of an experiment where an external electron beam is directed on an oxygen gas jet target; the forward scattered electron is detected in a magnetic spectrometer and the coincident, low-energy, recoil $\alpha$-particle is detected in a large acceptance detector centered around the direction of three-momentum transfer. We assume what we believe are reasonable experimental parameters to carry out such an experiment at the upcoming ERLs.  With an electron beam of energy 114 MeV and beam current of 40 mA incident on a hydrogen gas target of $5 \times 10^{18}$ cm$^{-2}$, we estimate that $S_{E1}$- and $S_{E2}$- factors can be determined at $E^{c.m.}_{\alpha}$ = 0.75 MeV to of order $\pm$20\% and $\pm$5\%, respectively, in 100 days of continuous data taking.

Assuming that the multi-Megawatt ERLs are realized with electron energy of about 100 MeV, a key technical challenge is to realize efficient, large solid-angle, low-energy $\alpha$-particle detection that is blind to the large rate of electromagnetic background. We note that previous work has shown~\cite{Alarcon2013} that the electron beam quality of 100 MeV Megawatt ERLs is high, with $\sim$50 $\mu$m 1$\sigma$ spatial size and with minimal halo. 
To reach high precision, the experiment must be efficient and stable over months of data taking.  However, we stress that the initial, key experiment to validate our proposed approach should focus on higher $E^{c.m.}_{\alpha}$ where the coincident electrodisintegration rates are significantly higher than in the astrophysical region and accordingly the running time is a more modest several weeks.  Such an experiment should elucidate the multipole structure of the electrodisintegration reaction, whose understanding is essential for extrapolation to the photodisintegration reaction.  If our approach is validated experimentally, one can then embark on the more ambitious measurement to determine the S-factors in the astrophysically interesting region at low $\alpha$-particle energies, where the electrodisintegration count rate drops precipitously.   

In the present study, we have focused on electrodisintegration of ${}^{16}$O into the ground states of ${}^4$He and ${}^{12}$C in the low momentum transfer $q$ region near threshold for the reaction. We have provided the bridge to photodisintegration and radiative capture (real-photon) reactions through the limit where the virtual photon involved in electron scattering becomes close to the real-photon line. Measurements of electrodisintegration thereby have the potential to provide a new way to approach the real-$\gamma$ photodisintegration cross sections and hence, through detailed balance, the capture reaction cross section and ultimately the astrophysical S-factors involved in that process. As discussed in the Introduction, obtaining information on these last quantities provides one of the high-priority goals in nuclear astrophysics.

After developing the general formalism for the electrodisintegration reaction ${}^{16}$O$(e,e'\alpha){}^{12}$C following past general treatments of such semi-inclusive reactions, together with some discussion of the photodisintegration reaction ${}^{16}$O$(\gamma,\alpha){}^{12}$C and radiative capture reaction $\alpha + {}^{12}$C$\rightarrow \gamma + {}^{16}$O, we have proceeded to develop parametrizations for the dynamical content in the problem. Since our focus is the region near threshold for the photo- or electro-disintegration reactions we are assured that the energy of the photon, $E_{\gamma}$, is small. Additionally, we have limited our attention to kinematics where the three-momentum transfer carried by the virtual photon in electrodisintegration, $q$, is also small. Here one needs to state what is meant by ``small''. We expect that the nuclear dynamics involved in the reactions occur with a typical nuclear scale $q_0 \cong 200$ -- 250 MeV/c, and accordingly we measure both quantities versus $q_0$, taking both ratios $\mu_1 \equiv E_\gamma/q_0$ and $\mu_2 \equiv q/q_0$ to be small. This allows us to expect that the lowest multipoles will dominate over higher multipolarity contributions and also to expand each of the small number of remaining multipoles in powers of $\mu_{1,2}$. Ultimately, as we have shown in detail in the body of this work, there are only a few parameters left that determine the dynamical content of the problem in the kinematic region of interest. Such procedures are simply an extension of what is typically done for photodisintegration or radiative capture.

Of course, it would be valuable to have a microscopic model for the reactions of interest here, although this is far from realizable at present. Even relatively crude models might be of some interest as they could help set the scales in the problem. For example, a cluster model in which the ${}^{12}$C ground state might be taken to be a cluster of three $\alpha$-particles and the ground state of ${}^{16}$O might involve four $\alpha$-particles could be pursued. We have not done so in this initial study, but instead have limited our attention to the parametrizations discussed above. When measurements are made of the kinematical dependences on $q$ and the angular distributions of the $\alpha$-particles are determined for each energy above threshold there is ample information to fix all of the parameters involved experimentally.

\begin{acknowledgments}
We acknowledge valuable discussions with P. Fisher, E. Tsentalovich and H. Weller.  This research was supported by the U.S. Department of Energy Office of Nuclear Physics under grant No. DE-fFG02-94ER40818. 
\end{acknowledgments}


\appendix*

\section{Extended Angular Distributions} \label{app:ResOctu}

Following \cite{Raskin1989}, the responses in terms of Legendre 
polynomials up to octupole contributions may be written
\begin{widetext}
\begin{flalign} \label{eq:RLextend}
R_L &= P_0 (\cos \theta_{\alpha}) \Bigg( |t_{C0}|^2 + |t_{C1}|^2 +|t_{C2}|^2+|t_{C3}|^2 \Bigg) \notag \\
    &+ P_1 (\cos \theta_{\alpha})  \Bigg(  2\sqrt{3} |t_{C0}||t_{C1}|\cos(\delta_{C1}-\delta_{C0}) + 4\sqrt{\frac{3}{5}} |t_{C1}||t_{C2}|\cos(\delta_{C2} - \delta_{C1})
    + \frac{18}{\sqrt{35}} |t_{C2}||t_{C3}|\cos(\delta_{C3} - \delta_{C2}) \Bigg) \notag \\
    &+ P_2 (\cos \theta_{\alpha})  \Bigg(2|t_{C1}|^2  + \frac{10}{7} |t_{C2}|^2  +   2 \sqrt{5} |t_{C0}||t_{C2}|\cos(\delta_{C2}-\delta_{C0})
     + 6\sqrt{\frac{3}{7}} |t_{C1}||t_{C3}|\cos(\delta_{C3} - \delta_{C1})  + \frac{4}{3}|t_{C3}|^2  \Bigg) \notag \\
    &+ P_3 (\cos \theta_{\alpha})  \Bigg(6\sqrt{\frac{3}{5}} |t_{C1}||t_{C2}|\cos(\delta_{C2} - \delta_{C1})
    + \frac{8}{3}\sqrt{\frac{7}{5}} |t_{C2}||t_{C3}|\cos(\delta_{C3} - \delta_{C2})  +   2\sqrt{7}  |t_{C0}||t_{C3}|\cos(\delta_{C3}-\delta_{C0})  \Bigg) \notag \\
    &+ P_4 (\cos \theta_{\alpha})  \Bigg(\frac{18}{7} |t_{C2}|^2
     + 8\sqrt{\frac{3}{7}} |t_{C1}||t_{C3}|\cos(\delta_{C3} - \delta_{C1})  + \frac{18}{11} |t_{C3}|^2  \Bigg) \notag \\
    & + P_5 (\cos \theta_{\alpha}) \Bigg( \frac{20}{3}\sqrt{\frac{5}{7}} |t_{C2}||t_{C3}|\cos(\delta_{C3} - \delta_{C2}) \Bigg) \notag \\
    & + P_6 (\cos \theta_{\alpha}) \Bigg( \frac{100}{33}|t_{C3}|^2 \Bigg) &&
\end{flalign}

\begin{flalign} \label{eq:RTextend}
R_T &= P_0 (\cos \theta_{\alpha}) \Bigg(|t_{E1}|^2 +|t_{E2}|^2  +|t_{E3}|^2  \Bigg) \notag \\
    &+ P_1 (\cos \theta_{\alpha})  \Bigg(\frac{6}{\sqrt{5}} |t_{E1}||t_{E2}|\cos(\delta_{E2}-\delta_{E1})
     + 12\sqrt{\frac{2}{35}} |t_{E2}||t_{E3}|\cos(\delta_{E3} - \delta_{E2}) \Bigg) \notag \\
    &+ P_2 (\cos \theta_{\alpha})  \Bigg(-|t_{E1}|^2  + \frac{5}{7} |t_{E2}|^2 
      + 6\sqrt{\frac{2}{7}} |t_{E1}||t_{E3}|\cos(\delta_{E3} - \delta_{E1})  +|t_{E3}|^2   \Bigg) \notag \\
    &+ P_3 (\cos \theta_{\alpha})  \Bigg(-\frac{6}{\sqrt{5}} |t_{E1}||t_{E2}|\cos(\delta_{E2} - \delta_{E1})
     + \frac{2}{3}\sqrt{\frac{14}{5}} |t_{E2}||t_{E3}|\cos(\delta_{E3} - \delta_{E2})   \Bigg) \notag \\
    &+ P_4 (\cos \theta_{\alpha})  \Bigg(-\frac{12}{7} |t_{E2}|^2
     -6\sqrt{\frac{2}{7}} |t_{E1}||t_{E3}|\cos(\delta_{E3} - \delta_{E1})  + \frac{3}{11} |t_{E3}|^2   \Bigg) \notag \\
    & + P_5 (\cos \theta_{\alpha}) \Bigg( -\frac{10}{3}\sqrt{\frac{10}{7}} |t_{E2}||t_{E3}|\cos(\delta_{E3} - \delta_{E2}) \Bigg)  \notag \\
    & + P_6 (\cos \theta_{\alpha}) \Bigg( -\frac{25}{11}|t_{E3}|^2 \Bigg)  &&
\end{flalign}
\begin{flalign} \label{eq:RTTextend}
R_{TT} &= - R_T \cos(2\phi_{\alpha}) . &&
\end{flalign}
\begin{flalign} \label{eq:RTLextend}
R_{TL} &= \cos\phi_{\alpha} \Bigg \{ P^1_1(\cos \theta_{\alpha}) \Bigg(  2\sqrt{3}|t_{C0}||t_{E1}|\cos(\delta_{E1}-\delta_{C0}) - 2\sqrt{\frac{3}{5}}|t_{C2}||t_{E1}| \cos(\delta_{C2}- \delta_{E1}) \notag \\ &+ \frac{6}{\sqrt{5}}|t_{C1}||t_{E2}| \cos(\delta_{C1} - \delta_{E2}) 
  - 6\sqrt{\frac{3}{35}}|t_{C3}||t_{E2}| \cos(\delta_{C3} - \delta_{E2}) +6\sqrt{\frac{6}{35}}|t_{C2}||t_{E3}| \cos(\delta_{C2} - \delta_{E3}) \Bigg)\notag \\
&+ P^1_2(\cos \theta_{\alpha}) \Bigg( 2|t_{C1}||t_{E1}|\cos(\delta_{C1}-\delta_{E1})+  2\sqrt{\frac{5}{3}}|t_{C0}||t_{E2}|\cos(\delta_{E2}-\delta_{C0})+\frac{10}{7\sqrt{3}}|t_{C2}||t_{E2}|\cos(\delta_{C2}-\delta_{E2}) \notag \\
&   - 2\sqrt{\frac{3}{7}}|t_{C3}||t_{E1}| \cos(\delta_{C3} - \delta_{E1}) +4\sqrt{\frac{2}{7}}|t_{C1}||t_{E3}| \cos(\delta_{C1} - \delta_{E3}) +\frac{2}{3}\sqrt{\frac{2}{3}}|t_{C3}||t_{E3}| \cos(\delta_{C3} - \delta_{E3})  \Bigg)  \notag\\
&+ P^1_3(\cos \theta_{\alpha}) \Bigg( 2\sqrt{\frac{3}{5}}|t_{C2}||t_{E1}|\cos(\delta_{C2}-\delta_{E1})+\frac{4}{\sqrt{5}}|t_{C1}||t_{E2}|\cos(\delta_{C1}-\delta_{E2}) \notag \\ &   +  2\sqrt{\frac{7}{6}}|t_{C0}||t_{E3}| \cos(\delta_{E3}-\delta_{C0}) +\frac{2}{3}\sqrt{\frac{7}{15}}|t_{C3}||t_{E2}| \cos(\delta_{C3} - \delta_{E2}) +2\sqrt{\frac{7}{30}}|t_{C2}||t_{E3}| \cos(\delta_{C2} - \delta_{E3}) \Bigg) \notag\\
&+ P^1_4(\cos \theta_{\alpha}) \Bigg( \frac{6\sqrt{3}}{7}|t_{C2}||t_{E2}|\cos(\delta_{C2}-\delta_{E2})   +2\sqrt{\frac{3}{7}}|t_{C3}||t_{E1}| \cos(\delta_{C3} - \delta_{E1})  \notag \\ 
&   +\frac{6}{\sqrt{14}}|t_{C1}||t_{E3}| \cos(\delta_{C1} - \delta_{E3}) +\frac{6}{11}\sqrt{\frac{3}{2}}|t_{C3}||t_{E3}| \cos(\delta_{C3} - \delta_{E3}) \Bigg) \notag \\
&  + P^1_5(\cos\theta_{\alpha})\Bigg( \frac{40}{3}\frac{1}{\sqrt{105}}|t_{C3}||t_{E2}|\cos(\delta_{C3}-\delta_{E2})+10\sqrt{\frac{2}{105}}|t_{C2}||t_{E3}|\cos(\delta_{C2}-\delta_{E3}) \Bigg)  \notag\\
&  + P^1_6(\cos\theta_{\alpha})\Bigg( \frac{50}{33}\sqrt{\frac{2}{3}}|t_{C3}||t_{E3}|\cos(\delta_{C3}-\delta_{E3})\Bigg)  \Bigg \} &&
\end{flalign}
For completeness we can also evaluate the Legendre 
polynomials to write expressions involving only sines and cosines of $\theta_{\alpha}$:
\begin{flalign}
R_L &= \overline{X^{0,0}_{fi}} \notag  \\
    &= |t_{C0}|^2 + 3|t_{C1}|^2\cos^2\theta_{\alpha}  + \frac{5}{16} |t_{C2}|^2(1+3\cos2\theta_{\alpha})^2 + \frac{7}{64}|t_{C3}|^2 (3\cos\theta_{\alpha} + 5 \cos3\theta_{\alpha})^2  \notag \\
    & +   2 \sqrt{3} |t_{C0}||t_{C1}|\cos(\delta_{C1}-\delta_{C0})\cos\theta_{\alpha}   \notag \\
    & +   \frac{\sqrt{5}}{2}|t_{C0}||t_{C2}|\cos(\delta_{C2}-\delta_{C0})(1+3\cos2\theta_{\alpha})  \notag \\
    & +   \frac{\sqrt{7}}{4}|t_{C0}||t_{C3}|\cos(\delta_{C3}-\delta_{C0})(3\cos\theta_{\alpha}+ 5 \cos3\theta_{\alpha})  \notag \\
    & + \frac{\sqrt{15}}{4}|t_{C1}||t_{C2}|\cos(\delta_{C2}-\delta_{C1})(5\cos\theta_{\alpha}+3\cos3\theta_{\alpha}) \notag \\
    & + \frac{\sqrt{21}}{8}|t_{C1}||t_{C3}|\cos(\delta_{C3}-\delta_{C1})(3+ 8\cos2\theta_{\alpha} +5 \cos4\theta_{\alpha}) \notag \\
    & + \frac{\sqrt{35}}{32}|t_{C2}||t_{C3}|\cos(\delta_{C3}-\delta_{C2})(30\cos\theta_{\alpha} + 19\cos3\theta_{\alpha} +15 \cos5\theta_{\alpha}) &&
\end{flalign}
\begin{flalign}
R_T &= \overline{X^{1,1}_{fi}} + \overline{X^{-1,-1}_{fi}}  \notag \\
    &= \frac{3}{2} |t_{E1}|^2\sin^2\theta_{\alpha} +\frac{15}{8} |t_{E2}|^2\sin^22\theta_{\alpha} +\frac{21}{256} |t_{E3}|^2(\sin2\theta_{\alpha} +5 \sin3\theta_{\alpha})^2 \notag \\
    & + \frac{3\sqrt{5}}{2}|t_{E1}||t_{E2}|\cos(\delta_{E2}-\delta_{E1})(\sin\theta_{\alpha}\sin2\theta_{\alpha}) \notag \\
    & + \frac{3}{2} \sqrt{\frac{5}{2}}|t_{E1}||t_{E3}|\cos(\delta_{E3}-\delta_{E1})(\sin^22\theta_{\alpha} - \sin^4\theta_{\alpha}) \notag \\
    & + \frac{3}{4} \sqrt{\frac{35}{2}}|t_{E2}||t_{E3}|\cos(\delta_{E3}-\delta_{E2})\sin2\theta_{\alpha}(\sin\theta_{\alpha}+\sin3\theta_{\alpha} - \sin^3\theta_{\alpha}) &&
\end{flalign}
\begin{flalign}
R_{TT} &= \overline{X^{1,-1}_{fi}} + \overline{X^{-1,1}_{fi}} = - R_T \cos2\phi_\alpha && .
\end{flalign}
\begin{flalign}
R_{TL} &= -2 Re \bigg[ \overline{X^{0,1}_{fi}} - \overline{X^{0,-1}_{fi}} \bigg] \notag \\
       &= \cos\phi_\alpha\bigg\{  -   2\sqrt{3}|t_{C0}||t_{E1}|\cos(\delta_{E1}-\delta_{C0})\sin\theta_{\alpha}  -   \sqrt{15}|t_{C0}||t_{E2}|\cos(\delta_{E2}-\delta_{C0})\sin2\theta_{\alpha} \notag \\
       & -   \frac{1}{4}\sqrt{\frac{21}{2}} |t_{C0}||t_{E3}|\cos(\delta_{E3}-\delta_{C0})(\sin\theta_{\alpha}+5\sin3\theta_{\alpha}) - 3|t_{C1}||t_{E1}|\cos(\delta_{C1}-\delta_{E1})\sin2\theta_{\alpha}  \notag \\
       & - \frac{3\sqrt{5}}{2}|t_{C1}||t_{E2}|\cos(\delta_{C1}-\delta_{E2})(\sin\theta_{\alpha}+\sin3\theta_{\alpha})  - \frac{3}{8}\sqrt{\frac{7}{2}}|t_{C1}||t_{E3}|\cos(\delta_{C1}-\delta_{E3})(6\sin2\theta_{\alpha} +5\sin4\theta_{\alpha})  \notag \\
       & + \frac{\sqrt{15}}{4}|t_{C2}||t_{E1}|\cos(\delta_{C2}-\delta_{E1})(\sin\theta_{\alpha}-3\sin3\theta_{\alpha})  - \frac{5\sqrt{3}}{8}|t_{C2}||t_{E2}|\cos(\delta_{C2}-\delta_{E2})(2\sin2\theta_{\alpha}-3\sin4\theta_{\alpha})  \notag \\
 & +  \frac{1}{32}\sqrt{\frac{3}{70}}|t_{C2}||t_{E3}|\cos(\delta_{C2}-\delta_{E3})(278\sin\theta_{\alpha}-455\sin3\theta_{\alpha}-525\sin5\theta_{\alpha}) \notag \\
 & +  \frac{\sqrt{21}}{8}|t_{C3}||t_{E1}|\cos(\delta_{C3}-\delta_{E1})(2\sin2\theta_{\alpha}-5\sin4\theta_{\alpha}) \notag \\
 & +  \frac{\sqrt{105}}{16}|t_{C3}||t_{E2}|\cos(\delta_{C3}-\delta_{E2})(2\sin\theta_{\alpha}-3\sin3\theta_{\alpha}-5\sin5
\theta_{\alpha}) \notag \\  
& -  \frac{7}{64} \sqrt{\frac{3}{2}}|t_{C3}||t_{E3}|\cos(\delta_{C3}-\delta_{E3})(13\sin2\theta_{\alpha}+20\sin4\theta_{\alpha}+25\sin6\theta_{\alpha}) \bigg\}  &&
\end{flalign}
\end{widetext}




\bibliography{O16electrodisintegration}

\begin{thebibliography}{65}%
\makeatletter
\providecommand \@ifxundefined [1]{%
 \@ifx{#1\undefined}
}%
\providecommand \@ifnum [1]{%
 \ifnum #1\expandafter \@firstoftwo
 \else \expandafter \@secondoftwo
 \fi
}%
\providecommand \@ifx [1]{%
 \ifx #1\expandafter \@firstoftwo
 \else \expandafter \@secondoftwo
 \fi
}%
\providecommand \natexlab [1]{#1}%
\providecommand \enquote  [1]{``#1''}%
\providecommand \bibnamefont  [1]{#1}%
\providecommand \bibfnamefont [1]{#1}%
\providecommand \citenamefont [1]{#1}%
\providecommand \href@noop [0]{\@secondoftwo}%
\providecommand \href [0]{\begingroup \@sanitize@url \@href}%
\providecommand \@href[1]{\@@startlink{#1}\@@href}%
\providecommand \@@href[1]{\endgroup#1\@@endlink}%
\providecommand \@sanitize@url [0]{\catcode `\\12\catcode `\$12\catcode
  `\&12\catcode `\#12\catcode `\^12\catcode `\_12\catcode `\%12\relax}%
\providecommand \@@startlink[1]{}%
\providecommand \@@endlink[0]{}%
\providecommand \url  [0]{\begingroup\@sanitize@url \@url }%
\providecommand \@url [1]{\endgroup\@href {#1}{\urlprefix }}%
\providecommand \urlprefix  [0]{URL }%
\providecommand \Eprint [0]{\href }%
\providecommand \doibase [0]{http://dx.doi.org/}%
\providecommand \selectlanguage [0]{\@gobble}%
\providecommand \bibinfo  [0]{\@secondoftwo}%
\providecommand \bibfield  [0]{\@secondoftwo}%
\providecommand \translation [1]{[#1]}%
\providecommand \BibitemOpen [0]{}%
\providecommand \bibitemStop [0]{}%
\providecommand \bibitemNoStop [0]{.\EOS\space}%
\providecommand \EOS [0]{\spacefactor3000\relax}%
\providecommand \BibitemShut  [1]{\csname bibitem#1\endcsname}%
\let\auto@bib@innerbib\@empty
\bibitem [{\citenamefont {Rolfs}\ and\ \citenamefont
  {Barnes}(1990)}]{Rolfs1990}%
  \BibitemOpen
  \bibfield  {author} {\bibinfo {author} {\bibfnamefont {C.}~\bibnamefont
  {Rolfs}}\ and\ \bibinfo {author} {\bibfnamefont {C.}~\bibnamefont {Barnes}},\
  }\href {\doibase 10.1146/annurev.ns.40.120190.000401} {\bibfield  {journal}
  {\bibinfo  {journal} {Annu. Rev. Nucl. Part. Sci.}\ }\textbf {\bibinfo
  {volume} {40}},\ \bibinfo {pages} {45} (\bibinfo {year} {1990})}\BibitemShut
  {NoStop}%
\bibitem [{\citenamefont {Woosley}\ \emph {et~al.}(2003)\citenamefont
  {Woosley}, \citenamefont {Heger}, \citenamefont {Rauscher},\ and\
  \citenamefont {Hoffman}}]{Woosley2003}%
  \BibitemOpen
  \bibfield  {author} {\bibinfo {author} {\bibfnamefont {S.~E.}\ \bibnamefont
  {Woosley}}, \bibinfo {author} {\bibfnamefont {A.}~\bibnamefont {Heger}},
  \bibinfo {author} {\bibfnamefont {T.}~\bibnamefont {Rauscher}}, \ and\
  \bibinfo {author} {\bibfnamefont {R.~D.}\ \bibnamefont {Hoffman}},\ }\href
  {\doibase 10.1016/S0375-9474(03)00673-0} {\bibfield  {journal} {\bibinfo
  {journal} {Nucl. Phys. A}\ }\textbf {\bibinfo {volume} {718}},\ \bibinfo
  {pages} {3c} (\bibinfo {year} {2003})}\BibitemShut {NoStop}%
\bibitem [{\citenamefont {Donnelly}\ \emph {et~al.}(2017)\citenamefont
  {Donnelly}, \citenamefont {Formaggio}, \citenamefont {Holstein},
  \citenamefont {Milner},\ and\ \citenamefont {Surrow}}]{DFHMSa}%
  \BibitemOpen
  \bibfield  {author} {\bibinfo {author} {\bibfnamefont {T.~W.}\ \bibnamefont
  {Donnelly}}, \bibinfo {author} {\bibfnamefont {J.~A.}\ \bibnamefont
  {Formaggio}}, \bibinfo {author} {\bibfnamefont {B.~R.}\ \bibnamefont
  {Holstein}}, \bibinfo {author} {\bibfnamefont {R.~G.}\ \bibnamefont
  {Milner}}, \ and\ \bibinfo {author} {\bibfnamefont {B.}~\bibnamefont
  {Surrow}},\ }\href {\doibase 10.1017/97811390282} {\emph {\bibinfo {title}
  {Foundations of {N}uclear and {P}article {P}hysics}}},\ \bibinfo {edition}
  {1st}\ ed.\ (\bibinfo  {publisher} {Cambridge University Press},\ \bibinfo
  {year} {2017})\ p.\ \bibinfo {pages} {533}\BibitemShut {NoStop}%
\bibitem [{\citenamefont {Tilley}\ \emph {et~al.}(1993)\citenamefont {Tilley},
  \citenamefont {Weller},\ and\ \citenamefont {Cheves}}]{Tilley1993}%
  \BibitemOpen
  \bibfield  {author} {\bibinfo {author} {\bibfnamefont {D.~R.}\ \bibnamefont
  {Tilley}}, \bibinfo {author} {\bibfnamefont {H.~R.}\ \bibnamefont {Weller}},
  \ and\ \bibinfo {author} {\bibfnamefont {C.~M.}\ \bibnamefont {Cheves}},\
  }\href {\doibase 10.1016/0375-9474(93)90073-7} {\bibfield  {journal}
  {\bibinfo  {journal} {Nucl. Phys. A}\ }\textbf {\bibinfo {volume} {565}},\
  \bibinfo {pages} {1} (\bibinfo {year} {1993})}\BibitemShut {NoStop}%
\bibitem [{\citenamefont {Dyer}\ and\ \citenamefont
  {Barnes}(1974)}]{DyerBarnes1974}%
  \BibitemOpen
  \bibfield  {author} {\bibinfo {author} {\bibfnamefont {P.}~\bibnamefont
  {Dyer}}\ and\ \bibinfo {author} {\bibfnamefont {C.~A.}\ \bibnamefont
  {Barnes}},\ }\href {\doibase 10.1016/0375-9474(74)90470-9} {\bibfield
  {journal} {\bibinfo  {journal} {Nucl. Phys. A}\ }\textbf {\bibinfo {volume}
  {233}},\ \bibinfo {pages} {495} (\bibinfo {year} {1974})}\BibitemShut
  {NoStop}%
\bibitem [{\citenamefont {Redder}\ \emph {et~al.}(1987)\citenamefont {Redder},
  \citenamefont {Becker}, \citenamefont {Rolfs}, \citenamefont {Trautvetter},
  \citenamefont {Donoghue}, \citenamefont {Rinckel}, \citenamefont {Hammer},\
  and\ \citenamefont {Langanke}}]{Redder1987}%
  \BibitemOpen
  \bibfield  {author} {\bibinfo {author} {\bibfnamefont {A.}~\bibnamefont
  {Redder}}, \bibinfo {author} {\bibfnamefont {H.~W.}\ \bibnamefont {Becker}},
  \bibinfo {author} {\bibfnamefont {C.}~\bibnamefont {Rolfs}}, \bibinfo
  {author} {\bibfnamefont {H.~P.}\ \bibnamefont {Trautvetter}}, \bibinfo
  {author} {\bibfnamefont {T.~R.}\ \bibnamefont {Donoghue}}, \bibinfo {author}
  {\bibfnamefont {T.~C.}\ \bibnamefont {Rinckel}}, \bibinfo {author}
  {\bibfnamefont {J.~W.}\ \bibnamefont {Hammer}}, \ and\ \bibinfo {author}
  {\bibfnamefont {K.}~\bibnamefont {Langanke}},\ }\href {\doibase
  10.1016/0375-9474(87)90555-0} {\bibfield  {journal} {\bibinfo  {journal}
  {Nucl. Phys. A}\ }\textbf {\bibinfo {volume} {462}},\ \bibinfo {pages} {385}
  (\bibinfo {year} {1987})}\BibitemShut {NoStop}%
\bibitem [{\citenamefont {Kremer}\ \emph {et~al.}(1988)\citenamefont {Kremer},
  \citenamefont {Barnes}, \citenamefont {Chang}, \citenamefont {Evans},
  \citenamefont {Filippone}, \citenamefont {Hahn},\ and\ \citenamefont
  {Mitchell}}]{Kremer1988}%
  \BibitemOpen
  \bibfield  {author} {\bibinfo {author} {\bibfnamefont {R.~M.}\ \bibnamefont
  {Kremer}}, \bibinfo {author} {\bibfnamefont {C.~A.}\ \bibnamefont {Barnes}},
  \bibinfo {author} {\bibfnamefont {K.~H.}\ \bibnamefont {Chang}}, \bibinfo
  {author} {\bibfnamefont {H.~C.}\ \bibnamefont {Evans}}, \bibinfo {author}
  {\bibfnamefont {B.~W.}\ \bibnamefont {Filippone}}, \bibinfo {author}
  {\bibfnamefont {K.~H.}\ \bibnamefont {Hahn}}, \ and\ \bibinfo {author}
  {\bibfnamefont {L.~W.}\ \bibnamefont {Mitchell}},\ }\href {\doibase
  10.1103/PhysRevLett.60.1475} {\bibfield  {journal} {\bibinfo  {journal}
  {Phys. Rev. Lett.}\ }\textbf {\bibinfo {volume} {60}},\ \bibinfo {pages}
  {1475} (\bibinfo {year} {1988})}\BibitemShut {NoStop}%
\bibitem [{\citenamefont {Ouellet}\ \emph {et~al.}(1996)\citenamefont
  {Ouellet}, \citenamefont {Butler}, \citenamefont {Evans}, \citenamefont
  {Lee}, \citenamefont {Leslie}, \citenamefont {MacArthur}, \citenamefont
  {McLatchie}, \citenamefont {Mak}, \citenamefont {Skensved}, \citenamefont
  {Whitton}, \citenamefont {Zhao},\ and\ \citenamefont
  {Alexander}}]{Ouellet1996}%
  \BibitemOpen
  \bibfield  {author} {\bibinfo {author} {\bibfnamefont {J.~M.~L.}\
  \bibnamefont {Ouellet}}, \bibinfo {author} {\bibfnamefont {M.~N.}\
  \bibnamefont {Butler}}, \bibinfo {author} {\bibfnamefont {H.~C.}\
  \bibnamefont {Evans}}, \bibinfo {author} {\bibfnamefont {H.~W.}\ \bibnamefont
  {Lee}}, \bibinfo {author} {\bibfnamefont {J.~R.}\ \bibnamefont {Leslie}},
  \bibinfo {author} {\bibfnamefont {J.~D.}\ \bibnamefont {MacArthur}}, \bibinfo
  {author} {\bibfnamefont {W.}~\bibnamefont {McLatchie}}, \bibinfo {author}
  {\bibfnamefont {H.-B.}\ \bibnamefont {Mak}}, \bibinfo {author} {\bibfnamefont
  {P.}~\bibnamefont {Skensved}}, \bibinfo {author} {\bibfnamefont {J.~L.}\
  \bibnamefont {Whitton}}, \bibinfo {author} {\bibfnamefont {X.}~\bibnamefont
  {Zhao}}, \ and\ \bibinfo {author} {\bibfnamefont {T.~K.}\ \bibnamefont
  {Alexander}},\ }\href {\doibase 10.1103/PhysRevC.54.1982} {\bibfield
  {journal} {\bibinfo  {journal} {Phys. Rev. C}\ }\textbf {\bibinfo {volume}
  {54}},\ \bibinfo {pages} {1982} (\bibinfo {year} {1996})}\BibitemShut
  {NoStop}%
\bibitem [{\citenamefont {Roters}\ \emph {et~al.}(1999)\citenamefont {Roters},
  \citenamefont {Rolfs}, \citenamefont {Strieder},\ and\ \citenamefont
  {Trautvetter}}]{Roters1999}%
  \BibitemOpen
  \bibfield  {author} {\bibinfo {author} {\bibfnamefont {G.}~\bibnamefont
  {Roters}}, \bibinfo {author} {\bibfnamefont {C.}~\bibnamefont {Rolfs}},
  \bibinfo {author} {\bibfnamefont {F.}~\bibnamefont {Strieder}}, \ and\
  \bibinfo {author} {\bibfnamefont {H.}~\bibnamefont {Trautvetter}},\ }\href
  {\doibase 10.1007/s100500050369} {\bibfield  {journal} {\bibinfo  {journal}
  {Eur. Phys. J. A}\ }\textbf {\bibinfo {volume} {6}},\ \bibinfo {pages} {451}
  (\bibinfo {year} {1999})}\BibitemShut {NoStop}%
\bibitem [{\citenamefont {Gialanella}\ \emph {et~al.}(2001)\citenamefont
  {Gialanella}, \citenamefont {Rogalla}, \citenamefont {Strieder},
  \citenamefont {Theis}, \citenamefont {Gy{\"{u}}rki}, \citenamefont {Agodi},
  \citenamefont {Alba}, \citenamefont {Aliotta}, \citenamefont {Campajola},
  \citenamefont {Zoppo} \emph {et~al.}}]{Gialanella2001}%
  \BibitemOpen
  \bibfield  {author} {\bibinfo {author} {\bibfnamefont {L.}~\bibnamefont
  {Gialanella}}, \bibinfo {author} {\bibfnamefont {D.}~\bibnamefont {Rogalla}},
  \bibinfo {author} {\bibfnamefont {F.}~\bibnamefont {Strieder}}, \bibinfo
  {author} {\bibfnamefont {S.}~\bibnamefont {Theis}}, \bibinfo {author}
  {\bibfnamefont {G.}~\bibnamefont {Gy{\"{u}}rki}}, \bibinfo {author}
  {\bibfnamefont {C.}~\bibnamefont {Agodi}}, \bibinfo {author} {\bibfnamefont
  {R.}~\bibnamefont {Alba}}, \bibinfo {author} {\bibfnamefont {M.}~\bibnamefont
  {Aliotta}}, \bibinfo {author} {\bibfnamefont {L.}~\bibnamefont {Campajola}},
  \bibinfo {author} {\bibfnamefont {A.~D.}\ \bibnamefont {Zoppo}},  \emph
  {et~al.},\ }\href {\doibase 10.1007/s100500170075} {\bibfield  {journal}
  {\bibinfo  {journal} {Eur. Phys. J. A}\ }\textbf {\bibinfo {volume} {11}},\
  \bibinfo {pages} {357} (\bibinfo {year} {2001})}\BibitemShut {NoStop}%
\bibitem [{\citenamefont {Kunz}\ \emph {et~al.}(2001)\citenamefont {Kunz},
  \citenamefont {Jaeger}, \citenamefont {Mayer}, \citenamefont {Hammer},
  \citenamefont {Staudt}, \citenamefont {Harissopulos},\ and\ \citenamefont
  {Paradellis}}]{Kunz2001}%
  \BibitemOpen
  \bibfield  {author} {\bibinfo {author} {\bibfnamefont {R.}~\bibnamefont
  {Kunz}}, \bibinfo {author} {\bibfnamefont {M.}~\bibnamefont {Jaeger}},
  \bibinfo {author} {\bibfnamefont {A.}~\bibnamefont {Mayer}}, \bibinfo
  {author} {\bibfnamefont {J.~W.}\ \bibnamefont {Hammer}}, \bibinfo {author}
  {\bibfnamefont {G.}~\bibnamefont {Staudt}}, \bibinfo {author} {\bibfnamefont
  {S.}~\bibnamefont {Harissopulos}}, \ and\ \bibinfo {author} {\bibfnamefont
  {T.}~\bibnamefont {Paradellis}},\ }\href {\doibase
  10.1103/PhysRevLett.86.3244} {\bibfield  {journal} {\bibinfo  {journal}
  {Phys. Rev. Lett.}\ }\textbf {\bibinfo {volume} {86}},\ \bibinfo {pages}
  {3244} (\bibinfo {year} {2001})}\BibitemShut {NoStop}%
\bibitem [{\citenamefont {Kunz}(2002)}]{Kunz2002phd}%
  \BibitemOpen
  \bibfield  {author} {\bibinfo {author} {\bibfnamefont {R.~W.}\ \bibnamefont
  {Kunz}},\ }\href@noop {} {Ph.D. thesis},\ \bibinfo  {school} {Univ. of
  Stuttgart} (\bibinfo {year} {2002})\BibitemShut {NoStop}%
\bibitem [{\citenamefont {Fey}(2004)}]{Fey2004phd}%
  \BibitemOpen
  \bibfield  {author} {\bibinfo {author} {\bibfnamefont {M.}~\bibnamefont
  {Fey}},\ }\href@noop {} {Ph.D. thesis},\ \bibinfo  {school} {Univ. of
  Stuttgart} (\bibinfo {year} {2004})\BibitemShut {NoStop}%
\bibitem [{\citenamefont {Sch{\"u}rmann}\ \emph {et~al.}(2005)\citenamefont
  {Sch{\"u}rmann}, \citenamefont {Leva}, \citenamefont {Gialanella},
  \citenamefont {Rogalla}, \citenamefont {Strieder}, \citenamefont {Cesare},
  \citenamefont {D'Onofrio}, \citenamefont {Imbriani}, \citenamefont {Kunz},
  \citenamefont {Lubritto} \emph {et~al.}}]{Schurmann2005}%
  \BibitemOpen
  \bibfield  {author} {\bibinfo {author} {\bibfnamefont {D.}~\bibnamefont
  {Sch{\"u}rmann}}, \bibinfo {author} {\bibfnamefont {A.~D.}\ \bibnamefont
  {Leva}}, \bibinfo {author} {\bibfnamefont {L.}~\bibnamefont {Gialanella}},
  \bibinfo {author} {\bibfnamefont {D.}~\bibnamefont {Rogalla}}, \bibinfo
  {author} {\bibfnamefont {F.}~\bibnamefont {Strieder}}, \bibinfo {author}
  {\bibfnamefont {N.~D.}\ \bibnamefont {Cesare}}, \bibinfo {author}
  {\bibfnamefont {A.}~\bibnamefont {D'Onofrio}}, \bibinfo {author}
  {\bibfnamefont {G.}~\bibnamefont {Imbriani}}, \bibinfo {author}
  {\bibfnamefont {R.}~\bibnamefont {Kunz}}, \bibinfo {author} {\bibfnamefont
  {C.}~\bibnamefont {Lubritto}},  \emph {et~al.},\ }\href {\doibase
  10.1140/epja/i2005-10175-2} {\bibfield  {journal} {\bibinfo  {journal} {Eur.
  Phys. J. A}\ }\textbf {\bibinfo {volume} {26}},\ \bibinfo {pages} {301–305}
  (\bibinfo {year} {2005})}\BibitemShut {NoStop}%
\bibitem [{\citenamefont {Assun\c{c}{\~{a}}o}\ \emph
  {et~al.}(2006)\citenamefont {Assun\c{c}{\~{a}}o}, \citenamefont {Fey},
  \citenamefont {Lefebvre-Schuhl}, \citenamefont {Kiener}, \citenamefont
  {Tatischeff}, \citenamefont {Hammer}, \citenamefont {Beck}, \citenamefont
  {Boukari-Pelissie}, \citenamefont {Coc}, \citenamefont {Correia} \emph
  {et~al.}}]{Assuncao2006}%
  \BibitemOpen
  \bibfield  {author} {\bibinfo {author} {\bibfnamefont {M.}~\bibnamefont
  {Assun\c{c}{\~{a}}o}}, \bibinfo {author} {\bibfnamefont {M.}~\bibnamefont
  {Fey}}, \bibinfo {author} {\bibfnamefont {A.}~\bibnamefont
  {Lefebvre-Schuhl}}, \bibinfo {author} {\bibfnamefont {J.}~\bibnamefont
  {Kiener}}, \bibinfo {author} {\bibfnamefont {V.}~\bibnamefont {Tatischeff}},
  \bibinfo {author} {\bibfnamefont {J.~W.}\ \bibnamefont {Hammer}}, \bibinfo
  {author} {\bibfnamefont {C.}~\bibnamefont {Beck}}, \bibinfo {author}
  {\bibfnamefont {C.}~\bibnamefont {Boukari-Pelissie}}, \bibinfo {author}
  {\bibfnamefont {A.}~\bibnamefont {Coc}}, \bibinfo {author} {\bibfnamefont
  {J.~J.}\ \bibnamefont {Correia}},  \emph {et~al.},\ }\href {\doibase
  10.1103/PhysRevC.73.055801} {\bibfield  {journal} {\bibinfo  {journal} {Phys.
  Rev. C}\ }\textbf {\bibinfo {volume} {73}},\ \bibinfo {pages} {055801}
  (\bibinfo {year} {2006})}\BibitemShut {NoStop}%
\bibitem [{\citenamefont {Makii}\ \emph {et~al.}(2009)\citenamefont {Makii},
  \citenamefont {Nagai}, \citenamefont {Shima}, \citenamefont {Segawa},
  \citenamefont {Mishima}, \citenamefont {Ueda}, \citenamefont {Igashira},\
  and\ \citenamefont {Ohsaki}}]{Makii2009}%
  \BibitemOpen
  \bibfield  {author} {\bibinfo {author} {\bibfnamefont {H.}~\bibnamefont
  {Makii}}, \bibinfo {author} {\bibfnamefont {Y.}~\bibnamefont {Nagai}},
  \bibinfo {author} {\bibfnamefont {T.}~\bibnamefont {Shima}}, \bibinfo
  {author} {\bibfnamefont {M.}~\bibnamefont {Segawa}}, \bibinfo {author}
  {\bibfnamefont {K.}~\bibnamefont {Mishima}}, \bibinfo {author} {\bibfnamefont
  {H.}~\bibnamefont {Ueda}}, \bibinfo {author} {\bibfnamefont {M.}~\bibnamefont
  {Igashira}}, \ and\ \bibinfo {author} {\bibfnamefont {T.}~\bibnamefont
  {Ohsaki}},\ }\href {\doibase 10.1103/PhysRevC.80.065802} {\bibfield
  {journal} {\bibinfo  {journal} {Phys. Rev. C}\ }\textbf {\bibinfo {volume}
  {80}},\ \bibinfo {pages} {065802} (\bibinfo {year} {2009})}\BibitemShut
  {NoStop}%
\bibitem [{\citenamefont {Sch{\"u}rmann}\ \emph {et~al.}(2011)\citenamefont
  {Sch{\"u}rmann}, \citenamefont {Leva}, \citenamefont {Gialanella},
  \citenamefont {Kunz}, \citenamefont {Strieder}, \citenamefont {Cesare},
  \citenamefont {Cesare}, \citenamefont {D'Onofrio}, \citenamefont {Fortak},
  \citenamefont {Imbriani} \emph {et~al.}}]{Schurmann2011}%
  \BibitemOpen
  \bibfield  {author} {\bibinfo {author} {\bibfnamefont {D.}~\bibnamefont
  {Sch{\"u}rmann}}, \bibinfo {author} {\bibfnamefont {A.~D.}\ \bibnamefont
  {Leva}}, \bibinfo {author} {\bibfnamefont {L.}~\bibnamefont {Gialanella}},
  \bibinfo {author} {\bibfnamefont {R.}~\bibnamefont {Kunz}}, \bibinfo {author}
  {\bibfnamefont {F.}~\bibnamefont {Strieder}}, \bibinfo {author}
  {\bibfnamefont {N.~D.}\ \bibnamefont {Cesare}}, \bibinfo {author}
  {\bibfnamefont {M.~D.}\ \bibnamefont {Cesare}}, \bibinfo {author}
  {\bibfnamefont {A.}~\bibnamefont {D'Onofrio}}, \bibinfo {author}
  {\bibfnamefont {K.}~\bibnamefont {Fortak}}, \bibinfo {author} {\bibfnamefont
  {G.}~\bibnamefont {Imbriani}},  \emph {et~al.},\ }\href {\doibase
  10.1016/j.physletb.2011.08.061} {\bibfield  {journal} {\bibinfo  {journal}
  {Phys. Lett. B}\ }\textbf {\bibinfo {volume} {703}},\ \bibinfo {pages} {557}
  (\bibinfo {year} {2011})}\BibitemShut {NoStop}%
\bibitem [{\citenamefont {Plag}\ \emph {et~al.}(2012)\citenamefont {Plag},
  \citenamefont {Reifarth}, \citenamefont {Heil}, \citenamefont {K{\"a}ppeler},
  \citenamefont {Rupp}, \citenamefont {Voss},\ and\ \citenamefont
  {Wisshak}}]{Plag2012}%
  \BibitemOpen
  \bibfield  {author} {\bibinfo {author} {\bibfnamefont {R.}~\bibnamefont
  {Plag}}, \bibinfo {author} {\bibfnamefont {R.}~\bibnamefont {Reifarth}},
  \bibinfo {author} {\bibfnamefont {M.}~\bibnamefont {Heil}}, \bibinfo {author}
  {\bibfnamefont {F.}~\bibnamefont {K{\"a}ppeler}}, \bibinfo {author}
  {\bibfnamefont {G.}~\bibnamefont {Rupp}}, \bibinfo {author} {\bibfnamefont
  {F.}~\bibnamefont {Voss}}, \ and\ \bibinfo {author} {\bibfnamefont
  {K.}~\bibnamefont {Wisshak}},\ }\href {\doibase 10.1103/PhysRevC.86.015805}
  {\bibfield  {journal} {\bibinfo  {journal} {Phys. Rev. C}\ }\textbf {\bibinfo
  {volume} {86}},\ \bibinfo {pages} {015805} (\bibinfo {year}
  {2012})}\BibitemShut {NoStop}%
\bibitem [{\citenamefont {Buchmann}\ \emph {et~al.}(1993)\citenamefont
  {Buchmann}, \citenamefont {Azuma}, \citenamefont {Barnes}, \citenamefont
  {D'Auria}, \citenamefont {Dombsky}, \citenamefont {Giesen}, \citenamefont
  {Jackson}, \citenamefont {King}, \citenamefont {Korteling}, \citenamefont
  {McNeely} \emph {et~al.}}]{Buchmann1993}%
  \BibitemOpen
  \bibfield  {author} {\bibinfo {author} {\bibfnamefont {L.}~\bibnamefont
  {Buchmann}}, \bibinfo {author} {\bibfnamefont {R.~E.}\ \bibnamefont {Azuma}},
  \bibinfo {author} {\bibfnamefont {C.~A.}\ \bibnamefont {Barnes}}, \bibinfo
  {author} {\bibfnamefont {J.~M.}\ \bibnamefont {D'Auria}}, \bibinfo {author}
  {\bibfnamefont {M.}~\bibnamefont {Dombsky}}, \bibinfo {author} {\bibfnamefont
  {U.}~\bibnamefont {Giesen}}, \bibinfo {author} {\bibfnamefont {K.~P.}\
  \bibnamefont {Jackson}}, \bibinfo {author} {\bibfnamefont {J.~D.}\
  \bibnamefont {King}}, \bibinfo {author} {\bibfnamefont {R.~G.}\ \bibnamefont
  {Korteling}}, \bibinfo {author} {\bibfnamefont {P.}~\bibnamefont {McNeely}},
  \emph {et~al.},\ }\href {\doibase 10.1103/PhysRevLett.70.726} {\bibfield
  {journal} {\bibinfo  {journal} {Phys. Rev. Lett.}\ }\textbf {\bibinfo
  {volume} {70}},\ \bibinfo {pages} {726} (\bibinfo {year} {1993})}\BibitemShut
  {NoStop}%
\bibitem [{\citenamefont {Azuma}\ \emph {et~al.}(1994)\citenamefont {Azuma},
  \citenamefont {Buchmann}, \citenamefont {Barker}, \citenamefont {Barnes},
  \citenamefont {D'Auria}, \citenamefont {Dombsky}, \citenamefont {Giesen},
  \citenamefont {Jackson}, \citenamefont {King}, \citenamefont {Korteling}
  \emph {et~al.}}]{Azuma1994}%
  \BibitemOpen
  \bibfield  {author} {\bibinfo {author} {\bibfnamefont {R.~E.}\ \bibnamefont
  {Azuma}}, \bibinfo {author} {\bibfnamefont {L.}~\bibnamefont {Buchmann}},
  \bibinfo {author} {\bibfnamefont {F.~C.}\ \bibnamefont {Barker}}, \bibinfo
  {author} {\bibfnamefont {C.~A.}\ \bibnamefont {Barnes}}, \bibinfo {author}
  {\bibfnamefont {J.~M.}\ \bibnamefont {D'Auria}}, \bibinfo {author}
  {\bibfnamefont {M.}~\bibnamefont {Dombsky}}, \bibinfo {author} {\bibfnamefont
  {U.}~\bibnamefont {Giesen}}, \bibinfo {author} {\bibfnamefont {K.~P.}\
  \bibnamefont {Jackson}}, \bibinfo {author} {\bibfnamefont {J.~D.}\
  \bibnamefont {King}}, \bibinfo {author} {\bibfnamefont {R.~G.}\ \bibnamefont
  {Korteling}},  \emph {et~al.},\ }\href {\doibase 10.1103/PhysRevC.50.1194}
  {\bibfield  {journal} {\bibinfo  {journal} {Phys. Rev. C}\ }\textbf {\bibinfo
  {volume} {50}},\ \bibinfo {pages} {1194} (\bibinfo {year}
  {1994})}\BibitemShut {NoStop}%
\bibitem [{\citenamefont {Tang}\ \emph {et~al.}(2010)\citenamefont {Tang},
  \citenamefont {Rehm}, \citenamefont {Ahmad}, \citenamefont {Brune},
  \citenamefont {Champagne}, \citenamefont {Greene}, \citenamefont {Hecht},
  \citenamefont {Henderson}, \citenamefont {Janssens}, \citenamefont {Jiang}
  \emph {et~al.}}]{Tang2010}%
  \BibitemOpen
  \bibfield  {author} {\bibinfo {author} {\bibfnamefont {X.~D.}\ \bibnamefont
  {Tang}}, \bibinfo {author} {\bibfnamefont {K.~E.}\ \bibnamefont {Rehm}},
  \bibinfo {author} {\bibfnamefont {I.}~\bibnamefont {Ahmad}}, \bibinfo
  {author} {\bibfnamefont {C.~R.}\ \bibnamefont {Brune}}, \bibinfo {author}
  {\bibfnamefont {A.}~\bibnamefont {Champagne}}, \bibinfo {author}
  {\bibfnamefont {J.~P.}\ \bibnamefont {Greene}}, \bibinfo {author}
  {\bibfnamefont {A.}~\bibnamefont {Hecht}}, \bibinfo {author} {\bibfnamefont
  {D.~J.}\ \bibnamefont {Henderson}}, \bibinfo {author} {\bibfnamefont
  {R.~V.~F.}\ \bibnamefont {Janssens}}, \bibinfo {author} {\bibfnamefont
  {C.~L.}\ \bibnamefont {Jiang}},  \emph {et~al.},\ }\href {\doibase
  10.1103/PhysRevC.81.045809} {\bibfield  {journal} {\bibinfo  {journal} {Phys.
  Rev.C}\ }\textbf {\bibinfo {volume} {81}},\ \bibinfo {pages} {045809}
  (\bibinfo {year} {2010})}\BibitemShut {NoStop}%
\bibitem [{\citenamefont {Plaga}\ \emph {et~al.}(1987)\citenamefont {Plaga},
  \citenamefont {Becker}, \citenamefont {Redder}, \citenamefont {Rolfs},
  \citenamefont {Trautvetter},\ and\ \citenamefont {Langanke}}]{Plaga1987}%
  \BibitemOpen
  \bibfield  {author} {\bibinfo {author} {\bibfnamefont {R.}~\bibnamefont
  {Plaga}}, \bibinfo {author} {\bibfnamefont {H.~W.}\ \bibnamefont {Becker}},
  \bibinfo {author} {\bibfnamefont {A.}~\bibnamefont {Redder}}, \bibinfo
  {author} {\bibfnamefont {C.}~\bibnamefont {Rolfs}}, \bibinfo {author}
  {\bibfnamefont {H.~P.}\ \bibnamefont {Trautvetter}}, \ and\ \bibinfo {author}
  {\bibfnamefont {K.}~\bibnamefont {Langanke}},\ }\href {\doibase
  10.1016/0375-9474(87)90436-2} {\bibfield  {journal} {\bibinfo  {journal}
  {Nucl. Phys. A}\ }\textbf {\bibinfo {volume} {465}},\ \bibinfo {pages} {291}
  (\bibinfo {year} {1987})}\BibitemShut {NoStop}%
\bibitem [{\citenamefont {Tischhauser}\ \emph {et~al.}(2009)\citenamefont
  {Tischhauser}, \citenamefont {Couture}, \citenamefont {Detwiler},
  \citenamefont {G{\"o}rres}, \citenamefont {Ugalde}, \citenamefont {Stech},
  \citenamefont {Wiescher}, \citenamefont {Heil}, \citenamefont {K{\"a}ppeler},
  \citenamefont {Azuma},\ and\ \citenamefont {Buchmann}}]{Tischhauser2009}%
  \BibitemOpen
  \bibfield  {author} {\bibinfo {author} {\bibfnamefont {P.}~\bibnamefont
  {Tischhauser}}, \bibinfo {author} {\bibfnamefont {A.}~\bibnamefont
  {Couture}}, \bibinfo {author} {\bibfnamefont {R.}~\bibnamefont {Detwiler}},
  \bibinfo {author} {\bibfnamefont {J.}~\bibnamefont {G{\"o}rres}}, \bibinfo
  {author} {\bibfnamefont {C.}~\bibnamefont {Ugalde}}, \bibinfo {author}
  {\bibfnamefont {E.}~\bibnamefont {Stech}}, \bibinfo {author} {\bibfnamefont
  {M.}~\bibnamefont {Wiescher}}, \bibinfo {author} {\bibfnamefont
  {M.}~\bibnamefont {Heil}}, \bibinfo {author} {\bibfnamefont {F.}~\bibnamefont
  {K{\"a}ppeler}}, \bibinfo {author} {\bibfnamefont {R.~E.}\ \bibnamefont
  {Azuma}}, \ and\ \bibinfo {author} {\bibfnamefont {L.}~\bibnamefont
  {Buchmann}},\ }\href {\doibase 10.1103/PhysRevC.79.055803} {\bibfield
  {journal} {\bibinfo  {journal} {Phys. Rev. C}\ }\textbf {\bibinfo {volume}
  {79}},\ \bibinfo {pages} {055803} (\bibinfo {year} {2009})}\BibitemShut
  {NoStop}%
\bibitem [{\citenamefont {deBoer}\ \emph {et~al.}(2017)\citenamefont {deBoer},
  \citenamefont {G{\"o}rres}, \citenamefont {Wiescher}, \citenamefont {Azuma},
  \citenamefont {Best}, \citenamefont {Brune}, \citenamefont {Fields},
  \citenamefont {Jones}, \citenamefont {Pignatari}, \citenamefont {Sayre} \emph
  {et~al.}}]{deBoer2017}%
  \BibitemOpen
  \bibfield  {author} {\bibinfo {author} {\bibfnamefont {R.~J.}\ \bibnamefont
  {deBoer}}, \bibinfo {author} {\bibfnamefont {J.}~\bibnamefont {G{\"o}rres}},
  \bibinfo {author} {\bibfnamefont {M.}~\bibnamefont {Wiescher}}, \bibinfo
  {author} {\bibfnamefont {R.}~\bibnamefont {Azuma}}, \bibinfo {author}
  {\bibfnamefont {A.}~\bibnamefont {Best}}, \bibinfo {author} {\bibfnamefont
  {C.}~\bibnamefont {Brune}}, \bibinfo {author} {\bibfnamefont
  {C.}~\bibnamefont {Fields}}, \bibinfo {author} {\bibfnamefont
  {S.}~\bibnamefont {Jones}}, \bibinfo {author} {\bibfnamefont
  {M.}~\bibnamefont {Pignatari}}, \bibinfo {author} {\bibfnamefont
  {D.}~\bibnamefont {Sayre}},  \emph {et~al.},\ }\href {\doibase
  10.1103/RevModPhys.89.035007} {\bibfield  {journal} {\bibinfo  {journal}
  {Rev. Mod. Phys.}\ }\textbf {\bibinfo {volume} {89}},\ \bibinfo {pages}
  {035007} (\bibinfo {year} {2017})}\BibitemShut {NoStop}%
\bibitem [{\citenamefont {DiGiovine}\ \emph {et~al.}(2015)\citenamefont
  {DiGiovine}, \citenamefont {Henderson}, \citenamefont {Holt}, \citenamefont
  {Raut}, \citenamefont {Rehm}, \citenamefont {Robinson}, \citenamefont
  {Sonnenschein}, \citenamefont {Rusev}, \citenamefont {Tonchev},\ and\
  \citenamefont {Ugalde}}]{DiGiovine2015}%
  \BibitemOpen
  \bibfield  {author} {\bibinfo {author} {\bibfnamefont {B.}~\bibnamefont
  {DiGiovine}}, \bibinfo {author} {\bibfnamefont {D.}~\bibnamefont
  {Henderson}}, \bibinfo {author} {\bibfnamefont {R.~J.}\ \bibnamefont {Holt}},
  \bibinfo {author} {\bibfnamefont {R.}~\bibnamefont {Raut}}, \bibinfo {author}
  {\bibfnamefont {K.~E.}\ \bibnamefont {Rehm}}, \bibinfo {author}
  {\bibfnamefont {A.}~\bibnamefont {Robinson}}, \bibinfo {author}
  {\bibfnamefont {A.}~\bibnamefont {Sonnenschein}}, \bibinfo {author}
  {\bibfnamefont {G.}~\bibnamefont {Rusev}}, \bibinfo {author} {\bibfnamefont
  {A.~P.}\ \bibnamefont {Tonchev}}, \ and\ \bibinfo {author} {\bibfnamefont
  {C.}~\bibnamefont {Ugalde}},\ }\href {\doibase 10.1016/j.nima.2015.01.060}
  {\bibfield  {journal} {\bibinfo  {journal} {Nucl. Instrum. Methods Phys. Res.
  A}\ }\textbf {\bibinfo {volume} {781}},\ \bibinfo {pages} {96} (\bibinfo
  {year} {2015})}\BibitemShut {NoStop}%
\bibitem [{\citenamefont {Holt}\ \emph {et~al.}(2018)\citenamefont {Holt},
  \citenamefont {Filippone},\ and\ \citenamefont {Pieper}}]{Holt2018}%
  \BibitemOpen
  \bibfield  {author} {\bibinfo {author} {\bibfnamefont {R.~J.}\ \bibnamefont
  {Holt}}, \bibinfo {author} {\bibfnamefont {B.~W.}\ \bibnamefont {Filippone}},
  \ and\ \bibinfo {author} {\bibfnamefont {S.~C.}\ \bibnamefont {Pieper}},\
  }\href@noop {} {\  (\bibinfo {year} {2018})},\ \Eprint
  {http://arxiv.org/abs/1809.10176} {arXiv:1809.10176 [nucl-ex]} \BibitemShut
  {NoStop}%
\bibitem [{\citenamefont {Ugalde}\ \emph {et~al.}(2013)\citenamefont {Ugalde},
  \citenamefont {DiGiovine}, \citenamefont {Holt}, \citenamefont {Henderson},
  \citenamefont {Rehm}, \citenamefont {Suleiman}, \citenamefont {Freyberger},
  \citenamefont {Grames}, \citenamefont {Kazim}, \citenamefont {Poelker} \emph
  {et~al.}}]{Ugalde2013}%
  \BibitemOpen
  \bibfield  {author} {\bibinfo {author} {\bibfnamefont {C.}~\bibnamefont
  {Ugalde}}, \bibinfo {author} {\bibfnamefont {B.}~\bibnamefont {DiGiovine}},
  \bibinfo {author} {\bibfnamefont {R.~J.}\ \bibnamefont {Holt}}, \bibinfo
  {author} {\bibfnamefont {D.}~\bibnamefont {Henderson}}, \bibinfo {author}
  {\bibfnamefont {K.~E.}\ \bibnamefont {Rehm}}, \bibinfo {author}
  {\bibfnamefont {R.}~\bibnamefont {Suleiman}}, \bibinfo {author}
  {\bibfnamefont {A.}~\bibnamefont {Freyberger}}, \bibinfo {author}
  {\bibfnamefont {J.}~\bibnamefont {Grames}}, \bibinfo {author} {\bibfnamefont
  {R.}~\bibnamefont {Kazim}}, \bibinfo {author} {\bibfnamefont
  {M.}~\bibnamefont {Poelker}},  \emph {et~al.},\ }\href
  {{https://www.jlab.org/exp_prog/proposals/13/PR12-13-005.pdf}} {\bibfield
  {journal} {\bibinfo  {journal} {Jefferson Lab Proposal}\ }\textbf {\bibinfo
  {volume} {PR12-13-005}} (\bibinfo {year} {2013})}\BibitemShut {NoStop}%
\bibitem [{\citenamefont {Gai}\ \emph {et~al.}(2010)\citenamefont {Gai},
  \citenamefont {Ahmed}, \citenamefont {Stave}, \citenamefont {Zimmerman},
  \citenamefont {Breskin}, \citenamefont {Bromberger}, \citenamefont {Chechik},
  \citenamefont {Dangendorf}, \citenamefont {Delbar}, \citenamefont {France},
  \citenamefont {Henshaw} \emph {et~al.}}]{Gai2010}%
  \BibitemOpen
  \bibfield  {author} {\bibinfo {author} {\bibfnamefont {M.}~\bibnamefont
  {Gai}}, \bibinfo {author} {\bibfnamefont {M.~W.}\ \bibnamefont {Ahmed}},
  \bibinfo {author} {\bibfnamefont {S.~C.}\ \bibnamefont {Stave}}, \bibinfo
  {author} {\bibfnamefont {W.~R.}\ \bibnamefont {Zimmerman}}, \bibinfo {author}
  {\bibfnamefont {A.}~\bibnamefont {Breskin}}, \bibinfo {author} {\bibfnamefont
  {B.}~\bibnamefont {Bromberger}}, \bibinfo {author} {\bibfnamefont
  {R.}~\bibnamefont {Chechik}}, \bibinfo {author} {\bibfnamefont
  {V.}~\bibnamefont {Dangendorf}}, \bibinfo {author} {\bibfnamefont
  {T.}~\bibnamefont {Delbar}}, \bibinfo {author} {\bibfnamefont {R.~H.}\
  \bibnamefont {France}}, \bibinfo {author} {\bibfnamefont {S.~S.}\
  \bibnamefont {Henshaw}},  \emph {et~al.},\ }\href {\doibase
  10.1088/1748-0221/5/12/P12004} {\bibfield  {journal} {\bibinfo  {journal}
  {JINST}\ }\textbf {\bibinfo {volume} {5}},\ \bibinfo {pages} {P12004}
  (\bibinfo {year} {2010})}\BibitemShut {NoStop}%
\bibitem [{\citenamefont {Azuma}\ \emph {et~al.}(2010)\citenamefont {Azuma},
  \citenamefont {Uberseder}, \citenamefont {Simpson}, \citenamefont {Brune},
  \citenamefont {Costantini}, \citenamefont {de~Boer}, \citenamefont
  {J.~G{\"o}rres}, \citenamefont {LeBlanc}, \citenamefont {Ugalde},\ and\
  \citenamefont {Wiescher}}]{Azuma2010}%
  \BibitemOpen
  \bibfield  {author} {\bibinfo {author} {\bibfnamefont {R.~E.}\ \bibnamefont
  {Azuma}}, \bibinfo {author} {\bibfnamefont {E.}~\bibnamefont {Uberseder}},
  \bibinfo {author} {\bibfnamefont {E.~C.}\ \bibnamefont {Simpson}}, \bibinfo
  {author} {\bibfnamefont {C.~R.}\ \bibnamefont {Brune}}, \bibinfo {author}
  {\bibfnamefont {H.}~\bibnamefont {Costantini}}, \bibinfo {author}
  {\bibfnamefont {R.~J.}\ \bibnamefont {de~Boer}}, \bibinfo {author}
  {\bibfnamefont {M.~H.}\ \bibnamefont {J.~G{\"o}rres}}, \bibinfo {author}
  {\bibfnamefont {P.~J.}\ \bibnamefont {LeBlanc}}, \bibinfo {author}
  {\bibfnamefont {C.}~\bibnamefont {Ugalde}}, \ and\ \bibinfo {author}
  {\bibfnamefont {M.}~\bibnamefont {Wiescher}},\ }\href {\doibase
  10.1103/PhysRevC.81.045805} {\bibfield  {journal} {\bibinfo  {journal} {Phys.
  Rev. C.}\ }\textbf {\bibinfo {volume} {81}},\ \bibinfo {pages} {045805}
  (\bibinfo {year} {2010})}\BibitemShut {NoStop}%
\bibitem [{\citenamefont {Tsentalovich}\ \emph {et~al.}(2000)\citenamefont
  {Tsentalovich}, \citenamefont {Turchinetz}, \citenamefont {Zwart},
  \citenamefont {Calarco}, \citenamefont {Butler}, \citenamefont {Nikolenko},
  \citenamefont {Shestakov},\ and\ \citenamefont
  {Toporkov}}]{Tsentalovich2000}%
  \BibitemOpen
  \bibfield  {author} {\bibinfo {author} {\bibfnamefont {E.}~\bibnamefont
  {Tsentalovich}}, \bibinfo {author} {\bibfnamefont {W.}~\bibnamefont
  {Turchinetz}}, \bibinfo {author} {\bibfnamefont {T.}~\bibnamefont {Zwart}},
  \bibinfo {author} {\bibfnamefont {J.}~\bibnamefont {Calarco}}, \bibinfo
  {author} {\bibfnamefont {M.}~\bibnamefont {Butler}}, \bibinfo {author}
  {\bibfnamefont {D.}~\bibnamefont {Nikolenko}}, \bibinfo {author}
  {\bibfnamefont {{\relax Yu}.}~\bibnamefont {Shestakov}}, \ and\ \bibinfo
  {author} {\bibfnamefont {D.}~\bibnamefont {Toporkov}},\ }\href@noop {} {}
  (\bibinfo {year} {2000}),\ \bibinfo {note} {{M}IT-Bates PAC proposal
  00-01}\BibitemShut {NoStop}%
\bibitem [{\citenamefont
  {Fri{{\v{s}}}{{\v{c}}}i{{\'{c}}}}(2017)}]{Friscic2017}%
  \BibitemOpen
  \bibfield  {author} {\bibinfo {author} {\bibfnamefont {I.}~\bibnamefont
  {Fri{{\v{s}}}{{\v{c}}}i{{\'{c}}}}},\ }\href
  {{http://einnconference.org/2017/wp-content/uploads/2017/10/EINN-2017-Conference-Abstracts-2810.pdf}}
  {\enquote {\bibinfo {title} {Electrodisintegration of {$^{16}$}{O} as a tool
  for investigating the {$^{12}$}{C}{($\alpha$,$\gamma$)$^{16}$}{O}
  reaction},}\ } (\bibinfo {year} {2017}),\ \bibinfo {note} {conf. Abstracts
  EINN2017}\BibitemShut {NoStop}%
\bibitem [{\citenamefont {Lunkenheimer}(2017)}]{Lunkenheimer2017}%
  \BibitemOpen
  \bibfield  {author} {\bibinfo {author} {\bibfnamefont {S.}~\bibnamefont
  {Lunkenheimer}},\ }\href
  {{http://wwwa1.kph.uni-mainz.de/WE-Heraeus-Seminar/650-Program1.pdf}}
  {\enquote {\bibinfo {title} {Studies of the nucleosynthesis
  {$^{12}$}{C}{($\alpha$,$\gamma$)$^{16}$}{O} in inverse kinematics for the
  {MAGIX} experiment at {MESA}},}\ } (\bibinfo {year} {2017}),\ \bibinfo {note}
  {650. WE-Heraeus-Seminar}\BibitemShut {NoStop}%
\bibitem [{\citenamefont {Raskin}\ and\ \citenamefont
  {Donnelly}(1989)}]{Raskin1989}%
  \BibitemOpen
  \bibfield  {author} {\bibinfo {author} {\bibfnamefont {A.~S.}\ \bibnamefont
  {Raskin}}\ and\ \bibinfo {author} {\bibfnamefont {T.~W.}\ \bibnamefont
  {Donnelly}},\ }\href {\doibase 10.1016/0003-4916(89)90337-0} {\bibfield
  {journal} {\bibinfo  {journal} {Ann. Phys.}\ }\textbf {\bibinfo {volume}
  {191}},\ \bibinfo {pages} {78} (\bibinfo {year} {1989})}\BibitemShut
  {NoStop}%
\bibitem [{\citenamefont {Hug}\ \emph {et~al.}(2017)\citenamefont {Hug},
  \citenamefont {Aulenbacher}, \citenamefont {Heine}, \citenamefont {Ledroit},\
  and\ \citenamefont {Simon}}]{Hug2017}%
  \BibitemOpen
  \bibfield  {author} {\bibinfo {author} {\bibfnamefont {F.}~\bibnamefont
  {Hug}}, \bibinfo {author} {\bibfnamefont {K.}~\bibnamefont {Aulenbacher}},
  \bibinfo {author} {\bibfnamefont {R.}~\bibnamefont {Heine}}, \bibinfo
  {author} {\bibfnamefont {B.}~\bibnamefont {Ledroit}}, \ and\ \bibinfo
  {author} {\bibfnamefont {D.}~\bibnamefont {Simon}},\ }in\ \href {\doibase
  10.18429/JACoW-LINAC2016-MOP106012} {\emph {\bibinfo {booktitle} {Proc. of
  Linear Accelerator Conference (LINAC'16), East Lansing, MI, USA, 25-30
  September 2016}}},\ \bibinfo {series and number} {\bibinfo {series} {Linear
  Accelerator Conference}\ No.~\bibinfo {number} {28}}\ (\bibinfo  {publisher}
  {JACoW},\ \bibinfo {address} {Geneva, Switzerland},\ \bibinfo {year} {2017})\
  pp.\ \bibinfo {pages} {313--315},\ \bibinfo {note} {{ISBN:}
  978-3-95450-169-4}\BibitemShut {NoStop}%
\bibitem [{\citenamefont {Hoffstaetter}\ \emph {et~al.}(2017)\citenamefont
  {Hoffstaetter}, \citenamefont {Trbojevic}, \citenamefont {Mayes},
  \citenamefont {Banerjee}, \citenamefont {Barley}, \citenamefont {Bazarov},
  \citenamefont {Bartnik}, \citenamefont {Berg}, \citenamefont {Brooks},
  \citenamefont {Burke} \emph {et~al.}}]{CBETAHoffstaetter2017}%
  \BibitemOpen
  \bibfield  {author} {\bibinfo {author} {\bibfnamefont {G.~H.}\ \bibnamefont
  {Hoffstaetter}}, \bibinfo {author} {\bibfnamefont {D.}~\bibnamefont
  {Trbojevic}}, \bibinfo {author} {\bibfnamefont {C.}~\bibnamefont {Mayes}},
  \bibinfo {author} {\bibfnamefont {N.}~\bibnamefont {Banerjee}}, \bibinfo
  {author} {\bibfnamefont {J.}~\bibnamefont {Barley}}, \bibinfo {author}
  {\bibfnamefont {I.}~\bibnamefont {Bazarov}}, \bibinfo {author} {\bibfnamefont
  {A.}~\bibnamefont {Bartnik}}, \bibinfo {author} {\bibfnamefont {J.~S.}\
  \bibnamefont {Berg}}, \bibinfo {author} {\bibfnamefont {S.}~\bibnamefont
  {Brooks}}, \bibinfo {author} {\bibfnamefont {D.}~\bibnamefont {Burke}},
  \emph {et~al.},\ }\href@noop {} {\  (\bibinfo {year} {2017})},\ \Eprint
  {http://arxiv.org/abs/1706.04245} {arXiv:1706.04245 [physics.acc-ph]}
  \BibitemShut {NoStop}%
\bibitem [{\citenamefont {Grieser}\ \emph {et~al.}(2018)\citenamefont
  {Grieser}, \citenamefont {Bonaventura}, \citenamefont {Brand}, \citenamefont
  {Hargens}, \citenamefont {Hetz}, \citenamefont {Le{\ss}mann}, \citenamefont
  {Westph{\"a}linger},\ and\ \citenamefont {Khoukaz}}]{GrieserKhoukaz2018}%
  \BibitemOpen
  \bibfield  {author} {\bibinfo {author} {\bibfnamefont {S.}~\bibnamefont
  {Grieser}}, \bibinfo {author} {\bibfnamefont {D.}~\bibnamefont
  {Bonaventura}}, \bibinfo {author} {\bibfnamefont {P.}~\bibnamefont {Brand}},
  \bibinfo {author} {\bibfnamefont {C.}~\bibnamefont {Hargens}}, \bibinfo
  {author} {\bibfnamefont {B.}~\bibnamefont {Hetz}}, \bibinfo {author}
  {\bibfnamefont {L.}~\bibnamefont {Le{\ss}mann}}, \bibinfo {author}
  {\bibfnamefont {C.}~\bibnamefont {Westph{\"a}linger}}, \ and\ \bibinfo
  {author} {\bibfnamefont {A.}~\bibnamefont {Khoukaz}},\ }\href {\doibase
  10.1016/j.nima.2018.07.076} {\bibfield  {journal} {\bibinfo  {journal} {Nucl.
  Instrum. Methods Phys. Res. A}\ }\textbf {\bibinfo {volume} {906}},\ \bibinfo
  {pages} {120} (\bibinfo {year} {2018})}\BibitemShut {NoStop}%
\bibitem [{MIT()}]{MIT2013}%
  \BibitemOpen
  \href@noop {} {\enquote {\bibinfo {title} {Workshop to {E}xplore {P}hysics
  {O}pportunities with {I}ntense, {P}olarized {E}lectron {B}eams up to 300
  {M}e{V}, {C}ambridge, {MA}, {M}arch 14-16, 2013, {E}dited by {R}oger
  {C}arlini, {F}rank {M}aas, and {R}ichard {M}ilner, {AIP} {C}onference
  {P}roceedings {V}ol. 1563.}}\ }\BibitemShut {NoStop}%
\bibitem [{\citenamefont {Donnelly}(2002)}]{Donnelly2002}%
  \BibitemOpen
  \bibfield  {author} {\bibinfo {author} {\bibfnamefont {T.~W.}\ \bibnamefont
  {Donnelly}},\ }\enquote {\bibinfo {title} {Electron scattering and the
  nuclear many-body problem},}\ in\ \href {\doibase
  10.1007/978-94-010-0460-2_3} {\emph {\bibinfo {booktitle} {The Nuclear
  Many-Body Problem 2001}}},\ \bibinfo {editor} {edited by\ \bibinfo {editor}
  {\bibfnamefont {W.}~\bibnamefont {Nazarewicz}}\ and\ \bibinfo {editor}
  {\bibfnamefont {D.}~\bibnamefont {Vretenar}}}\ (\bibinfo  {publisher}
  {Springer Netherlands},\ \bibinfo {address} {Dordrecht},\ \bibinfo {year}
  {2002})\ pp.\ \bibinfo {pages} {19--26}\BibitemShut {NoStop}%
\bibitem [{\citenamefont {Bjorken}\ and\ \citenamefont
  {Drell}(1964)}]{Bjorken1964}%
  \BibitemOpen
  \bibfield  {author} {\bibinfo {author} {\bibfnamefont {J.~D.}\ \bibnamefont
  {Bjorken}}\ and\ \bibinfo {author} {\bibfnamefont {S.~D.}\ \bibnamefont
  {Drell}},\ }\href@noop {} {\emph {\bibinfo {title} {Relativistic Quantum
  Mechanics}}}\ (\bibinfo  {publisher} {McGraw-Hill, New York},\ \bibinfo
  {year} {1964})\BibitemShut {NoStop}%
\bibitem [{\citenamefont {Donnelly}\ and\ \citenamefont
  {Raskin}(1986)}]{Donnelly1986}%
  \BibitemOpen
  \bibfield  {author} {\bibinfo {author} {\bibfnamefont {T.~W.}\ \bibnamefont
  {Donnelly}}\ and\ \bibinfo {author} {\bibfnamefont {A.~S.}\ \bibnamefont
  {Raskin}},\ }\href {\doibase 10.1016/0003-4916(86)90173-9} {\bibfield
  {journal} {\bibinfo  {journal} {Ann. Phys.}\ }\textbf {\bibinfo {volume}
  {169}},\ \bibinfo {pages} {247} (\bibinfo {year} {1986})}\BibitemShut
  {NoStop}%
\bibitem [{\citenamefont {Blatt}\ and\ \citenamefont
  {Weisskopf}(1979)}]{BlattWeisskopf1979}%
  \BibitemOpen
  \bibfield  {author} {\bibinfo {author} {\bibfnamefont {J.~M.}\ \bibnamefont
  {Blatt}}\ and\ \bibinfo {author} {\bibfnamefont {V.~F.}\ \bibnamefont
  {Weisskopf}},\ }\href {\doibase 10.1007/978-1-4612-9959-2} {\emph {\bibinfo
  {title} {Theoretical Nuclear Physics}}}\ (\bibinfo  {publisher} {Springer,
  New York, NY},\ \bibinfo {year} {1979})\BibitemShut {NoStop}%
\bibitem [{\citenamefont {Sick}\ \emph {et~al.}(1969)\citenamefont {Sick},
  \citenamefont {Hughes}, \citenamefont {Donnelly}, \citenamefont {Walecka},\
  and\ \citenamefont {Walker}}]{Sick1969}%
  \BibitemOpen
  \bibfield  {author} {\bibinfo {author} {\bibfnamefont {I.}~\bibnamefont
  {Sick}}, \bibinfo {author} {\bibfnamefont {E.~B.}\ \bibnamefont {Hughes}},
  \bibinfo {author} {\bibfnamefont {T.~W.}\ \bibnamefont {Donnelly}}, \bibinfo
  {author} {\bibfnamefont {J.~D.}\ \bibnamefont {Walecka}}, \ and\ \bibinfo
  {author} {\bibfnamefont {G.~E.}\ \bibnamefont {Walker}},\ }\href {\doibase
  10.1103/PhysRevLett.23.1117} {\bibfield  {journal} {\bibinfo  {journal}
  {Phys. Rev. Lett.}\ }\textbf {\bibinfo {volume} {23}},\ \bibinfo {pages}
  {1117} (\bibinfo {year} {1969})}\BibitemShut {NoStop}%
\bibitem [{\citenamefont {Rose}(1953)}]{Rose1953}%
  \BibitemOpen
  \bibfield  {author} {\bibinfo {author} {\bibfnamefont {M.~E.}\ \bibnamefont
  {Rose}},\ }\href {\doibase 10.1103/PhysRev.91.610} {\bibfield  {journal}
  {\bibinfo  {journal} {Phys. Rev.}\ }\textbf {\bibinfo {volume} {91}},\
  \bibinfo {pages} {610} (\bibinfo {year} {1953})}\BibitemShut {NoStop}%
\bibitem [{\citenamefont {Barker}\ and\ \citenamefont
  {Kajino}(1991)}]{Barker1991}%
  \BibitemOpen
  \bibfield  {author} {\bibinfo {author} {\bibfnamefont {F.~C.}\ \bibnamefont
  {Barker}}\ and\ \bibinfo {author} {\bibfnamefont {T.}~\bibnamefont
  {Kajino}},\ }\href {\doibase 10.1071/PH910369} {\bibfield  {journal}
  {\bibinfo  {journal} {Aust. J. Phys.}\ }\textbf {\bibinfo {volume} {44}},\
  \bibinfo {pages} {369} (\bibinfo {year} {1991})}\BibitemShut {NoStop}%
\bibitem [{\citenamefont {Barker}()}]{Barker1974}%
  \BibitemOpen
  \bibfield  {author} {\bibinfo {author} {\bibfnamefont {F.~C.}\ \bibnamefont
  {Barker}},\ }\href@noop {} {}\bibinfo {note} {{p}rivate communication to P.
  Dyer and C. A. Barnes, Nucl. Phys. A 233, 495 (1974)}\BibitemShut {NoStop}%
\bibitem [{\citenamefont {Knutson}(1999)}]{Knutson1999}%
  \BibitemOpen
  \bibfield  {author} {\bibinfo {author} {\bibfnamefont {L.~D.}\ \bibnamefont
  {Knutson}},\ }\href {\doibase 10.1103/PhysRevC.59.2152} {\bibfield  {journal}
  {\bibinfo  {journal} {Phys. Rev. C}\ }\textbf {\bibinfo {volume} {59}},\
  \bibinfo {pages} {2152} (\bibinfo {year} {1999})}\BibitemShut {NoStop}%
\bibitem [{\citenamefont {Watson}(1954)}]{Watson1954}%
  \BibitemOpen
  \bibfield  {author} {\bibinfo {author} {\bibfnamefont {K.~M.}\ \bibnamefont
  {Watson}},\ }\href {\doibase 10.1103/PhysRev.95.228} {\bibfield  {journal}
  {\bibinfo  {journal} {Phys. Rev.}\ }\textbf {\bibinfo {volume} {95}},\
  \bibinfo {pages} {228} (\bibinfo {year} {1954})}\BibitemShut {NoStop}%
\bibitem [{\citenamefont {Brune}(2001)}]{Brune2001}%
  \BibitemOpen
  \bibfield  {author} {\bibinfo {author} {\bibfnamefont {C.~R.}\ \bibnamefont
  {Brune}},\ }\href {\doibase 10.1103/PhysRevC.64.055803} {\bibfield  {journal}
  {\bibinfo  {journal} {Phys. Rev. C}\ }\textbf {\bibinfo {volume} {64}},\
  \bibinfo {pages} {055803} (\bibinfo {year} {2001})}\BibitemShut {NoStop}%
\bibitem [{\citenamefont {Lane}\ and\ \citenamefont {Thomas}(1958)}]{Lane1958}%
  \BibitemOpen
  \bibfield  {author} {\bibinfo {author} {\bibfnamefont {A.~M.}\ \bibnamefont
  {Lane}}\ and\ \bibinfo {author} {\bibfnamefont {R.~G.}\ \bibnamefont
  {Thomas}},\ }\href {\doibase 10.1103/RevModPhys.30.257} {\bibfield  {journal}
  {\bibinfo  {journal} {Rev. Mod. Phys.}\ }\textbf {\bibinfo {volume} {30}},\
  \bibinfo {pages} {257} (\bibinfo {year} {1958})}\BibitemShut {NoStop}%
\bibitem [{\citenamefont {Meyer}\ \emph {et~al.}(2001)\citenamefont {Meyer},
  \citenamefont {Aschenauer}, \citenamefont {Blok}, \citenamefont {Groep},
  \citenamefont {Hicks}, \citenamefont {Holvoet}, \citenamefont {Jans},
  \citenamefont {Lapik{\'a}s}, \citenamefont {Lannoy}, \citenamefont {Nooren}
  \emph {et~al.}}]{DeM2001}%
  \BibitemOpen
  \bibfield  {author} {\bibinfo {author} {\bibfnamefont {G.~D.}\ \bibnamefont
  {Meyer}}, \bibinfo {author} {\bibfnamefont {E.~C.}\ \bibnamefont
  {Aschenauer}}, \bibinfo {author} {\bibfnamefont {H.~P.}\ \bibnamefont
  {Blok}}, \bibinfo {author} {\bibfnamefont {D.}~\bibnamefont {Groep}},
  \bibinfo {author} {\bibfnamefont {K.}~\bibnamefont {Hicks}}, \bibinfo
  {author} {\bibfnamefont {H.}~\bibnamefont {Holvoet}}, \bibinfo {author}
  {\bibfnamefont {E.}~\bibnamefont {Jans}}, \bibinfo {author} {\bibfnamefont
  {L.}~\bibnamefont {Lapik{\'a}s}}, \bibinfo {author} {\bibfnamefont
  {B.}~\bibnamefont {Lannoy}}, \bibinfo {author} {\bibfnamefont {G.~J.~L.}\
  \bibnamefont {Nooren}},  \emph {et~al.},\ }\href {\doibase
  10.1016/S0370-2693(01)00750-X} {\bibfield  {journal} {\bibinfo  {journal}
  {Phys. Lett. B}\ }\textbf {\bibinfo {volume} {513}},\ \bibinfo {pages} {258}
  (\bibinfo {year} {2001})}\BibitemShut {NoStop}%
\bibitem [{\citenamefont {Meija}\ \emph {et~al.}(2016)\citenamefont {Meija},
  \citenamefont {Coplen}, \citenamefont {Berglund}, \citenamefont {Brand},
  \citenamefont {Bi{\`e}vre}, \citenamefont {Gr{\"o}ning}, \citenamefont
  {Holden}, \citenamefont {Irrgeher}, \citenamefont {Loss}, \citenamefont
  {Walczyk},\ and\ \citenamefont {Prohaska}}]{Meija2016}%
  \BibitemOpen
  \bibfield  {author} {\bibinfo {author} {\bibfnamefont {J.}~\bibnamefont
  {Meija}}, \bibinfo {author} {\bibfnamefont {T.~B.}\ \bibnamefont {Coplen}},
  \bibinfo {author} {\bibfnamefont {M.}~\bibnamefont {Berglund}}, \bibinfo
  {author} {\bibfnamefont {W.~A.}\ \bibnamefont {Brand}}, \bibinfo {author}
  {\bibfnamefont {P.~D.}\ \bibnamefont {Bi{\`e}vre}}, \bibinfo {author}
  {\bibfnamefont {M.}~\bibnamefont {Gr{\"o}ning}}, \bibinfo {author}
  {\bibfnamefont {N.~E.}\ \bibnamefont {Holden}}, \bibinfo {author}
  {\bibfnamefont {J.}~\bibnamefont {Irrgeher}}, \bibinfo {author}
  {\bibfnamefont {R.~D.}\ \bibnamefont {Loss}}, \bibinfo {author}
  {\bibfnamefont {T.}~\bibnamefont {Walczyk}}, \ and\ \bibinfo {author}
  {\bibfnamefont {T.}~\bibnamefont {Prohaska}},\ }\href {\doibase
  10.1515/pac-2015-0503} {\bibfield  {journal} {\bibinfo  {journal} {Pure Appl.
  Chem.}\ }\textbf {\bibinfo {volume} {88(3)}},\ \bibinfo {pages} {293–306}
  (\bibinfo {year} {2016})}\BibitemShut {NoStop}%
\bibitem [{Pho()}]{PhotoNucCross}%
  \BibitemOpen
  \href@noop {} {}\bibinfo {note}
  {\url{https://wiki.jlab.org/ciswiki/index.php/Simulations_and_Backgrounds\#Relevant_Theoretical_Cross_Sections}}\BibitemShut
  {NoStop}%
\bibitem [{\citenamefont {Ziegler}\ \emph {et~al.}(2010)\citenamefont
  {Ziegler}, \citenamefont {Ziegler},\ and\ \citenamefont
  {Biersack}}]{Ziegler2010}%
  \BibitemOpen
  \bibfield  {author} {\bibinfo {author} {\bibfnamefont {J.~F.}\ \bibnamefont
  {Ziegler}}, \bibinfo {author} {\bibfnamefont {M.~D.}\ \bibnamefont
  {Ziegler}}, \ and\ \bibinfo {author} {\bibfnamefont {J.~P.}\ \bibnamefont
  {Biersack}},\ }\href {\doibase 10.1016/j.nimb.2010.02.091} {\bibfield
  {journal} {\bibinfo  {journal} {Nucl. Instrum. Methods Phys. Res. B}\
  }\textbf {\bibinfo {volume} {268}},\ \bibinfo {pages} {1818–1823} (\bibinfo
  {year} {2010})}\BibitemShut {NoStop}%
\bibitem [{SRI()}]{SRIM}%
  \BibitemOpen
  \href@noop {} {}\bibinfo {note} {\url{http://www.srim.org/}}\BibitemShut
  {NoStop}%
\bibitem [{\citenamefont {K{\"o}hler}\ \emph {et~al.}(2013)\citenamefont
  {K{\"o}hler}, \citenamefont {Bonaventura}, \citenamefont {Grieser},
  \citenamefont {Hergem{\"o}ller}, \citenamefont {Ortjohann}, \citenamefont
  {T{\"a}schner}, \citenamefont {Zannotti},\ and\ \citenamefont
  {Khoukaz}}]{Kohler2012}%
  \BibitemOpen
  \bibfield  {author} {\bibinfo {author} {\bibfnamefont {E.}~\bibnamefont
  {K{\"o}hler}}, \bibinfo {author} {\bibfnamefont {D.}~\bibnamefont
  {Bonaventura}}, \bibinfo {author} {\bibfnamefont {S.}~\bibnamefont
  {Grieser}}, \bibinfo {author} {\bibfnamefont {A.-K.}\ \bibnamefont
  {Hergem{\"o}ller}}, \bibinfo {author} {\bibfnamefont {H.-W.}\ \bibnamefont
  {Ortjohann}}, \bibinfo {author} {\bibfnamefont {A.}~\bibnamefont
  {T{\"a}schner}}, \bibinfo {author} {\bibfnamefont {A.}~\bibnamefont
  {Zannotti}}, \ and\ \bibinfo {author} {\bibfnamefont {A.}~\bibnamefont
  {Khoukaz}},\ }\enquote {\bibinfo {title} {{C}luster-{J}et {B}eam
  {V}isualisation with {M}icro {C}hannel {P}lates},}\ in\ \href
  {http://repository.gsi.de/record/51925} {\emph {\bibinfo {booktitle} {GSI
  Scientific Report 2012}}},\ Vol.\ \bibinfo {volume} {2013-1}\ (\bibinfo
  {publisher} {GSI Helmholtzzentrum f{\"u}r Schwerionenforschung},\ \bibinfo
  {address} {Darmstadt},\ \bibinfo {year} {2013})\ p.~\bibinfo {pages}
  {18}\BibitemShut {NoStop}%
\bibitem [{\citenamefont {Kordyasz}\ \emph {et~al.}(2015)\citenamefont
  {Kordyasz}, \citenamefont {Neindre}, \citenamefont {Parlog}, \citenamefont
  {Casini}, \citenamefont {Bougault}, \citenamefont {Poggi}, \citenamefont
  {Bednarek}, \citenamefont {Kowalczyk}, \citenamefont {Lopez}, \citenamefont
  {Merrer} \emph {et~al.}}]{Kordyasz2015}%
  \BibitemOpen
  \bibfield  {author} {\bibinfo {author} {\bibfnamefont {A.~J.}\ \bibnamefont
  {Kordyasz}}, \bibinfo {author} {\bibfnamefont {N.~L.}\ \bibnamefont
  {Neindre}}, \bibinfo {author} {\bibfnamefont {M.}~\bibnamefont {Parlog}},
  \bibinfo {author} {\bibfnamefont {G.}~\bibnamefont {Casini}}, \bibinfo
  {author} {\bibfnamefont {R.}~\bibnamefont {Bougault}}, \bibinfo {author}
  {\bibfnamefont {G.}~\bibnamefont {Poggi}}, \bibinfo {author} {\bibfnamefont
  {A.}~\bibnamefont {Bednarek}}, \bibinfo {author} {\bibfnamefont
  {M.}~\bibnamefont {Kowalczyk}}, \bibinfo {author} {\bibfnamefont
  {O.}~\bibnamefont {Lopez}}, \bibinfo {author} {\bibfnamefont
  {Y.}~\bibnamefont {Merrer}},  \emph {et~al.},\ }\href {\doibase
  10.1140/epja/i2015-15015-2} {\bibfield  {journal} {\bibinfo  {journal} {Eur.
  Phys. J. A}\ }\textbf {\bibinfo {volume} {51}},\ \bibinfo {pages} {5}
  (\bibinfo {year} {2015})}\BibitemShut {NoStop}%
\bibitem [{\citenamefont {Friedman}\ \emph {et~al.}(1988)\citenamefont
  {Friedman}, \citenamefont {Bertsche}, \citenamefont {Michel}, \citenamefont
  {Morris}, \citenamefont {Muller},\ and\ \citenamefont {Tans}}]{Friedman1988}%
  \BibitemOpen
  \bibfield  {author} {\bibinfo {author} {\bibfnamefont {P.~G.}\ \bibnamefont
  {Friedman}}, \bibinfo {author} {\bibfnamefont {K.~J.}\ \bibnamefont
  {Bertsche}}, \bibinfo {author} {\bibfnamefont {M.~C.}\ \bibnamefont
  {Michel}}, \bibinfo {author} {\bibfnamefont {D.~E.}\ \bibnamefont {Morris}},
  \bibinfo {author} {\bibfnamefont {R.~A.}\ \bibnamefont {Muller}}, \ and\
  \bibinfo {author} {\bibfnamefont {P.~P.}\ \bibnamefont {Tans}},\ }\href
  {\doibase 10.1063/1.1139973} {\bibfield  {journal} {\bibinfo  {journal} {Rev.
  Sci. Instrum.}\ }\textbf {\bibinfo {volume} {59}},\ \bibinfo {pages} {98}
  (\bibinfo {year} {1988})}\BibitemShut {NoStop}%
\bibitem [{\citenamefont {Kumagai}\ \emph {et~al.}(2013)\citenamefont
  {Kumagai}, \citenamefont {Ohnishi}, \citenamefont {Fukuda}, \citenamefont
  {Takeda}, \citenamefont {Kameda}, \citenamefont {Inabe}, \citenamefont
  {Yoshida},\ and\ \citenamefont {Kubo}}]{Kumagai2013}%
  \BibitemOpen
  \bibfield  {author} {\bibinfo {author} {\bibfnamefont {H.}~\bibnamefont
  {Kumagai}}, \bibinfo {author} {\bibfnamefont {T.}~\bibnamefont {Ohnishi}},
  \bibinfo {author} {\bibfnamefont {N.}~\bibnamefont {Fukuda}}, \bibinfo
  {author} {\bibfnamefont {H.}~\bibnamefont {Takeda}}, \bibinfo {author}
  {\bibfnamefont {D.}~\bibnamefont {Kameda}}, \bibinfo {author} {\bibfnamefont
  {N.}~\bibnamefont {Inabe}}, \bibinfo {author} {\bibfnamefont
  {K.}~\bibnamefont {Yoshida}}, \ and\ \bibinfo {author} {\bibfnamefont
  {T.}~\bibnamefont {Kubo}},\ }\href {\doibase 10.1016/j.nimb.2013.08.050}
  {\bibfield  {journal} {\bibinfo  {journal} {Nucl. Instr. Meth. Phys. Res. B}\
  }\textbf {\bibinfo {volume} {317}},\ \bibinfo {pages} {717} (\bibinfo {year}
  {2013})}\BibitemShut {NoStop}%
\bibitem [{\citenamefont {Iguaz}\ \emph {et~al.}(2014)\citenamefont {Iguaz},
  \citenamefont {Panebianco}, \citenamefont {Axiotis}, \citenamefont
  {Druillole}, \citenamefont {Fanourakis}, \citenamefont {Geralis},
  \citenamefont {Giomataris}, \citenamefont {Harissopulos}, \citenamefont
  {Lagoyannis},\ and\ \citenamefont {Papaevangelou}}]{Iguaz2014}%
  \BibitemOpen
  \bibfield  {author} {\bibinfo {author} {\bibfnamefont {F.~J.}\ \bibnamefont
  {Iguaz}}, \bibinfo {author} {\bibfnamefont {S.}~\bibnamefont {Panebianco}},
  \bibinfo {author} {\bibfnamefont {M.}~\bibnamefont {Axiotis}}, \bibinfo
  {author} {\bibfnamefont {F.}~\bibnamefont {Druillole}}, \bibinfo {author}
  {\bibfnamefont {G.}~\bibnamefont {Fanourakis}}, \bibinfo {author}
  {\bibfnamefont {T.}~\bibnamefont {Geralis}}, \bibinfo {author} {\bibfnamefont
  {I.}~\bibnamefont {Giomataris}}, \bibinfo {author} {\bibfnamefont
  {S.}~\bibnamefont {Harissopulos}}, \bibinfo {author} {\bibfnamefont
  {A.}~\bibnamefont {Lagoyannis}}, \ and\ \bibinfo {author} {\bibfnamefont
  {T.}~\bibnamefont {Papaevangelou}},\ }\href {\doibase
  10.1016/j.nima.2013.09.061} {\bibfield  {journal} {\bibinfo  {journal} {Nucl.
  Instr. Meth. Phys. Res. A}\ }\textbf {\bibinfo {volume} {735}},\ \bibinfo
  {pages} {399} (\bibinfo {year} {2014})}\BibitemShut {NoStop}%
\bibitem [{\citenamefont {Deaconu}\ \emph {et~al.}(2017)\citenamefont
  {Deaconu}, \citenamefont {Leyton}, \citenamefont {Corliss}, \citenamefont
  {Druitt}, \citenamefont {Eggleston}, \citenamefont {Guerrero}, \citenamefont
  {Henderson}, \citenamefont {Lopez}, \citenamefont {Monroe},\ and\
  \citenamefont {Fisher}}]{Deaconu2017}%
  \BibitemOpen
  \bibfield  {author} {\bibinfo {author} {\bibfnamefont {C.}~\bibnamefont
  {Deaconu}}, \bibinfo {author} {\bibfnamefont {M.}~\bibnamefont {Leyton}},
  \bibinfo {author} {\bibfnamefont {R.}~\bibnamefont {Corliss}}, \bibinfo
  {author} {\bibfnamefont {G.}~\bibnamefont {Druitt}}, \bibinfo {author}
  {\bibfnamefont {R.}~\bibnamefont {Eggleston}}, \bibinfo {author}
  {\bibfnamefont {N.}~\bibnamefont {Guerrero}}, \bibinfo {author}
  {\bibfnamefont {S.}~\bibnamefont {Henderson}}, \bibinfo {author}
  {\bibfnamefont {J.}~\bibnamefont {Lopez}}, \bibinfo {author} {\bibfnamefont
  {J.}~\bibnamefont {Monroe}}, \ and\ \bibinfo {author} {\bibfnamefont
  {P.}~\bibnamefont {Fisher}},\ }\href {\doibase 10.1103/PhysRevD.95.122002}
  {\bibfield  {journal} {\bibinfo  {journal} {Phys. Rev. D}\ }\textbf {\bibinfo
  {volume} {95}},\ \bibinfo {pages} {122002} (\bibinfo {year}
  {2017})}\BibitemShut {NoStop}%
\bibitem [{\citenamefont {Astabatyan}\ \emph {et~al.}(2012)\citenamefont
  {Astabatyan}, \citenamefont {Ivanov}, \citenamefont {Lukyanov}, \citenamefont
  {Markaryan}, \citenamefont {Maslov}, \citenamefont {Penionzhkevich},\ and\
  \citenamefont {Revenko}}]{Astabatyan2012}%
  \BibitemOpen
  \bibfield  {author} {\bibinfo {author} {\bibfnamefont {R.~A.}\ \bibnamefont
  {Astabatyan}}, \bibinfo {author} {\bibfnamefont {M.~P.}\ \bibnamefont
  {Ivanov}}, \bibinfo {author} {\bibfnamefont {S.~M.}\ \bibnamefont
  {Lukyanov}}, \bibinfo {author} {\bibfnamefont {E.~R.}\ \bibnamefont
  {Markaryan}}, \bibinfo {author} {\bibfnamefont {V.~A.}\ \bibnamefont
  {Maslov}}, \bibinfo {author} {\bibfnamefont {{\relax Yu}.~E.}\ \bibnamefont
  {Penionzhkevich}}, \ and\ \bibinfo {author} {\bibfnamefont {R.~V.}\
  \bibnamefont {Revenko}},\ }\href {\doibase 10.1134/S0020441212020017}
  {\bibfield  {journal} {\bibinfo  {journal} {Instrum. Exp. Tech.}\ }\textbf
  {\bibinfo {volume} {55}},\ \bibinfo {pages} {335} (\bibinfo {year}
  {2012})}\BibitemShut {NoStop}%
\bibitem [{\citenamefont {Breskin}\ and\ \citenamefont
  {Chechik}(1984)}]{Breskin1984}%
  \BibitemOpen
  \bibfield  {author} {\bibinfo {author} {\bibfnamefont {A.}~\bibnamefont
  {Breskin}}\ and\ \bibinfo {author} {\bibfnamefont {R.}~\bibnamefont
  {Chechik}},\ }\href {\doibase 10.1016/0168-9002(84)90096-2} {\bibfield
  {journal} {\bibinfo  {journal} {Nucl. Instr. Meth. Phys. Res. A}\ }\textbf
  {\bibinfo {volume} {221}},\ \bibinfo {pages} {363} (\bibinfo {year}
  {1984})}\BibitemShut {NoStop}%
\bibitem [{\citenamefont {Trbojevic}\ \emph {et~al.}(2017)\citenamefont
  {Trbojevic} \emph {et~al.}}]{Trbojevic2017}%
  \BibitemOpen
  \bibfield  {author} {\bibinfo {author} {\bibfnamefont {D.}~\bibnamefont
  {Trbojevic}} \emph {et~al.},\ }in\ \href {\doibase
  10.18429/JACoW-IPAC2017-TUOCB3} {\emph {\bibinfo {booktitle} {Proc. of
  International Particle Accelerator Conference (IPAC'17), Copenhagen, Denmark,
  May, 2017}}},\ \bibinfo {series and number} {\bibinfo {series} {International
  Particle Accelerator Conference}\ No.~\bibinfo {number} {8}}\ (\bibinfo
  {publisher} {JACoW},\ \bibinfo {address} {Geneva, Switzerland},\ \bibinfo
  {year} {2017})\ pp.\ \bibinfo {pages} {1285--1289},\ \bibinfo {note} {{ISBN:}
  978-3-95450-182-3}\BibitemShut {NoStop}%
\bibitem [{\citenamefont {Lafferty}\ and\ \citenamefont
  {Wyatt}(1995)}]{Lafferty1995}%
  \BibitemOpen
  \bibfield  {author} {\bibinfo {author} {\bibfnamefont {G.~D.}\ \bibnamefont
  {Lafferty}}\ and\ \bibinfo {author} {\bibfnamefont {T.~R.}\ \bibnamefont
  {Wyatt}},\ }\href {\doibase 10.1016/0168-9002(94)01112-5} {\bibfield
  {journal} {\bibinfo  {journal} {Nucl. Instrum. Methods Phys. Res. A}\
  }\textbf {\bibinfo {volume} {355}},\ \bibinfo {pages} {541} (\bibinfo {year}
  {1995})}\BibitemShut {NoStop}%
\bibitem [{\citenamefont {Alarcon}\ \emph {et~al.}(2013)\citenamefont
  {Alarcon}, \citenamefont {Balascuta}, \citenamefont {Benson}, \citenamefont
  {Bertozzi}, \citenamefont {Boyce}, \citenamefont {Cowan}, \citenamefont
  {Douglas}, \citenamefont {Evtushenko}, \citenamefont {Fisher}, \citenamefont
  {Ihloff} \emph {et~al.}}]{Alarcon2013}%
  \BibitemOpen
  \bibfield  {author} {\bibinfo {author} {\bibfnamefont {R.}~\bibnamefont
  {Alarcon}}, \bibinfo {author} {\bibfnamefont {S.}~\bibnamefont {Balascuta}},
  \bibinfo {author} {\bibfnamefont {S.~V.}\ \bibnamefont {Benson}}, \bibinfo
  {author} {\bibfnamefont {W.}~\bibnamefont {Bertozzi}}, \bibinfo {author}
  {\bibfnamefont {J.~R.}\ \bibnamefont {Boyce}}, \bibinfo {author}
  {\bibfnamefont {R.}~\bibnamefont {Cowan}}, \bibinfo {author} {\bibfnamefont
  {D.}~\bibnamefont {Douglas}}, \bibinfo {author} {\bibfnamefont
  {P.}~\bibnamefont {Evtushenko}}, \bibinfo {author} {\bibfnamefont
  {P.}~\bibnamefont {Fisher}}, \bibinfo {author} {\bibfnamefont
  {E.}~\bibnamefont {Ihloff}},  \emph {et~al.},\ }\href {\doibase
  10.1103/PhysRevLett.111.164801} {\bibfield  {journal} {\bibinfo  {journal}
  {Phys. Rev. Lett.}\ }\textbf {\bibinfo {volume} {111}},\ \bibinfo {pages}
  {164801} (\bibinfo {year} {2013})}\BibitemShut {NoStop}%
\end{thebibliography}%

\end{document}